\documentclass[11pt]{article}
\usepackage{jheppub} 
\usepackage{graphicx}
\usepackage{psfrag}
\usepackage{caption}
\usepackage{subcaption}
\usepackage{enumerate}
\usepackage{framed}
\usepackage{amssymb}
\usepackage[svgnames]{xcolor}
\usepackage{comment}
\usepackage[normalem]{ulem}
\usepackage{amsmath}
\usepackage{fancyhdr}
\usepackage{physics}
\usepackage{siunitx}
\usepackage{float}
\usepackage[toc]{appendix}
\usepackage{tikz}
\usepackage{tkz-tab}
\usepackage{soul}
\usepackage[compat=1.1.0]{tikz-feynman}
\usetikzlibrary{shapes.geometric, arrows}
\numberwithin{equation}{section}
\renewcommand\vec{\boldsymbol}

\definecolor{darkgreen}{rgb}{0.0, 0.5, 0.0}

\newcommand{\hinvMpc}{\,h\, {\rm Mpc}^{-1}\,}
\newcommand{\fnl}{f_{\rm NL}}

\graphicspath{{Figures/}}

\begin{document}
	
\title{Efficiently evaluating loop integrals in the EFTofLSS using QFT integrals with massive propagators}

\author[a]{Charalampos Anastasiou,}
\author[b]{Diogo P.~L. Bragan\c{c}a,}
\author[a]{Leonardo Senatore,}
\author[b]{Henry Zheng,}

\affiliation[a]{Institute for Theoretical Physics, ETH Zurich, 8093 Z\"urich, Switzerland}
\affiliation[b]{SITP and KIPAC, Department of Physics and SLAC, Stanford University, Stanford, CA 94305}
\emailAdd{babis@phys.ethz.ch}
\emailAdd{braganca@stanford.edu}
\emailAdd{lsenatore@phys.ethz.ch}
\emailAdd{henryzheng@stanford.edu}

\date{\today}

\abstract{
	We develop a new way to analytically calculate loop integrals in the Effective Field Theory of Large Scale-Structure. Previous available methods show severe limitations beyond the one-loop power spectrum due to analytical challenges and computational and memory costs. Our new method is based on fitting the linear power spectrum with cosmology-independent functions that resemble integer powers of quantum field theory massive propagators with complex masses. 
	A remarkably small number of them is sufficient to reach enough accuracy. Similarly to former approaches, the cosmology dependence is encoded in the coordinate vector of the expansion of the linear power spectrum in our basis. We first produce cosmology-independent tensors where each entry is the loop integral evaluated on a given combination of basis vectors. For each cosmology, the evaluation of a loop integral amounts to contracting this tensor with the coordinate vector of the linear power spectrum. The 3-dimensional loop integrals for our basis functions can be evaluated using techniques familiar to particle physics, such as recursion relations and Feynman parametrization. We apply our formalism to evaluate the one-loop bispectrum of galaxies in redshift space. The final analytical expressions are quite simple and can be evaluated with little computational and memory cost. {We show that the same expressions resolve the integration of all one-loop $N$-point function in the EFTofLSS.} This method, which is originally presented here, has already been applied in the first one-loop bispectrum analysis of the BOSS data to constraint $\Lambda$CDM parameters and primordial non-Gaussianities~\cite{DAmico:2022osl,DAmico:2022gki}.
}
	
\maketitle
\clearpage

\section{Introduction and Conclusions}
	
The Effective Field Theory of Large Scale Structure (EFTofLSS)~\cite{Baumann:2010tm,Carrasco:2012cv} describes the dynamics of the large-scale structures of the universe in the mildly non-linear regime. 
The development of the theory, from its initial formulation to the application to data, has been a decade long effort, where several important developments have been obtained at all stages. 
Very schematically, it has been necessary to develop the description of dark matter, biased tracers and redshift space distortions, as well as a non perturbative treatment of some infrared effects. 
	We provide a summary of some important results obtained prior the application to data in this footnote~\footnote{The initial formulation of the EFTofLSS was performed in Eulerian space in~\cite{Baumann:2010tm,Carrasco:2012cv}, and subsequently extended to Lagrangian space in~\cite{Porto:2013qua}.
	The dark matter power spectrum has been computed at one-, two- and three-loop orders in~\cite{Carrasco:2012cv, Carrasco:2013sva, Carrasco:2013mua, Carroll:2013oxa, Senatore:2014via, Baldauf:2015zga, Foreman:2015lca, Baldauf:2015aha, Baldauf:2015zga, Baldauf:2015aha, Cataneo:2016suz, Lewandowski:2017kes,Konstandin:2019bay}.
	These calculations were accompanied by some theoretical developments of the EFTofLSS, such as a careful understanding of renormalization~\cite{Carrasco:2012cv,Pajer:2013jj,Abolhasani:2015mra} (including rather-subtle aspects such as lattice-running~\cite{Carrasco:2012cv} and a better understanding of the velocity field~\cite{Carrasco:2013sva,Mercolli:2013bsa}), of several ways for extracting the value of the counterterms from simulations~\cite{Carrasco:2012cv,McQuinn:2015tva}, and of the non-locality in time of the EFTofLSS~\cite{Carrasco:2013sva, Carroll:2013oxa,Senatore:2014eva}.
	These theoretical explorations also include an enlightening study in 1+1 dimensions~\cite{McQuinn:2015tva}.
	An IR-resummation of the long displacement fields had to be performed in order to reproduce the Baryon Acoustic Oscillation (BAO) peak, giving rise to the so-called IR-Resummed EFTofLSS~\cite{Senatore:2014vja,Baldauf:2015xfa,Senatore:2017pbn,Lewandowski:2018ywf,Blas:2016sfa}. 
	Accounts of baryonic effects were presented in~\cite{Lewandowski:2014rca,Braganca:2020nhv}. The dark-matter bispectrum has been computed at one-loop in~\cite{Angulo:2014tfa, Baldauf:2014qfa}, the one-loop trispectrum in~\cite{Bertolini:2016bmt}, and the displacement field in~\cite{Baldauf:2015tla}.
	The lensing power spectrum has been computed at two loops in~\cite{Foreman:2015uva}.
	Biased tracers, such as halos and galaxies, have been studied in the context of the EFTofLSS in~\cite{ Senatore:2014eva, Mirbabayi:2014zca, Angulo:2015eqa, Fujita:2016dne, Perko:2016puo, Nadler:2017qto,Donath:2020abv} (see also~\cite{McDonald:2009dh}), the halo and matter power spectra and bispectra (including all cross correlations) in~\cite{Senatore:2014eva, Angulo:2015eqa}. Redshift space distortions have been developed in~\cite{Senatore:2014vja, Lewandowski:2015ziq,Perko:2016puo}. 
	Neutrinos have been included in the EFTofLSS in~\cite{Senatore:2017hyk,deBelsunce:2018xtd}, clustering dark energy in~\cite{Lewandowski:2016yce,Lewandowski:2017kes,Cusin:2017wjg,Bose:2018orj}, and primordial non-Gaussianities in~\cite{Angulo:2015eqa, Assassi:2015jqa, Assassi:2015fma, Bertolini:2015fya, Lewandowski:2015ziq, Bertolini:2016hxg}.
	Faster evaluation schemes for the calculation of some of the loop integrals have been developed in~\cite{Simonovic:2017mhp}.
	Comparison with high-quality $N$-body simulations to show that the EFTofLSS can accurately recover the cosmological parameters have been performed in~\cite{DAmico:2019fhj,Colas:2019ret,Nishimichi:2020tvu,Chen:2020zjt}.}.

With the results of~\cite{Perko:2016puo} the EFTofLSS became ready to be applied to data, in particular to the power spectrum of galaxies in redshift space. 
Ref.~\cite{DAmico:2019fhj,Ivanov:2019pdj,Colas:2019ret} provided the first application of the EFTofLSS to data, by being able to extract the cosmological parameters from the analysis of the full shape of the galaxy bispectrum of BOSS observations~\cite{BOSS:2016wmc}. 
Since then, many applications to data have followed. 
We summarize some of the main results concerning the application to data in this footnote~\footnote{
	The EFTofLSS prediction at one-loop order has been used to analyze the BOSS galaxy Power Spectrum~\cite{DAmico:2019fhj,Ivanov:2019pdj,Colas:2019ret}, and Correlation Function~\cite{Zhang:2021yna,Chen:2021wdi}. 
	This was extended to eBOSS in~\cite{Simon:2022csv}.
	The BOSS galaxy-clustering bispectrum monopole was analyzed in~\cite{DAmico:2019fhj,Philcox:2021kcw} using the EFTofLSS prediction at tree-level. All $\Lambda$CDM cosmological parameters have been measured from these data by only imposing a prior from Big Bang Nucleosynthesis {(BBN)}, reaching quite a remarkable precision. For example, the present amount of matter, $\Omega_m$, and the Hubble constant (see also~\cite{Philcox:2020vvt,DAmico:2020kxu} for subsequent refinements) have error bars that are similar to the ones obtained from the Cosmic Microwave Background (CMB)~\cite{Planck:2018vyg}.
	For clustering and smooth quintessence models, limits on the equation of state $w$ of  dark energy  of $\lesssim 5\%$ have been set using only late-time measurements~\cite{DAmico:2020kxu,DAmico:2020tty,Simon:2022csv}, similar to the ones from CMB~\cite{Planck:2018vyg}.  These measurements establish a new, CMB-independent, method for determining the Hubble constant~\cite{DAmico:2019fhj}, with precision comparable to one from the cosmic ladder~\cite{Riess:2019cxk,Freedman:2019jwv} and CMB. Some models that were proposed to alleviate the tension in the Hubble measurements between the CMB and cosmic ladder (see e.g.~\cite{Verde:2019ivm}) have also been compared to data~\cite{DAmico:2020ods,Ivanov:2020ril,Niedermann:2020qbw,Smith:2020rxx,Simon:2022adh}. 
	}. 
	
One application to data that is very relevant for this paper is the one where the one-loop EFTofLSS prediction for the bispectrum was compared to BOSS to measure the $\Lambda$CDM parameters~\cite{DAmico:2022osl} or to set limits on some parameters related to primordial non-Gaussianities~\cite{DAmico:2022gki} (see also~\cite{Cabass:2022wjy,Cabass:2022ymb} for a contemporary and a subsequent paper which constrain the same parameters but using the EFTofLSS tree-level prediction)~\footnote{The non-Gaussianity parameters that were constrained are $\fnl^{\rm equil.}\,,\ \fnl^{\rm orth.}$, and $\fnl^{\rm loc.}$, which are predicted to be produced by some single-clock~\cite{Creminelli:2005hu,Senatore:2009gt} or multiple fields~\cite{{Bernardeau:2002jy,Lyth:2002my,Zaldarriaga:2003my,Babich:2004gb,Senatore:2010wk}} inflationary models.}. Ref.~\cite{DAmico:2022osl,DAmico:2022gki} are important for this paper because the computational tool to evaluate the one-loop bispectrum in the analysis is the one we originally present here.
	
Let us first explain the challenge in performing a data analysis using the EFTofLSS. In practice, one needs to evaluate the model predictions as a function of the cosmological and EFT parameters, and determine what are the parameter regions allowed by the data. Since the EFTofLSS equations are typically solved perturbatively, evaluating the prediction requires the computation of loop integrals. 
In principle, these depend on the cosmology and on the EFTofLSS parameters, which are being scanned over as we compare theory and data. Certainly, evaluating the loop integrals in the EFTofLSS takes computational time, and therefore it might be challenging to analyze the data scanning over thousands of combinations of cosmological parameters and EFTofLSS parameters.
The problem of scanning over the EFTofLSS parameters has been solved in~\cite{DAmico:2019fhj,Colas:2019ret} by defining them as prefactors of the loop expressions that they multiply. 
	So, at the cost of increasing the number of the loop integrals to perform, one does not need to recompute the loop as the EFTofLSS parameters are changed~\footnote{Additionally, some of these parameters are analytically marginalized over, so they are not even scanned numerically.}. 
A first way in which this problem of re-evaluating the loops for each cosmological parameter configuration was solved was to precompute the observables in a grid of the cosmological parameters that are being sampled. This was the approach of the first analysis of~\cite{DAmico:2019fhj}. A second approach has been to use the fact that, typically, one is interested in cosmologies close to each other, where the cosmological parameters change by a few percent; so, one can approximate the Likelihood of the data given the model parameters, with a small-order Taylor expansion around a given reference cosmology. 
	This approach was initially developed in~\cite{Cataneo:2016suz}, and applied to data in~\cite{Colas:2019ret}~\footnote{The only parameter that can scan for a few tens of per cent is the amplitude of the linear power spectrum,~$A_s$. But in this case the dependence is practically analytical, and so, in principle, there would be no computational cost in changing this parameter, or, alternatively, the Taylor expansion works very well also for large variations of the parameter.}.
	
A useful development was the one of~\cite{Simonovic:2017mhp}. 
By expanding in the linear power-spectrum for each cosmology in a cosmology-independent basis of power laws with complex exponent (the so-called FFTLog basis), Ref.~\cite{Simonovic:2017mhp} realized that the loop integrals for each combination of basis functions could be done analytically (see also~\cite{Schmittfull:2016jsw,McEwen:2016fjn} for some partial implementation of closely related ideas). 
In this way, evaluating the loop integral for a given power spectrum amounts to contracting the loop evaluated for a generic combination of basis vectors (which is a tensor), with the cosmology-dependent coefficients ({\it i.e} the coordinate vectors). 
Since the tensors are cosmology independent, and need just to be computed once, scanning over cosmology corresponds to expanding the linear power spectrum in the FFTLog basis and performing the contraction, which is naively computationally very efficient. 
This was the technique used in~\cite{Ivanov:2019pdj}.
	
However, the FFTLog formalism has a major limitation: to accurately fit the power spectrum one needs a number of basis vectors $N={\cal{O}}(64)$ or larger~\cite{Philcox:2022frc}. 
For example, the one-loop bispectrum calculation requires a sum of $N^3$ terms. 
It is therefore computationally very demanding to compute the tensors even once, both in terms of memory, and in terms of CPU time. 
Perhaps even worse, given the many bias coefficients, the resulting tensors require very large memory (${\cal{O}}$(50)Gb~\cite{Philcox:2022frc}),  and perhaps, if loadable in a CPU, a non-negligible running time in performing the tensor contraction. 
This severely restricts the capabilities of this formalism to be used to run Markov Chain Monte Carlo (MCMC), which is the way the likelihood of the data is usually sampled. 
	
On top of this, the analytic complexity of the FFTLog formalism rises steeply beyond the computation of the one-loop power spectrum. 
The formalism requires loop integrals with propagators raised to a power which is not an integer and has an imaginary part~\cite{Simonovic:2017mhp}. 
While for the one-loop power spectrum the resulting integrals can be readily expressed in terms of Beta functions that are computed in a straightforward manner, the corresponding integrals for the one-loop bispectrum give rise to significantly more intricate Appel hypergeometric functions~\cite{Simonovic:2017mhp,Boos:1990rg,Anastasiou:1999ui}. 
The latter  satisfy recurrence relations which have been exploited in  Ref.~\cite{Simonovic:2017mhp} to reduce the number of  independent  integrals. 
However, this reduction is significantly less effective for complex powers of propagators than for integers. Beyond one-loop, in the generic case of complex  propagator powers, the multiloop integrals which are required in the FFTLog approach belong to unclassified species of generalized hypergeometric functions with poorly explored recurrence and convergence properties.
It appears to us extremely challenging to perform the loop integrals with the FFTLog basis of functions, in general. 
In fact, as of today, we are unaware of a complete evaluation of the two-loop integrals of the power spectrum in the EFTofLSS with this method.
	
Therefore, a method requiring less basis functions and which allows the application of  efficient methods for their evaluation appears to be very useful. 
This is the purpose of this paper. We propose a new decomposition of the linear  power spectrum, using analytical functions  
consisting of Quantum Field Theory (QFT) propagators for massive particles, with  masses which are in general complex. 
These functions are needed in a smaller number to fit sufficiently well the linear power spectrum. 
We obtain a good accuracy only with 16 functions, which is a significant reduction, and reduces the number of terms in the one-loop bispectrum calculation by $(64/16)^3\sim 64$ with respect to the FFTLog method (if we take as a reference FFTLog with 64 basis functions). 
This solves the problem of speed and memory requirement. 
In fact, the matrices that are used to perform the BOSS bispectrum analysis of~\cite{DAmico:2022osl,DAmico:2022gki} can be evaluated on 2.5 CPU hours using 800 Mb of storage, and cost negligible time in the tensor contraction in the MCMC.

By introducing mass parameters, we can afford to restrict the powers of propagators to be integers. 
This brings several advantages.  
The integrals satisfy recurrence identities which can be derived simply with the method of integration by parts~\cite{Tkachov:1981wb,Chetyrkin:1981qh}. 
For integer powers of propagators, the reduction identities admit a very constraining set of boundary conditions emerging from the fact that, in dimensional regularization, loop integrals with all powers of propagators being negative integers vanish. Thus, they  reduce to a smaller  set  of basis integrals (master integrals) than in the general case of analytic powers of propagators.  
At one-loop, as we will explicitly show, the system of integration by parts identities can be easily brought in a diagonal form which, in turn,  can be implemented as a computer algorithm to reduce  integrals with arbitrary integer powers  of propagators to master integrals.
At two-loops and beyond, the diagonalization of the integration by parts identities and the reduction are intertwined and they can be performed  with an ordered Gauss elimination algorithm  introduced by Laporta~\cite{Laporta:2000dsw}.
As we will show in Sec.~\ref{sec:all-N-point},  {\it all} one-loop master integrals with massive or massless propagators in three space dimensions can be solved with the method of integrating Feynman parameters. 
At two-loops, powerful methods developed for QFT master integrals can be used, in principle, for their evaluation. Concretely,  massive multi-loop integrals are very likely amenable to the differential equations~\cite{Remiddi:1997ny,Kotikov:1990kg,Gehrmann:1999as} method which has been successfully employed in most modern QFT amplitude computations of the last two decades. 
	
	Using the new basis functions,  we derive analytical expressions for the loop integrals appearing in the one-loop power spectrum and bispectrum (of massive tracers in redshift space), by generalizing QFT results in three spacetime dimensions for complex masses~\footnote{Due to the ultralocality in space of the dark matter equations of motion, the time integrals of the loops are factorized and can be performed once and for all, leaving the loop integrals to depend only on the spatial momenta.}. 
	The evaluation is based on the following procedure: loop integrals are reduced to master integrals with recursion, based on diagonalized integration by parts relations. The master integrals are then evaluated analytically by introducing Feynman parameters.  
	The integration over Feynman parameters needs to be performed carefully due to the complex values of the masses in the propagators and the presence of branch cuts in the antiderivatives. 
	However, the final expressions that we obtain for the one-loop master integrals are simple logarithmic and trigonometric functions, perhaps further confirming the goodness of this approach even at the analytical level (on top of the small number of fitting functions that are needed).
	Concretely, we find three one-loop master  integrals:  the one-loop tadpole, the one-loop  
	bubble and the  one-loop triangle, all with unit powers of propagators. 
	We provide expressions for the master integrals which are valid for the complex values of internal masses that we use in our parametrization and real values of external momenta. We should remark, that our list of master integrals is exhaustive for arbitrary correlation functions in EFTofLSS and, in general, QFT in $D=3$ Euclidean dimensions. 
	In fact, it has been shown~\cite{vanNeerven:1983vr}  that all $N-$point one-loop integrals with $N > D$ are always reducible to $D-$point integrals, which, for $D=3$,  one needs in the bispectrum and that we have solved here.  {We provide a discussion about this in Sec.~\ref{sec:all-N-point}}. 
	
	In comparison to typical loop amplitude calculations in QFT, we encounter integrals with higher integer powers of propagators. 
	We therefore require a relatively high number of recursive steps for reducing them all  to master integrals. 
	This poses a potential challenge for calculating the master integral coefficients analytically, as it is typically done for QFT amplitude computations in four spacetime dimensions.  
	However, in three dimensions, a numerical reduction is a very well suited approach. 
	In particular, the mass parameters that we have introduced have predefined numerical values which can be substituted in the  recurrence identities.  
	A further simplification occurs for one-loop three-dimensional integrals, as the master integrals, within dimensional regularization, do not exhibit  $1/(D-3)$ poles and the reduction of integrals with higher than unity powers of propagators can be performed entirely numerically, setting the space dimension to exactly $D=3$. 
	We optimize the efficiency of the reduction with a memoization technique~\cite{Memo}.  
	A numerical reduction naturally entails a risk of loss of arithmetic precision and rounding errors. 
	We mitigate this risk with the implementation of variable arithmetic precision.   
	We have authored Python and Rust codes for the evaluation of the integrals with numerical reduction and computation of the master integrals.
	We have also made a toy implementation of integral reduction within MAPLE, which has a native variable numerical precision, memoization and it also allows for obtaining some benchmark symbolic results. 
	As mentioned, the output of the Python and Rust codes has been already used to set the first and strong limits on primordial non-Gaussianities from Large-Scale Structure by using BOSS data~\cite{DAmico:2022gki} as well as to perform the first analysis of the $\Lambda$CDM model using the one-loop bispectrum on the same data~\cite{DAmico:2022osl}.
	
	The paper is organized as follows. 
	In Sec.~\ref{sec:decomp}, we present the expansion of the power spectrum in the basis of functions similar to massive propagators. 
	In Sec.~\ref{sec:PandB1loop}  we present the comparison of our analytical integrations against numerical integration, validating in this way the formalism. In Sec.~\ref{sec:l_func}, we present the recursion relation to reduce the loop integrals to the master integrals. 
	In Sec.~\ref{sec:master} we present the integration of the master integrals. {In Sec. 6, we show how all one-loop $N$-point functions integrals can be reduced to the master integrals we compute in this paper.}

	\paragraph{Public Codes:}  
The Python code is publicly available on GitHub~\footnote{
\url{https://github.com/dbraganca/python-integer-powers}}, and the Rust code is available upon request.

	\section{Decomposition of the power spectrum}
	\label{sec:decomp}
	
	We approximate the linear power spectrum $P_{\rm lin}$ by a fitting function $P_{\rm fit}(k)$ given by  
	\begin{equation}\label{eq:Pfit1}
		P_{\rm fit}(k) = \frac{\alpha_0}{1+\frac{k^2}{k_{{\rm UV},0}^2}}  + \sum_{n=1}^{N-1} \alpha_n f(k^2, k^2_{{\rm peak},n},k^2_{{\rm UV}, n}, i_n, j_n) = \sum_{n=0}^{N-1} \alpha_n f_n(k^2)\,.
	\end{equation}
	The function $f$ is given by
	\begin{equation}
		\label{eq:f}
		f(k^2, k^2_{\rm peak},k^2_{\rm UV}, i, j) \equiv \frac{\left(k^2 / k_0^2\right)^i}{\left(1 + \frac{(k^2 - k^2_{\rm peak})^2}{k^4_{\rm UV}}\right)^j}\,,
	\end{equation}
	where $k_0$, $k^2_{\rm peak}$ and $k^2_{\rm UV}$ are predetermined cosmology independent parameters, and $i$ and $j$ are positive integers, with $i\leq j$. We define $f_n(k^2) \equiv f(k^2, k^2_{{\rm peak},n},k^2_{{\rm UV}, n}, i_n, j_n)$ for $n\geq 1$ and $f_0(k^2) \equiv 1/(1+k^2/k_{{\rm UV},0}^2)$, and use $k_0 = \frac{1}{20} h/$Mpc. 
	The cosmology dependence is encoded in the fitting coefficients $\alpha_n$. 
	$N$ is the number of fitting functions used (throughout this paper, we use $N=16$). 
	We also denote $\vec{\alpha}$ and $\vec{f}$ as vectors whose $n$-th entry is given, respectively, by the elements $\alpha_n$ and $f_n$. 
	Note that $\alpha_n$ has the same dimensions as $1/k^3$.
	
	We can select a number of points $N_p$ of $P_{\rm lin}$ and determine $\vec{\alpha}$ using a least squares regression:
	\begin{equation}
		\label{eq:alpha}
		\vec{\alpha} = (X^T X)^{-1} X^T \vec{P}_{\rm lin}\,,
	\end{equation}
	where $\vec{P}_{\rm lin}$ is a $N_p$ dimensional vector and $X$ is a $N_p \times N$ matrix. $\vec{P}_{\rm lin}$ and $X$ are given by
	\begin{equation}
		\begin{split}
			P_{{\rm lin},i} = P_{\rm lin}(k_i)\,,\\
			X_{ij} = f_j(k^2_i)\,,
		\end{split}
	\end{equation}
	$k_i$'s being the $N_p$ wave-numbers of each fitting point. 
	In this paper, we use $N_p = 100$ logarithmically spaced points from $k_{\rm min}= 10^{-3}$ to $k_{\rm max}=  0.6$.
	
\begin{figure}[H]
		\centering
		\includegraphics[width=0.7\textwidth]{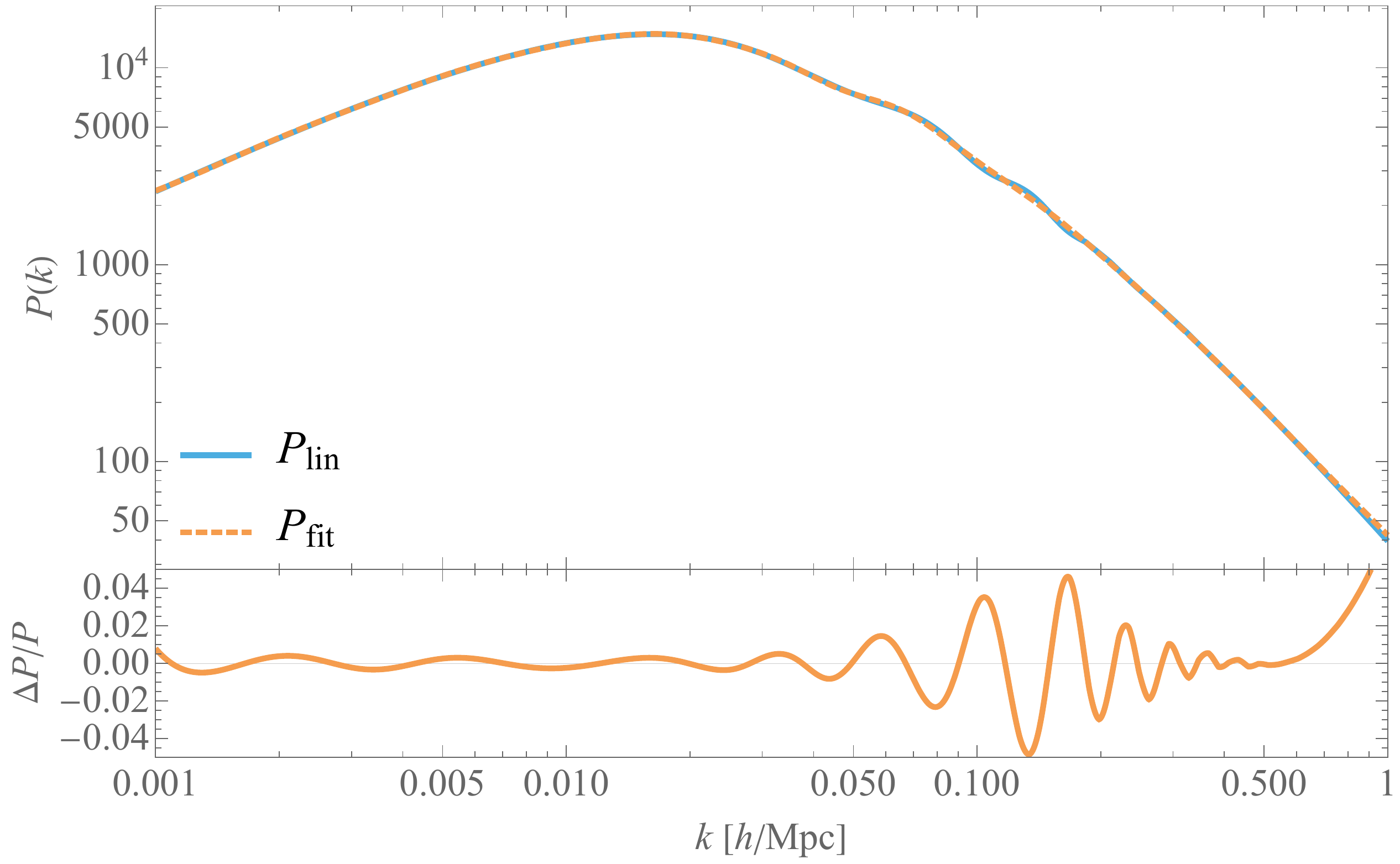}
		\caption{\label{fig:Pfitplot} Comparison of $P_{\rm lin}$ with $P_{\rm fit}$, from $k = 10^{-3} \hinvMpc$ to $k = 1 \hinvMpc$. {Note that even if the fit is only performed up to $0.6 \hinvMpc$, the error is within 5\% up to $1\hinvMpc$.}}
	\end{figure}

	It is useful to write our specific decomposition by grouping terms as follows:
	\begin{equation}
		\begin{split}
			P_{\rm fit}(k) = & \frac{a_0}{1+\frac{k^2}{k_{{\rm UV},0}^2}}  + \sum_{i=1}^3  a_i\; f(k^2, k^2_{\rm peak, 1},k^2_{\rm UV, 1}, 0, i)\\
			& + \sum^3_{i = 0}\left(b_i\; f(k^2, k^2_{\rm peak, 2},k^2_{\rm UV, 2}, 1, i + 1) + c_i \; f(k^2, k^2_{\rm peak, 3},k^2_{\rm UV, 3}, 0, i + 2)\right.\\
			& \left.+ d_i\; f(k^2, k^2_{\rm peak, 4},k^2_{\rm UV, 4}, 0, i + 1)\right)\,,
		\end{split}
	\end{equation}	
	where our parameters are:
	\begin{align*}
		& k^2_{\rm UV, 0} = 1 \times 10^{-4} \ h^2/{\rm Mpc}^2\,, \\
		& k^2_{\rm peak, 1} = -3.4 \times 10^{-2} \ h^2/{\rm Mpc}^2 \,,\quad k^2_{\rm UV, 1} = 6.9 \times 10^{-2} \ h^2/{\rm Mpc}^2 \,, \\
		& k^2_{\rm peak, 2} =-1 \times 10^{-3}  \ h^2/{\rm Mpc}^2\,,\quad k^2_{\rm UV, 2} = 8.2 \times 10^{-3} \ h^2/{\rm Mpc}^2 \,, \\
		& k^2_{\rm peak, 3} = -7.6 \times 10^{-5}  \ h^2/{\rm Mpc}^2\,,\quad k^2_{\rm UV, 3} = 1.3 \times 10^{-3}   \ h^2/{\rm Mpc}^2 \,,\\
		& k^2_{\rm peak, 4} = -1.56 \times 10^{-5} \ h^2/{\rm Mpc}^2 \,,\quad k^2_{\rm UV, 4} = 1.35 \times 10^{-5} \ h^2/{\rm Mpc}^2\,.
	\end{align*}	
	The fitting function $P_{\rm fit}$ is accurate up to 5\% at all relevant scales, as shown in Fig.~\ref{fig:Pfitplot}. 
	We can change the precision of $P_{\rm fit}$ by increasing or decreasing $N$, which we can use to better capture the BAO wiggles, as is shown in Appendix~\ref{sec:appfit}.
	There is one aspect that makes the residuals of the fit sufficient, as we will show. 
	We are interested only in using $ P_{\rm fit}$ to make predictions for the EFTofLSS. Therefore, any mismatch in the UV, {\it i.e.} $k\gtrsim 0.5\hinvMpc$,  can be absorbed in the counterterms, and, if the difference is just ${\cal{O}}(1)$, the order of magnitude of the counterterms is not affected by this.
	Additionally, we only use $P_{\rm fit}$ inside the loop integrals, and $P_{\rm lin}$ outside the integrals.
	Therefore only momenta comparable to the external ones strongly affect the result. 
	Since loops are quantitatively irrelevant at low wavenumbers, and modes much longer than the ones of interest do not contribute due to IR safety~\cite{Carrasco2014,Lewandowski2017}, we do not need an accurate fit for $k\lesssim 0.001\hinvMpc$ (and in fact this is even more than what is really needed in the IR).  
	On top of this, the effect of residuals that are highly oscillating tends to be suppressed upon integration.
	Therefore,  despite the residuals being greater than 1\%, the error of the integrals will be much smaller, as we will show.
	
	Each one of our fitting functions $f$ in Eq.~\eqref{eq:f} can itself be expressed as a sum of QFT propagator-like functions by decomposing the denominator in the following way
	\begin{equation}
		\label{eq:denf}
		\frac{1}{{\left(1 + \frac{(k^2 - k^2_{\rm peak})^2}{k^4_{\rm UV}}\right)^j}} = \frac{k^{4j}_{\rm UV}}{\left(k^2 - k^2_{\rm peak} -i \, k^2_{\rm UV}\right)^j\left(k^2 - k^2_{\rm peak} + i \, k^2_{\rm UV}\right)^j}\,,
	\end{equation}
	and then noticing that 
	\begin{equation}
		\frac{k^2_{\rm UV}}{\left(k^2 - k^2_{\rm peak} -i \, k^2_{\rm UV}\right)\left(k^2 - k^2_{\rm peak} + i \, k^2_{\rm UV}\right)} = -\frac{i/2}{k^2 - k^2_{\rm peak} -i \, k^2_{\rm UV}} + \frac{i/2}{k^2 - k^2_{\rm peak} + i \, k^2_{\rm UV}}\,.
		\label{eq:func_decomp}
	\end{equation}
	The last term is indeed a sum of two propagators with complex masses.
	We can then proceed iteratively to decompose the right hand side (r.h.s.) of Eq.~\eqref{eq:denf}.
	Therefore each $f$ gets schematically decomposed in 
	\begin{equation}
		\label{eq:decomp}
		f(k^2, k^2_{\rm peak}, k^2_{\rm UV}, i, j) = \sum_{n=1}^j  k_{\rm UV}^{2(n-i)} \,k^{2 i} \left( \frac{\kappa_n }{(k^2 + M)^n} + \frac{\kappa^*_n}{(k^2 + M^*)^n}\right)\,,
	\end{equation}
	where each $\kappa_n$ is a complex constant, and $M \equiv - k^2_{\rm peak} + i \, k^2_{\rm UV}$, which is $k$-independent. 
	Note that we have an exception $M_0 = k^2_{\rm UV, 0}$ which is purely real positive such that $f_0(k^2) =k^2_{\rm UV,0}/(k^2_{\rm UV,0} + k^2)= M_0/(M_0 + k^2)$.
	Note {also} that $M$ has the same dimensions as $k^2$. We do not write $M^2$ to avoid clutter.
	For example, we have
	\begin{align}
		& f(k^2, k^2_{\rm peak}, k^2_{\rm UV}, 0, 1) = \frac{i k_{\rm UV}^2}{2}\left( \frac{1}{k^2 + M} -\frac{1}{k^2 + M^*}\right) \,, \\
		& f(k^2, k^2_{\rm peak}, k^2_{\rm UV}, 0, 2) = \frac{i k_{\rm UV}^2}{4}\left( \frac{1}{k^2 + M} -\frac{1}{k^2 + M^*}\right) - \frac{k_{\rm UV}^4}{4} \left( \frac{1}{(k^2 + M)^2} +\frac{1}{(k^2 + M^*)^2}\right)\,.
	\end{align}
	This way, we can decompose each fitting function $f_n$ in \eqref{eq:Pfit1}, and for each term in the decomposition calculate the corresponding loop integrals coming from cosmological perturbation theory. 
	After that, we can sum the contributions to obtain the loop integral for a single $f_n$, a pair $(f_{n_1},f_{n_2})$ or a triplet $(f_{n_1},f_{n_2},f_{n_3})$, depending on whether the loop integrals involve one, two or three power spectra. 
	Then, we just have to multiply each loop integral by the corresponding $\alpha_n$ coefficients and sum. In Section \ref{sec:PandB1loop} this will be demonstrated with the example of the one-loop power spectrum.
	
	So our scheme is the following:
	\begin{itemize}
		\item Perform a linear regression on the power spectrum to obtain $P_{\rm fit}$;
		\item Calculate for each diagram the corresponding cosmology independent tensor for each {needed} combination of fitting functions $f_n$;
		\item Contract tensor with cosmology dependent fitting coefficients $\alpha_n$ to obtain the full loop integral.
	\end{itemize}
	
	\section{One-loop power spectrum and one-loop bispectrum}
	\label{sec:PandB1loop}
	
	\subsection{One-loop power spectrum in real space}
	\label{sec:P1loop}
	
	In this paper, we will focus on the loop diagrams, and limit ourselves to one-loop order. 
	We will therefore always neglect the counterterms of the EFTofLSS, as at this order they are tree-level diagrams. 
	Therefore, the one-loop contribution to the power spectrum, $P_{\rm 1-loop}$, consists of two diagrams,
	\begin{align}
		P_{\rm 1-loop}(k,\tau) = D(\tau)^4\left[P_{22}(k) + P_{13}(k)\right] \,,
	\end{align}
	where $\tau$ is conformal time and $D(\tau)$ is the linear growth factor. 
	The individual terms are given by,
	\begin{align}
		P_{22}(k) &= 2\int_q \left[F_2(\vec{q},\vec{k}-\vec{q})\right]^2 P_{\rm lin}(q)P_{\rm lin}(|\vec{k}-\vec{q}|) {\ ,}\\ 
		P_{13}(k) &= 6P_{\rm lin}(k)\int_q F_3(\vec{q},-\vec{q},\vec{k})P_{\rm lin}(q)\,,
	\end{align}
	where $\int_q \equiv \int \frac{dq^3}{(2\pi)^3}$, and $F_i$ are the standard perturbation theory kernels. 
	
	Similarly, we define 
	\begin{align}
		\label{eq:P22fit}
		\bar{P}_{22}(k) &= 2\int_q \left[F_2(\vec{q},\vec{k}-\vec{q})\right]^2 P_{\rm fit}(q)P_{\rm fit}(|\vec{k}-\vec{q}|){ \ ,}\\ 
		\label{eq:P13fit}
		\bar{P}_{13}(k) &= 6P_{\rm lin}(k)\int_q F_3(\vec{q},-\vec{q},\vec{k})P_{\rm fit}(q)\,,
	\end{align}
	that are calculated using the $P_{\rm fit}$ in the integrands instead of $P_{\rm lin}$.

	\begin{figure}[H]
		\centering
		\includegraphics[scale = 0.6]{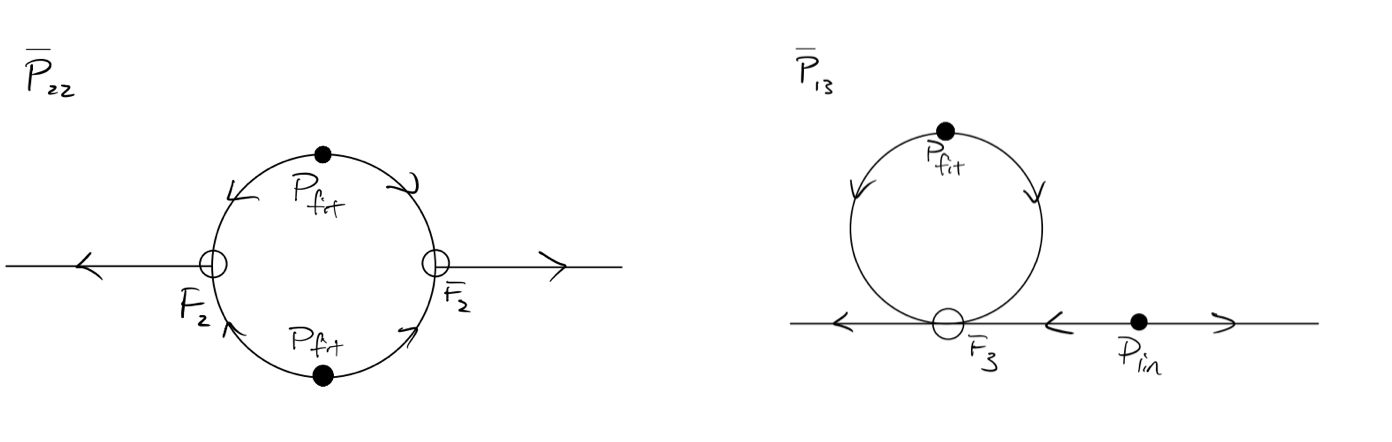}
		\caption{Diagrams contributing to the one-loop power spectrum. $P_{\rm lin}$ are contributions that are not being integrated. $P_{\rm fit}$ are always integrated. For the exact numerical result, $P_{\rm fit}$ should be replaced by $P_{\rm lin}$. Left: $P_{22}$ diagram, given in Eq.~\eqref{eq:P22fit}. Right: $P_{13}$ diagram, given in Eq.~\eqref{eq:P13fit}.}
		\label{fig:p1loop}
	\end{figure}

	The kernels $F_2$ and $F_3$ are decomposed into integer powers of $q^2, k^2$ and $|\vec{k} - \vec{q}|^2$. 
	For example, the kernel $F_2$ has the form,
	\begin{equation}
		\begin{split}
			\label{eq:F2decomp}
			F_2(\vec{q}, \vec{k} - \vec{q}) &= \frac{5}{14} + \frac{3k^2}{28q^2} + \frac{3k^2}{28 |\vec{k} - \vec{q}|^2}- \frac{5q^2}{28|\vec{k} - \vec{q}|^2}-\frac{5|\vec{k} - \vec{q}|^2}{28q^2}+\frac{k^4}{14|\vec{k} - \vec{q}|^2q^2} \\
			&= \sum_i c_i |\vec{k} - \vec{q}|^{2 m_{i}}\,q^{2 n_{i}} k^{-2 (m_{i} + n_{i})} \,,
		\end{split}
	\end{equation}
	where $n_i$ and $m_i$ are integers, and $c_i$ are the corresponding coefficients~\footnote{In this paper, we use the perturbation theory kernels up to $F_4$. The full expressions are too long to include in this paper, and we refer the reader to~\cite{Bernardeau:2001qr} for a derivation}.
	
	Our goal now is to find expressions for $\bar{P}_{22}$ and $\bar{P}_{13}$. 
Let us start with $\bar{P}_{22}$.
The strategy is to find analytical expressions for the loop integral, with $P_{\rm fit}$ replaced by a generic fitting function $f_n$. 
In this way, we generate a $N \times N$ matrix $M^{(22)}$ whose elements are given by 
	\begin{equation}
		M^{(22)}_{ij}(k^2) \equiv  2 \int_q \left[F_2(\vec{q},\vec{k}-\vec{q})\right]^2 f_{i}(q^2) f_{j}(|\vec{k} - \vec{q}|^2) \,.
	\end{equation}
Given $M^{(22)}_{ij}$, $\bar{P}_{22}$ is simply given by
\begin{equation}
	\bar{P}_{22}(k) = \vec{\alpha}^T  M^{(22)}(k^2) \vec{\alpha}\,,
\end{equation}
where $\vec{\alpha}$ is given in Eq.~\eqref{eq:alpha}.
With our decomposition of the kernels, $M^{(22)}_{ij}$ can be written as a sum of terms of the following form:
	\begin{equation}
		\label{eq:Ifunc}
		I_{ij}(m, n, k^2) \equiv \int_q q^{2m}|\vec{k} - \vec{q}|^{2n}   f_{i}(q^2) f_{j}(|\vec{k} - \vec{q}|^2)\,,
	\end{equation}
with $k$-dependent coefficients.
Next, decomposing the $f$ functions as in Eq.~\eqref{eq:decomp}, we find that each $I_{ij}$ can be written as a sum of terms like:
\begin{align}\label{eq:LBdefinition}
	L_B(n_1, d_1, n_2, d_2, k^2, M_1, M_2) \equiv  \int_q\frac{|\vec{k} - \vec{q}|^{2n_{1}}q^{2n_{2}}}{(|\vec{k} - \vec{q}|^2+M_{1})^{d_{1}}(q^2+M_{2})^{d_{2}}} \,,
\end{align}
where $M_1$ and $M_2$ are complex squared masses, and with $L_B$ being multiplied by $k$-dependent coefficients. 
The integral $ L_B$ can be calculated analytically as shown in Sec.~\ref{sec:bubrec}.
Following this procedure we analytically calculate $\bar{P}_{22}$. 

Similarly, for $\bar{P}_{\rm 13}$ we can define a $N$ dimensional vector
$M^{(13)}$  whose elements are given by
	\begin{equation}
		\label{eq:M13}
		M^{(13)}_{i}(k^2) \equiv  6 \int_q F_3(\vec{q},-\vec{q},\vec{k}) f_{i}(q^2)\,,
	\end{equation}
and $\bar{P}_{13}$ is given by 
\begin{equation}
	\bar{P}_{13}(k) =  P_{\rm lin}(k) M^{(13)}(k^2)\cdot \vec{\alpha}\,.
\end{equation}
As for $M^{(22)}$, the matrix $M^{(13)}$ can be written as a sum of terms of the form 
	\begin{equation}
		I_{i\emptyset}(m, n, k^2) \equiv \int_q q^{2m}  |\vec{k} - \vec{q}|^{2n} f_{i}(q^2)\,,
	\end{equation}
	where $\emptyset$ means $f_j$ is removed from the integrand of $I_{ij}$ defined in Eq.~\eqref{eq:Ifunc} (which can be done by setting some entries of $L_B$ to 0). 
	After decomposing the $f$ functions, each $I_{i\emptyset}$ can be written as {a sum} of $k$-dependent coefficients times integrals of the form (\ref{eq:LBdefinition}) with $d_2 = 0$.
	
	\paragraph{UV-correction to compare with numerical integration:}
	To test the accuracy of our approach, we need a way to correctly compare with numerical integration. Our numerical integration bounds are $q_{\rm IR} = 1.0 \times 10^{-4} \hinvMpc$ (lower bound) and $q_{\rm UV} = 1.0 \hinvMpc$ (upper bound). 
	For example, the numerical $P_{22}$ is given by
	\begin{equation}
		P_{22}(k) = \int_{q_{\rm IR}}^{q_{\rm UV}} \int_{\Omega_2} d\Omega_2 \, dq\, p_{22}(\vec{q},\vec{k}-\vec{q})\,,
	\end{equation}
	where $p_{22}$ includes everything in the integrand (including the integration measure).
	When we calculate the loop integrals analytically, the integration is performed across the whole space i.e. up to $q \to \infty$. 
	For example, the analytical $\bar{P}_{22}$ is given by
	\begin{equation}
		\bar{P}_{22}(k) = \int_{0}^{\infty} \int_{\Omega_2} d\Omega_2 \, dq \, \bar{p}_{22}(\vec{q},\vec{k}-\vec{q})\,,
	\end{equation}
	where $\bar{p}_{22}$ includes everything in the integrand (including the integration measure).
	So to precisely test the accuracy of our result, we need to subtract from the analytical result the UV contribution from $q=q_{\rm UV}$ to $q=\infty$~\footnote{For the purpose of simply matching the numerical integration, we can do the same procedure to correct for the IR contribution. However, in our decomposition, that contribution is very subdominant.}. 
	We do this for the sole purpose of matching the numerical integration: the difference we correct for here comes from the difference between $P_{\rm fit}$ and the exact linear power spectrum within the non-linear regime, and, on the physical analysis,  is degenerate with ${\cal{O}}(1)$ corrections to the counterterms (more on this point also later).
	
	To illustrate, for $P_{22}$ this UV contribution is  given by
	\begin{equation}
		\bar{P}_{22}^{\rm UV}(k) = \int_{q_{\rm UV}}^{\infty} \int_{\Omega_2} d\Omega_2 \, dq \, \bar{p}_{22}(\vec{q},\vec{k}-\vec{q})\,.
	\end{equation}
	In the $P_{22}$ case, this contribution does not matter significantly, because the kernel goes as $q^{-4}$ and in our basis $\lim_{q \rightarrow \infty}P_{\rm fit}(q) \sim 1/q^2$, giving a contribution of the order $\mathcal{O}(q_{\rm UV}^{-5})$~\footnote{Note that since we fit up to $0.6 \hinvMpc$ there could be in principle a significant difference between $P_{\rm fit}(q)$ and $P_{\rm lin}(q)$ for $q \in [0.6,1]\hinvMpc$, which could influence the matching with numerical integration. 
		But, as shown in Fig.~\ref{fig:Pfitplot}, that difference is small, and in particular subdominant compared with the UV contribution from $1\hinvMpc$ to $\infty$.}. 
	
\begin{figure}[h]
		\centering
		\includegraphics[width=\textwidth]{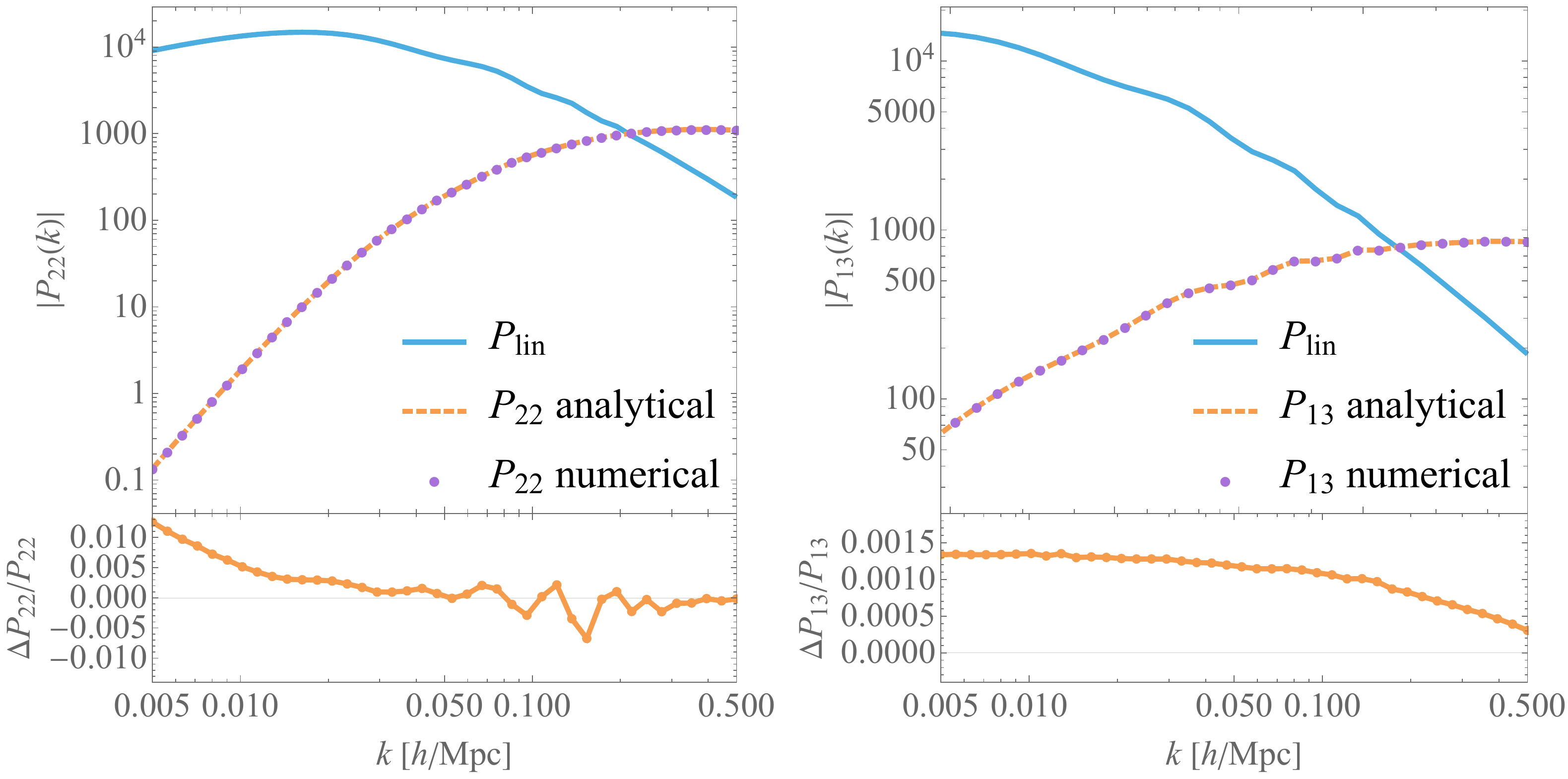}
		\caption{\label{fig:P13P22}Two contributions for the one-loop power spectrum. Left: comparison between analytical result, $\bar{P}_{22}$, and exact numerical result, $P_{22}$. Right: comparison between analytical result, $\bar{P}^{{\rm comp}}_{13}$, and exact numerical result, $P_{13}$. These result were obtained using $N=16$ fitting functions.}
	\end{figure}

	However, in the $P_{13}$, the contribution is more significant.
	To estimate that contribution, we expand the $P_{13}$ integrand in Eq.~\eqref{eq:M13} for each  function $f_i$ up to second order in $1/q$ (if we wanted more precision we could go to higher orders, but in our case that is not necessary). 
	We define $m^{(13)}_{\rm UV}$ and $M^{(13)}_{\rm UV}$ as the following, 
	\begin{align}
		\label{eq:m13}
		m^{(13)}_{{\rm UV}, i} &\equiv  \int_{\Omega_2} d\Omega_2 \lim_{q\to \infty} q^2\,  6 F_3(\vec{q},-\vec{q},\vec{k}) f_{i}(q^2) \leq {\cal{O}}\left(\frac{k^2}{q^2}\right)\,,\\
		M^{(13)}_{{\rm UV}, i} &\equiv \int_{q_{\rm UV}}^{\infty} \frac{dq}{(2\pi)^3}\, m^{(13)}_{{\rm UV}, i} \, .
	\end{align}
	So, we obtain a simple vector $m^{(13)}_{\rm UV}$ where each component is $c_i k^2/q^2$, $c_i$ being a constant which depends on the function parameters.
	We then integrate $m^{(13)}_{\rm UV}$ from $q=q_{\rm UV}$ to $q=\infty$, obtaining the $\bar{P}_{13}$ UV contribution for each function $M^{(13)}_{\rm UV}$. 
	The elements of $M^{(13)}_{\rm UV}$ are given by 
	\begin{equation}
		M^{(13)}_{{\rm UV},i} = {\frac{c_i}{(2\pi)^3}} k^2/q_{\rm UV}\,.
	\end{equation}
	In order to estimate the UV contribution $\bar{P}_{13}^{\rm UV}$, we just have to calculate
	\begin{equation}
		\bar{P}_{13}^{\rm UV} = P_{\rm lin}\, M^{(13)}_{\rm UV} \cdot \alpha\,,
	\end{equation}
	where $ M^{(13)}_{\rm UV}$ is cosmology independent.
	So, to accurately compare with the numerically calculated $P_{13}$, we should use
	\begin{equation}
		\bar{P}^{\rm comp}_{13} = \bar{P}_{13} - \bar{P}_{13}^{\rm UV}\,.
	\end{equation}
	Note again that, when doing the full EFTofLSS analysis, this contribution can be absorbed in the effective terms {by a ${\cal{O}}(1)$ shift}. 
	Here, we just show that we do not need any extra degree of freedom to correctly match the numerical result.
	The comparison of the $P_{13}$ and $P_{22}$ diagrams analytically evaluated with our code using $P_{\rm fit}$ to numerical integration using the exact $P_{\rm lin}$ is shown in Fig.~\ref{fig:P13P22}.
	
\begin{figure}
		\centering
		\includegraphics[width=\textwidth]{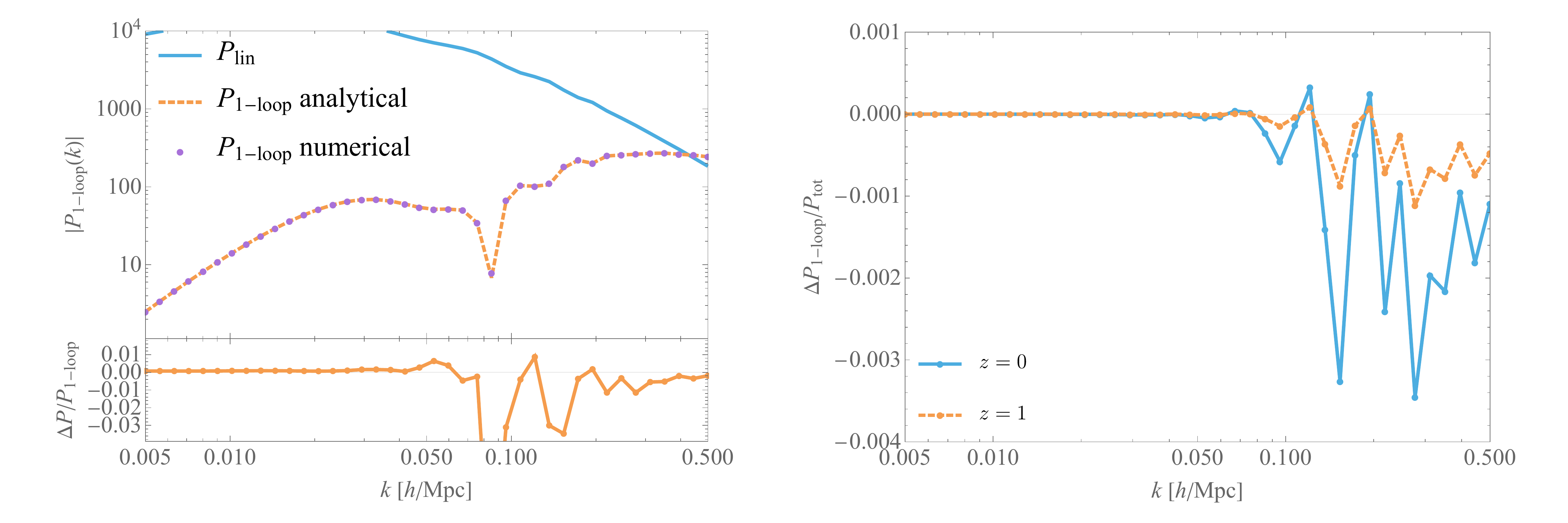}
		\caption{\label{fig:P1loop} Left: Comparison between analytical result, $\bar{P}_{\rm 1-loop}$, and exact numerical result, $P_{\rm 1-loop}$, obtained by summing the contributions in Fig.~\ref{fig:P13P22}. Right: 1-loop error relative to the full power spectrum including the linear contribution at redshifts $z=0$ and $z=1$.  } 
	\end{figure}

	It is actually more interesting to compare the accuracy of our approximation against the numerical integration of the one-loop contribution to the power spectrum, and, even more importantly, with the power spectrum {\it up to} one loop~\footnote{Here we neglect the counterterms, which tend to make the loop even smaller, so that the reported accuracy is probably an underestimate of the actual accuracy.}. 
	In fact, this is the observable quantity that gets compared against data. 
	This comparison is shown in Fig.~\ref{fig:P1loop}. We present results both at
	redshift $z=0$, where the loop, for a given wavenumber, is the largest, and
	with $z=1$, which is perhaps more relevant observationally, where the errors are safely below per mill.
	
	\subsection{One-loop bispectrum in real space}
	\label{sec:bispec_rs}
	
	The matter one-loop order contribution to the bispectrum $B_{\rm 1-loop}$ consists of four diagrams,
	\begin{equation}
		\begin{split}
			& B_{\rm 1-loop}(k_1,k_2,k_3,\tau) = \\  &D(\tau)^6\left[B_{222}(k_1,k_2,k_3) + B^I_{321}(k_1,k_2,k_3) + B^{II}_{321}(k_1,k_2,k_3) + B_{411}(k_1,k_2,k_3) \right]\,,
		\end{split}
	\end{equation}
	where $\vec{k}_1 + \vec{k}_2 + \vec{k}_3 = 0$. 
	The terms are given by
	\begin{align}
		&B_{222}(k_1,k_2,k_3) = 8 \int_{\vec{q}} F_2 (\vec{q},\vec{k}_1-\vec{q}) F_2(\vec{k}_1-\vec{q},\vec{k}_2+\vec{q}) F_2(\vec{k}_2+\vec{q},-\vec{q}) \nonumber \\
		& \qquad \qquad \times P_{\rm lin}(q) P_{\rm lin}(|\vec{k}_1-\vec{q}|) P_{\rm lin}(|\vec{k}_2+\vec{q}|)\ , \\
		& B_{321}^I(k_1,k_2,k_3) = 6 P_{\rm lin}(k_1) \int_{\vec{q}} F_3(-\vec{q},-\vec{k}_2+\vec{q},-\vec{k}_1) F_2(\vec{q},\vec{k}_2-\vec{q}) P_{\rm lin}(q) P_{\rm lin}(|\vec{k}_2-\vec{q}|) \nonumber \\
		&\qquad \qquad + 5\;{\rm perms}\ , \\
		& B_{321}^{II}(k_1,k_2,k_3) = F_2(\vec{k}_1,\vec{k}_2) P_{\rm lin}(k_1) P_{13}(k_2) + 5\;{\rm perms}\ , \\
		& B_{411}(k_1,k_2,k_3) = 12 P_{\rm lin}(k_1) P_{\rm lin}(k_2) \int_{\vec{q}} F_4(\vec{q},-\vec{q},-\vec{k}_1,-\vec{k}_2) P_{\rm lin}(q) + 2\;{\rm cyclic\; perms}\,.
	\end{align}
	In the same way as in Sec.~\ref{sec:P1loop}, we define $\bar{B}_{222}$, $\bar{B}^I_{321}$,
	$\bar{B}^{II}_{321}$, and $\bar{B}_{411}$ using $P_{\rm fit}$ instead of $P_{\rm lin}$. 
	We can also expand the kernels in integer powers of $q^2$, $|\vec{k}_1-\vec{q}|^2$, and $|\vec{k}_2+\vec{q}|^2$, e.g. for $B_{222}$:
	\begin{equation}
		\begin{split}
			&F_2 (\vec{q},\vec{k}_1-\vec{q}) F_2(\vec{k}_1-\vec{q},\vec{k}_2+\vec{q}) F_2(\vec{k}_2+\vec{q},-\vec{q}) = \\
			& \qquad \qquad \sum_{n_1,n_2,n_3}C_{n_1,n_2,n_3}(k^2_1,k^2_2,k^2_3)q^{2n_{1}}|\vec{k}_1 - \vec{q}|^{2n_{2}}|\vec{k}_2 + \vec{q}|^{2n_{3}}\ ,
		\end{split}
	\end{equation}
	where $C_{n_1,n_2,n_3}(k^2_1,k^2_2,k^2_3)$ are coefficients of each term and $n_1, n_2$ and $n_3$ are integers.
	When terms like $|\vec{k}_2-\vec{q}|^2$ or $|\vec{k}_3-\vec{q}|^2$ appear, it is always possible to isolate the term and perform a change of variable to obtain a dependence {only on the variables} $q^2$, $|\vec{k}_1-\vec{q}|^2$, and $|\vec{k}_2+\vec{q}|^2$.
	To compare with the numerical integration, we use the following parametrizations:
	\begin{align}
		&\vec{q} = q\left(\sqrt{1-\nu ^2} \cos (\phi),\sqrt{1-\nu ^2} \sin
		(\phi),\nu  \right)\,, \\ 
		&\vec{k}_1 = k_1(0,0,1)\,, \\ 
		&\vec{k}_2 = k_2\left(0,\sqrt{1-y^2},y\right)\,,
	\end{align}
	where $y$ is the cosine of the angle between $\vec{k}_1$ and $\vec{k}_2$, given by
	\begin{equation}
		y \equiv \frac{-k_1^2-k_2^2+k_3^2}{2 k_1 k_2}\,,
	\end{equation}
	and $\nu$ is the cosine of the angle between $\vec{q}$ and $\vec{k}_1$.
	For this case we proceed as explained for the one-loop power spectrum, but now we generalize the function $L_B$ to a new function $L$ defined as
	\begin{equation}
		\label{eq:Lfunc}
		\begin{split}
			& L(n_1, d_1, n_2, d_2, n_3, d_3, k_1^2, k_2^2, k_3^2, M_1, M_2, M_3) \equiv \\
			& \qquad \int_q\frac{|\vec{k}_1 - \vec{q}|^{2n_{1}}q^{2n_{2}}|\vec{k}_2 + \vec{q}|^{2n_{3}}}{(|\vec{k}_1 - \vec{q}|^2+M_{1})^{d_{1}}(q^2+M_{2})^{d_{2}}(|\vec{k}_2 + \vec{q}|^2+M_{3})^{d_{3}}}   \,,
		\end{split}
	\end{equation}
	which can be analytically calculated as well.   
	Note that we have
	\begin{equation}
		L_B(n_1, d_1, n_2, d_2, k^2, M_1, M_2) = L(n_1, d_1, n_2, d_2, 0, 0, k^2, 0, 0, M_1, M_2, 0)\,.
	\end{equation}
	In the rest of the paper, we will sometimes omit the arguments in $L$ that are not exponents.
	Similarly, we generalize the function $I_{ij}$ to a function $J_{ijk}$ defined as
	\begin{equation}
		\label{eq:Jfunc}
		J_{ijk}(l, m, n, k_1^2, k_2^2, k_3^2) \equiv \int_q q^{2l}|\vec{k}_1 - \vec{q}|^{2m} \,  |\vec{k}_2 + \vec{q}|^{2n} f_{i}(q^2) f_{j}(|\vec{k}_1 - \vec{q}|^2)   f_{k}(|\vec{k}_2 + \vec{q}|^2)\,.
	\end{equation}
We have 
\begin{equation}
	I_{ij}(l,m,k^2) = J_{ij\emptyset}(l, m, 0, k^2, 0, 0)\,,
\end{equation}
where $\emptyset$ means $f_k$ is removed from the integrand of $J_{ijk}$ defined in Eq.~\eqref{eq:Jfunc}.
We can then use the function $J$ to calculate the different contributions for the one-loop bispectrum where each term involves the function $J$ multiplied by $\vec{k}_i$ dependent coefficients, i.e. $M^{(\rm diagram)} 
\supset C(k^2_1, k^2_2, k^2_3)J_{ijk}(l,m,n,k^2_1,k^2_2,k^2_3)$. 
Notice that, as $I_{ij}$ can be represented by the $J_{ijk}$ function with $f_k$ removed or $I_{i\emptyset}$ with $f_j$ and $f_k$ removed, we can evaluate every diagram as a sum of $J_{ijk}$ functions with $k$-dependent coefficients $C(k^2_1, k^2_2, k^2_3)$ and setting indices to $\emptyset$ when necessary (this can be done by setting some entries of $L$ to 0).
	
	\begin{figure}[H]
		\centering
		\includegraphics[width=.6\textwidth]{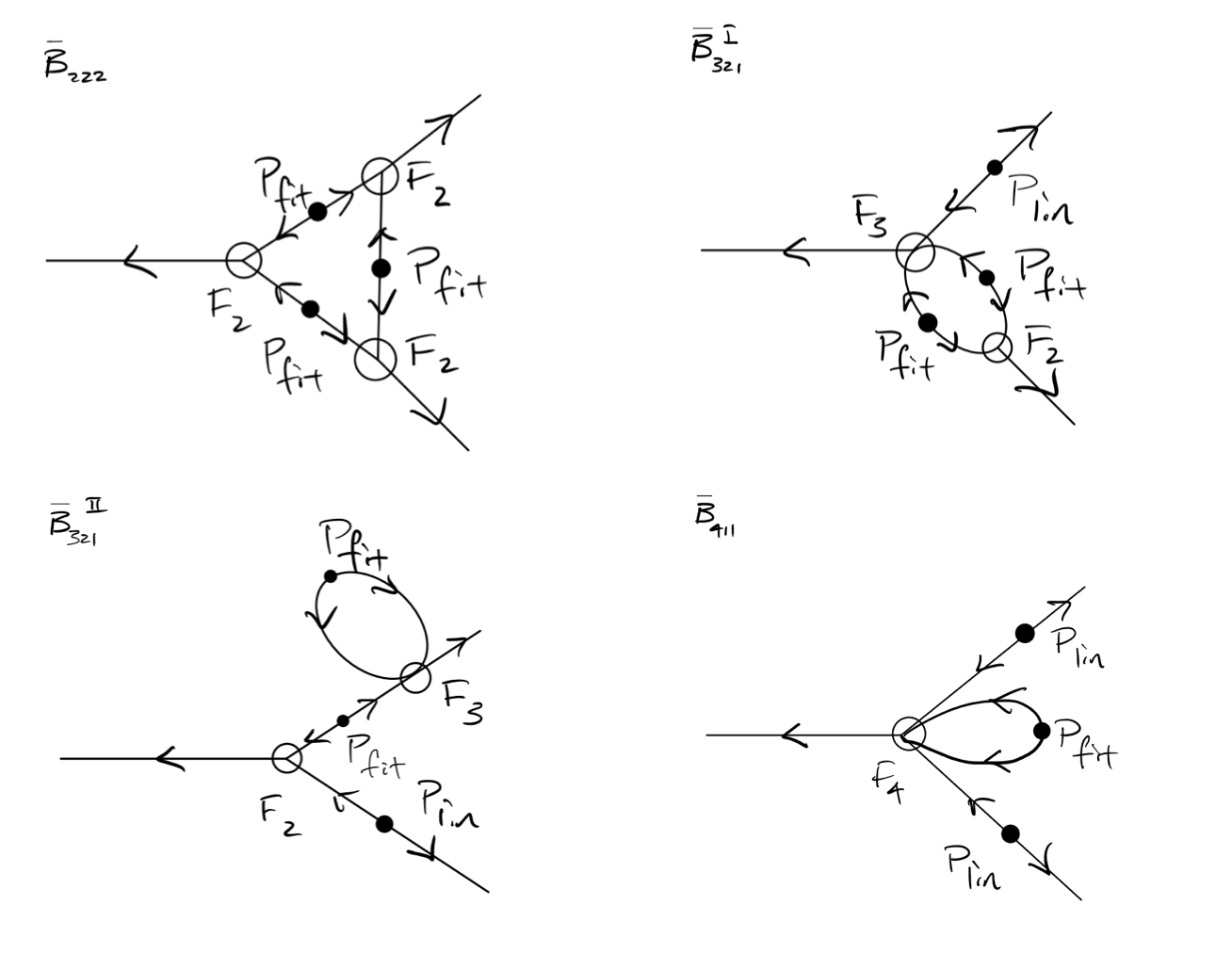}
		\caption{\label{fig:bispec_diagram} 
			Diagrams contributing to the one-loop bispectrum. 
			$P_{\rm lin}$ are contributions that are not being integrated. $P_{\rm fit}$ are always integrated. 
			For the exact numerical result, $P_{\rm fit}$ should be replaced by $P_{\rm lin}$. 
			Arrows represent the flow of time, lines without crossed are retarded Green's function, while lines with crosses are correlation functions of the initial fields. }
	\end{figure}
\begin{figure}[H]
		\centering
		\includegraphics[width=\textwidth]{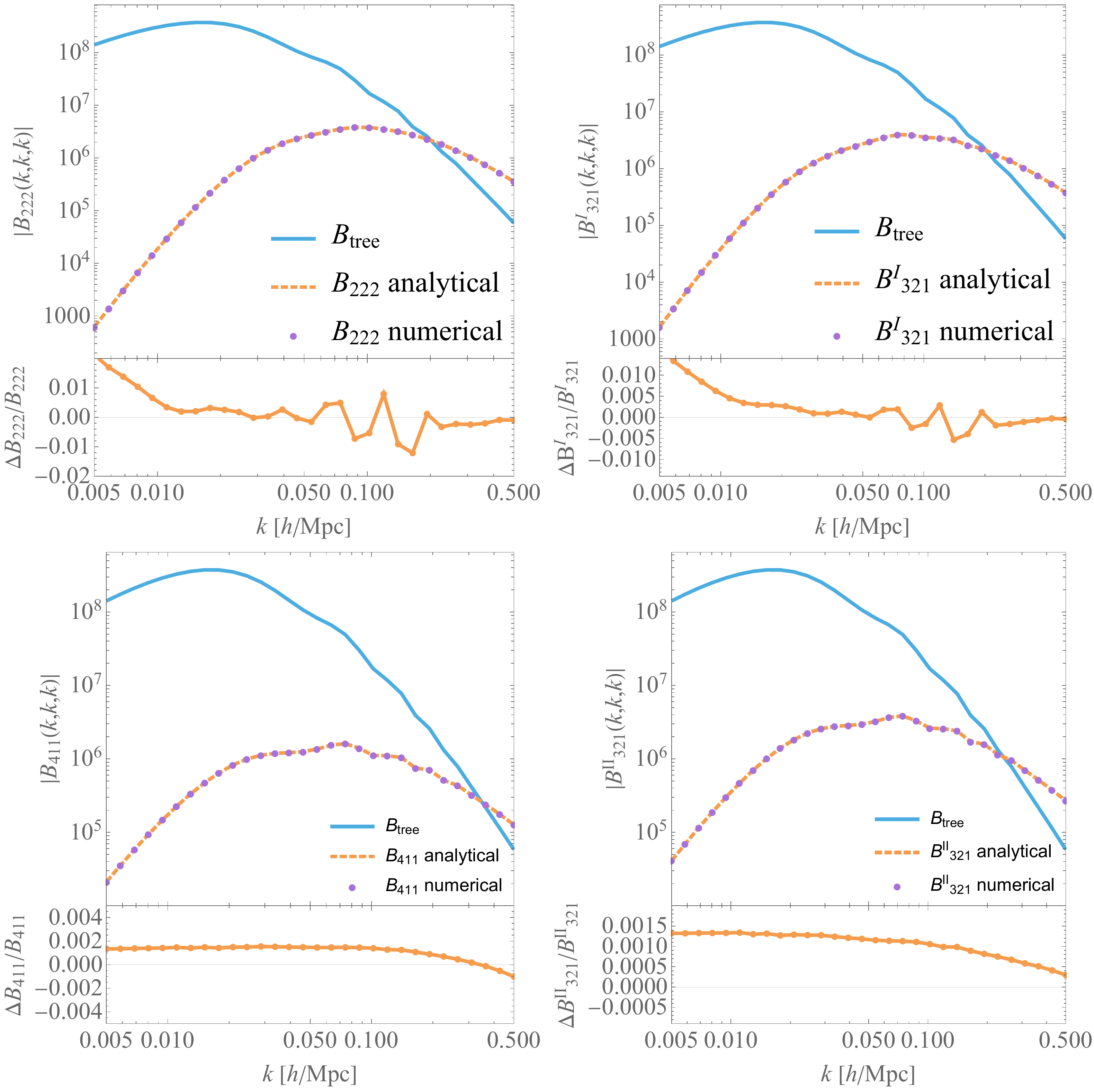}
		\caption{\label{fig:BAll} Comparison of the analytical result with the numerical integration for each one of the 4 diagrams contributing to the 1-loop bispectrum.}
	\end{figure}

	Defining the $N \times N \times N$ tensor $M^{(222)}$ as 
	\begin{equation}
		\begin{split}
			M^{(222)}_{ijk} = 8 \int_q &F_2 (\vec{q},\vec{k}_1-\vec{q}) F_2(\vec{k}_1-\vec{q},\vec{k}_2+\vec{q}) F_2(\vec{k}_2+\vec{q},-\vec{q}) \\ &\times 
			f_{i}(q^2) f_{j}(|\vec{k}_1 - \vec{q}|^2)  f_{k}(|\vec{k}_2 + \vec{q}|^2)\,,
		\end{split}
	\end{equation}
	where we dropped the arguments to alleviate the notation, one obtains $\bar{B}_{222}$ as follows:
	\begin{equation}
		\bar{B}_{222} = \sum_{ijk} M^{(222)}_{ijk} \alpha_i \alpha_j \alpha_k\,.
	\end{equation}
	Note that $M^{(222)}$ can easily be calculated if one has a formula for $J$. 
	The comparison with the numerical integration can be seen in Fig.~\ref{fig:BAll}.
	
	$\bar{B}_{321}^I$ can be calculated by defining the $N \times N$ matrix
	$M^{(321)}$ as 
	\begin{equation}
		M^{(321)}_{ij} = 6 \int_q  F_3(-\vec{q},-\vec{k}_2+\vec{q},-\vec{k}_1) F_2(\vec{q},\vec{k}_2-\vec{q}) f_{i}(q^2) f_{j}(|\vec{k}_2 - \vec{q}|^2)\,,
	\end{equation}
	where $M^{(321)}$ is evaluated again with our $J$ function. 
	The full $\bar{B}^I_{321}$ diagram is,
	\begin{equation}
		\bar{B}^I_{321} = P_{\rm lin}(k_1) \left(\vec{\alpha}^T \cdot M^{(321)} \cdot \vec{\alpha}\right)  + 5\;{\rm perms} \,.
	\end{equation}
	where $\vec{\alpha}$ are our fitting coefficients of $P_{\rm lin}$. 
	The comparison the the numerical integration can be seen in Fig.~\ref{fig:BAll}.
	
\begin{figure}[H]
		\centering
		\includegraphics[width=\textwidth]{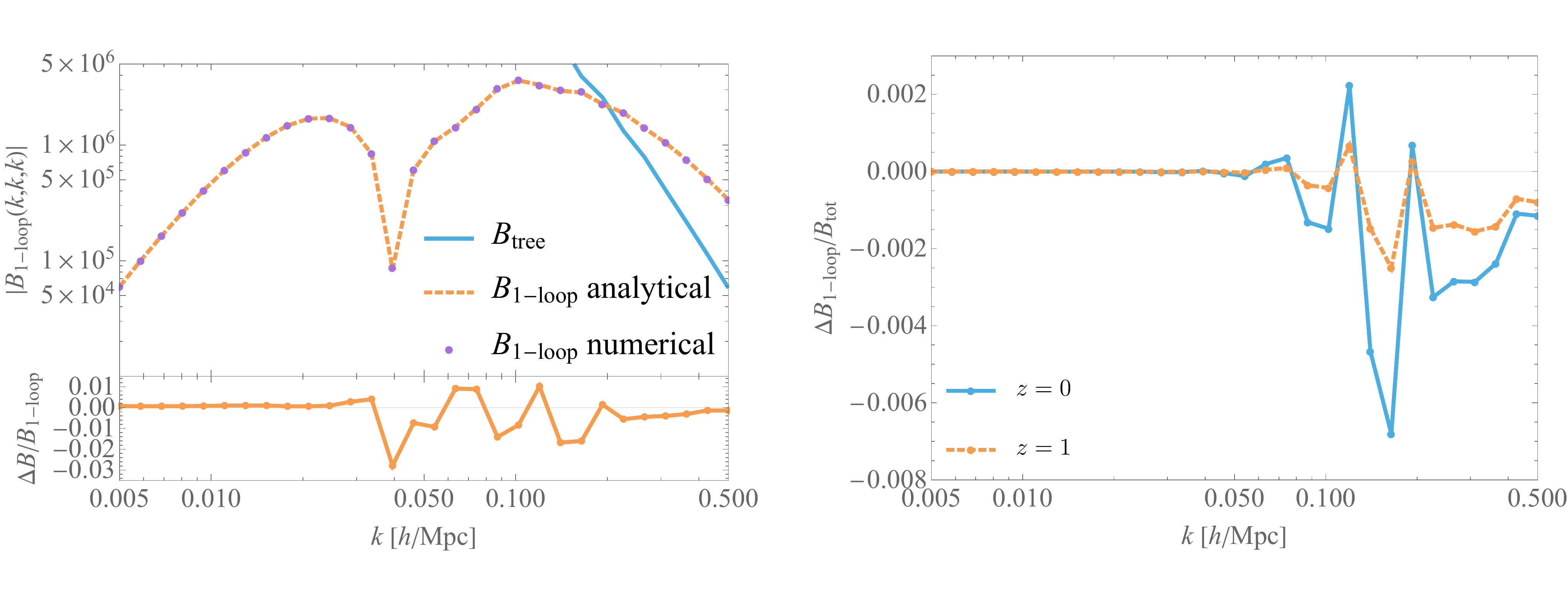}
		\caption{\label{fig:B1loop}Left: Comparison between analytical result $\bar{B}_{\rm 1-loop}$ and exact numerical result $B_{\rm 1-loop}$, obtained by summing the contributions in Fig.~\ref{fig:BAll}. 
			Right: 1-loop error relative to the full bispectrum including the linear contribution at redshifts $z=0$ and $z=1$.  }
	\end{figure}

	$\bar{B}^{II}_{321}$ is just calculated using the formalism for $\bar{P}_{13}$.
	Finally, $\bar{B}_{411}$ is calculated using the vector $M^{(411)}$ defined by
	\begin{equation}
		M^{(411)}_i = 12\int_{\vec{q}} F_4(\vec{q},-\vec{q},-\vec{k}_1,-\vec{k}_2) f_{i}(q^2)\,,
	\end{equation}
	and then doing the dot product
	\begin{equation}
		\bar{B}_{411} = P_{\rm lin}(k_1) P_{\rm lin}(k_2) ( M^{411}\cdot \vec{\alpha}) + 2\;{\rm cyclic\; perms}\,,
	\end{equation}
	where again $M^{(411)}$ is evaluated using our $J$ function.
	For both $\bar{B}^{II}_{321}$ and $\bar{B}_{411}$ we performed the UV-correction outline in Sec.~\ref{sec:P1loop} in order to more accurately compare with the numerical integration.
	The comparison with the numerical integration can be seen in Fig.~\ref{fig:BAll}.
	Summing all the contributions, we can obtain the full 1-loop bispectrum and compare it with the numerical integration. 
	Notice that, as shown in the right plot of Fig.~\ref{fig:B1loop}, we obtain sub-percent agreement on $B_{\rm tot} = B_{\rm tree} + B_{1-\rm loop}$ up to $k_{\rm max} = 0.5$ and the relative error is further suppressed by $D(z)^2$ for higher redshift. 
	Since we are not adding the counterterms, which tend to make the loops smaller, this is most probably an overestimate of the error.
	
	\paragraph{Calculation for scalene triangles.}
	
\begin{figure}[H]
		\centering
		\includegraphics[width=\textwidth]{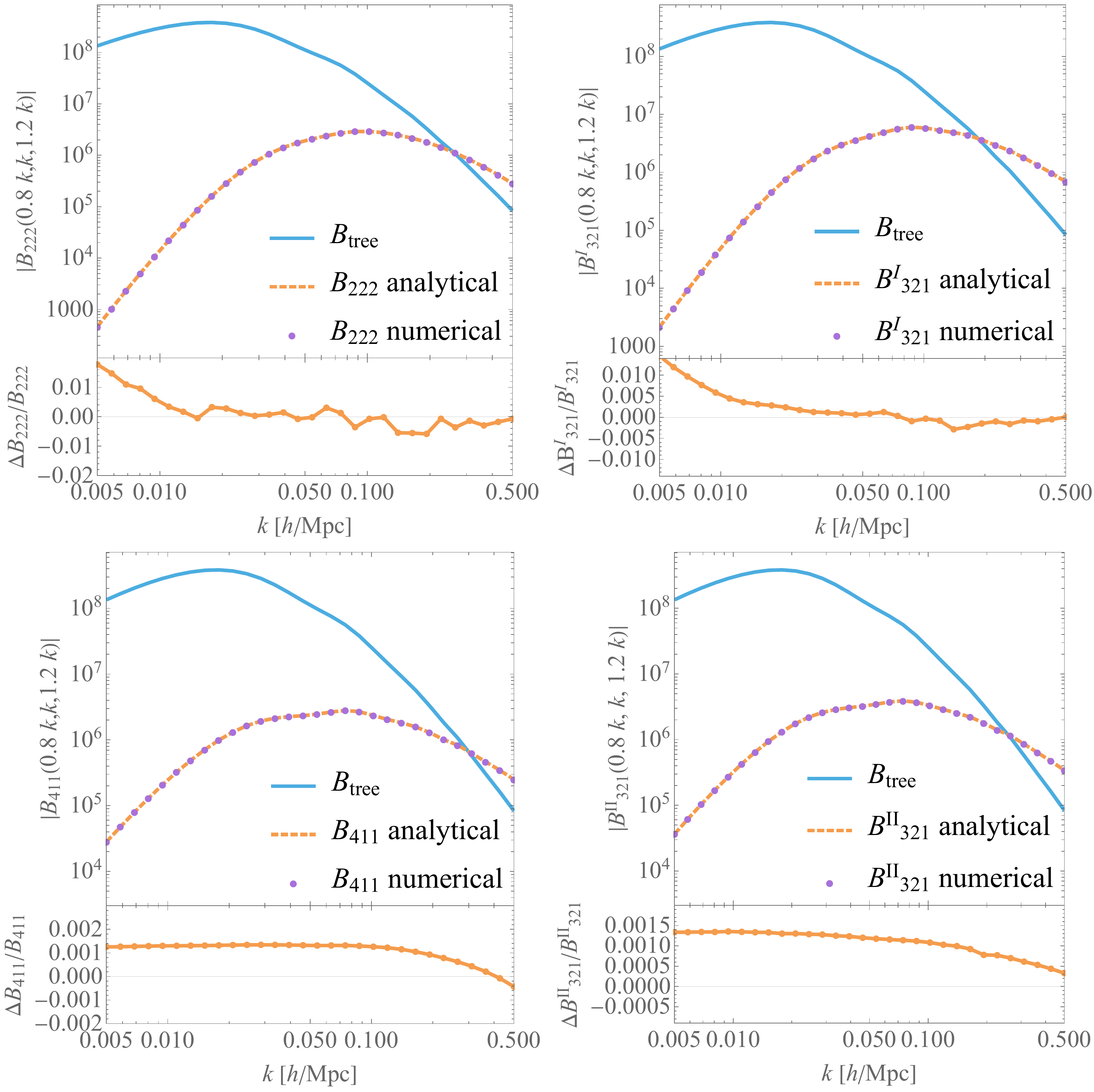}
		\caption{\label{fig:BAllScalene} Comparison of the analytical result with the numerical integration for each one of the 4 diagrams contributing to the 1-loop bispectrum. 
			This time we use scalene triangles with sides $0.8k$, $k$, and $1.2k$.}
	\end{figure}

\begin{figure}[H]
		\centering
		\includegraphics[width=\textwidth]{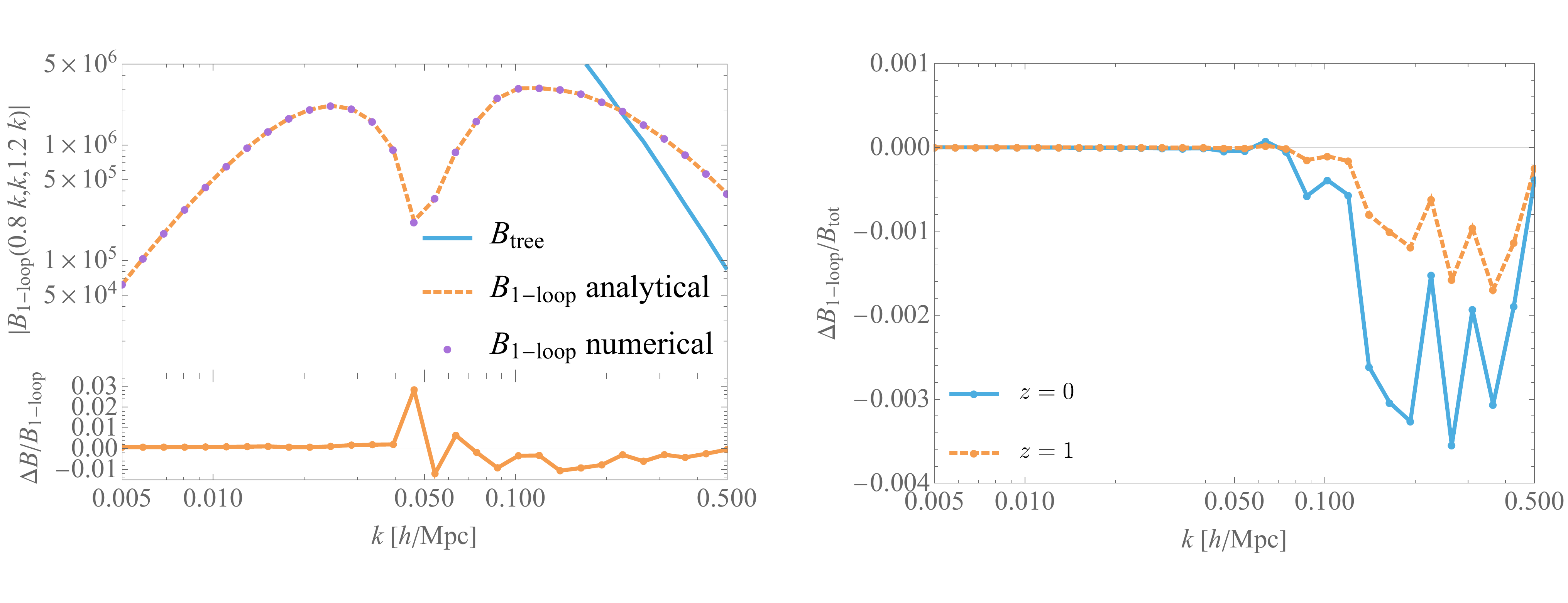}
		\caption{\label{fig:B1loopScalene}Left: Comparison between analytical result $\bar{B}_{\rm 1-loop}$ and exact numerical result $B_{\rm 1-loop}$, obtained by summing the contributions in Fig.~\ref{fig:BAllScalene}. 
			Right: 1-loop error relative to the full bispectrum including the linear contribution at redshifts $z=0$ and $z=1$. }
	\end{figure}

	We have so far calculated the diagrams for equilateral triangles. 
	We also compare for general scalene triangle, and verify that a similar precision is achieved, as can be seen in Figs.~\ref{fig:BAllScalene} and \ref{fig:B1loopScalene}.
	
	\subsection{Redshift space matter one-loop bispectrum}
	\label{sec:bispec_reds}
	We provide a brief description of the galaxy bispectrum in redshift space at one-loop order. 
	The equations for the kernels are derived and provided in~\cite{DAmico:2022ukl}. 
	We recommend the reader to cite that paper for the original derivation.
	In redshift space, we must make the following change of coordinates,
	\begin{align}
		\vec{x}_r = \vec{x} + \frac{\hat{z}\cdot \vec{v}}{aH}\hat{z} \,,
	\end{align}
	where $\vec{x}$ is the real space position vector, $\hat{z}$ is the direction of the line of sight, $\vec{v}$ is the peculiar velocity, $a$ is the scale factor, and $H$ is the Hubble-Lema\^{i}tre parameter. Under this change of coordinates, the halo density field is,
	\begin{align}
		1 + \delta_{r}(\vec{x}_r) =  (1 + \delta(\vec{x}))\left| \frac{\partial \vec{x}_r}{\partial \vec{x}}\right|^{-1}\ .
	\end{align}
	Thus, in Fourier space, this becomes,
	\begin{align}
		\delta_{r}(\vec{k}) = \delta(\vec{k}) + \int d^3x e^{-i\vec{k}\cdot\vec{x}}\left( {\rm exp} \left[ -i\frac{k_z}{aH}v_{z}(\vec{x}) \right] -1
		\right)\left( 1 + \delta(\vec{x}) \right),
	\end{align}
	where $k_z$ is the wave-number along the line of sight and $v_z$ is the peculiar velocity along the line of sight.
	The redshift-space field $\delta^{(n)}_{r}$ can then be expressed in terms of integrals of products of $\delta^{(1)}$ and the redshift-space kernel $K^{(n)}_{r}$,
	\begin{align}
		\delta^{(n)}_{r} = \int\prod^n_i d^3q_iK^{(n)}_{r}(\vec{q}_1, \dots, \vec{q}_n)\delta^3_D(\vec{k}-\sum^n_i\vec{q}_i)\prod^n_i\delta^{(1)}(\vec{q}_i).
	\end{align}
	The full expressions for the redshift-space kernels can be found in~\cite{DAmico:2022ukl}. 
	
\begin{figure}[H]
		\centering
		\includegraphics[width=\textwidth]{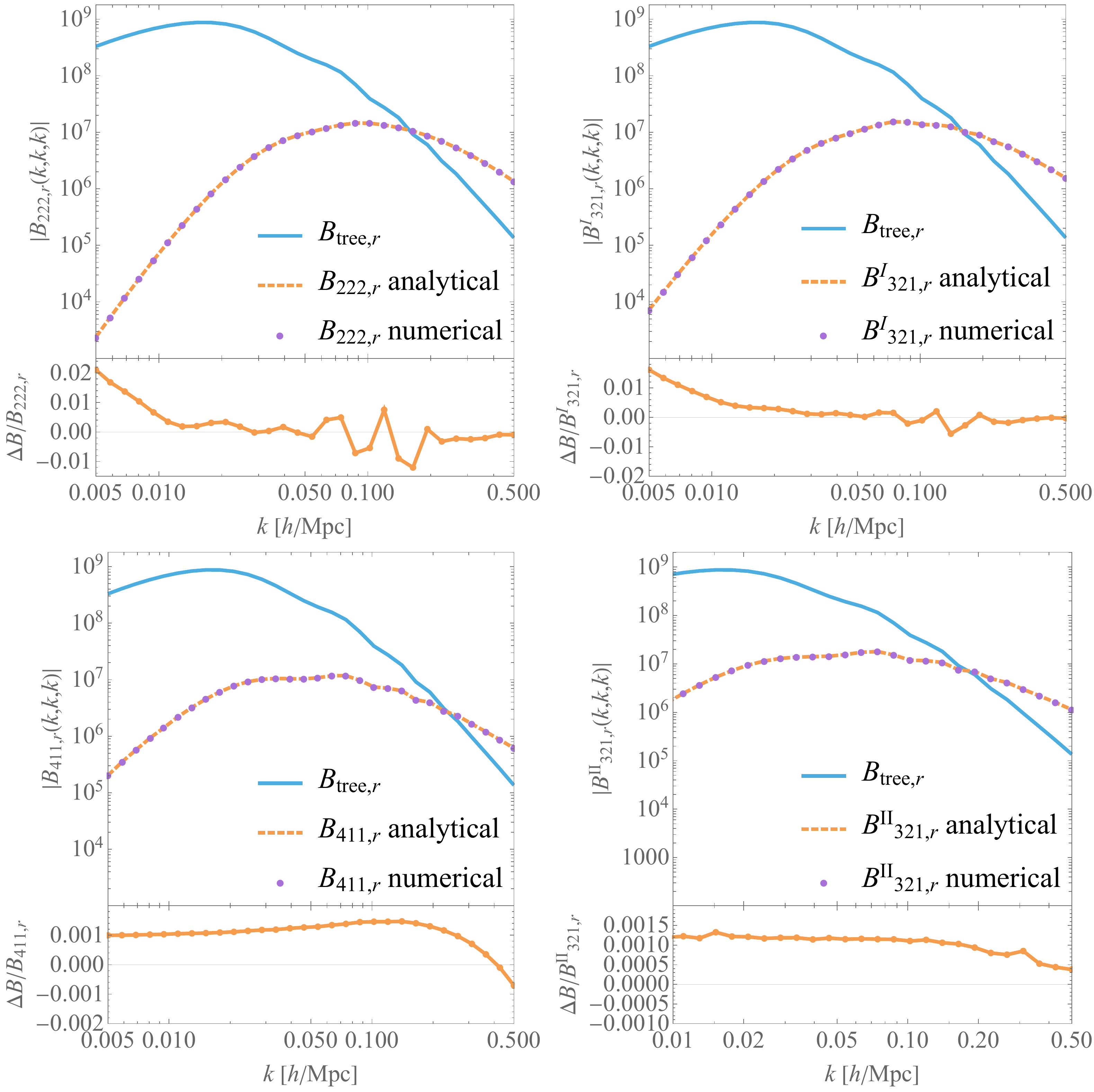}
		\caption{\label{fig:BAllred} Comparison of the analytical result with the numerical integration for each one of the 4 diagrams contributing to the 1-loop bispectrum.}
	\end{figure}

\begin{figure}[H]
		\centering
		\includegraphics[width=\textwidth]{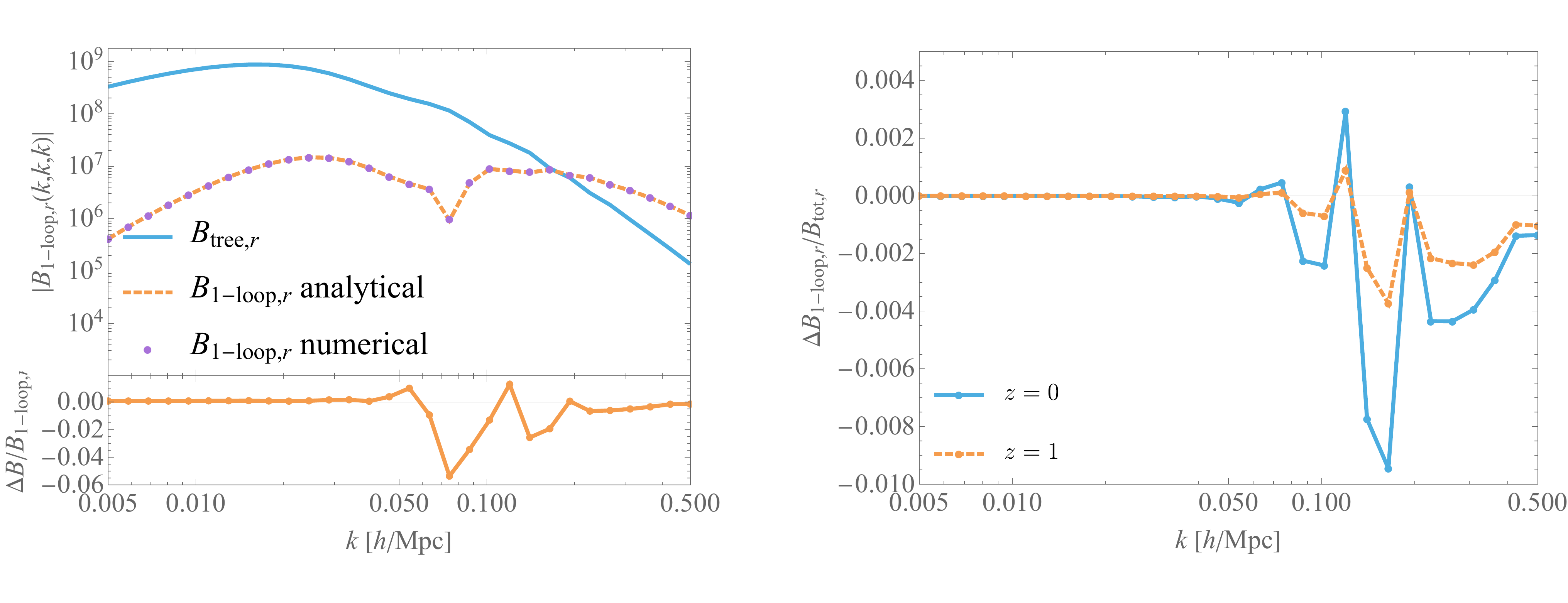}
		
		\caption{\label{fig:B1loopred} Left: 
			Comparison of $\bar{B}_{\rm 1-loop, r}$ using $P_{\rm fit}$ evaluated with our code ($B_{\rm
				1-loop, r}$ analytical) and numerical integration using exact $P_{\rm lin}$
			($B_{\rm 1-loop, r}$ numerical), both obtained by summing the contributions in Fig.~\ref{fig:BAllred}. 
			Right: 1-loop error relative to the tree level redshift space
			bispectrum including the linear contribution at redshifts $z=0$ and $z=1$, {whose ratio is just the ratio of the growth factors at the two redshifts squared}. 	
		} 
	\end{figure}

	The contributions to the one-loop bispectrum in redshift-space in terms of the redshift space kernels are given by,
	\begin{align}
		B_{222,r}(\vec{k}_1,\vec{k}_2,\vec{k}_3) &= \int_q K^{(2)}_{r}(-\vec{q},\vec{q}+\vec{k}_1)K^{(2)}_{r}(\vec{q}+\vec{k}_1,k_2-\vec{q})K^{(2)}_{r}(\vec{k}_2-\vec{q},\vec{q}) \\
		& \qquad \times P_{\rm lin}(q)P_{\rm lin}(|\vec{k}_1-\vec{q}|)P_{\rm lin}(|\vec{k}_2+\vec{q}|)\nonumber\ , \\
		B^I_{321,r}(\vec{k}_1,\vec{k}_2,\vec{k}_3) &= K^{(1)}_{r}(\vec{k}_1) P_{\rm lin}(k_1) 
		\int_q K^{(2)}_{r}(\vec{q},\vec{k}_2-\vec{q})K^{(3)}_{r}(-\vec{q},\vec{k}_2-\vec{q},-\vec{k}_1)  \\
		& \qquad \times P_{\rm lin}(q)P_{\rm lin}(|\vec{k}_2 + \vec{q}|) +\, 5\,\textrm{perms} \nonumber\ , \\
		B^{II}_{321,r}(\vec{k}_1,\vec{k}_2,\vec{k}_3) &= K^{(1)}_{r}(\vec{k}_2)K^{(2)}_{r}(\vec{k}_1,\vec{k}_2)P_{\rm lin}(k_1)P_{\rm lin}(k_2)\\
		& \qquad \times \int_q K^{(3)}_{r}(\vec{k}_1,\vec{q},-\vec{q})P_{\rm lin}(\vec{q}) + \,5\,\textrm{perms}\nonumber\ , \\
		B_{411,r}(\vec{k}_1,\vec{k}_2,\vec{k}_3) &= P_{\rm lin}(k_1)P_{\rm lin}(k_2) K^{(1)}_{r}(\vec{k}_1)K^{(1)}_{r}(\vec{k}_2)
		\int_q K^{(4)}_{r}(\vec{q},-\vec{q},-\vec{k}_2,\vec{k}_1)P_{\rm lin}(q) \nonumber \\
		+ \,2\,\textrm{cyclic perms}.
	\end{align}
	The redshift space bispectrum depends on the angles of $\vec{k}_i$ relative to $\hat{z}$. 
	It can be decomposed into a sum of multipoles. 
	In this section, we only consider the monopole, which is obtained by simply averaging over the angles.
	We proceed as in the real space case, defining first the corresponding approximations $\bar{B}_{\rm diagram}$. 
	Each kernel can be decomposed in terms dependent on $q^2$, $(\vec{k}_1-\vec{q})^{2}$, and $(\vec{k}_2 + \vec{q})^{2}$ as well as line of sight terms which depend on $\hat{\vec{k}}_1\cdot \hat{\vec{z}}$, $\hat{\vec{k}}_2\cdot \hat{\vec{z}}$, $\hat{\vec{k}}_3\cdot \hat{\vec{z}}$, and $\hat{\vec{q}}\cdot \hat{\vec{z}}$. 
	The integration of the non-rotationally invariant parts associated to the projections along the line of sight will be handled by reducing them to linear combinations of rotationally invariant integrands multiplied by suitable, non-rotationally invariant tensors, similar to the approach in Sec.~\ref{sec:tensor} (see also~\cite{DAmico:2022ukl}). 
	At that point, the ingredient needed to complete the integration is still just the function $J_{ijk}$ defined in Eq.~\eqref{eq:Jfunc}. 
	We use it to construct the new matrices $M^{(222),r}$ (corresponding to $\bar{B}_{222,r}^I$), $M^{(3211),r}$ (corresponding to $\bar{B}_{321,r}^I$), $M^{(3212),r}$ (corresponding to $\bar{B}_{321,r}^{II}$), and $M^{(411),r}$ (corresponding to $\bar{B}_{411,r}^I$).
	Then, comparing with the numerical integration, we obtain again a very good agreement, as shown in Fig.~\ref{fig:BAllred}. 
	For the full one-loop bispectrum in redshift space, the result is shown in Fig.~\ref{fig:B1loopred}. 
	As usual, since we are not adding the counterterms, this is probably an overestimate of errors.
	
	\paragraph{UV-subtraction of the $q^0$ part of the kernel.} 
	We are presenting results for dark matter, but our procedure holds unaltered also for biased tracers. 
	In the case of dark matter, the UV behavior of the kernels is smaller or equal than $1/q^2$, for $q\to\infty$. 
	Combined with the fact that in our basis $P_{\rm fit}(q)$ goes like $1/q^2$ as $q\to\infty$, this means that all integrals are convergent, making the comparison with numerical integration straightforward.
	However, for biased tracers, the kernels go as $q^0$ in the same limit, and it is useful to perform the following procedure in order to compare with numerical results.
	Using $P_{\rm fit}(q)$ within the loop integrals, one subtracts from the redshift space kernels of $B^{II}_{321, r}$ and $B_{411, r}$ terms that scale as $q^0$.
	In fact such terms will introduce UV divergences in the integral, since these two diagrams contain only one $P_{\rm fit}(q)$. 
	In our analytical integration procedure, these $q^0$ terms will give rise to potentially-large finite parts proportional to the mass squared $M_i$'s. 
	While these contributions are degenerate with counterterms, they make the comparison with the numerical results less straightforward.
	We can define 
	\begin{align}
		K^{(3), \rm UV}_{r} &= \int d\Omega_2\lim_{|q|\rightarrow \infty} K^{(3)}_{r} \,,\\
		K^{(4), \rm UV}_{r} &= \int d\Omega_2\lim_{|q|\rightarrow \infty}K^{(4)}_{r}\,,\\
		K^{(3), \rm UVsub}_{r} &= K^{(3)}_{r} - K^{(3), \rm UV}_{r}\,,\\
		K^{(4), \rm UVsub}_{r} &= K^{(4)}_{r} - K^{(4), \rm UV}_{r}\,,
	\end{align}
	and we will use $K^{(3), \rm UVsub}_{r}$ and $K^{(4), \rm UVsub}_{r}$
	inside the one-loop integrals of $B^{II}_{321,r}$ and $B_{411,r}$.
	As mentioned above, such procedure is useful to accurately compare with the numerical integration, but when considering the full EFTofLSS, the justification for this comes from the fact that some counterterms are degenerate with these UV subtractions.
	
	{Additional checks, where we compare our {analytical} integration directly with numerical integration of $P_{\rm fit}$ (so that, with infinite numerical precision, the results should agree), are presented in App.~\ref{app:syst}.} 
	These results validate this formalism to quickly calculate loop integrals in the EFTofLSS. In the remaining part of the paper, we will present a detailed calculation of the function $L$ introduced in Eq.~\eqref{eq:Lfunc}.

	\section{$L$-function evaluation}
	\label{sec:l_func}
	
	With our power spectrum decomposition given in Eq.~\eqref{eq:Pfit1}, we remind readers that the evaluation of the 1-loop bispectrum involves integrals of the type shown in Eq.~\eqref{eq:Lfunc}. 
	For clarity, we rewrite the expression here with some arguments dropped, which is a notation that we will use in this section.
	\begin{align}
		\label{eq:Lgen}
		L(n_1,d_1,n_2,d_2,n_3,d_3) = \int_q\frac{(\vec{k}_1-\vec{q})^{2n_1}\vec{q}^{2n_2}(\vec{k}_2+\vec{q})^{2n_3}}{((\vec{k}_1-\vec{q})^2+M_1)^{d_1}(\vec{q}^2+M_2)^{d_2}((\vec{k}_2+\vec{q})^2+M_3)^{d_3}}
	\end{align}
	where $n_1,n_2,n_3$ can be positive or negative integers and $d_1,d_2,d_3 \geq 0$. 
	We call the expression in Eq.~\eqref{eq:Lgen} the general triangle integral named after the shape of the corresponding Feynman diagram (see Fig.~\ref{fig:bispec_diagram}).
	The procedure for calculating a given $L$ will be to perform several recursion steps to reduce the powers of $n_i$ and $d_i$.
	
\begin{figure}
		\centering
		\includegraphics[scale = 0.5]{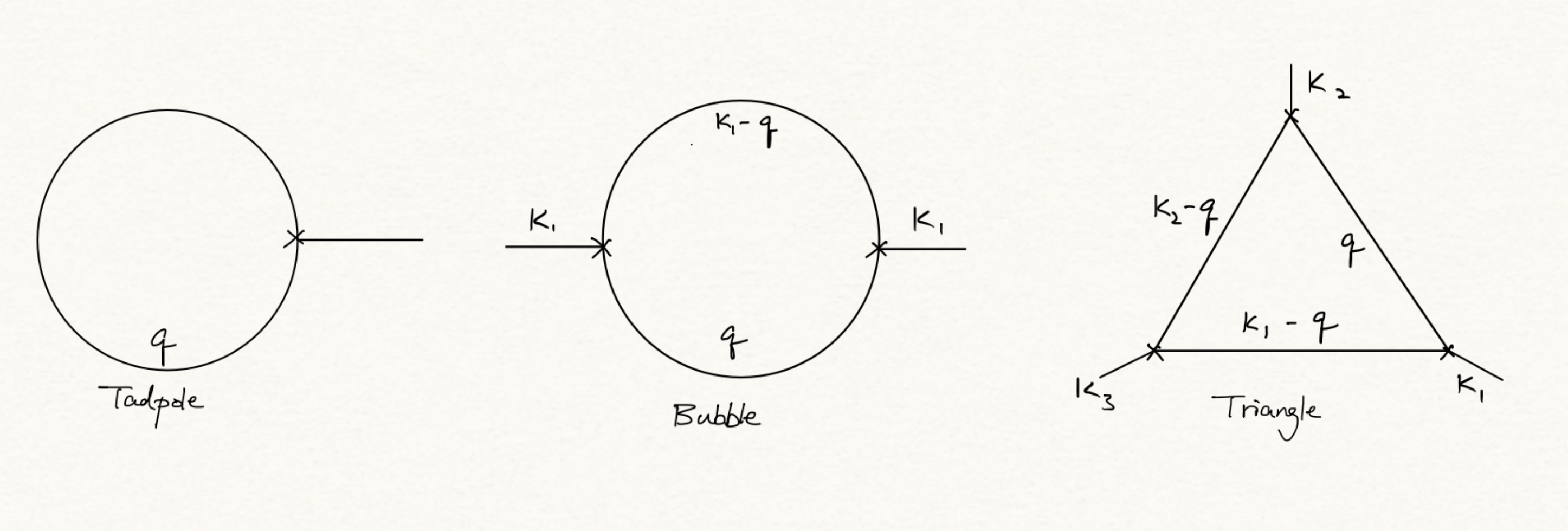}
		\caption{Tadpole, bubble, and triangle master integrals represented by Feynman diagrams.		}
		\label{fig:feynman}
	\end{figure}

	The recursions eventually terminate resulting in $L$ being a sum of master integrals, which we call Tadpole, Bubble, and Triangle master integrals, given by: 
	\begin{enumerate}
		\item Tadpole: 
		\begin{align}
			\label{eq:tadpolefamily}
			\text{Tad}(M_j, {n}, d) = \int \frac{d^3\vec{q}}{\pi^{3/2}}\frac{{(\vec{p}_i^2)^n}}{(\vec{p}^2_i + M_j)^{d}}  \,,
		\end{align}
		where $\vec{p_i} = \{\vec{k}_1 - \vec{q}, \vec{q}, \vec{k}_2 + \vec{q}\}$ and $M_j = \{M_1, M_2, M_3\}$. 
		
		\item Bubble:
		\begin{align}
			\label{eq:bubmaster}
			B_{\rm master}(k^2,M_1,M_2) =  \int \frac{d^3\vec{q}}{\pi^{3/2}}\frac{1}{(q^2 + M_1)( |\vec{k}-\vec{q}|^2 + M_2)} \,.
		\end{align}
		
		\item Triangle:
		\begin{align}
			\begin{split}
				T_{\rm master}(k_1^2, k_2^2, k_3^2, M_1,& M_2, M_3) = \\
				&\int \frac{d^3\vec{q}}{\pi^{3/2}}\frac{1}{(q^2 + M_1)( |\vec{k}_1-\vec{q}|^2 + M_2)( |\vec{k}_2+\vec{q}|^2 + M_3)},
			\end{split}
		\end{align}
		where $\vec{k}_1 + \vec{k}_2 + \vec{k}_3 = 0$.
	\end{enumerate}
	These master integrals are evaluated in closed form, as explained in~Sec.~\ref{sec:master}.  
	The name of the master integrals comes from the number of propagators of the associated Feynman diagram as shown in Fig.~\ref{fig:feynman}.
	
	Formally, the family of Tadpole master integrals of Eq.~(\ref{eq:tadpolefamily}) is reducible with recurrence identities to just one of them, $\text{Tad}(M_j, 0, 1)$. 
	However, as we will shortly remind, all one-loop tadpoles admit a simple analytic expression in terms of Gamma functions and a reduction for them is, in practice, unnecessary. 
	
	While the integrals that we actually evaluate analytically are the master integrals defined just above, we point out that, in intermediate steps, we may find that one of the $n_i$ and $d_i$ are 0 or two of the $n_i$ and $d_i$ are 0. 
	We call these intermediate integrals the general bubble integrals, also named after the shape of the corresponding Feynman diagrams. 
	Note that the general integrals are different from the master integrals.
	The general bubble integral has the form,
	\begin{align}
		L_B(n_1, d_1, n_2, d_2) = \int_q \frac{\vec{p}^{2n_1}_i\vec{p}^{2n_2}_j}{(\vec{p}_i^2 + M_i)^{d_1}(\vec{p}_j^2 + M_j)^{d_2}},
		\label{eq:bubblegen}
	\end{align}
	where $\vec{p_i} = \{\vec{k_1} - \vec{q}, \vec{q}, \vec{k}_2 + \vec{q} \}$ and $i \neq j$. 	
	These general integrals will be also themselves then further reduced to master integrals as defined above, as will later be explained.
	
	{		The calculations of all master integrals and reduction coefficients will be carried out setting the number of dimensions to $D=3$. The sum of $L$-functions of Eq.~\eqref{eq:Lgen} in the expressions for the power spectrum and the bispectrum are finite and numerically integrable in $D=3$ dimensions. Paradoxically, we will be still using dimensional regularization subtly, as it is common in QFT integral computations, and thinking of the dimension  as a regulator of ultraviolet divergences, $D=3-2\epsilon$. While the limit of $\epsilon \to 0$ is smooth for the sum of contributions to our physical results,  we will need to analytically continue $\epsilon$ to values for which every integral emerging in intermediate expressions should also be convergent. The effect of this analytic continuation will be that infinite parts in $D=3$  get dropped within dimensional regularization, {but this does not affect the final result}.
		
		As a {concrete} example, we will compute here the tadpole master integrals, which we have announced above to appear in our final results, as functions of both the dimension $D$ and an ultraviolet cutoff $\Lambda$.
		We define,
		\begin{equation}
			\text{Tad}_\Lambda(M, {n}, d) \equiv \int \frac{d^D\vec q}{\pi^{\frac D 2} } \frac{\left( \vec q^2 \right)^n}{ \left( \vec q^2 +M \right)^d}
			\Theta\left( \vec q^2 < \Lambda \right)
		\end{equation}
		which, by using spherical coordinates and rescaling the integration variable $| \vec q | = x \sqrt{\Lambda}$, we recognize to be, up to factors, the integral representation of a hypergeometric function,
			{
			\begin{eqnarray}
				\text{Tad}_\Lambda(M, {n}, d) &=& \frac{{ 2 }\Lambda^{\frac D 2 +n} M^{-d}}{\Gamma\left( \frac D 2 \right)} \int_0^1 dx  \frac{x^{D  +2 n -1}}{
					{ \left(1+ \frac{\Lambda}{M} x^2\right)^d}} \nonumber \\
				&=&
				\frac{{ \Lambda^{\frac D 2 +n} M^{-d}}
					\Gamma\left( \frac D 2 +n\right)
				}{\Gamma\left( \frac D 2 \right) \Gamma\left( \frac D 2+n+1 \right)}                           {}_2F_1\left(d, \frac D 2+n, \frac D 2 +n +1, -\frac \Lambda M \right) \,.
			\end{eqnarray}
		}
		As we are interested in the behaviour at large values of the cutoff $\Lambda$, we transform~\footnote{
			The analytic continuation of the hypergeometric function is given by the identity
			\begin{eqnarray}
				{}_2F_1\left(a,b,c, x \right) &=&
				(-x)^{-a} \,        \frac{\Gamma(c) \Gamma(b-a)}{\Gamma(b) \Gamma(c-a)}                {}_2F_1\left(a,1+a-c,1+a-b, 1/x \right)
				\nonumber \\
				&&+                                      (-x)^{-b} \,        \frac{\Gamma(c) \Gamma(a-b)}{\Gamma(a) \Gamma(c-b)}                {}_2F_1\left(b,1+b-c,1+b-a, 1/x \right).
			\end{eqnarray}
			It is also useful to recall that $ {}_2F_1\left(a,0,c, x \right)={}_2F_1\left(0,b,c, x \right)=1$.
		} the hypergeometric function to an equivalent form appropriate for an expansion in $M/\Lambda$. We find 
		\begin{eqnarray}
			\label{eq:TadpoleCutoff}
			&& \text{Tad}_\Lambda(M, {n}, d)=                                                                                                                      M^{\frac D 2 +n - d}           \frac {\Gamma\left( \frac D 2 + n\right)}{\Gamma\left( \frac D 2\right) }\,                                                                                                                            \frac{
				\Gamma\left( d-n- \frac D 2 \right)                                                                                                                            }{
				\Gamma\left( d \right)                                                                                                                    }
			\nonumber \\
			&& \hspace{0cm}
			+ \Lambda^{\frac D 2 +n - d}
			\frac {\Gamma\left( \frac D 2 + n\right)}{\Gamma\left( \frac D 2\right) }     \frac{
				\Gamma\left(\frac D 2+n -d \right)
			}
			{
				\Gamma\left(\frac D 2+n \right) \Gamma\left(1+\frac D 2+n -d \right)
			}
			\;     {}_2F_1\left(d, d-n - \frac D 2, 1+d-n - \frac D 2, -\frac{M}{\Lambda}  \right)
			\nonumber \\ &&
		\end{eqnarray}  
		where the hypergeometric function behaves at the $\Lambda \to \infty $ limit as
		\begin{equation}
			{}_2F_1\left(d, d-n - \frac D 2, 1+d-n - \frac D 2, -\frac{M}{\Lambda}  \right) =  1  + {\cal O}\left( \frac{M}{\Lambda}\right).  
		\end{equation}
		
		As can be seen from the factor multiplying the hypergeometric function in Eq.~(\ref{eq:TadpoleCutoff}),  the $\Lambda \to \infty$ limit converges only for $\frac{D}{2} + n < d$.   By analytic continuation, within dimensional regularization, and anticipating that in the sum of integrals in physical results the {cutoff} dependent terms cancel, they are set to zero.    
		We will thus be using the dimensional regularization expression
		\begin{eqnarray}
			\label{eq:TadpoleDREG}
			\text{Tad}(M, {n}, d)
			&=&
			\left. \lim_{\Lambda \to \infty}
			\text{Tad}_\Lambda(M, {n}, d)  \right|_{\rm dim. \, reg.}
			\nonumber \\
			&=&
			M^{\frac D 2 +n - d}   \,
			\frac {\Gamma\left( \frac D 2 + n\right)}{\Gamma\left( \frac D 2\right) }   \, 
			\frac{
				\Gamma\left( d-n- \frac D 2 \right)                                                                                                                            }{
				\Gamma\left( d \right)                                                                                                                    } \,,
		\end{eqnarray}  
		Importantly, the part of the integral which is accounted for in dimensional regularization in the last equality of Eq.~(\ref{eq:TadpoleDREG}) has a smooth limit as $\epsilon \to 0$ and we can set $D=3$ directly in our expressions.~\footnote{This feature is specific to one-loop, as a two-loop tadpole integral has an $1/\epsilon$ poles and the $\epsilon$ regulator cannot be taken to zero directly from the start.}  
	}
	
	\subsection{$L$-function calculation flowchart} 
	In this section we outline the procedure used to compute the function $L$ in Eq.~\eqref{eq:Lgen}.  
	\begin{figure}
		\centering
		\begin{tikzpicture}[node distance = 2cm] 
			\tikzstyle{startstop} = [rectangle, rounded corners, minimum width=3cm, minimum height=1cm,text centered, draw=black, fill=red!30]
			\tikzstyle{io} = [trapezium, trapezium left angle=70, trapezium right angle=110, minimum width=3cm, minimum height=1cm, text centered, draw=black, fill=blue!30]
			\tikzstyle{io2} = [trapezium, trapezium left angle=70, trapezium right angle=110, minimum width=2cm, minimum height=1cm, text centered, draw=black, fill=blue!30]
			\tikzstyle{process} = [rectangle, minimum width=3cm, minimum height=1cm, text centered, draw=black, fill=orange!30]
			\tikzstyle{decision} = [diamond, minimum width=3cm, minimum height=1cm, text centered, draw=black, fill=green!30]
			\tikzstyle{arrow} = [thick,->,>=stealth]
			\centering
			\node (start) [startstop] {Input: $n_i, d_i, M_j, k_i$ for $i = 1\dots 3$, $j = 1\dots 4$};
			\node (in0) [io, below of=start] {$L(n_1,d_1,n_2,d_2,n_3,d_3)$};
			\node (in1) [io, below of=in0] {$T(d_1,d_2,d_3)$};
			\node (in3) [io2, right of=in1, xshift = 7 cm] {$\textrm{Tensor reduction}$};
			\node (pro1) [process, below of=in1, yshift=-2cm] {$T_{\rm master}(M_j,k_i)$};
			\node (in2) [io, right of=pro1, xshift = 4cm] {$B(d_1,d_2)$};
			\node (pro2) [process, below of=pro1, xshift = 3cm] {$B_{\rm master}(M_j,k_i)$};
			\node (pro3) [process, right of=pro2, xshift = 4cm] {$\textrm{Tad}(d_1,M_j,k_i)$};
			\draw[arrow] (start) --  (in0);
			\draw[arrow] (in0) -- node[anchor = west] {$n_1 = n_2 = n_3 = 0$} (in1);
			\draw[arrow] (in1) -- node[anchor = west, yshift = -0.5cm] {$d_1=d_2=d_3=1$} (pro1);
			\draw[arrow] (in1) -- node[anchor = west, xshift = 0.2cm] {$\exists i: d_i = 0$} (in2);
			\draw[arrow] (in1) -- node[anchor = west, xshift=-0.7cm, yshift = -0.5cm] {$\exists i: d_i < 0$} (in3);
			\draw[arrow] (in3) -- node[anchor = north, xshift = 1cm, yshift = 0.2cm] {} (in2);
			\draw[arrow] (in2) -- node[anchor = west, xshift=0.4cm, yshift=-0.2cm]{$\exists i: d_i = 0$} (pro3);
			\draw[arrow] (in2) -- node[anchor = west] {$d_1= d_2 = 1$} (pro2);
			\draw [arrow] (in0.east)arc(-160:160:1)node[anchor=west, xshift = 2cm] {\small $L$-recursion};
			\draw [arrow] (in1.west)arc(20:340:1)node[anchor=south, xshift = -0.4cm, yshift = -1.3cm] {\small $T$-recursion};
			\draw [arrow] (in2.east)arc(-160:160:1)node[anchor=west, xshift = 1.3cm, yshift=0.8cm] {\small $B$-recursion};
		\end{tikzpicture}
		\caption{\label{fig:code_flowchart}Procedural flowchart of calculation used to evaluate $L$ defined in Eq.~\eqref{eq:Lgen} in the one-loop bispectrum. 
			Blue textboxes represent intermediate integrals that require further reduction via recursion procedures of previous sections. 
			Yellow textboxes represent final integrals that are evaluated analytically.  
		} 
	\end{figure}
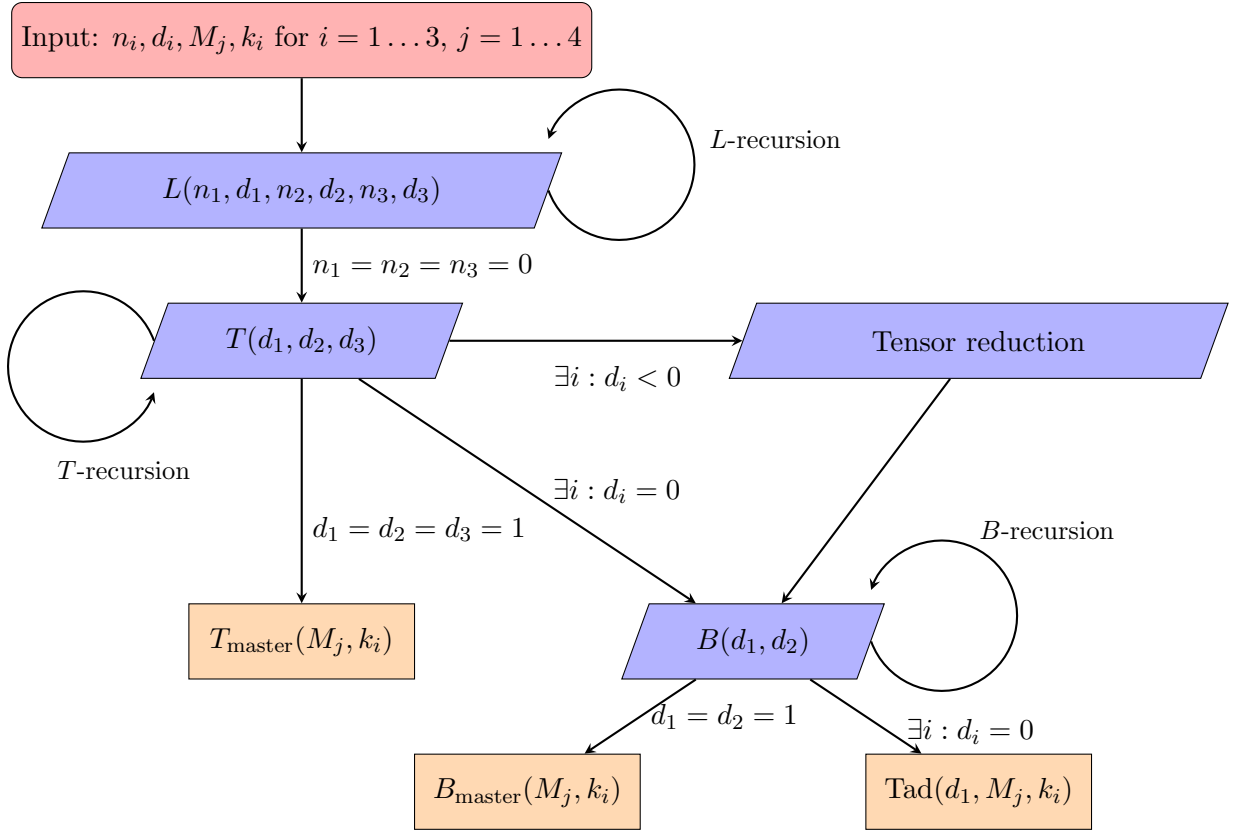
	
	For the bispectrum, the longest path of decomposition starts with a term in which all $n_i \neq 0, d_i \neq 0$, in which case we perform the triangle integral recursions discussed in Sec.~\ref{sec:L-recursion}. 
	At the end of this step of recursion, we are left with a reduced triangle integral we call $T(d_1,d_2,d_3)$, which has the form of~\eqref{eq:tri_reduc}. 
	We then proceed to further reduce $T(d_1,d_2,d_3)$ using the procedure outlined in Sec.~\ref{sec:T_reduc}, by the end of which, depending on the boundary values of $n_i$'s, we are either left with the final triangle master integral, $T_{\rm master}(M_j,k_i)$, or a bubble integral that requires further reduction. 
	Should a bubble term remain after the reduced triangle integral reduction, we perform the recursions outlined in Sec.~\ref{sec:bubrec}, where we will be left with either a bubble master integral, $B_{\rm master}(M_j,k_i)$, or a tadpole integral, $\textrm{Tad}(d_1,M_j,k_i)$. 
	In Fig.~\ref{fig:code_flowchart}, we summarize the code procedure pictorially.

	\subsection{Simplifying $L$ using recursion relations}
	\label{sec:trirec}
	
	\subsubsection{$L$-recursion}{\label{sec:L-recursion}}
	
	The first step in our calculation is to decompose $L$ into terms where all $n_i = 0$.
	To do so, as described in Fig.~\ref{fig:code_flowchart}, we use the so-called $L$-recursion.
	This recursion is different depending on whether $n_i > 0$ or $n_i < 0$. 
	Notice that at the starting point of the recursion, $d_i\geq0$.
	\begin{enumerate}
		\item Case $n_i > 0$:
		Let us outline the procedure to simplify $L$ when some $n_i$ is positive. Consider for example $n_1 > 0$. 
		First, we define
		\begin{align}
			I_+(n_1,d_1) = \frac{(\vec{k}_1-\vec{q})^{2n_1}}{((\vec{k}_1-\vec{q})^2+M_1)^{d_1}}\,.
		\end{align}
		Note that $I_+$ is a factor of the integrand in Eq.~\eqref{eq:Lgen}.
		We can reduce the numerator's exponent with the following simple manipulation,
		\begin{align}
			\label{eq:numrecursion}
			I_+(n_1,d_1) &= \frac{(\vec{k}_1-\vec{q})^{2(n_1-1)}((\vec{k}_1-\vec{q})^2+M_1-M_1)}{((\vec{k}_1-\vec{q})^2+M_1)^{d_1}}\\
			&=I_+(n_1 - 1,d_1-1) - M_1 \, I_+(n_1-1,d_1)\,.
		\end{align}
		In this case, the recursion terminates when $d_1 = 0$ or when $n_1 = 0$.
		We can use this simple relation to reduce the numerator exponents in the integrand of $L(n_1,d_1,n_2,d_2,n_3,d_3)$ if any $n_i$ is positive.
		In fact, if e.g. $n_1>0$, $L(n_1,d_1,n_2,d_2,n_3,d_3)$ can be expressed as,
		\begin{equation}
			\label{eq:Lrec1}
			\begin{split}
				&L(n_1 > 0, d_1, n_2, d_2, n_3, d_3) \\
				&= L(n_1-1,d_1-1,n_2,d_2,n_3,d_3) - M_1 \, L(n_1-1,d_1,n_2,d_2,n_3,d_3)\,.
			\end{split}
		\end{equation}
		
		\item Case $n_i < 0$:
		In this case, we use a slightly different procedure to simplify $L$.
		Consider for example $n_1 < 0$. 
		Similarly as the previous case, we can define
		\begin{align}
			I_{-}(n_1,d_1) = \frac{1}{(\vec{k}_1-\vec{q})^{-2n_1}((\vec{k}_1-\vec{q})^2+M_1)^{d_1}}\,,
		\end{align}
		where all factors in the denominator have positive exponents.
		We can reduce the absolute value of $n_1$ with the following manipulation,
		\begin{align}
			\begin{split}
				I_{-}(n_1,d_1) &= \frac{1}{M_1} \left(\frac{1}{(\vec{k}_1-\vec{q})^{-2n_1}((\vec{k}_1-\vec{q})^2+M_1)^{d_1-1}} \right.\\
				&\qquad \qquad \left.- \frac{1}{(\vec{k}_1-\vec{q})^{-2(n_1+1)}((\vec{k}_1-\vec{q})^2+M_1)^{d_1}}\right)\\
				&=\frac{1}{M_1}\left(I_{-}(n_1,d_1-1) - I_{-}(n_1+1,d_1)\right)\,,
				\label{I_func}
			\end{split}
		\end{align}
		where again the recursion terminates when $n_1 = 0$ or $d_1 = 0$ as each iteration through the recursion reduces the absolute powers of either $n_1$ or $d_1$ by 1. 
		Therefore in the case of $n_1 < 0$, the general triangle integral can be expressed as,
		\begin{equation}
			\label{eq:Lrec2}
			\begin{split}
				&L(n_1 < 0, n_2, n_3, d_1, d_2, d_3) \\
				&= \frac{1}{M_1}\left(L(n_1, n_2, n_3, d_1 -1, d_2, d_3) - L(n_1+1, n_2, n_3, d_1, d_2, d_3)\right)\,.
			\end{split}
		\end{equation}
	\end{enumerate}
	We can use Eqs.~\eqref{eq:Lrec1} and \eqref{eq:Lrec2} to decrease the absolute value of $n_i$ (or the corresponding $d_i$) for both $n_i>0$ and $n_i<0$. 
	
	There are four possibilities to end the recursion for a specific $i$:
	\begin{itemize}
		\item If $n_i=0$ but $d_i\geq0$, we can also redo this recursion for a different $i$ for which $n_i \neq 0$. 
		If there are no more $n_i \neq 0$ and all $d_i > 0$, we can continue simplifying using integration by parts, with what we call $T$-recursion, as described in Sec.~\ref{sec:T_reduc}. 
		In the case of all $n_1 = n_2 = n_3 = 0$ and one $d_i = 0$, then we proceed to $B$-recursion as described in Sec.~\ref{sec:bubrec}.  
		
		\item If $d_i = 0$ and $n_i>0$, we redefine $d_i = -n_i <0$ and use a tensor reduction method to simplify the expression, as detailed in Sec.~\ref{sec:tensor}. 
		\item If $d_i = 0$ and $n_i<0$, we are left with a simpler $L$ function with $M_i = 0$, $d_i=-n_i$, and $n_i=0$. We can then redo this recursion for a different $i$ provided that $n_i\neq 0$.
	\end{itemize}

	\subsubsection{$T$-recursion}
	\label{sec:T_reduc}
	In the previous subsection, we were able to reduce $L$ in Eq.~(\ref{eq:Lgen}) to a form where there are no $n_i$'s.
	If one has $d_i=0$ for some $i$, then we go to the $B$-recursion in Sec.~\ref{sec:bubrec}.
	In this section we outline the recursion relation used to reduce $L$ when all $n_i=0$ and all $d_i>0$.
	We define the following integral
	\begin{align}
		T(d_1,d_2,d_3) = \int_q\frac{1}{((\vec{k}_1-\vec{q})^2+M_1)^{d_1}(\vec{q}^2+M_2)^{d_2}((\vec{k}_2+\vec{q})^2+M_3)^{d_3}}\,,
		\label{eq:tri_reduc}
	\end{align}
	which is just $L$ with all $n_i = 0$.
	We can define 
	\begin{equation}
		t(d_1,d_2,d_3) \equiv \frac{1}{((\vec{k}_1-\vec{q})^2+M_1)^{d_1}(\vec{q}^2+M_2)^{d_2}((\vec{k}_2+\vec{q})^2+M_3)^{d_3}}\,,
	\end{equation}
	and from the divergence theorem, we have 
	\begin{align}
		\int_q\frac{\partial}{\partial q_{\mu}}\cdot \left( q_{\mu}t(d_1,d_2,d_3) \right) &=0\ ,\\
		\int_q\frac{\partial}{\partial q_{\mu}}\cdot \left( k_{1\mu}t(d_1,d_2,d_3) \right)&=0\ ,\\\
		\int_q\frac{\partial}{\partial q_{\mu}}\cdot \left( k_{2\mu}t(d_1,d_2,d_3)\right) &=0\ .   
	\end{align}
	{The above identities are derived in exactly three dimensions, assuming convergent behaviour of the integrals in the ultraviolet. 
		More generally, this is guaranteed for Integration By Parts~(IBP) identities within dimensional regularization~\cite{Tkachov:1981wb,Chetyrkin:1981qh}, which, as we have explained earlier, we subtly employ. In this {article}, all IBP identities that we will employ have a regular $\epsilon=0$ limit and we  can set the dimension to its physical value from the beginning. 
		The role of dimensional regularization is important for the reduction, furnishing a  terminating condition by setting non-convergent integrals in $D=3$ with no denominators to zero. 
		This can also be seen from the dimensional regularization expression of Eq.~(\ref{eq:TadpoleDREG}) which gives
		\begin{equation}
			\text{Tad}(M, {n}, 0)
			=   \text{Tad}(0, {n}, 0) =0.
		\end{equation}
	}      
	Noticing that 
	\begin{align}
		\frac{\partial}{\partial q_{\mu}}\cdot q_{\mu} &= 3 + q_{\mu}\frac{\partial}{\partial q_{\mu}}\ ,\\
		\frac{\partial}{\partial q_{\mu}}\cdot k_{i\mu} &= k_{i\mu} \cdot \frac{\partial}{\partial q_{\mu}}\,,
	\end{align}
	and calculating the derivative, we get
	\begin{align}
		(3-d_{1223})\hat{0}+d_1k_{1s}\widehat{1^+} + d_3(k_{2s})\widehat{3^+} + 2M_2d_2\widehat{2^+} -d_1\widehat{1^+}\widehat{2^-}-d_3\widehat{2^-}\widehat{3^+} = 0\ ,\\
		(d_1-d_2)\hat{0} + d_1(k_{1s}-2 M_1)\widehat{1^+} - \widehat{2^-}(d_1\widehat{1^+} + d_3\widehat{3^+})-d_2(k_{1s}-2M_2)\widehat{2^+} + \nonumber \\
		\widehat{1^-}(d_2\widehat{2^+} + d_3\widehat{3^+})-d_3(k_{3s}-k_{2s})\widehat{3^+} = 0\ ,\\
		(d_2 - d_3)\hat{0} + d_1(k_{3s}-k_{1s})\widehat{1^+} - \widehat{3^-}(d_1\widehat{1^+} + d_2\widehat{2^+}) + \widehat{2^-}(d_3\widehat{3^+} + d_1\widehat{1^+})+\nonumber\\
		d_2(k_{2s}-2M_2)\widehat{2^+} - d_3(k_{2s}-2M_3)\widehat{3^+} = 0\ ,
	\end{align}
	where
	\begin{align}
		k_{1s} &= k^2_1 + M_2 + M_1 \ ,\\
		k_{2s} &= k^2_2 + M_2 + M_3\ , \\
		k_{3s} &= k^2_3 + M_3 + M_1\ , \\
		d_{1223} &= d_1 + 2 d_2 + d_3\ , 
	\end{align}
	and we also defined ladder operators $\hat{0}$, $\widehat{1^\pm}$, $\widehat{2^\pm}$, and $\widehat{3^\pm}$, that act on $T$ as
	\begin{align}
		&\widehat{0}\,T(d_1, d_2, d_3)= T(d_1, d_2, d_3)\,,\\
		&\widehat{1^{\pm}}\,T(d_1, d_2, d_3) =T(d_1 \pm 1,d_2,d_3)\,,\\
		&\widehat{2^{\pm}}\,T(d_1, d_2, d_3) =T(d_1 ,d_2 \pm 1,d_3)\,,\\
		&\widehat{3^{\pm}}\,T(d_1, d_2, d_3) =T(d_1 ,d_2,d_3 \pm 1)\,.
	\end{align}
	The action of two ladder operators on $T$ is for example
	\begin{equation}
		\label{eq:ladderaction}
		\widehat{1^{+}}\widehat{2^{-}} [T(d_1,d_2,d_3)] =  T(d_1+1,d_2-1,d_3)\,.
	\end{equation}
	
	Solving for terms only involving $\widehat{1^+}, \widehat{2^+}, \widehat{3^+}$, we obtain, 
	\begin{align}
		\widehat{1^{+}} &= {-k_{s,23}\widehat{1^{+}}\widehat{2^{-}} + \frac{k_{s,22} d_2}{d_1}\widehat{2^{+}}\widehat{1^{-}} - k_{s,12}\widehat{ 1^{+}}\widehat{3^{-}} + \frac{k_{s,22} d_3}{d_1}\widehat{3^{+}}\widehat{1^{-}} - \frac{k_{s,12} d_2}{d_1}\widehat{2^{+}}\widehat{3^{-}} }\nonumber \\
		&-\frac{k_{s,23} d_3}{d_1}\widehat{3^{+}}\widehat{2^{-}} + \left(k_{s,12}\frac{3 - d_{1233}}{d_1}-k_{s,22}\frac{3 - d_{1123}}{d_1}+k_{s,23}\frac{3 - d_{1223}}{d_1}\right)\hat{0}\,,\\
		\widehat{2^{+}} &= {\frac{k_{s,33} d_1}{d_2}\widehat{1^{+}}\widehat{2^{-}} -k_{s,23}\widehat{2^{+}}\widehat{1^{-}}-\frac{k_{s,31} d_1}{d_2}\widehat{1^{+}}\widehat{3^{-}} -\frac{k_{s,23} d_3}{d_2}\widehat{3^{+}}\widehat{1^{-}} -k_{s,31}\widehat{2^{+}}\widehat{3^{-}} }\nonumber \\
		&+\frac{k_{s,33} d_3}{d_2}\widehat{3^{+}}\widehat{2^{-}} + \left(k_{s,23}\frac{3 - d_{1123}}{d_2}+k_{s,31}\frac{3 - d_{1233}}{d_2}-k_{s,33}\frac{3 - d_{1223}}{d_2}\right)\hat{0}\,,\\
		\widehat{3^+} &= -\frac{k_{s,31} d_1}{d_3}\widehat{1^+}\widehat{2^{-}} -\frac{k_{s,12} d_2}{d_3}\widehat{2^+}\widehat{1^{-}} -k_{s,12} \widehat{1^+}\widehat{3^{-}} + \frac{k_{s,11} d_1}{d_3}\widehat{3^+}\widehat{1^{-}} + \frac{k_{s,11} d_2}{d_3}\widehat{2^+}\widehat{3^{-} }\nonumber\\
		&-k_{s,31}\widehat{3^+}\widehat{2^-} + \left(-k_{s,11}\frac{3 - d_{1233}}{d_3}+k_{s,12}\frac{3 - d_{1123}}{d_3}+k_{s,31}\frac{3 - d_{1223}}{d_3}\right)\hat{0}\,,
	\end{align}
	where
	\begin{align}
		k_{s,11} &= \frac{-4 M_1 M_2 + k_{1s}^2}{\rm jac}\ ,\\
		k_{s,12} &= \frac{-2k_{3s} M_2 + k_{1s} k_{2s}}{\rm jac}\ ,\\
		k_{s,22} &= \frac{-4M_2 M_3 + k_{2s}^2}{\rm jac}\ , \\
		k_{s,23} &= \frac{-2 k_{1s} M_3 + k_{2s} k_{3s}}{\rm jac}\ , \\
		k_{s,31} &= \frac{-2 k_{2s} M_1 + k_{1s} k_{3s}}{\rm jac}\ , \\
		k_{s,33} &= \frac{-4M_1M_3 + k_{3s}^2}{\rm jac}\ , \\
		\textrm{jac} &= -8M_1M_2M_3 + 2k_{1s}^2M_3 + 2k_{2s}^2M_1 + 2k_{3s}^2M_2 -2k_{1s}k_{2s}k_{3s}\, , \\
		d_{ijkl} &= d_i + d_j + d_k + d_l\,.
	\end{align}
	The important point of this solution is that we can express an operator that raises the overall value of $d_1+d_2+d_3$ by one unit as a combination of operators that do not raise  $d_1+d_2+d_3$.
	Therefore, if $d_1 > 0$, let $\widehat{1^+}$ act on $T(d_1-1,d_2,d_3)$, and use the solution we found for $\widehat{1^+}$ in terms of the other raising or lowering operators:
	\begin{align}
		\begin{split}\label{eq:rec_solved}
			T(d_1,d_2,d_3) &= -k_{s,23}T(d_1,d_2-1,d_3) + \frac{k_{s,22}d_2}{d_1 -1}T(d_1-2,d_2+1,d_3)  \\
			&  -k_{s,12}T(d_1,d_2,d_3-1) + \frac{k_{s,22}d_3}{d_1 -1}T(d_1-2,d_2,d_3+1) \\
			&  -\frac{k_{s,12}d_2}{d_1-1}T(d_1-1,d_2+1,d_3-1)-\frac{k_{s,23}d_3}{d_1-1}T(d_1-1,d_2-1,d_3+1) \\
			& +\left(k_{s,12}\frac{2-d_1-d_2-2d_3}{d_1-1}-k_{s,22}\frac{1-2d_1-d_2-d_3}{d_1-1}\right.\\
			&\left.+k_{s,23}\frac{2-d_1-2d_2-d_3}{d_1-1} \right) T(d_1-1,d_2,d_3)\,,
		\end{split}
	\end{align}
	Similarly, if $d_2 > 0$, let $\widehat{2^+}$ act on $T(d_1,d_2-1,d_3)$, and use the solution we found for $\widehat{2^+}$ in terms of the other raising or lowering operators:
	\begin{align}
		\begin{split}
			T(d_1,d_2,d_3) &= \frac{k_{s,33} d_1}{d_2-1}T(d_1+1,d_2-2,d_3) -k_{s,23}T(d_1-1,d_2,d_3)\\
			& -\frac{k_{s,31} d_1}{d_2-1}T(d_1+1,d_2-1,d_3-1) -\frac{k_{s,23} d_3}{d_2-1}T(d_1-1,d_2-1,d_3+1) \\
			& -k_{s,31}T(d_1,d_2,d_3-1) +\frac{k_{s,33} d_3}{d_2-1}T(d_1,d_2-2,d_3+1)\\
			& + \left(k_{s,23}\frac{2-2d_1 -d_2-d_3}{d_2-1}+k_{s,31}\frac{2-d_1-d_2-2d_3}{d_2-1}\right.\\
			& \left.-k_{s,33}\frac{1-d_1-2d_2-d_3}{d_2-1}\right)T(d_1,d_2-1,d_3)\,.
		\end{split}
	\end{align}
	Similarly, if $d_3 > 0$, let $\widehat{3^+}$ act on $T(d_1,d_2,d_3-1)$, and use the solution we found for $\widehat{3^+}$ in terms of the other raising or lowering operators:
	\begin{align}
		\begin{split}
			T(d_1,d_2,d_3)&=-\frac{k_{s,31} d_1}{d_3-1}T(d_1+1,d_2-1,d_3-1) -\frac{k_{s,12} d_2}{d_3-1}T(d_1-1,d_2+1,d_3-1)\\ &-k_{s,12}T(d_1+1,d_2,d_3-2) + \frac{k_{s,11} d_1}{d_3-1}T(d_1-1,d_2,d_3)\\
			&+ \frac{k_{s,11} d_2}{d_3-1}T(d_1,d_2+1,d_3-2) -k_{s,31}T(d_1,d_2-1,d_3)\\
			&+ \left(-k_{s,11}\frac{1-d_1-d_2-2d_3}{d_3-1}+k_{s,12}\frac{3 -2d_1-d_2-d_3-1}{d_3-1}\right.\\
			&\left.+k_{s,31}\frac{d-d_1-2d_2-d_3-1}{d_3-1}\right)T(d_1,d_2,d_3-1)\,.
		\end{split}
	\end{align}
	The recursion terminates either when one of the $d_i$ vanishes, in which case we evaluate the bubble integral that will be analyzed in Sec.~\ref{sec:bubrec}, or when $d_1=d_2=d_3=1$, in which case we evaluate a master integral as shown in Sec.~\ref{sec:master}. 	 
	We stress again that in each of these recursion relations, the sum $d_1 + d_2 + d_3$ decreases by one from the l.h.s. to the r.h.s., e.g. $T(d_1 + 1, d_2 - 1, d_3 - 1)$, thus the recursion relation is guaranteed to terminate. 
	
	\subsubsection{$B$-recursion}
	\label{sec:bubrec}
	
	The general bubble integral is given by Eq.~(\ref{eq:bubblegen}). 
	Once we obtain a bubble integral from the recursion relation of the former section, we can use the same recursion relations as in Sec.~\ref{sec:trirec} to reduce the $n_i$ exponents. 
	We then obtain one of three possible expressions:
	\begin{equation}
		\label{eq:brecurse}
		\begin{split}
			& \int_q\frac{1}{((\vec{k}_1-\vec{q})^2 + M_1)^{d_1}(\vec{q}^2 + M_2)^{d_2}}\, , \\
			& \int_q\frac{1}{((\vec{k}_1-\vec{q})^2 + M_1)^{d_1}((\vec{k}_2+\vec{q})^2 + M_3)^{d_3}}\, , \\
			& \int_q\frac{1}{(\vec{q}^2 + M_2)^{d_2}((\vec{k}_2+\vec{q})^2 + M_3)^{d_3}}\,.
		\end{split}
	\end{equation}
	
	To simplify the reduction procedure, we can make a shift in $\vec{q}$ to put the last two terms into the same form as the first,
	\begin{align}
		\int_q\frac{1}{((\vec{k}_1-\vec{q})^2 + M_1)^{d_1}((\vec{k}_2+\vec{q})^2 + M_3)^{d_3}}&=\int_q\frac{1}{((\vec{k}_3-\vec{q})^2+M_3)^{d_3}(\vec{q}^2+M_1)^{d_1}} \,, \\
		\int_q\frac{1}{(\vec{q}^2 + M_2)^{d_2}((\vec{k}_2+\vec{q})^2 + M_3)^{d_3}} &= \int_q\frac{1}{((\vec{k}_2-\vec{q})^2+M_2)^{d_2}(\vec{q}^2+M_3)^{d_3}}\,,
	\end{align}
	where we did the shift $\vec{q}\rightarrow \vec{k}_1-\vec{q}$ and $\vec{q}\rightarrow -\vec{k}_2+\vec{q}$ respectively. Now we define the function,
	\begin{align}
		B(a,b,k^2,M_i,M_j)=\int_q\frac{1}{((\vec{k}-\vec{q})^2 + M_i)^{a}(\vec{q}^2 + M_j)^{b}}\,,
	\end{align}
	where $a$ and $b$ are integers, $M_i = \{M_1, M_2, M_3\}$ and $i \neq j$.
	We define 
	\begin{equation}
		B_{\rm int}(a,b,k^2,M_i,M_j) \equiv \frac{1}{((\vec{k}-\vec{q})^2 + M_i)^{a}(\vec{q}^2 + M_j)^{b}}\,.
	\end{equation}
	Following the same approach as in the $T$-recursion, we can use the divergence theorem to obtain
	\begin{align}
		\int_q\frac{\partial}{\partial q_{\mu}}\cdot \left( q_{\mu}B_{\rm int}(a,b,k^2,M_i,M_j) \right) &=0\ ,\\
		\int_q\frac{\partial}{\partial q_{\mu}}\cdot \left( k_{\mu}B_{\rm int}(a,b,k^2,M_i,M_j) \right)&=0\,.
	\end{align}
	This yields the following relations: 
	\begin{align}
		(3-a-2b)\hat{0} + a(k^2 + M_i + M_j)\widehat{a^+} - a\, \widehat{a^+} \widehat{b^-} + 2b\,M_j\,\widehat{b^+}&=0\ , \\
		(a-b)\hat{0}-a\, \widehat{a^+}\widehat{b^-} +a(k^2-M_i+M_j) \widehat{a^+} + b\,\widehat{a^{-}}\widehat{b^+} - b(k^2+M_i-M_j)\widehat{b^+} &=0\ , 
	\end{align}
	where we define ladder operators $\hat{0}$, $\widehat{a^\pm}$, $\widehat{b^\pm}$ that act on $B$ as
	\begin{align}
		&\widehat{0}\,B(a, b, k, M_i, M_j)= B(a, b, k, M_i, M_j)\,,\\
		&\widehat{a^{\pm}}\,B(a, b, k, M_i, M_j) =B(a\pm1, b, k, M_i, M_j)\,, \\
		&\widehat{b^{\pm}}\,B(a, b, k, M_i, M_j) =B(a, b\pm1, k, M_i, M_j)\,.
	\end{align}
	The action of the ladder operators on $B$ is analogous to the one outlined in Eq.~\eqref{eq:ladderaction}.
	Solving for $\widehat{a^{+}}, \widehat{b^+}$, 
	\begin{align}
		\widehat{a^+} &= \frac{\left[\left((3-a-b)\frac{2M_j-k_s}{a} -2M_j + b\frac{k_s}{a}\right)\hat{0} + k_s \widehat{a^+}\widehat{b^-} -b\frac{2M_j}{a}\widehat{a^-}\widehat{b^+}\right]}{k_s^2-4M_iM_j}\,,\\
		\widehat{b^+} &= \frac{\left[\left((3-a-b)\frac{2M_i - k_s}{b}+a\frac{k_s}{b} -2M_i \right)\hat{0}-\frac{2aM_i}{b}\widehat{a^+}\widehat{b^-}+k_s \widehat{a^-}\widehat{b^+}\right]}{k_s^2-4M_iM_j}\,,
	\end{align}
	where $k_s = k^2 + M_i + M_j$.
	Notice that the recursive solutions we found are such that $\widehat{a^+}$ and $\widehat{b^+}$ raise the value of $a+b$ by one unit. 
	Therefore, if $a>1$, let $\widehat{a^+}$ act on $B(a-1, b)$ and use the recursive relation we found:
	\begin{align}
		\begin{split}
			B(a, b) &= \bigg[\left((3-a-b)\frac{2M_j-k_s}{a} -2M_j + b\frac{k_s}{a}\right)B(a-1, b) \\
			& + k_s B(a, b-1)- b\frac{2M_j}{a}B(a-2, b+1)\bigg]/(k_s^2-4M_iM_j)\,,
		\end{split}
	\end{align}
	where we have omitted the arguments of $B$ that do not change for clarity. 
	We will use this lighter notation in the rest of the paper as well. 
	Similarly, if $b>1$, let $\widehat{b^+}$ act on $B(a, b-1)$ and use the recursive relation we found:
	\begin{align}
		\begin{split}
			B(a, b) &= \bigg[\left((3-a-b)\frac{2M_i- k_s}{b}+a\frac{k_s}{b} -2M_i \right)B(a, b-1) \\
			&-\frac{2aM_i}{b}B(a+1, b-2)+k_s B(a-1, b)\bigg]/(k_s^2-4M_iM_j)\,.
		\end{split}
	\end{align}
	The recursion relation terminates when $a = 0$ or $b = 0$, in which case a tadpole integral remains whose calculation we derive in Sec.~\ref{sec:master}, or when $a = b = 1$, in which case we evaluate the bubble master integral, also calculated in Sec.~\ref{sec:master}. 
	
	\subsubsection{Tensor Reduction}
	\label{sec:tensor}
	
	Following the numerator exponent reduction in Eq.~(\ref{eq:numrecursion}) for $n_3 > 0$ and $d_3 > 0$, the recursion may terminate when $n_3 > 0$ and $d_3 = 0$, at which point we may be left with an integral of the form
	\begin{align}
		\int_q\frac{(\vec{k}_2+\vec{q})^{2n_3}}{(\vec{k}_1-\vec{q})^2+M_1)^{d_1}(\vec{q}^2+M_2)^{d_2}}.
	\end{align}
	Such term requires further reduction as our goal is to reduce all denominator exponents $d_i \rightarrow 1$ or $0$, and all numerator exponents $n_i \rightarrow 0$ if $d_i \neq 0$ and $n_i \rightarrow -1$ if $d_i = 0$. 
	At that point we proceed to evaluate the triangle, bubble, or tadpole master integral.
	In the evaluation of the 1-loop bispectrum, the power $n_3$ satisfies $n_3 \leq 4$. In this section, we outline the procedure to evaluate such terms. 
	Consider the simplest case of $n_3 = 1$, 
	\begin{align}
		& \int_q\frac{(\vec{k}_2+\vec{q})^{2}}{(\vec{k}_1-\vec{q})^2+M_1)^{d_1}(\vec{q}^2+M_2)^{d_2}} = \int_q\frac{k^2_2 + q^2 + M_2 - M_2 + 2\vec{k}_2\cdot \vec{q}}{(\vec{k}_1-\vec{q})^2+M_1)^{d_1}(\vec{q}^2+M_2)^{d_2}} \nonumber \\
		&  = (k^2_2-M_2)B(d_1,d_2,k_2,M_1,M_2) + B(d_1,d_2-1,k_2,M_1,M_2) \nonumber \\
		& \qquad \qquad \qquad +2\int_q\frac{\vec{k}_2\cdot \vec{q}}{(\vec{k}_1-\vec{q})^2+M_1)^{d_1}(\vec{q}^2+M_2)^{d_2}}\,.
	\end{align}
	The first two terms can be evaluated following the bubble recursion formulas of the previous section. To evaluate the last term, we use a tensor reduction method~\cite{Passarino:1978jh} by considering the integral,
	\begin{align}
		B_{\mu}(d_1,d_2,k_1,M_1,M_2)=\int_q\frac{q_{\mu}}{(\vec{k}_1-\vec{q})^2+M_1)^{d_1}(\vec{q}^2+M_2)^{d_2}} = k_{1\mu}A\,,
	\end{align}
	where in the last step we noticed that the integral must be proportional to $k_{1\mu}$. 
	We can then contract both sides with $k^{\mu}_1$ to obtain
	\begin{align}
		\begin{split}
			&\int_q\frac{\vec{k}_1\cdot \vec{q}}{(\vec{k}_1-\vec{q})^2+M_1)^{d_1}(\vec{q}^2+M_2)^{d_2}}=\int_q\frac{\frac{1}{2}\left(-(\vec{k}_1 - \vec{q})^2+k^2_1+q^2\right)}{(\vec{k}_1-\vec{q})^2+M_1)^{d_1}(\vec{q}^2+M_2)^{d_2}}\\
			&= \frac{1}{2}\left(-B(d_1-1,d_2)+B(d_1,d_2-1)+(M_1-M_2)B(d_1,d_2) \right) = k^2_1A
		\end{split}\\
		\Rightarrow \,  A &= \frac{1}{2k^2_1}\left(-B(d_1-1,d_2)+B(d_1,d_2-1)+(M_1-M_2)B(d_1,d_2) \right).
	\end{align}
	Hence,
	\begin{align}
		\begin{split}
			& B_{\mu}(d_1,d_2,k_1,M_1,M_2)=\frac{k_{1\mu}}{2k^2_1}\left(-B(d_1-1,d_2)+B(d_1,d_2-1)\right.\\
			& \hspace{9cm}\left.+(M_1-M_2)B(d_1,d_2)\right)\ ,
		\end{split}\\
		\begin{split}
			\Rightarrow \, &\qquad  {2 k_{2\mu} B_{\mu}} = 2\int_q\frac{\vec{k}_2\cdot \vec{q}}{(\vec{k}_1-\vec{q})^2+M_1)^{d_1}(\vec{q}^2+M_2)^{d_2}}  \\
			& \qquad \qquad = \frac{\vec{k}_2\cdot \vec{k}_1}{k^2_1}\left(-B(d_1-1,d_2)+B(d_1,d_2-1)+(M_1-M_2)B(d_1,d_2) \right)\,.
		\end{split}
	\end{align}
	For $n_3 = 2$, we will additionally need to evaluate terms proportional to
	\begin{align}
		B_{\mu\nu}(d_1,d_2,k_1,M_1,M_2) =\int_q\frac{q_{\mu}q_{\nu}}{(\vec{k}_1-\vec{q})^2+M_1)^{d_1}(\vec{q}^2+M_2)^{d_2}} = k_{1\mu}k_{1\nu}A_1 + \delta_{\mu\nu}A_2 \,.
	\end{align}
	We can again contract both sides with $k_{1\mu}k_{1\nu}$ and $\delta_{\mu\nu}$ to obtain two equations allowing us to solve for $A_1$ and $A_2$. 
	The general procedure is then to construct all possible symmetric tensors from $k_{1\mu}$ and $\delta_{\mu\nu}$ multiplied by arbitrary coefficients $A_i$. 
	Taking contractions of these tensors, one may solve the system of equations to obtain $A_i$ as functions of bubble integrals. 
	We will require expressions  for $n_3 = 2, 3, 4$, but they are straightforward but too long to include in the paper. 
	
	Lastly, we encounter terms of the form,
	\begin{align}
		\int_q\frac{(\vec{k}_1 - \vec{q})^{2n_1}(\vec{k}_2+\vec{q})^{2n_3}}{(\vec{q}^2+M_2)^{d_2}}.    
	\end{align}
	Consider the case of $n_1 = n_3 = 1$, expanding the numerator we find terms schematically of the form 
	\begin{align}
		\begin{split}
			&\int_q\frac{(\vec{k}_1 - \vec{q})^{2}(\vec{k}_2+\vec{q})^{2}}{(\vec{q}^2+M_2)^{d_2}} \supset \\ 
			&\left\{\int_q\frac{1}{(\vec{q}^2+M_2)^{d_2}},\int_q\frac{q_{\mu}}{(\vec{q}^2+M_2)^{d_2}},\int_q\frac{q_{\mu}q_{\nu}}{(\vec{q}^2+M_2)^{d_2}},\right.\\
			& \hspace{7cm} \left.\int_q\frac{q_{\mu}q_{\nu}q_{\rho}}{(\vec{q}^2+M_2)^{d_2}},\int_q\frac{q_{\mu}q_{\nu}q_{\rho}q_{\sigma}}{(\vec{q}^2+M_2)^{d_2}} \right\}.
		\end{split}
	\end{align}
	Integrals involving odd powers of $q$ vanish as the integrand is odd under $\vec{q}\rightarrow -\vec{q}$. 
	Integrands with even powers of $q$ can be evaluated with the standard dimensional regularization formula, e.g.,
	\begin{align}
		\int_q\frac{q_{\mu}q_{\nu}}{(\vec{q}^2+M_2)^{d_2}} = \frac{3\delta_{\mu\nu}}{2}M^{-d_2+5/2}_2\Gamma(d_2-5/2)/\Gamma(d_2).
	\end{align}
	The full decomposition for the case of $n_1 = n_3 = 1$ is,
	\begin{align}
		\begin{split}
			&\int_q\frac{(\vec{k}_1 - \vec{q})^{2}(\vec{k}_2+\vec{q})^{2}}{(\vec{q}^2+M_2)^{d_2}} = \int_q \frac{(k^2_1 - 2 \vec{k}_1 \cdot \vec{q} + q^2)(k^2_2 + 2\vec{k}_2 \cdot q + q^2)}{(\vec{q}^2+M_2)^{d_2}}\ ,\\
			&= \int_q\frac{k^2_1k^2_2 + (k^2_1 + k^2_2)q^2 - 4 (\vec{k}_1 \cdot \vec{q}) (\vec{k}_2 \cdot \vec{q}) + q^4}{(\vec{q}^2+M_2)^{d_2}}\ ,\\
			&= k^2_1k^2_2\Gamma(d_2 - 3/2)M^{-d_2 + 3/2}_2/\Gamma(d_2) + (k^2_1 + k^2_2)\frac{3}{2}\Gamma(d_2 - 5/2)M^{-d_2 + 5/2}_2/\Gamma(d_2) -  \\
			&\qquad 4\vec{k}_1\cdot\vec{k}_2\frac{1}{2}\Gamma(d_2 - 5/2)M^{-d_2 + 5/2}_2/\Gamma(d_2) + \frac{15}{4}\Gamma(d_2 - 7/2)M^{-d_2 + 7/2}_2/\Gamma(d_2)\,.
		\end{split}
	\end{align}
	For higher powers of $n_1, n_3$ we expand the numerator momenta and keep even powers of $q$ and evaluate each term with dim reg..
	At the level of the one loop bispectrum in redshift space we encounter at most the case $n_1 = n_3 = 4$ in the $B_{411,r}$ diagram, which we have implemented in our code.
	
	We now point out that these kind of manipulations are used also in another context. 
	In evaluating the loop integrals in redshift space, we often encounter integrals where the internal momenta are contracted with the line of sight direction $\hat z$. 
	The resulting non-rotationally-invariant integrals are evaluated in a very similar manner as explained in this section. 
	That is, first we expand the integrals in the most general tensor structure allowed by the momenta dependence of the loop integral. 
	This is made out of all the possible tensors ${\cal T}^{i_1,\ldots,i_n}$ built out of $\delta^{ij}, \vec k_1,\vec k_2$ (there is no need to use $\vec k_3$ by momentum conservation) with the right number of indexes. 
	Then, we perform suitable contractions with the same tensors ${\cal T}^{i_1,\ldots,i_n}$, each one leading to a rotationally invariant integral not involving $\hat z$, and which can be evaluated with the techniques presented here.
	Finally, we solve the linear system to find the coefficients of the expansion in ${\cal T}^{i_1,\ldots,i_n}$ of the original integral we wished to evaluate. 
	The procedure is conceptually straightforward, but perhaps cumbersome, and we do not present the details here (see also~\cite{DAmico:2022ukl}).
	
	\subsubsection{Recursion example}
	
	After explaining the main recursions used to evaluate $L$ and reduce it to master integrals, we give here an example of how it works.
	Let us consider the following specific $L$ with $n_1=1,d_1 = 2,n_2 = 0,d_2 = 1,n_3 = 0$, and $d_3 = 1$:
	\begin{equation}
		\begin{split}
			&L(n_1=1,d_1 = 2,n_2 = 0,d_2 = 1,n_3 = 0,d_3 = 1) = \\
			& \qquad \int_{\vec{q}}\frac{|\vec{k}_1-\vec{q}|^{2}}{((\vec{k}_1 - \vec{q})^2+M_1)^{2}(\vec{q}^2+M_2)((\vec{k}_2 + \vec{q})^2 + M_3)}\,.
		\end{split}
	\end{equation}
	Such a term can arise in the evaluation of $\bar{B}_{222}$,
	\begin{align}
		\bar{B}_{222} &\supset C(k^2_1,k^2_2,k^2_3)\int_q |\vec{k}_1 - q|^2P_{\rm fit}(|\vec{k}_1 - q|)P_{\rm fit}(q)P_{\rm fit}(|\vec{k}_2 + q|)\\
		P_{\rm fit}(k) &\supset  \alpha_1 \frac{1}{(k^2 + M_1)^2}\ ,\alpha_2 \frac{1}{(k^2 + M_2)}\ ,\alpha_3 \frac{1}{(k^2 + M_3)} 
	\end{align}
	We now demonstrate the procedure of our code to evaluate such a term.
	
	\begin{enumerate}
		\item Step 1: Reduction of $n_1 = 1$ using $L$-reduction from Sec.~\ref{sec:trirec}: 
		\begin{align}
			\begin{split}
				L(n_1 = 1,2,0,1,0,1) &= L(n_1 = 0,1,0,1,0,1) - M_1L(n_1 = 0,2,0,1,0,1)\ \\
				&= T_{\rm master}(k^2_1,k^2_2,k^2_3,M_1,M_2,M_3) - M_1T(2,1,1).
			\end{split}
		\end{align}
		\item Step 2: Reduction of $d_1$ using $T$-reduction from Sec.~\ref{sec:T_reduc}:
		\begin{align}
			\begin{split}
				T(2,1,1) &= -k_{s, 23}T(2,0,1) + k_{s, 22}T(0,2,1) + k_{s, 22}T(2,1,0) - k_{s,12}T(0,1,2) \\
				& - k_{s,12}T(1,2,0) - k_{s,23}T(1,0,2)+ \left(-3k_{s,12} -5k_{s,22} - 3k_{s,23}\right)T_{\rm master}\\
				&= -k_{s, 23}B(d_1 = 2,d_3 = 1) + k_{s, 22}B(d_2 = 2,d_3 = 1)  \\
				&+ k_{s, 22}B(d_1 = 2,d_2 = 1) - k_{s,12}B(d_2 = 1,d_3 = 2) \\
				&- k_{s,12}B(d_1 = 1,d_2 = 2) - k_{s,23}B(d_1 = 1,d_3 = 2) \\
				&+ \left(-3k_{s,12} -5k_{s,22} - 3k_{s,23}\right)T_{\rm master}\,.
			\end{split}
		\end{align}
		\item Step 3: Reduction of Bubble integrals using $B$-reduction from Sec.~\ref{sec:bubrec} (we outline the reduction of only the first term, $B(d_1 = 2, d_3 = 1)$, for simplicity):
		\begin{align}
			\begin{split}
				&B(d_1 = 2, d_3 = 1) = \\
				&=\frac{\left((-2M_3 + \frac{k_s}{2})B_{\rm master}(k^2_2,M_1,M_3) + k_s B(2,0) - M_3 B(0,2)/2\right)}{k^2_s - 4M_1M_3}\ ,\\
				&= \frac{\Big((-2M_3 + \frac{k_s}{2})B_{\rm master}(k^2_2,M_1,M_3) + k_s\textrm{Tad}(M_1,n_1 = 0, d_1 = 2) - M_3\textrm{Tad}(M_3,n_3 = 0, d_3 = 2)/2\Big)}{k^2_s - 4M_1M_3}\,.
			\end{split}
		\end{align}
		We perform the $B$-reduction for every bubble integral appearing after $T$-reduction. 
		\item Step 4: Evaluating all master integrals:\\
		The general $L$-function,  $L(1,2,0,1,0,1)$ has now been decomposed into various $T_{\rm master}$, $B_{\rm master}$ and $\textrm{Tad}$ master integrals, which we evaluate analytically and sum.
	\end{enumerate}	
	
	\section{Master integrals calculation}
	\label{sec:master}
	
	\subsection{Expressions}
	The final result of the recursion relations leads to three types of irreducible integrals which we call the tadpole, bubble, and triangle master integrals, named after the shape of the corresponding Feynman diagrams. 
	We remind their form here for convenience.
	\begin{enumerate}
		\item Tadpole: 
		\begin{align}
			\text{Tad}(M_j, {n}, d) = \int \frac{d^3\vec{q}}{\pi^{3/2}}\frac{{(\vec{p}_i^2)^n}}{(\vec{p}^2_i + M_j)^{d}} = \frac{2}{\sqrt{\pi}}{\Gamma(n + 3/2)}\Gamma(d - n - 3/2)M^{n -d + 3/2}_j/\Gamma(d).
		\end{align}
		where $\vec{p_i} = \{\vec{k}_1 - \vec{q}, \vec{q}, \vec{k}_2 + \vec{q}\}$ and $M_j = \{M_1, M_2, M_3\}$.
		The expression above holds within dimensional regularization for $D=3$, as we discussed in Eq.~(\ref{eq:TadpoleDREG}).
		
		\item Bubble:
		\begin{align}
			B_{\rm master}(k^2,M_1,M_2) =  \int \frac{d^3\vec{q}}{\pi^{3/2}}\frac{1}{(q^2 + M_1)( |\vec{k}-\vec{q}|^2 + M_2)}
		\end{align}
		
		\item Triangle:
		\begin{align}
			\begin{split}
				T_{\rm master}(k_1^2, k_2^2, k_3^2, M_1,& M_2, M_3) = \\
				&\int \frac{d^3\vec{q}}{\pi^{3/2}}\frac{1}{(q^2 + M_1)( |\vec{k}_1-\vec{q}|^2 + M_2)( |\vec{k}_2+\vec{q}|^2 + M_3)},
			\end{split}
		\end{align}
		where $\vec{k}_1 + \vec{k}_2 + \vec{k}_3 = 0$.
	\end{enumerate}
	
	To compute the Bubble and Triangle master integrals, we will use Schwinger and Feynman parametrizations within dimensional regularization. 
	However, there are subtleties that have to be dealt with in the case of complex masses. 
	
	\subsection{Calculation tools}
	\label{sec:prelim}
	
	To find closed form expressions for the master integrals, we will use some tools that we now outline.
	
	\paragraph{Generalized multiplicative formulas for logs and square roots.}
	
	Following \cite{tHooft:1978jhc}, we use logs and square roots that have a branch cut in the negative real axis, with their value on the negative real axis taken by continuity from the upper part~\footnote{So $\log(-1)=i\pi$ and $\sqrt{-1}=i$.}.	
	The rule for the log of a product is then:
	\begin{equation}
		\begin{split}
			& \log(a b) = \log(a) + \log(b) + \eta(a,b)\,, \\
			&\eta(a,b) = 2 \pi i \left[ \theta(-a) \theta(-b) \theta(ab) - \theta(a) \theta(b) \theta(-ab) \right]\,,
			\label{eq:eta}
		\end{split}
	\end{equation}
	where $a$ and $b$ are complex numbers, and $\theta$ is a function defined as
	\begin{equation}
		\label{eq:theta}
		\theta(a) = 
		\begin{cases}
			1, \,& \textrm{if } \Im a > 0 \,,\\ 
			0, \, &\textrm{if } \Im a < 0 \,,\\
			1, \, &\textrm{if } \Im a = 0 \textrm{ and } \Re a<0 \,,\\
			0, \, &\textrm{if } \Im a = 0 \textrm{ and } \Re a\geq0\,.
		\end{cases}
	\end{equation}
	Specifically, if $a$ is a positive real number, we have $\theta(a)=0$ and $\theta(-a)=1$.  
	Notice that in the real line this is the opposite of the standard definition of the Heaviside function.
	Useful particular cases are:
	\begin{align}
		& \log(a b) = \log(a) + \log(b) \,, \text{ if $\Im a$ and $\Im b$ have different sign,}\\
		& \log(\frac{a}{b}) = \log(a) - \log(b) \,, \text{ if $\Im a$ and $\Im b$ have the same sign.}
	\end{align}
	Defining the square root as $\sqrt{a} \equiv \exp(\log(a)/2)$, we obtain the analogous expressions:
	\begin{equation}
		\begin{split}
			& \sqrt{a b} = s(a,b)\sqrt{a}\sqrt{b}\,, \\
			& s(a,b) \equiv \exp{\eta(a,b)/2} = (-1)^{\theta(-a) \theta(-b) \theta(ab)} (-1)^{\theta(a) \theta(b) \theta(-ab)}
			\label{eq:s}
			\,,
		\end{split}
	\end{equation}
	and specifically:
	\begin{align}
		\label{eq:ssigncondition}
		& \sqrt{a b} = \sqrt{a}  \sqrt{b} \,, \text{ if $\Im a$ and $\Im b$ have different sign,}\\
		\label{eq:srealposcondition}	
		& {\sqrt{a b} = \sqrt{a}  \sqrt{b} \,, \text{ if $a$ is real and positive}}\\
		& \sqrt{\frac{a}{b}} = \frac{\sqrt{a}}{\sqrt{b}}\,, \text{ if $\Im a$ and $\Im b$ have the same sign.}
	\end{align}	
	As a quick example, let us look at $\sqrt{-z}$.
	Applying Eq.~\eqref{eq:s}, we obtain: 
	\begin{equation}
		\sqrt{-z} = s(-1,z)\, i \sqrt{z} =  (-1)^{\theta(1) \theta(- z)} (-1)^{\theta(-1) \theta(z)} i \sqrt{z}\,,
	\end{equation}
	where we used $\theta(z)^2 = \theta(z)$. 
	Using our prescription, we get $\theta(-1) = 1$ and $\theta(1) = 0$, and our result simplifies to 
	\begin{equation}
		\sqrt{-z} = s(-1,z)\, i \sqrt{z} =  (-1)^{\theta(z)} i \sqrt{z}\,,
	\end{equation}
	which means that $\sqrt{-z} = i\sqrt{z}$ only in the lower complex plane and in the positive real axis, while $\sqrt{-z} = -i\sqrt{z}$ in the upper complex plane and in the negative real axis. 
	Note also that $\Re \sqrt{z}\geq 0$ for any complex number $z$.
	
	Furthermore, we use the following conventions for simplifying exponents $(a^b)^c=a^{b c}$ only if $c$ is an integer. 
	So, for example, $a^{3/2}$ can be interpreted as $(a^{1/2})^3$ but not as $(a^{3})^{1/2}$~(\footnote{These conventions are the same as in Mathematica.}).
	
	\paragraph{Schwinger parametrization.}
	Let us derive the Feynman parametrization for the bubble integral using Schwinger parameters~\cite{Weinzierl:2006qs}. 
	We will use two important (equivalent) identities.
	First, if $\Im A >0$, we use
	\begin{equation}
		\label{eq:schwinger1}
		\frac{i}{A} = \int_0^\infty ds (1 + i\epsilon)\, \exp(i A (1 + i \epsilon)s)\,,
	\end{equation}
	where $\epsilon > 0$ and can be arbitrarily small. 
	Notice that if $\Im A < 0$, the integrand of Eq.~\eqref{eq:schwinger1} is divergent as the imaginary part of the exponent becomes positive. 
	Thus in the case of $\Im A < 0$ we use another analogous result (which is just the complex conjugate of Eq.~\eqref{eq:schwinger1})
	\begin{equation}
		\label{eq:schwinger2}
		-\frac{i}{A} = \int_0^\infty ds (1 - i\epsilon)\, \exp(-i A(1 - i\epsilon) s)\,,
	\end{equation}	
	If $\Im A = 0$ and $\Re A > 0$, both equations are valid.
	
	\subsection{Calculation of the bubble master integral}
	\label{sec:bubmaster}
	
	We now have the necessary ingredients to detail the calculation of the bubble master integral defined in Eq.~\eqref{eq:bubmaster}. 
	The calculation is slightly different depending on whether the imaginary part of the masses in the denominator have the same sign or not. 
	Both cases are relevant for our decomposition of $P_{\rm lin}$. 
	
	\paragraph{Masses with the same imaginary part sign.}
	Looking at the bubble integral, we note that the imaginary parts of each term in the denominator is the same as the imaginary part of the corresponding mass. 
	Let us then first look at the case where both masses have a positive imaginary part. Using our identities, we obtain straightforwardly: 
	\begin{equation}
		\label{eq:bubble}
		\begin{split}
			&B_{\rm master}(k^2,M_1,M_2)= \\
			&- \int_0^\infty ds_1 \int_0^\infty ds_2\,\int (1 + i\epsilon_1)(1 + i\epsilon_2)\frac{d^3\vec{q}}{\pi^{3/2}} e^{i(q^2+M_1)(1 + i\epsilon_1)s_1}e^{i((\vec{k}-\vec{q})^2+M_2)(1 + i\epsilon_2)s_2}\,.
		\end{split}
	\end{equation}
	simplifying, we obtain: 
	\begin{equation}
		\begin{split}
			&B_{\rm master}(k^2,M_1,M_2)= - \int_0^\infty ds_1 \int_0^\infty ds_2\,\int \frac{d^3\vec{q}}{\pi^{3/2}} (1 + i\epsilon_1)(1 + i\epsilon_2)\\
			&\exp{i S_+ (-\vec{q}+ \frac{s_2(1 + i\epsilon_2)}{S_+}\vec{k})^2+i\frac{s_1 s_2(1 + i\epsilon_1)(1 + i\epsilon_2)}{S_+}k^2 + i (M_1 s_1(1 + i\epsilon_1) +M_2 s_2(1 + i\epsilon_2))}\,,
		\end{split}
	\end{equation}
	where $S_+= (1 + i\epsilon_1)s_1+(1 + i\epsilon_2)s_2$.
	
	We can now continue in two ways that give the same result:
	\begin{itemize}
		\item We do directly the Gaussian integral in $q$, then do a change of variable $s_1 = \tau x$, $s_2 = \tau (1-x)$ and then integrate in $\tau$, or alternatively 
		
		\item We do a change of variable $s_1 = \tau x$, $s_2 = \tau (1-x)$, integrate in $\tau$, and then do the $q$ integral in the standard dim reg. way.
	\end{itemize} 
	
	Let us pick the first way for concreteness.
	We can now set $\epsilon \equiv \epsilon_1s_1 + \epsilon_2s_2 > 0$, and we are guaranteed that the Gaussian integral converges (because we have $S_+ = s_1 + s_2 + i \epsilon$, as $\epsilon$ will be taken to zero at the end of the calculation).
	After doing the Gaussian $q$ integral, and setting $\epsilon \to 0$ where no poles show up, we get
	\begin{equation}
		\begin{split}
			&B_{\rm master}(k^2,M_1,M_2)= \\
			&  - \int_0^\infty ds_1 \int_0^\infty ds_2\,\frac{1}{(- i(s_1+s_2+ i \epsilon))^{3/2}} \exp{i\frac{s_1 s_2}{s_1+s_2+i\epsilon}k^2+ i (M_1 s_1 +M_2 s_2)}\,,
		\end{split}
	\end{equation}
	where we used the result
	\begin{equation}
		\label{eq:gauss}
		\int d^d q \exp(i a q^2) = \frac{\pi^{d/2}}{(-i a)^{d/2}}\,, \text{ if } \Im a > 0\, ,
	\end{equation}
	with $a = S_+$, and in particular $\Im(a) = \epsilon >0$.
	
	Now, doing the change of variables described before: $s_1 = \tau x$, $s_2 = \tau (1-x)$, and noting that the Jacobian of the transformation is $\tau$, we get
	\begin{equation}
		\begin{split}
			&B_{\rm master}(k^2,M_1,M_2)= \\
			& \qquad - \frac{1}{(- i)^{3/2}} \int_0^1 dx \int_0^\infty d\tau\, \tau (\tau+i \epsilon)^{-3/2}  \exp{i \frac{\tau^2}{\tau+i \epsilon} x (1-x)k^2+ i \tau (M_1 x +M_2 (1-x))}\,,
		\end{split}
	\end{equation}
	and since the integral is convergent when $\epsilon \to 0$, we can set $\epsilon=0$ and get:
	\begin{equation}
		\begin{split}
			& B_{\rm master}(k^2,M_1,M_2)= \\
			& \qquad - \frac{1}{(- i)^{3/2}} \int_0^1 dx \int_0^\infty d\tau\, \tau^{-1/2}  \exp{i \tau x (1-x)k^2+ i \tau (M_1 x +M_2 (1-x))}\,,
		\end{split}
	\end{equation}
	and finally performing the $\tau$ integral yields the integral:
	\begin{align}
		\label{eq:Bsame}
		\begin{split}
			B_{\rm master}(k^2,M_1,M_2)&= \frac{\Gamma(1/2)}{(- i)^{3/2}} \int_0^1 dx \frac{(-i)^{3/2}}{\sqrt{x (1-x)k^2+ M_1 x +M_2 (1-x)}}\\
			& = \sqrt{\pi}\int_0^1 dx \frac{1}{\sqrt{x (1-x)k^2+ M_1 x +M_2 (1-x)}} \,, 
		\end{split}
	\end{align}
	which is a standard Feynman integral.  
	Note that in this case the square root does not have any branch cut, because its argument always has a positive imaginary part, from our assumption on the masses $M_1$ and $M_2$. 
	We were thus able to find the Feynman parameter integral using Schwinger parameters for this case. 
	For two masses with negative imaginary parts, the exact same steps apply, and we obtain the same result.
	
	Solving this integral yields:
	\begin{equation}
		\label{eq:Bfinalsame}
		\begin{split}
			& B_{\rm master}(k^2,M_1,M_2) = \\
			& \frac{\sqrt{\pi}}{k} \left[ i \log \left(2 \sqrt{x (1-x) + m_1 x + m_2 (1-x)}+i (m_1-m_2-2 x+1)\right)\right]_{x=0}^{x=1} \\
			& \hspace{3cm}  \, - {\rm discontinuities}\,,
		\end{split}
	\end{equation}
	where $m_1 = M_1/k^2$ and $m_2 = M_2/k^2$. {Here and in the rest of the paper, the term `{\rm discontinuities}' means that, in order to use a given expression, we need to check if the argument of the antiderivative (in the case the argument of the $\log$ as a function of $x$) crosses any branch cut in the integration region (in this case the negative real axis). 
		If it does, using the antiderivative would erroneously add the size of the discontinuity to the integral, and so we need to add/subtract this amount from the formula of the integral in terms of the antiderivative (in this case $2 \pi i$), depending on the direction of the crossing.}

	\paragraph{Branch cut crossings.}
	We can analyze under what conditions a log branch cut crossing happens.
	Let us define the argument of the $\log$ as
	\begin{equation}
		\label{eq:A}
		A(x, m_1, m_2) \equiv 2 \sqrt{x (1-x) + m_1 x + m_2 (1-x)}+i (m_1-m_2-2 x+1)\,.
	\end{equation}
	Then, we have a branch cut crossing when $A(x, m_1, m_2) = -t$, where $t>0$, for some $x \in ]0,1[$.
	We now want to prove two statements: first, that there can be at most one branch cut crossing, and second that (there is a single branch cut crossing) iff ($\Im A(1,m_1,m_2) > 0$ and $\Im A(0,m_1,m_2) < 0$). 
	
\begin{figure}[H]
		\centering
		\includegraphics[width=.6\textwidth]{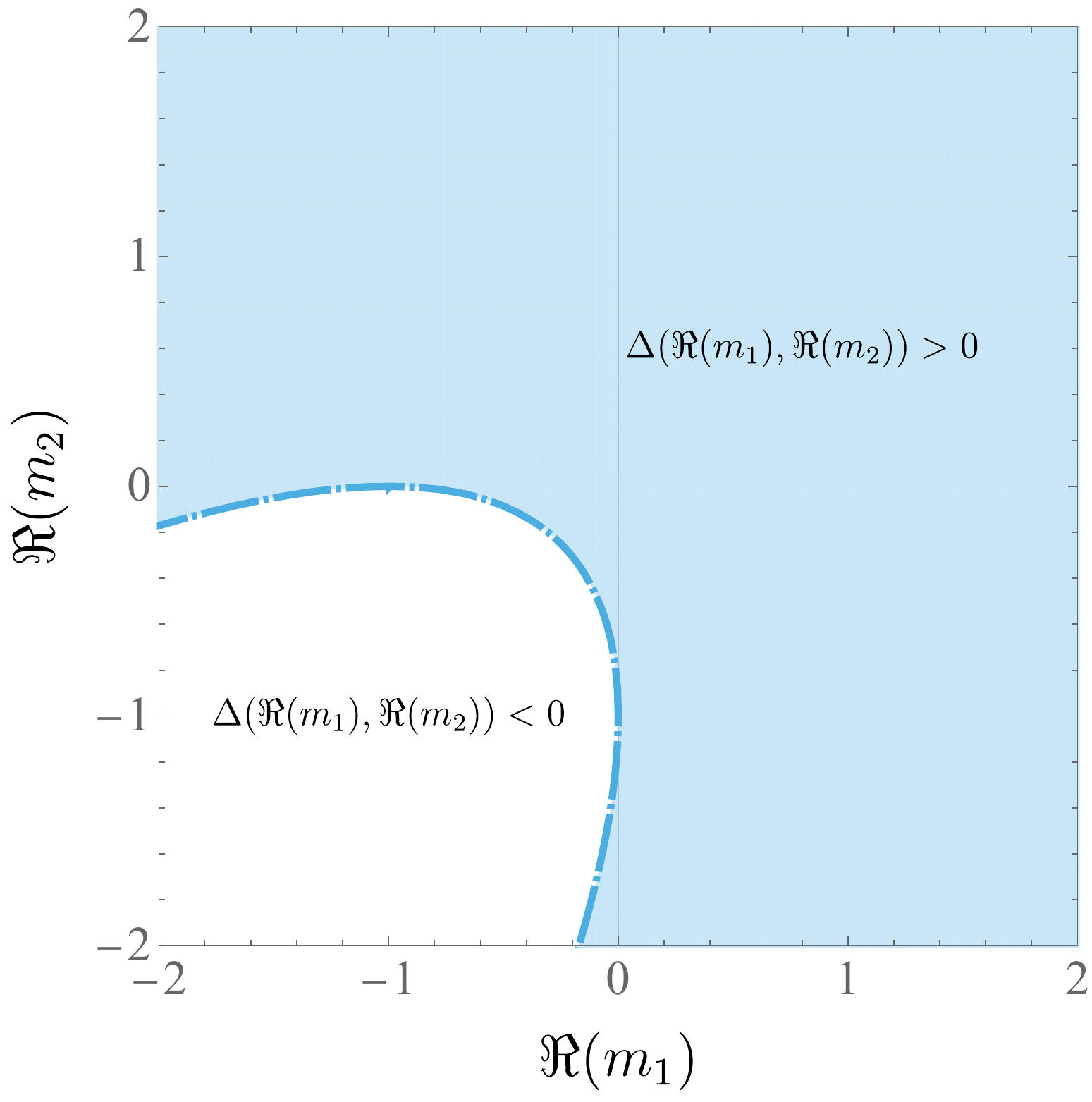}
		\caption{Region plot showing where $\Delta(\Re(m_1), \Re(m_2))>0$ as a function of $\Re(m_1)$ and $\Re(m_2)$. 
			All masses used in this paper satisfy this condition.}
		\label{fig:contour}
	\end{figure}

	Let us first prove that there can be at most only one branch cut crossing.
	Solving $A(x, m_1, m_2) = -t$ for $x$ yields
	\begin{align}
		& 2\sqrt{x (1-x) + m_1 x + m_2 (1-x)}+i (m_1-m_2-2 x+1) = -t \\
		\label{eq:At}\Rightarrow\, & 2\sqrt{x (1-x) + m_1 x + m_2 (1-x)} = -t - i (m_1 - m_2 - 2 x +1) \\
		\begin{split}
			\Rightarrow\, & 4 x (1-x) + 4 m_1 x + 4 m_2 (1-x) = \\
			& \qquad \qquad t^2 + 2 i t (m_1 - m_2 - 2 x +1) - (m_1 - m_2 - 2 x +1)^2 
		\end{split}
		\\
		\label{eq:deltaeq}
		\Rightarrow\, & \Delta(m_1, m_2) = t^2 + 2 i t (m_1 - m_2 + 1) - 4 i t x  \,,
	\end{align}
	where
	\begin{equation}
		\begin{split}
			\Delta(m_1, m_2) &\equiv m_1^2-2 m_1 m_2+2 m_1+m_2^2+2 m_2+1 \\
			& = (m_1-m_2)^2 + 2(m_1 + m_2) + 1\\
			& = (m_2 - m_1 + 1)^2 + 4 m_1 = (m_1 - m_2 + 1)^2 + 4 m_2\,.
			\label{eq:Delta}
		\end{split}
	\end{equation}
	Eq.~\eqref{eq:deltaeq} gives two constraints: one for the real part, and another for the imaginary part. 
	The real part equation gives us directly $t$
	\begin{equation}
		\label{eq:t}
		t_\pm = \Im(m_1)-\Im(m_2)\pm\sqrt{\Delta(\Re(m_1), \Re(m_2))}\,,
	\end{equation}
	where we need to impose $\Delta(\Re(m_1), \Re(m_2))>0$ in order for $t$ to be real.
	In Fig.~\ref{fig:contour}, we show the region where this condition is satisfied. 
	In particular, it is always satisfied if both $\Re m_1>0$ and $\Re m_2>0$, which is the case for the masses that we use in our decomposition. 
	
	Taking the real part of Eq.~\eqref{eq:At}, we see that $\Re(l.h.s.) > 0$ and that, plugging in $t_\pm$, that $\Re(r.h.s.) = \mp \sqrt{\Delta(\Re(m_1), \Re(m_2))}$, corresponding to $t_\pm$ respectively. 
	Therefore only the solution $t_-$ corresponds to a positive real part, and can satisfy the equation. We therefore discard $t_+$.

	The imaginary part equation gives us $x$ as a function of $t$:
	\begin{equation}
		x_t = \frac{1}{2} + \frac{\Re(m_1-m_2) (-\Im(m_1)+\Im(m_2)+t)-\Im(m_1+m_2)}{2 t}\,,
	\end{equation}
	and plugging $t=t_-$ yields
	\begin{align}
		& x_- = \frac{- 2 \Im(m_2) - (\Re(m_1)-\Re(m_2)+1) \sqrt{\Delta(\Re(m_1), \Re(m_2))}}{2 \left(\Im(m_1)-\Im(m_2)-\sqrt{\Delta(\Re(m_1), \Re(m_2))}\right)}\,. \label{eq:xminus}
	\end{align} 
	This means that there is at most one branch cut crossing of the $\log$ in Eq.~\eqref{eq:Bfinalsame}. 
	In fact, the branch cut crossing happens if $x_- \in ]0,1[$ and $t_- > 0$. 
	The explicit conditions on $m_1$ and $m_2$ for the branch cut can be found in Appendix~\ref{sec:appmasscond}. 
	However, there is a simpler summary formula that allows us to automatically account for the branch crossing, as we now discuss.
	
	Having proved that there can be at most one branch cut crossing, we now want to prove that (there is a single branch cut crossing) iff ($\Im A(1,m_1,m_2) > 0$ and $\Im A(0,m_1,m_2) < 0$).
	We divide the proof in two parts:
	\begin{enumerate}
		\item \label{proof1} (there is a branch cut crossing) $\Rightarrow$ ($\Im A(1,m_1,m_2) > 0$ and $\Im A(0,m_1,m_2) < 0$),
		\item \label{proof2} ($\Im A(1,m_1,m_2) > 0$ and $\Im A(0,m_1,m_2) < 0$) $\Rightarrow$ (there is a branch cut crossing).
	\end{enumerate}
	Before proceeding with the proof, let us derive an auxiliary result. 
	Assume that $A(x, m_1, m_2)$ defined in Eq.~\eqref{eq:A} is crossing the real axis, which implies $\Im A = 0$. 
	Writing $A(x, m_1, m_2)=-t$, where $t \in \mathbb{R}$, we can differentiate $A$ with respect to $x$, obtaining:
	\begin{equation}
		\label{eq:dAdx}
		\begin{split}
			\frac{dA}{dx} & = \frac{m_1-m_2-2 x+1}{\sqrt{x (m_1-m_2-x+1)+m_2}}-2 i \\
			& = \frac{m_1-m_2-2 x+1 - 2 i \sqrt{x (m_1-m_2-x+1)+m_2}}{\sqrt{x (m_1-m_2-x+1)+m_2}} \\
			& = -i \frac{A(x, m_1, m_2)}{\sqrt{x (m_1-m_2-x+1)+m_2}} \\
			& =  \frac{i t}{\sqrt{x (m_1-m_2-x+1)+m_2}}\,.
		\end{split}
	\end{equation}
	We can now proceed and prove statement \ref{proof1}.
	Assume that there is a branch cut crossing i.e. that $A(x, m_1, m_2)=-t$ and $t>0$. 
	In this case, since $\Re(\sqrt{z}) \geq 0$, we have that $\Re(t/\sqrt{x (m_1-m_2-x+1)+m_2}) \geq 0$. 
	Now, using Eq.~\eqref{eq:dAdx}, we get $\Im(dA/dx) > 0$ at the branch cut crossing, which shows that $A$ crosses the branch cut always from the negative imaginary plane to the positive imaginary plane, and thus by continuity that $\Im(A(1,m_1,m_2)) > 0$ and $\Im(A(0,m_1,m_2)) < 0$~(\footnote{{We remind the reader that since the masses have the same imaginary part, there is no branch-cut crossing for the square root that defines $A(x)$, and so $A(x)$ is continuous.}}).
	
	It remains to prove the converse: statement \ref{proof2}.
	If $\Im(A(1,m_1,m_2)) > 0$ and $\Im(A(0,m_1,m_2)) < 0$, that means by continuity that there was a crossing of the imaginary axis, at $x_{\rm cross}$.
	Moreover, at the branch cut, $\Im(dA/dx) > 0$. 
	Let us call $-t \equiv A(x_{\rm cross},m_1,m_2)$.  If $t>0$, there is a branch cut crossing. 
	Let us prove that this is the case. At $x=x_{\rm cross}$, and taking the imaginary part of Eq.~\eqref{eq:dAdx}, we get
	\begin{equation}
		\begin{split}
			& \Im\left(\frac{dA}{dx}\right) = t \Re\left(\frac{1}{\sqrt{x (m_1-m_2-x+1)+m_2}} \right) \\
			\Rightarrow\; & t = \frac{\Im\left(\frac{dA}{dx}\right)}{\Re\left(\frac{1}{\sqrt{x (m_1-m_2-x+1)+m_2}} \right)} >0\,.
		\end{split}
	\end{equation}
	Therefore, in this case $A$ crosses the real axis at the branch cut.
	We have then proven statement \ref{proof2}.
	
	This concludes the proof that 
	(there is a single branch cut crossing) $\Leftrightarrow$ ($\Im A(1,m_1,m_2) > 0$ and $\Im A(0,m_1,m_2) < 0$).
	
	This allows us to simply incorporate the branch cut crossings in Eq.~\eqref{eq:Bfinalsame}. 
	In fact, the correct result of the integration is that, for an infinitesimal path crossing the negative real axis from below to above, the result should be infinitesimally small. 
	But using our formula \eqref{eq:Bfinalsame} for the indefinite integral we would get $2\pi i$. 
	Given that we cross the branch-cut always from below to above, Eq.~\eqref{eq:Bfinalsame} can be made to take into account branch cut crossings by modifying it as
	\begin{align}
		\label{eq:Bfinalsamecut}
		\begin{split}
			B_{\rm master}(k^2,M_1,M_2)& =\frac{\sqrt{\pi}}{k} i [\log \left(A(1,m_1,m_2)\right) - \log \left(A(0,m_1,m_2)\right)\\
			& \qquad  -2\pi i H(\Im A(1, m_1, m_2)) H(-\Im A(0, m_1, m_2))]
			\,,
		\end{split}
	\end{align}
	where 
	\begin{align}
		& A(0, m_1, m_2) = 2 \sqrt{m_2}+i (m_1-m_2+1)\,, \\
		& A(1, m_1, m_2) = 2 \sqrt{m_1}+i (m_1-m_2-1)\,,
	\end{align}
	and $H(x)$ is the Heaviside step function extended at $x=0$ so that $H(0)=0$.	
	
	\paragraph{Masses with opposite imaginary part sign}
	
	Let us focus now on the case where the two masses have a different sign in the imaginary part. For concreteness, we assume $\Im M_1 > 0$ and $\Im M_2 < 0$.
	
	In this case we need non-zero $\epsilon$ insertions with Eqs.~\eqref{eq:schwinger1} and \eqref{eq:schwinger2} since we {would} develop poles without them.  
	We now need to write
	\begin{equation}
		\label{eq:bubblediff}
		\begin{split}
			&B_{\rm master}(k^2,M_1,M_2)= \\
			& \int_0^\infty ds_1 \int_0^\infty ds_2\,\int (1 + i\epsilon_1)(1 - i\epsilon_2)\frac{d^3\vec{q}}{\pi^{3/2}} e^{i(q^2+M_1)(1 + i\epsilon_1)s_1}e^{-i((\vec{k}-\vec{q})^2+M_2)(1 - i\epsilon_2)s_2}\, ,
		\end{split}
	\end{equation}
	which can be rearranged as 
	\begin{equation}
		\begin{split}
			&B_{\rm master}(k^2, M_1, M_2) = \int_0^\infty ds_1 ds_2\frac{d^3q}{\pi^{3/2}}(1 + i\epsilon_1)(1 - i\epsilon_2) \\
			\times &\exp\Big( i S \left( q + \frac{k(1 - i\epsilon_2)s_2}{S} \right)^2 
			- i\frac{k^2(1+i\epsilon_1) s_1 (1-i\epsilon_2) s_2}{S} + i\left( (1 + i\epsilon_1)s_1 M_1 \right.\\
			& \left. \qquad \qquad - M_2(1 - i\epsilon_2)s_2 \right) \Big),
		\end{split}
	\end{equation}
	where $S = (1 + i\epsilon_1)s_1 - (1-i\epsilon_2)s_2 \equiv s_1 - s_2 + i\epsilon${, where similarly to before we have redefined~$\epsilon$}. 
	Notice that without $\epsilon$ insertions, $S \rightarrow 0$ for $s_1 = s_2$. 
	Thus we retain $\epsilon$ insertions to shift the pole. 
	The $q$ integral will be convergent {for  $\epsilon>0$}. 
	Doing the momentum integral and taking $\epsilon \rightarrow 0$ where possible, we obtain
	\begin{align}
		B_{\rm master}(k^2, M_1,M_2) = \frac{1}{(-i)^{3/2}}\int ds_1ds_2\frac{1}{S^{3/2}}e^{iI}\,,
	\end{align}
	where $I = M_1 s_1 - M_2 s_2 - k^2 s_1s_2/S$. 
	Now we make the	change of variables, $s_1 = \tau x$ and $s_2 = \tau(1-x)$ and redefine $\epsilon \rightarrow \tau\epsilon$: 
	\begin{align}
		\begin{split}
			B_{\rm master}(k^2,M_1,M_2) =& \frac{1}{-i^{3/2}}\int^1_0dx\int^{\infty}_0d\tau \frac{\tau^{-1/2}}{(2x-1 + i\epsilon)^{3/2}} \\
			&\times \exp\left( i\tau\left(
			M_1 x - M_2(1-x) - \frac{k^2x(1-x)}{2x-1 + i\epsilon}
			\right)\right)\,.
		\end{split}
	\end{align}
	The $\tau$ integration then gives,
	\begin{align}
		\label{eq:preSP}
		B_{\rm master}(k^2,M_1,M_2) =   -\sqrt{\pi}\int^1_0dx \frac{1}{(2x - 1 + i\epsilon)^{3/2}}\frac{1}{\sqrt{-\frac{k^2x(1-x)}{2x-1+ i\epsilon} + M_1x - M_2(1-x)}}.
	\end{align}
	
	We will now make use of the Sokhotski-Plemelj (SP) theorem~\cite{Davies:1990fe}, which states,
	\begin{align}
		\label{eq:SPformula}
		\lim_{\epsilon \rightarrow 0^+}\int^b_a\frac{f(x,\epsilon)}{x - x_0 + i\epsilon} = P.V.\int^b_a\frac{\lim_{\epsilon \rightarrow 0^+}f(x,\epsilon)}{x - x_0}dx - \lim_{\epsilon \rightarrow 0^+}i\pi f(x_0,\epsilon)\, ,
	\end{align}
	for $f(x,\epsilon)$ being a continuous and non-singular function along the region of integration on the real line. 
	We choose $f(x,\epsilon) = \frac{1}{2}\frac{1}{\sqrt{2x - 1+ i\epsilon}\sqrt{-\frac{k^2x(1-x)}{2x-1+i\epsilon} + M_1x - M_2(1-x)}}$, and $x_0 = 1/2$. $f(x,\epsilon)$ satisfies the condition of being continuous and non-singular in the limit of $\epsilon \to 0^+$ even at $x = 1/2$. (in fact $\lim_{\epsilon \to 0^+}f(1/2,\epsilon) = -i$). We then obtain from Eq.~\eqref{eq:preSP},		
	\begin{align}
		\label{eq:PVintegral}
		\begin{split}
			&B_{\rm master}(k^2,M_1,M_2) = -\sqrt{\pi}\\
			&\times \left(P.V. \int^1_0dx \frac{1}{2x - 1}\frac{1}{\sqrt{2x - 1}\sqrt{-\frac{k^2x(1-x)}{2x-1} + M_1x - M_2(1-x)}}-\frac{i \pi}{2} \left(\frac{-2 i}{k}\right) \right)\,,
		\end{split}
	\end{align}
	where we have safely taken the limit $\epsilon \to 0$ in the second term of Eq.~\eqref{eq:PVintegral}. 	
	We can now take the expression in Eq.~\eqref{eq:PVintegral} and do a change of variable to put it in a more familiar form.
	Indeed, by introducing $\hat{x} \equiv \frac{x}{2 x - 1}$, and introducing another $\epsilon$ to explicitly implement the principal value, we get the mapping
	\begin{equation}
		x \in \left[0,\frac12 - \frac{\epsilon}{4}\right] \cup \left[\frac12 + \frac{\epsilon}{4}, 1\right] \Rightarrow \hat{x} \in \left[0, \frac12 - \frac{1}{\epsilon}\right] \cup \left[\frac12 + \frac{1}{\epsilon}, 1\right]\,,
	\end{equation}
	with $\pm\infty$ in $\hat{x}$ corresponding to $\frac{1}{2}^{\pm}$ in $x$ when we take the limit $\epsilon \rightarrow 0$~(\footnote{Notice that we write the interval as $ \left[\frac12 + \frac{1}{\epsilon}, 1\right]$ even if $\epsilon<1$ because this represents the mapping of the boundary of the one-dimensional integration: as an interval on the real axis, it should be written as $ \left[1,\frac12 + \frac{1}{\epsilon}\right]$. 
	}).  
	We will use this change of variable to avoid the pole at $x=1/2$.
	Applying this change of variable to Eq.~\eqref{eq:PVintegral} yields:
	\begin{equation}
		\label{eq: Bdiffsign}
		\begin{split}
			B_{\rm master}(k^2,M_1,M_2) = \sqrt{\pi}&\left(\int_0^{\frac12 - \frac{1}{\epsilon}} d\hat{x} \frac{i}{\sqrt{-\hat{x} (1-\hat{x})k^2 - M_1 \hat{x} - M_2 (1-\hat{x})}} + \right.\\ 
			&\left.\int_{\frac12 + \frac{1}{\epsilon}}^{1} d\hat{x} \frac{1}{\sqrt{\hat{x} (1-\hat{x})k^2+ M_1 \hat{x} +M_2 (1-\hat{x})}} + \frac{\pi}{k}\right)\,,
		\end{split}
	\end{equation}
	which is a very familiar integrand (see Eq.~\eqref{eq:Bsame}).
	
	We can further simplify this by noticing that, recalling the function $s$ defined in Eq.~\eqref{eq:s},
	\begin{equation}
		s(-1, \hat{x} (1-\hat{x})k^2+ M_1 \hat{x} +M_2 (1-\hat{x})) = 1
	\end{equation}
	if ${\hat{x}} < 0$, $\Im(M_1) > 0$, and $\Im(M_2) < 0$ (because the second argument of $s$ always has a negative imaginary part in this case).
	Therefore, we can simplify \eqref{eq: Bdiffsign}, obtaining
	\begin{equation}
		\label{eq:Bdiffsign2}
		\begin{split}
			B_{\rm master}(k^2,M_1,M_2) = \sqrt{\pi}&\left(\int_0^{\frac12 - \frac{1}{\epsilon}}  \frac{d\hat{x}}{\sqrt{\hat{x} (1-\hat{x})k^2 + M_1 \hat{x} + M_2 (1-\hat{x})}} +\right.\\
			&\left.\int_{\frac12 + \frac{1}{\epsilon}}^{1}  \frac{d\hat{x}}{\sqrt{\hat{x} (1-\hat{x})k^2+ M_1 \hat{x} + M_2 (1-\hat{x})}} + \frac{\pi}{k} \right)\,.
		\end{split}
	\end{equation}
	Now both components have the same integrand. Let us define
	\begin{align}
		\label{eq:I}
		&I \equiv \int_0^{\frac12 - \frac{1}{\epsilon}}  \frac{d\hat{x}}{\sqrt{\hat{x} (1-\hat{x}) + m_1 \hat{x} + m_2 (1-\hat{x})}} \,,\\ 
		\label{eq:II}
		&II \equiv \int_{\frac12 + \frac{1}{\epsilon}}^{1}  \frac{d\hat{x}}{\sqrt{\hat{x} (1-\hat{x}) + m_1 \hat{x} + m_2 (1-\hat{x})}}\,.
	\end{align}
	where again $m_1 = M_1/k^2$ and $m_2 = M_2/k^2$.
	Notice that we have
	\begin{equation}
		B_{\rm master}(k^2,M_1,M_2) = \frac{\sqrt{\pi}}{k}\left(I + II + \pi \right)\,.
	\end{equation}
	To alleviate the notation, we now redefine the integration variable $\hat{x}$ as $x$.
	As written above in Eq.~\eqref{eq:Bfinalsame}, the antiderivative corresponding to $I$ and $II$ is
	\begin{equation}
		\begin{split}
			G(x) &\equiv i \log \left(2 \sqrt{x (1-x) + m_1 x + m_2 (1-x)}+i (m_1-m_2-2 x+1)\right)\\
			&=i \log A(x, m_1, m_2)\,,
		\end{split}
	\end{equation}
	where we defined $G(x)$ {and $A(x, m_1, m_2)$ was defined in~(\ref{eq:A})}. 
	Then, we have for $I + II$:
	\begin{equation}
		\label{eq:Bdiffsign3}
		\begin{split}
			&I + II = G(1) - G(0) + i \lim_{\epsilon \to 0} \left(\log \left(\frac{1}{4} i \epsilon  \Delta(m_1, m_2) \right) - \log \left(-\frac{1}{4} i \epsilon  \Delta(m_1, m_2) \right)\right){- \,  {\rm discontinuities}}\,,
		\end{split}
	\end{equation}
	where $\Delta(m_1,m_2)$ is given by Eq.~\eqref{eq:Delta}. 
	{We remind that the meaning of `- discontinuities' is given below Eq.~(\ref{eq:Bfinalsame}).}
	Simplifying Eq.~\eqref{eq:Bdiffsign3} using Eq.~\eqref{eq:eta} with $a=-1, b=\frac{1}{4} i \epsilon  \Delta(m_1, m_2)$, yields
	\begin{equation}
		\label{eq:Bdiffsign4}
		\begin{split}
			I + II &= G(1) - G(0) + i \left(- \log(-1) -\eta(-1,\frac{1}{4} i \epsilon  \Delta(m_1, m_2)) \right){ - \,  {\rm discontinuities}}\\
			&= G(1) - G(0) + \pi - i \, \eta(-1, i \Delta(m_1, m_2)){ - \,  {\rm discontinuities}}  \,,
		\end{split}
	\end{equation}
	where $\eta$ is defined in Eq.~\eqref{eq:eta}. 
	So Eq.~\eqref{eq:Bdiffsign2} simplifies to
	\begin{equation}
		\label{eq:Bfinaldiff}
		B_{\rm master}(k^2,M_1,M_2) = \frac{\sqrt{\pi}}{k}\left[G(1) - G(0) + 2\pi - i \, \eta(-1, i \Delta(m_1, m_2)) \right]{- \,  {\rm discontinuities}}\,.
	\end{equation}
	The simplified expression for $i \eta(-1, i \Delta(m_1, m_2))$ is:
	\begin{equation}
		i \eta(-1, i \Delta(m_1, m_2)) = {+} 2  \pi \times \theta\left(i \Delta(m_1,m_2) \right)\,,
	\end{equation}
	where $\theta$ is defined in Eq.~\eqref{eq:theta}, and used that $\theta(a)^2=\theta(a)$. 
	In this way, we can rewrite Eq.~\eqref{eq:Bfinaldiff} as
	\begin{equation}
		\label{eq:Bfinaldiff2}
		B_{\rm master}(k^2,M_1,M_2) = {\frac{\sqrt{\pi}}{k}}\left[G(1) - G(0) + 2\pi (1 - \theta\left(i \Delta(m_1,m_2) \right)) \right]{ - \,  {\rm discontinuities}}\,.
	\end{equation}
	Note that Eq.~\eqref{eq:Bfinaldiff2} is very similar to Eq.~\eqref{eq:Bfinalsamecut}. 
	We will see next that if we take into account branch cut crossings the two equations are exactly identical.
	
	\paragraph{Branch cut crossings.}
	Now, as in the previous case, let us consider the branch cut crossings of the antiderivative that can occur in the integration region. 
	Following the steps of the previous case, where the imaginary part of the masses have the same sign, we verify that there can only be at most one branch cut crossing, happening at $x_-$ given by Eq.~\eqref{eq:xminus}.
	
	Knowing that there can be at most one branch cut crossing, we want to prove as in the last section that:
	(there is a single branch cut crossing) \textbf{iff} ($\Im A(1,m_1,m_2) > 0$ and $\Im A(0,m_1,m_2) < 0$).
	To prove this statement, let us look at how the imaginary part of $A(x,m_1,m_2)$ varies. 
	We first look at the $\pm \infty$ limits, corresponding to $\frac{1}{2} \pm \frac{1}{\epsilon}$ integration limits in Eqs.~\eqref{eq:I} and \eqref{eq:II}.
	Direct computation shows that (keeping in mind that $\Im(m_1)>0$ and $\Im(m_2)<0$)
	\begin{equation}
		\begin{split}
			&A\left(\frac{1}{2} - \frac{1}{\epsilon},m_1,m_2\right) = \frac{1}{4} i \Delta(m_1,m_2) \epsilon + {\cal{O}}(\epsilon^2) \,, \\
			&A\left(\frac{1}{2} + \frac{1}{\epsilon},m_1,m_2\right) = -\frac{1}{4} i \Delta(m_1,m_2) \epsilon + {\cal{O}}(\epsilon^2)\ .
		\end{split}
	\end{equation} 
	Thus, $\Im\left(A\left(\frac{1}{2} \mp \frac{1}{\epsilon},m_1,m_2\right)\right) = \pm \frac{\epsilon}{4} \Re(\Delta(m_1,m_2)) $. 
	We now want to analyze the sign of $t_-$ defined in Eq.~\eqref{eq:t} to check when branch cuts happen. 
	Recall that there is a branch cut crossing somewhere in the real line (but not necessarily in the integration region) when $t_->0$. 
	To investigate when $t_->0$, using the definition of $t_-$ given in Eq.~\eqref{eq:t}, we find that the condition to $t_->0$ is equivalent to the following: 			
	\begin{equation}
		\label{eq:branchcond}
		\begin{split}
			&t_- > 0 \\
			\Leftrightarrow \, & 0 < \sqrt{\Delta(\Re(m_1), \Re(m_2))} < \Im(m_1) - \Im(m_2) \\
			\Leftrightarrow \, & 0 < \sqrt{\Re(\Delta(m_1,m_2)) + (\Im(m_1-m_2))^2} < \Im(m_1) - \Im(m_2) \\
			\Leftrightarrow \, & 0 < \Re(\Delta(m_1,m_2)) + (\Im(m_1-m_2))^2 < (\Im(m_1-m_2))^2 \\
			\Leftrightarrow \, & -(\Im(m_1-m_2))^2 < \Re(\Delta(m_1,m_2))  < 0\,.
		\end{split}
	\end{equation}		
	
	Our next step is now to show that, if we carefully take branch cut crossings into account, that Eq.~\eqref{eq:Bfinalsamecut} is valid in this situation as well.
	We will consider two cases separately: $\Re(\Delta(m_1,m_2)) > 0$ and $\Re(\Delta(m_1,m_2)) < 0$.
	For each case, we will check both if there are branch cut crossings and how the $\theta$ function in Eq.~\eqref{eq:Bfinaldiff2} simplifies, so that we get a definite expression for $B_{\rm master}$.
	While considering both cases, it is important to have the following auxiliary result in mind: using the definition of $\Delta$ in Eq.~\eqref{eq:Delta}, we have
	\begin{equation}
		\Delta(\Re(m_1), \Re(m_2)) = \Re(\Delta(m_1,m_2)) + (\Im(m_1-m_2))^2\,.
	\end{equation}
	
	In the following, we will use the property $\Delta(\Re(m_1), \Re(m_2))>0$, which is satisfied for all our masses. 
	In Fig.~\ref{fig:contour}, we show the region where this condition is satisfied. 
	As mentioned before, in particular, it is always satisfied if both $\Re m_1>0$ and $\Re m_2>0$, which is the case in our decomposition.
	\begin{itemize}
		\item Case $\Re(\Delta(m_1,m_2)) > 0$: in this case
		$\Im\left(A\left(-\infty, m_1, m_2\right)\right) > 0$ and $\Im\left(A\left(+\infty, m_1, m_2\right)\right) < 0$. 
		Also, the condition in Eq.~\eqref{eq:branchcond} is not satisfied, meaning that there is no branch cut crossing. The sign of $\Im A$ is given by the following table:
		\begin{figure}[H]
			\centering
			\begin{tikzpicture}
				\tkzTabInit[lgt=3]{$x$ / 1 , ${\Im A(x,m_1,m_2)}$ / 1}{$-\infty$, $x_-$, $+\infty$}
				\tkzTabLine{, +, z, -, }
			\end{tikzpicture}
		\end{figure}
		
		Thus, since $\Re(\Delta(m_1,m_2)) > 0 \Rightarrow \theta(i \Delta(m_1,m_2)) = 1$,
		Eq.~\eqref{eq:Bfinaldiff2} is simply given by 
		\begin{equation}
			B_{\rm master}(k^2,M_1,M_2) = {\frac{\sqrt{\pi}}{k}}\left[G(1) - G(0) \right]\,.
		\end{equation}
		Therefore, since in this case, from looking at the sign of $\Im A$ displayed in the table, we can never have simultaneously $\Im A(1, m_1, m_2)>0$ and $\Im A(0, m_1, m_2)<0$, Eq.~\eqref{eq:Bfinalsamecut} is valid.
		
		\item Case $\Re(\Delta(m_1,m_2)) < 0$: in this case
		$\Im\left(A\left(-\infty, m_1, m_2\right)\right) < 0$ and $\Im\left(A\left(+\infty, m_1, m_2\right)\right) > 0$. 
		Since we are assuming $\Delta(\Re(m_1), \Re(m_2))>0$, the condition in Eq.~\eqref{eq:branchcond}  is satisfied, implying that there is a branch cut crossing in the real line (but not necessarily in the integration region).
		The sign of $\Im A$ is given by the following table:
		\begin{figure}[H]
			\centering
			\begin{tikzpicture}
				\tkzTabInit[lgt=3]{$x$ / 1 , ${\Im A(x,m_1,m_2)}$ / 1}{$-\infty$, $x_-$, $+\infty$}
				\tkzTabLine{, -, z, +, }
			\end{tikzpicture}
		\end{figure}
		There are three possibilities for the position of $x_-$: $x_- \leq 0$, $0 < x_- < 1$, $x_- \geq 1$. 
		For each one, remembering that $\Im\left(A\left(x_-, m_1, m_2\right)\right)=0$ and that $\Im\left(A\left(x, m_1, m_2\right)\right)$ only has one zero, we have: 
		\begin{align*}
			x_- \leq 0: &\quad \Im\left(A\left(0, m_1, m_2\right)\right) > 0 \text{ and } \Im\left(A\left(1, m_1, m_2\right)\right) > 0\,, \\
			0 < x_- < 1: &\quad \Im\left(A\left(0, m_1, m_2\right)\right) < 0 \text{ and } \Im\left(A\left(1, m_1, m_2\right)\right) > 0 \,,\\
			x_- \geq 1: &\quad \Im\left(A\left(0, m_1, m_2\right)\right) < 0 \text{ and } \Im\left(A\left(1, m_1, m_2\right)\right) < 0 \,.
		\end{align*}
		For $x_- \leq 0$ and $x_- \geq 1$, the branch cut crossing is in the integration region, so we need to take the crossing into account. 
		In these two cases, it amounts to subtracting $2\pi$ from the expression inside the square brackets in Eq.~\eqref{eq:Bfinaldiff2}. 
		Therefore, since $\Re(\Delta(m_1,m_2)) < 0 \Rightarrow \theta(i \Delta(m_1,m_2)) = 0$,  Eq.~\eqref{eq:Bfinaldiff2} for $x_- \leq 0$ and $x_- \geq 1$ is simply given by 
		\begin{equation}
			B_{\rm master}(k^2,M_1,M_2) = {\frac{\sqrt{\pi}}{k}}\left[G(1) - G(0)\right]\,.
		\end{equation}
		
		For $0 < x_- < 1$, there is no branch cut crossing in the integration region, so Eq.~\eqref{eq:Bfinaldiff} becomes
		\begin{equation}
			B_{\rm master}(k^2,M_1,M_2) = {\frac{\sqrt{\pi}}{k}}\left[G(1) - G(0) + 2\pi\right]\,.
		\end{equation}
	\end{itemize}
	Notice that in this particular case we have $\Im\left(A\left(0, m_1, m_2\right)\right) < 0$ and $\Im\left(A\left(1, m_1, m_2\right)\right) > 0$.
	
	Combining all these results, we find the remarkable result that we have the same expressions as in Eq.~\eqref{eq:Bfinalsamecut}, which is therefore valid regardless of the relative signs of $\Im(m_1)$ and $\Im(m_2)$:
	\begin{align}
		\label{eq:Bfinalcut}
		\begin{split}
			B_{\rm master}(k^2,M_1,M_2)& =\frac{\sqrt{\pi}}{k} i [\log \left(A(1,m_1,m_2)\right) - \log \left(A(0,m_1,m_2)\right)\\
			& \qquad  \qquad -2\pi i H(\Im A(1, m_1, m_2)) H(-\Im A(0, m_1, m_2))]
			\,,
		\end{split}
	\end{align}
	where 
	\begin{align}
		& A(0, m_1, m_2) = 2 \sqrt{m_2}+i (m_1-m_2+1)\,, \\
		& A(1, m_1, m_2) = 2 \sqrt{m_1}+i (m_1-m_2-1)\,,
	\end{align}
	provided that the condition $\Delta(\Re(m_1),\Re(m_2)) >0$ is satisfied.
	This interesting observation makes $B_{\rm master}(k^2,M_1,M_2)$ extremely efficient to evaluate numerically. 
	This last expression Eq.~\eqref{eq:Bfinalcut} hints, by its simplicity, at some closer relation between the case where the masses have the same sign of the imaginary part and where they have opposite signs. 
	Indeed, it can be proven using contour integration that the two cases are closely related. 
	This proof is given in the Appendix~\ref{sec:appcontour}.

	\subsection{Calculation of the triangle master integral}
	\label{sec:triangle_master}
	
	Let us compute the triangle master integral,
	\begin{align}
		T_{\rm master}(k^2_1, k^2_2, k_3^2, M_1, M_2, M_3) = \int \frac{d^3\vec{q}}{\pi^{3/2}}\frac{1}{((\vec{k}_1 - \vec{q})^2 + M_1) (q^2 + M_2)( (\vec{k}_2 +
			\vec{q})^2 + M_3)}\ .
	\end{align}
	The procedure will be similar to the bubble integral. 
	
	\subsubsection{Simplifying the master integral}
	
	For this integral, it is much simpler to work under the assumption that all masses have a positive real part (which is the case for our masses).
	So, in this section, we will consider $\left\{ \Re(M_1),\Re(M_2),\Re(M_3) \right\} > 0$~(\footnote{Note that we could also have considered this assumption from the beginning in the bubble integration. 
		However, the procedure presented there allows for masses that do not necessarily have a positive real part.
	}).
	The most general case is considered in the App.~\ref{sec:trianglediff} in order to accommodate {a more accurate} parametrization of the BAO wiggles.
	
	Let us then assume $\left\{ \Re(M_1),\Re(M_2),\Re(M_3) \right\} > 0$. 
	In this case, we can perform the integrations without having to keep track of the signs of the imaginary parts. 
	In fact, we can choose the following Schwinger parametrization,
	\begin{align}
		\frac{1}{A} = \int^{\infty}_0ds \exp(-As)
	\end{align}
	for $\Re(A)>0$. 
	In this case, the triangle master integral can be written as,
	\begin{align}
		\begin{split}
			&T_{\rm master}(k^2_1,k^2_2,M_1,M_2,M_3) = \int_0^{+\infty} ds_1 ds_2 ds_3 \\ 
			&\qquad \int \frac{d^3q}{\pi^{3/2}} \exp\left[ -\left((\vec{k}_1 - \vec{q})^2 + M_1\right)s_1 - (q^2 + M_2)s_2 - \left(
			(\vec{k}_2 - \vec{q})^2 + M_3 \right)s_3 \right]\ ,
		\end{split}
	\end{align}
	since if $ \Re(M_i)> 0$, then $\Re(A)>0$.  
	Expanding the exponent, we obtain
	\begin{equation}
		\begin{split}
			&-(s_1 + s_2 + s_3)\left( \vec{q}^2 - 2\vec{q}\cdot\frac{s_1\vec{k}_1 - s_3\vec{k}_2}{s_1 + s_2 + s_3} \right) - s_1k^2_1 - s_1M_1 - s_2M_2 - s_3k^2_2 - s_3M_3\\
			&= -(s_1 + s_2 + s_3)\left( \vec{q} - \frac{s_1\vec{k}_1 - s_3\vec{k}_2}{s_1 + s_2 + s_3} \right)^2 - \frac{s_1s_3k^2_3 + s_1s_2k^2_1 +
				s_2s_3k^2_2}{s_1 + s_2 + s_3} - s_1M_1 - s_2M_2 - s_3M_3\,.
		\end{split}
	\end{equation}
	
	We can perform the Gaussian integral in $\vec{q}$ as $s_1 + s_2 + s_3 > 0$, obtaining
	\begin{equation}
		T_{\rm master} = \int_0^{+\infty} ds_1 ds_2 ds_3\frac{e^{- I}}{(s_1 + s_2 + s_3)^{3/2}}\,,
	\end{equation}
	where $I = \frac{s_1 s_2 k^2_1 + s_1 s_3 k^2_3 + s_2 s_3 k^2_2}{s_1 + s_2 + s_3} + s_1 M_1 + s_2 M_2 + s_3 M_3$.
	
	Changing integration variables to $s_1 = \tau x_1, s_2 = \tau x_2, s_3 = \tau(1 - x_1 - x_2) = \tau x_3$, where $\tau \in [0,+\infty[$, $x_1 \in [0,1]$, $x_2 \in [0,1]$, and $x_1 + x_2 < 1$ and $x_3=1-x_2-x_1$, and observing that the Jacobian is $\tau^2$, we obtain
	\begin{equation}
		\label{eq:tausame}
		T_{\rm master} = \int_0^1 dx_1 \int_0^{1-x_1} dx_2 \int_0^{+\infty}d\tau \tau^{1/2}e^{-\tau \tilde{I}}\,,
	\end{equation}
	where we defined $\tilde{I} = \frac{x_1 x_2 k^2_1 + x_1 x_3 k^2_3 + x_2 x_3 k^2_2}{x_1 + x_2 + x_3} + x_1 M_1 + x_2 M_2 + x_3 M_3$. 
	Since $\Re \tilde{I}>0$, we can safely do the $\tau$ integration in Eq.~\eqref{eq:tausame}, getting
	\begin{equation}
		\label{eq:Tbeforesame}
		T_{\rm master} = \frac{\sqrt{\pi}}{2}\int_0^1 dx_1 \int_0^{1-x_1} dx_2 \tilde{I}^{-3/2}\,.
	\end{equation}
	
	Next, we make another change of variables $x_2  = (1 - x)y,\, x_1 = x$, obtaining
	\begin{equation}
		\label{eq:Tsame}
		\begin{split}
			T_{\rm master} &= {\frac{\sqrt{\pi}}{2}\int_0^1 dx dy (1-x)\hat{I}^{-3/2}}\\
			&= \frac{\sqrt{\pi}}{2}\int_0^1 \frac{dx dy (1-x)}{\left(-ay^2 + by + c \right)^{3/2}}\ , 
		\end{split}
	\end{equation}
	where 
	\begin{equation}
		\begin{split}
			\hat{I} \equiv  &M_1 x +  M_2 y (1-x) + M_3 (1-y) (1-x) + \\
			&k^2_1 x y (1-x) + k^2_2 (1-x)^2 y (1-y) + k^2_3 x (1-x)(1-y)\,,
		\end{split}
	\end{equation}
	$a = k_2^2 (1-x)^2$, $b = (1-x)(k_2^2 + M_2 - M_3+x(k_1^2-k_2^2-k_3^2))$, and $c = M_3(1-x) + k_3^2 x (1-x) + M_1 x$.
	Notice that $\hat{I} = -ay^2 + by + c$.
	Performing the indefinite integral in $y$, we obtain:
	\begin{equation}
		T_{\rm master}(k^2_1, k^2_2, k^2_3, M_1, M_2, M_3) =\frac{\sqrt{\pi}}{2}\int_0^1 dx \left.\frac{2(1-x) (2 a y - b)}{\left(b^2 + 4 a c\right) \sqrt{-ay^2 + by + c}}\right|^{y = 1}_{y = 0}\,,
	\end{equation}
	valid if $b^2 + 4 a c \neq 0$. If $b^2 + 4 a c = 0$, we obtain
	\begin{equation}
		T_{\rm master}(k^2_1, k^2_2, k^2_3, M_1, M_2, M_3) =\frac{\sqrt{\pi}}{2}\int_0^1 dx \left.\frac{2 (x-1)}{(b-2 a y) \sqrt{-\frac{(b-2 a y)^2}{a}}}\right|^{y = 1}_{y = 0}\,.
	\end{equation}
	{Here there are no discontinuities as we explain below.}
	Let us just consider the case $b^2 + 4 a c \neq 0$ as the case  $b^2 + 4 a c = 0$ has measure 0 in the $x$ integration, and so can be neglected.
	Replacing the values of $a$, $b$, and $c$, and rearranging, we obtain:
	\begin{align}
		\label{eq:param}
		T_{\rm master} = \frac{\sqrt{\pi}}{2}\int_0^1 dx \left.\frac{N_1 x + N_0}{\sqrt{R_2 x^2 + R_1 x + R_0}\left(S_2 x^2 + S_1 x + S_0\right)}\right|^{y = 1}_{y = 0}\,,
	\end{align}
	where $N_1, N_0, R_2, R_1, R_0, S_2, S_1, S_0$ are functions of $y$ evaluated at $ y = 1$ and $y = 0$, and independent of $x$, given by:
	\begin{align}
		\label{eq:trimaster_params}
		\begin{split}
			N_1 &= - 2 k_1^2+ 2 (1-2y) k_2^2 + 2 k_3^2 \,,\\
			N_0 &= -2 M_2 + 2 M_3 + 2 k_2^2 (-1+2y) \,,\\
			R_2 &= k_3^2 (-1 + y) - k_1^2 y + k_2^2 (1 - y) y\,,\\
			& = k_3^2 (-1 + y) - k_1^2 y \,,\\
			R_1 &= k_1^2 y+ 2 k_2^2 y\left(y-1\right)+(1-y) k_3^2 + M_1 - M_2 y+M_3 (y-1) \\
			&= k_1^2 y +(1-y) k_3^2 + M_1 - M_2 y+M_3 (y-1)\,,\\
			R_0 &= M_3 (1-y) + k_2^2 y (1-y) + M_2 y \\
			&= M_3 (1-y) + M_2 y \,,\\
			S_2 &= k_1^4-2 k_1^2 (k_2^2+k_3^2)+(k_2^2-k_3^2)^2 \,,\\
			S_1 &= 2 \left(k_1^2 (k_2^2+M_2-M_3)-k_2^4+k_2^2 (k_3^2+2 M_1-M_2-M_3)+k_3^2 (M_3-M_2)\right) \,,\\
			S_0 &= k_2^4+2 k_2^2 (M_2+M_3)+(M_2-M_3)^2\,,
		\end{split}
	\end{align}
	where for the second equality in $R_2$, $R_1$, and $R_0$ we have used the fact that $y$ is either 1 or 0 (so we eliminate terms that vanish in both cases).
	Notice that, before eliminating terms that vanish for $y=0$ and $y=1$, one has $\hat{I} = R_2 x^2 + R_1 x + R_0$.
	Note {also} that the integrand in Eq.~\eqref{eq:param} contains $\sqrt{\hat{I}}$, and so does not cross any branch cut in each integration region as $\textrm{sign}(\Im(\hat{I}))$ is constant as $\textrm{sign}\Im(\tilde{I})$ is constant.
	So, to calculate the $y$ integral for a general $x$ in each region, it suffices to take the difference in the $y$ antiderivative.
	
	Before integrating $T$ we can put it in a simplified form, by factoring the second order polynomials: 
	\begin{align}
		\label{eq:Tsamefinal}
		\begin{split}
			T_{\rm master} &=  \frac{\sqrt{\pi}}{2}\int_0^1 dx \left. \frac{N_1 x + N_0}{\sqrt{R_2(x - z_+)(x - z_-)}S_2(x - x_+)(x - x_-)} \right|^{y = 1}_{y = 0}\\
			&= \frac{\sqrt{\pi}}{2}\int_0^1 dx \left. \frac{\left(\frac{N_0+N_1 x_+}{S_2(x - x_+)(x_+ - x_-)} - \frac{N_0+N_1 x_-}{S_2(x - x_-)(x_+ - x_-)}\right)}{\sqrt{R_2(x - z_+)(x - z_-)}} \right|^{y = 1}_{y = 0} \\
			&= \frac{\sqrt{\pi}}{2}\int_0^1 dx \left( \frac{c_1}{\sqrt{R_2(x - z_+)(x - z_-)}(x-x_+)} \right.\\
			&\hspace{5cm} \left.\left. +\frac{c_2}{\sqrt{R_2(x - z_+)(x - z_-)}(x-x_-)} \right)\right|^{y = 1}_{y = 0} \,, 
		\end{split}
	\end{align}
	where $z_{\pm} = -\frac{R_1}{2R_2} \pm \frac{\sqrt{R^2_1 - 4 R_0R_2}}{2R_2}$, $x_{\pm} = -\frac{S_1}{2S_2} \pm \frac{\sqrt{S^2_1 - 4 S_0S_2}}{2S_2}$, $c_1 = \frac{N_0+N_1 x_+}{S_2(x_+ - x_-)}$, and $c_2 = - \frac{N_0+N_1 x_-}{S_2(x_+ - x_-)}$. 
	We define $F_{\rm int}$ as 
	\begin{equation}
		\label{eq:Fint}
		F_{\rm int}(R_2, z_+, z_-, x_0) = \frac{\sqrt{\pi}}{2}\int_0^1 dx \frac{1}{\sqrt{R_2(x - z_+)(x - z_-)}(x-x_0)}\,,
	\end{equation}
	where $x_0$ can take on $x_\pm$.
	We can then write $T_{\rm master}$ as 
	\begin{align}  
		T_{\rm master} &= \left[c_1 F_{\rm int}(R_2, z_+, z_-, x_+) + c_2 F_{\rm int}(R_2, z_+, z_-, x_-)\right]^{y = 1}_{y = 0}\,.
	\end{align}
	
	Note that this approach is only valid for $S_2 \neq 0$. The case $S_2=0$ corresponds to totally flat triangles satisfying $|k_3| = |k_1| \pm |k_2|$, which are observationally uninteresting because any binning will force the evaluation of non-flat triangles~(\footnote{In the code that we publicly release with this paper, the case of flat triangles is in practice included by adding $0.00001\hinvMpc$ to $k_3$.}).
	The evaluation of $F_{\rm int}$ is discussed in Section~\ref{sec:Fint}. 
	
	\subsubsection{Derivation of $F_{\rm int}$}
	\label{sec:Fint}
	
	We now derive a closed form expression for the function $F_{\rm int}$ defined in Eq.~\eqref{eq:Fint}, that we rewrite here for convenience:
	\begin{equation}
		F_{\rm int}(R_2, z_+, z_-, x_0) = \frac{\sqrt{\pi}}{2}\int_0^1 dx \frac{1}{\sqrt{R_2(x - z_+)(x - z_-)}(x-x_0)}\,,
	\end{equation}
	where $R_2$ is a negative real, $z_+$, $z_-$, and $x_0$ are in general complex numbers. 
	
	This integral only makes sense if the square root in the integrand does not cross any branch cut. Thus, we will separate the square root using our formula. 
	\begin{equation}
		\label{eq:Fint2}
		F_{\rm int}(R_2, z_+, z_-, x_0) = \frac{\sqrt{\pi}}{2} \int_0^1 dx \frac{s(z_+-x,x-z_-)}{\sqrt{|R_2|}\sqrt{(z_+ - x)}\sqrt{(x - z_-)}(x-x_0)}\,.
	\end{equation}
	Under our parametrization of the masses, $s(z_+-x,x-z_-)$ is constant which means we can take $s(z_+-x,x-z_-) = s(z_+,-z_-)$.
	This can be seen the following way: on the one hand, since $\Re M_i >0$, we have $\Re (R_2(x - z_+)(x - z_-)) > 0$ if $0<x<1$ both for $y=0$ and $y=1$. 
	This means that $\sqrt{R_2(x - z_+)(x - z_-)}$ cannot cross any branch cut.
	On the other hand, for fixed $y$ (and so fixed $z_\pm$), both $(z_+ - x)$ and $(x-z_-)$ have a constant imaginary part sign and so do not cross any branch cut.
	Since $\sqrt{(z_+ - x)(x - z_-)} = s(z_+-x,x-z_-) \sqrt{(z_+ - x)}\sqrt{(x - z_-)}$, these observations imply that $s(z_+-x,x-z_-)$ is constant~\footnote{ 
		Noticing that $R_2(x-z_-)(x-z_+) = \hat{I}$, we have that
		\begin{align}
			\Re(R_2(x-z_-)(x-z_+)) &=\Re(M_1x + M_3(1-y)(1-x) + k^2_3(1-x)(1-y)x \nonumber \\
			&+ M_2 y (1-x) + (1-x)y(k^2_1 x + k^2_2(1-x) (1-y))) > 0\,.
		\end{align}
		Remembering that we are using masses that have a positive real part, each term in the equation above is positive for $0<x<1$ and $0<y<1$. 
		Thus, $\sqrt{R_2(x-z_-)(x-z_+)}$ cannot cross any branch cut in the region $0<x<1$
		and $0<y<1$. 
		Since $\sqrt{R_2(x-z_-)(x-z_+)}=\sqrt{|R_2|}\sqrt{(z_+-x)(x-z_-)}$, also $\sqrt{(z_+-x)(x-z_-)}$ has no branch cut crossing. 
		Now, since there are no branch cut crossing for any $0<y<1$, for the purpose of evaluating the function $s(,)$, we can fix $y$ such that $z_{\pm}$ is fixed. 
		If $z_{\pm}$ is fixed, the imaginary part of $z_+ - x$ and $x - z_-$ are also fixed. 
		Hence, since $s(,)$ only depends on the imaginary parts of the arguments, $s(z_+ - x, x-z_-)$ is
		also constant. 
	}.
	Integrating yields:
	\begin{equation}
		\label{eq:Fintexpr}
		F_{\rm int}(R_2, z_+, z_-, x_0) = s(z_+,-z_-)\frac{\sqrt{\pi} }{\sqrt{|R_2|}} \left.\frac{\arctan\left(\frac{\sqrt{z_+-x}\sqrt{x_0 - z_-}}{\sqrt{x_0-z_+}\sqrt{x-z_-}}\right)}{\sqrt{x_0 - z_+}\sqrt{x_0 - z_-}}\right|_{x=0}^{x=1}{ - \,  {\rm discontinuities}}\,,
	\end{equation}
	where we remind that the definition of `$-$discontinuities' is given below Eq.~\eqref{eq:Bfinalsame}.
	This would be the final result if $\arctan$ did not have any branch cuts, and if $F_{\rm int}$ had no indeterminacies. 
	We now outline how to incorporate possible branch cut crossings, and later how to incorporate possible indeterminate results in Eq.~\eqref{eq:Fintexpr}.
	
	\subsubsection{Branch cut crossings}
	\label{sec:Fintcut}
\begin{figure}[ht]
		\centering
		\includegraphics[scale = 0.5]{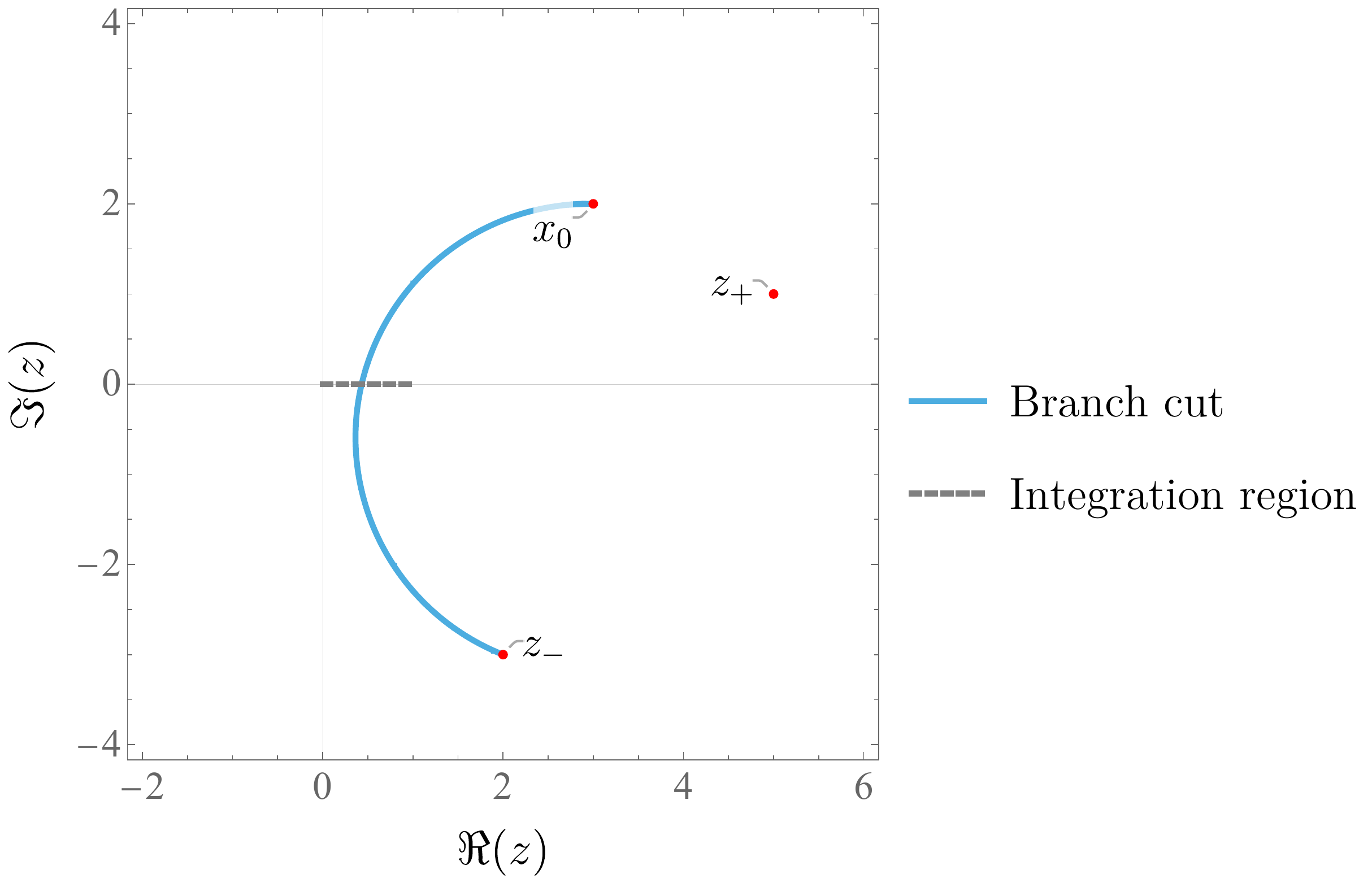}
		\caption{Arc described by $B\leq 0$, where $B$ is defined in Eq.~\eqref{eq:B}, and integration region of Eq.~\eqref{eq:Fint2}. A branch cut crossing happens at the intersection of the two lines. In this plots, we used $z_+ = 1 + i$, $z_- = 2 - 3i$, and $x_0 = 3+2i$.}
		\label{fig:Bcircle}
	\end{figure}

	Let us start by analyzing the branch cuts of $\arctan$. 
	There are two of them, both in the imaginary axis. 
	The first goes from $i$ to $+ i \infty$ and the second goes from $-i$ to $- i \infty$. The discontinuity works as follows: 
	\begin{equation}
		\begin{split}
			\lim_{\epsilon\to 0} \arctan(x\, i ) -\arctan(x\, i - \epsilon) = \pi\,, \, |x|>1\,,\\
			\lim_{\epsilon\to 0} \arctan(x\, i +\epsilon) -\arctan(x\, i - \epsilon) = \frac{\pi}{2}\,, \, |x|=1\,,
		\end{split}
	\end{equation}
	where we use continuity from the positive real direction.
	So we need to find when the argument of the $\arctan$ intersects a branch cut. 
	Let us define, for complex $z$,
	\begin{equation}
		\label{eq:Adef}
		A(z, z_+, z_-, x_0) \equiv \frac{\sqrt{z_+-z}\sqrt{x_0 - z_-}}{\sqrt{x_0-z_+}\sqrt{z-z_-}}\,.
	\end{equation}
	As was said above, the branch cuts are $A \in (-i\infty,-i] \cup [i, i\infty) $, which is equivalent to $A^2 \leq -1$.
	Defining $B \equiv A^2 + 1$, the condition for a branch cut crossing can be written $B \leq 0 \Leftrightarrow (\arg(B) = \pi) \cup (B=0)$.
	$B$ is given by 
	\begin{equation}
		\label{eq:B}
		B = \frac{(x_0 - z)(z_+ - z_-)}{(x_0-z_+)(z-z_-)} \,,
	\end{equation}
	so that 
	\begin{align}
		\label{eq:argcrit}
		\arg(B) = \pi \Leftrightarrow  \arg\left(\frac{z-x_0}{z-z_-}\right) = \pi + \arg\left(\frac{z_+-x_0}{z_+-z_-}\right)\; \mod 2\pi\,,
	\end{align}
	which, as we now explain, describes the arc of the circle defined by $z_+$, $z_-$, and $x_0$, that ends in $x_0$ and $z_-$ and does {\it not} include $z_+$, as shown in Fig.~\ref{fig:Bcircle}.
	
	In fact, a result from geometry is that an arc of a circle that contains points $(ABC)$ where $A$ and $B$ are the end points and $C$ belongs to the arc has a property that the angle between the oriented segment $\overrightarrow{AC}$ and $\overrightarrow{BC}$ (let us call it $\alpha$) is constant along the arc.
	Moreover, the arc of the same circle that does not contain $C$ has a property that the angle 
	between $\overrightarrow{AD}$ and $\overrightarrow{BD}$, with $D$ belonging to this arc, is constant and equal to $\pi + \alpha$.
	Eq.~\eqref{eq:argcrit} {indicates} that the angle between the segment $\overrightarrow{ x_0\,z}$ and the segment $\overrightarrow{ z_-\,z}$ is constant and equal to $\pi+\beta$, with $\beta$ being the angle between the segment $\overrightarrow{x_0\,z_+}$ and the segment $\overrightarrow{ z_-\,z_+}$.
	Therefore, $z$ describes the arc of the circle defined by the points $x_0$, $z_-$ and $z_+$ that does not contain $z_+$.  
	
	A branch cut crossing happens if the arc intersects the region of integration in Eq.~\eqref{eq:Fint2}, which is the segment $\Re(z) \in [0,1] \cap \Im(z)=0$.
	To calculate the possible intersection points with this segment, we obtain the equation of the full circle (not just the arc), using the following expression~\footnote{
		Eq.~\eqref{eq:circledet} is a standard result to write the equation of a circle as a function of three points that constitute that circle. 
		You can see it the following way:
		\begin{itemize}
			\item The expression $a  (x^2 + y^2) + b\,x + c\,y + d = 0$ uniquely defines a circle for $a \neq 0$.
			\item The determinant expression in Eq.~\eqref{eq:circledet} represents such an expression.
			\item Points $z_+$, $z_-$ and $x_0$ satisfy the equation by construction (the determinant is zero because two rows are identical).
			\item Therefore, Eq.~\eqref{eq:circledet} represents the equation of a circle that passes by $z_+$, $z_-$ and $x_0$.
		\end{itemize}
	}
	\begin{equation}
		\label{eq:circledet}
		\begin{vmatrix}
			x^2 + y^2 & x & y & 1 \\
			|z_+|^2 & \Re z_+ & \Im z_+ & 1  \\
			|z_-|^2 & \Re z_- & \Im z_- & 1  \\
			|x_0|^2 & \Re x_0 & \Im x_0 & 1  \\
		\end{vmatrix}
		=0\,,
	\end{equation}
	where $x=\Re z$ and $y=\Im z$.
	Then, we set $y=0$ and get a quadratic equation for $x$, $a x^2 + b x + c = 0$, where 
	\begin{align}
		\begin{split}
			a =& \Im(z_-) (\Re(z_+)-\Re(x_0))+\Im(z_+) (\Re(x_0)-\Re(z_-)) \\
			& \hspace{8cm} +\Im(x_0) (\Re(z_-)-\Re(z_+))\,,
		\end{split}\\
		\begin{split}
			b =& \Im(x_0) \left(\Im(z_+)^2-\Im(z_-)^2+\Re(z_+)^2-\Re(z_-)^2\right) \\
			&+\Im(z_-) \left(\Re(x_0)^2-\Re(z_+)^2\right) +\Im(z_+) \left(\Im(z_-)^2-\Re(x_0)^2+\Re(z_-)^2\right)\\
			&+\Im(x_0)^2 (\Im(z_-)-\Im(z_+))-\Im(z_+)^2 \Im(z_-) \,,
		\end{split}\\
		\begin{split}
			c =& \Im(z_+)^2 (\Im(z_-) \Re(x_0)-\Im(x_0) \Re(z_-)) \\
			&+\Im(z_+) \left(\Re(z_-) \left(\Im(x_0)^2-\Re(x_0) \Re(z_-)+\Re(x_0)^2\right)-\Im(z_-)^2 \Re(x_0)\right)\\
			&+\Re(z_+) (\Im(x_0) (\Im(z_-)^2-\Re(z_+) \Re(z_-)+\Re(z_-)^2)\\
			&+\Im(z_-) \Re(x_0) (\Re(z_+)-\Re(x_0))-\Im(x_0)^2 \Im(z_-))\,.
		\end{split}
	\end{align}
	Next, we calculate the discriminant $\textrm{Disc} = b^2 - 4 a c$. 
	If $\textrm{Disc} \leq 0$, there are no branch cut crossings for real $x$. 
	If $\textrm{Disc}>0$, there are two crossings $x_1$ and $x_2$ between the full circle and the real line. 
	Then, we check if $x_1$ and $x_2$ are between 0 and 1. 
	If $x_1$ and/or $x_2$ lie in [0,1], then we check Eq.~\eqref{eq:argcrit} to verify that the point belongs to the correct arc.
	Now, after this, we can have 3 cases:
	\begin{itemize}
		\item No branch cut crossings: then we just use Eq.~\eqref{eq:Fintexpr} to evaluate $F_{\rm int}$. 
		\item 1 branch cut crossing: we add/subtract $\pi$ to the $\arctan$ in Eq.~\eqref{eq:Fintexpr}, if the sign of $\Re \frac{dA}{dx}$ at the crossing is negative/positive, respectively. 
		\item 2 branch cut crossings: in this case, we always return to the original branch of the $\arctan$, so there is no need to add anything to Eq.~\eqref{eq:Fintexpr}.
	\end{itemize}
	
	Finally, there is still the case $B=0$, which holds only for $z=x_0$. 
	This means that if $x_0$ is real and  $0<x_0<1$, then we have an extra branch cut crossing (that only adds/subtracts $\frac{\pi}{2}$ instead of $\pi$, if the sign of $\Re \frac{dA}{dx}$ is negative/positive respectively). 
	
	Now that we have analyzed branch cut crossings, let us look at possible indeterminate expression in Eq.~\eqref{eq:Fintexpr}.
	
	\paragraph{Indeterminate expressions in Eq.~\eqref{eq:Fintexpr}.}
	Let us consider the following part of Eq.~\eqref{eq:Fintexpr}:
	\begin{equation}
		Z(x,z_+,z_-,x_0) \equiv \frac{\arctan\left(\frac{\sqrt{z_+-x}\sqrt{x_0 - z_-}}{\sqrt{x_0-z_+}\sqrt{x-z_-}}\right)}{\sqrt{x_0 - z_+}\sqrt{x_0 - z_-}}\,.
	\end{equation}
	In our parametrization of the masses, it is possible that the argument of $\arctan$, $A(x,z_+,z_-,x_0)$ as defined in Eq.~\eqref{eq:Adef}, diverges, or that the whole expression gives an indeterminate or otherwise problematic expression. 
	Here, we will only consider problems that appear for our masses.
	Let us see how to handle these cases.
	
	First, we look at divergences in $A(x,z_+,z_-,x_0)$.
	Divergences can appear in two situations: when $x=z_-$ (which is relevant only for $z_-=0$ or $1$ because this is where we evaluate the antiderivative) and when $x_0 = z_+$.
	Regarding the first case, we know that the integration variable $x$ runs from 0 to 1 in the $F_{\rm int}$ integral, and we know that, for a general complex $z$: 
	\begin{align}
		&\lim_{X \to 0^+} \arctan\left(\frac{z}{X}\right) = \frac{\sqrt{z^2}\pi}{2 z} \label{eq:limit1}\\
		&\lim_{X \to 0^+} \arctan\left(\frac{z}{i X}\right) = i \frac{\sqrt{-z^2}\pi}{2 z}\,. \label{eq:limit2}
	\end{align} 
	The problematic values of $z_-$ are 0 and 1 because they are the values at which we calculate the antiderivative. 
	Let us consider the two cases.
	When $z_- = 0$, we have for a general function $f$, $\lim_{x \to 0^+} f(\sqrt{x-z_-}) = \lim_{X \to 0^+} f(X)$. 
	When $z_- = 1$, we have $\lim_{x \to 1^-} f(\sqrt{x-z_-}) = \lim_{X \to 0^+} f(i X)$. 
	Therefore when $z_- = 0$ we use Eq.~\eqref{eq:limit1} with $z = \frac{\sqrt{z_+} \sqrt{x_0}}{\sqrt{x_0-z_+}}$, giving 
	\begin{equation}
		\begin{split}
			\lim_{x\to 0^+} Z(x,z_+,0,x_0) &= \frac{\pi}{2} \sqrt{\left(\frac{\sqrt{z_+} \sqrt{x_0}}{\sqrt{x_0-z_+}}\right)^2}\frac{\sqrt{x_0-z_+}}{\sqrt{z_+} \sqrt{x_0}}\frac{1}{\sqrt{x_0-z_+}\sqrt{x_0}} \\
			&= \frac{\pi}{2} \sqrt{\frac{z_+\, x_0}{x_0-z_+}}\frac{1}{\sqrt{z_+} x_0}\,.
		\end{split}	
	\end{equation}
	
	When $z_- = 1$ we use Eq.~\eqref{eq:limit2}, with $z = \frac{\sqrt{z_+-1} \sqrt{x_0-1}}{\sqrt{x_0-z_+}}$, obtaining
	\begin{equation}
		\begin{split}
			\lim_{x\to 1^-} Z(x,z_+,1,x_0) &= i\frac{\pi}{2} \sqrt{-\left(\frac{\sqrt{z_+-1} \sqrt{x_0-1}}{\sqrt{x_0-z_+}}\right)^2}\frac{\sqrt{x_0-z_+}}{\sqrt{z_+-1} \sqrt{x_0-1}} \frac{1}{\sqrt{x_0-z_+}\sqrt{x_0-1}}\\
			&= i\frac{\pi}{2} \sqrt{-\frac{(z_+-1) (x_0-1)}{(x_0-z_+)}}\frac{1}{\sqrt{z_+-1} (x_0-1)} \,.
		\end{split}
	\end{equation}
	
	Before discussing the case $x_0 = z_+$, let us analyze the case $x_0 = z_-$.
	We can notice an indetermination of the type $0/0$ in $Z(x,z_+,z_-,x_0)$.
	We can use L'H\^{o}pital's rule in this case (which here is equivalent to using Taylor approximation for $\arctan$ for a small argument $z$: $\arctan(z) \approx z$), obtaining
	\begin{align}\nonumber
		\lim_{x_0\to z_-}Z &=  \lim_{x_0\to z_-} \frac{\arctan\left( \frac{\sqrt{z_+ - x}\sqrt{x_0 -
					z_-}}{\sqrt{x_0 - z_+}\sqrt{x - z_-}}\right)}{\sqrt{x_0 - z_+}\sqrt{x_0 -
				z_-}}=  \lim_{x_0\to z_-}\frac{\frac{\sqrt{z_+ - x}\sqrt{x_0 - z_-}}{\sqrt{x_0 -
					z_+}\sqrt{x - z_-}}}{\sqrt{x_0 - z_+}\sqrt{x_0 - z_-}}\\
		&=\lim_{x_0 \to z_-} \frac{\sqrt{z_+ - x}}{{(x_0 - z_+)}\sqrt{x-z_-}}
		= \frac{\sqrt{z_+ - x}}{(z_- - z_+)\sqrt{x-z_-}}
	\end{align}
	where in the second step we use $\lim_{z\to 0}\arctan(z) = z$. 
	
	Finally, let us discuss the case $x_0 = z_+$. 
	Note that, in this case, $Z(x,z_+,z_-,x_0)$ diverges, so it is unusable to calculate the integral in Eq.~\eqref{eq:Fint2}. 
	The solution starts by noticing that $F_{\rm int}$ in Eq.~\eqref{eq:Fint2} is symmetric under $z_+ \leftrightarrow z_-$. 
	In fact, notice that $z_+$ and $z_-$ are just roots of the quadratic equation that shows up in the integrand in $x$.
	Therefore, all of our derivation remains the same if we defined
	$\tilde{z}_+ = z_-$ and $\tilde{z}_- = z_+$, i.e. just swapping which root
	we call $z_{\pm}$. 
	Hence, the $Z$ function is still valid if we chose to swap and call $z_+$ what we called $z_-$ previously and vice versa for $z_-$.   
	Naively the $\arctan$ function will diverge in the case of $x_0 = z_+$,
	but we can swap $z_+$ with $z_-$ such that we end up with the case of $x_0 = z_-$ instead and then proceed with the L'H\^{o}pital as before. 
	
	\paragraph{Note on $s(z_+,-z_-)$ function in $F_{\rm int}$.}
	In the expression for $F_{\rm int}$ in Eq.~\eqref{eq:Fintexpr}, there is the factor $s(z_+,-z_-)$. 
	This factor is undefined if either $z_+$ and/or $z_-$ vanish. 
	In that case, we use Eq.~\eqref{eq:Fint2} to set the value of $s$, taking an infinitesimally small $x>0$. 
	For example, when $z_+=0$, and $z_- \neq 0$, we get $s(-x, -z_-)$. 
	When $z_+=z_-=0$, we obtain $s(-x, x)=1$.

	\section{One-loop integrals for all $N$-point correlation functions in the EFTofLSS\label{sec:all-N-point}} 
	
	In the previous sections, we demonstrated how to map EFTofLSS correlation functions to QFT integrals with massive propagators raised
	to integer powers. 
	For the one-loop power spectrum and the one-loop bispectrum, we demonstrated how to reduce these QFT
	integrals with recurrence identities to master  integrals that we computed analytically. 
	The master integral reduction techniques that
	we employed can be applied more generally, beyond the first important physical examples that we explicitly discussed earlier, in order to
	compute higher point correlation functions. 
	In this section, we explain how to compute algorithmically the one-loop QFT integrals that
	our method generates in generic $N$-point one-loop correlators of the EFTofLSS.
	
	We will show that, within dimensional regularization,
	the reduction  of $N$-point one-loop integrals to
	master integrals is completely analogous to the procedure that we detailed for the power spectrum and  the bispectrum.  
	What is more, we will show that, in the physical limit of exactly $D=3$ dimensions, no additional master integrals beyond what we have already presented and computed are needed.  
	As we will review here, box, pentagon,
	hexagon, etc. master integrals are fully determined from three and lower point master integrals~\cite{vanNeerven:1983vr} (the one-loop
	tadpole, bubble and triangle master integrals of Section~\ref{sec:master}) through finite parts ${\cal O}\left(\epsilon^0 \right)$ in the dimensional regulator $\epsilon = (3 - D)/2$ expansion.
	
	Our method maps one-loop integrals in EFTofLSS correlators
	to generic $N$-point scalar one-loop integrals of the form,    
	\begin{eqnarray}
		\label{eq:genNpoint}
		I\left( \nu_1, \ldots \nu_M\right) =
		\int d^Dq \prod_{i=1}^M \frac{1}{{\cal A}_i^{\nu_i}}\ ,
	\end{eqnarray}
	with $M\leq N$.
	The loop momentum $q$ is Euclidean and, in dimensional
	regularization,  has $D=3-2\epsilon$ space dimensions.
	The terms
	\begin{eqnarray}
		{\cal A}_i = \left( q+p_i \right)^2 + M_i \, ,
	\end{eqnarray}
	correspond to denominators of scalar propagators with
	complex mass terms $M_i$.
	These factors are raised to positive or negative integer powers
	$\nu_i$ (in the notation of the previous sections, $\nu_i=d_i$ or
	$\nu_i=-n_i$ with $n_i, d_i \in \mathbb{N}$).
	The momenta  $p_i$ are real linear combinations 
	\begin{equation}
		p_{i} =\sum_{m=1}^i k_m \, ,  
	\end{equation}
	of the $N$ external momenta $k_i$, which satisfy the momentum conservation condition
	\begin{equation}
		p_N=\sum_{m=1}^{N} k_m =0.
	\end{equation}
	
	As a first step, we may apply the Passarino-Veltman
	technique~\cite{Passarino:1978jh}  in order to eliminate propagator factors with negative powers  $\nu_r=-n_r, \; n_r
	\in \mathbb{N}, \{r\}  \subset \{1,\ldots , M\}$. Such integrals have
	a polynomial of the loop momentum $q$ in the numerator. Using
	simple identities, for example 
	\[
	(\vec q^2)^2 = \delta_{\mu_1 \mu_2} \delta_{\mu_3 \mu_4} \, q^{\mu_1}
	q^{\mu_2} q^{\mu_3} q^{\mu_4} , \quad \vec q^2  \left( \vec q \cdot \vec
	p_1  \right) = \delta_{\mu_1 \mu_2} p_{1\mu_3} \, q^{\mu_1}
	q^{\mu_2} q^{\mu_3}, \quad \ldots,
	\]
	we write the numerator as a contraction of tensors. This operation casts the
	original scalar integral, which includes negative powers $\nu_r$ of
	propagators, as a superposition of tensor integrals,  with a tensor which is a product
	of loop momenta in the numerator and propagator
	factors which are now raised to exclusively positive powers ($\nu_i=d_i
	\geq 0$)
	\begin{eqnarray}
		\label{eq:tensorNpoint}
		I[q^{\mu_1}q^{\mu_2}  \ldots q^{\mu_R} ] \equiv \int d^Dq
		\frac{q^{\mu_1}q^{\mu_2}  \ldots  q^{\mu_R}}{\prod_{i \not \in \{ r\}}   {\cal A}_i^{d_i}}
		\quad . 
	\end{eqnarray}
	Above, we have introduced a short notation where, in the bracket, at the left-hand side, we label the integral with its
	numerator. The integration in Eq.~(\ref{eq:tensorNpoint}) yields a
	superposition of  tensors which are  
	products of the metric and the external momenta  $p_i$ found in the
	denominator factors ${\cal A}_i $, 
	\begin{equation}
		\label{eq:tensordecomposition}
		I[q^{\mu_1}q^{\mu_2}  \ldots q^{\mu_R}] = \sum_{{\bf t}} C_{\bf t} \, B^{\mu_1 \mu_2
			\ldots \mu_R}_{({\bf t})}\left( \delta^{\mu \nu} , \vec p_i \right), \quad
		i \not \in \{ r \}.
	\end{equation}
	The set of momenta $p_i$ which enter $B^{\mu_1 \mu_2
		\ldots}_{({\bf t})}$ above is not the full set of external momenta for the
	original integral of Eq.~(\ref{eq:genNpoint}).  It is a subset, as it
	does not contain the external momenta $p_r$ in propagator factors
	${\cal A}_r$ which were inverted to the numerator, by being raised to a negative power $\nu_r =-n_r$. This fact will be used next to eliminate the loop momentum vectors $q^\mu$ from the numerator, {as we will now show}. 
	The tensors $B^{\mu_1 \mu_2
		\ldots \mu_R}_{({\bf t})}$ can be constructed,  using standard
	linear algebra matrix diagonalization,  to form an orthonormal basis of
	rank-$R$ tensors, 
	\begin{equation}
		B^{\mu_1 \mu_2 \ldots \mu_R}_{({\bf t})}
		B_{({\bf t}^\prime) {\mu_1 \mu_2
				\ldots \mu_R}} = \delta_{tt^\prime}. 
	\end{equation}
	Then, the tensor integral decomposition of
	Eq.~(\ref{eq:tensordecomposition}) takes the explicit form,
	\begin{equation}
		\label{eq:tensordecompE}
		I[q^{\mu_1}q^{\mu_2}  \ldots q^{\mu_R}] = \sum_{{\bf t}}
		I\left[q_{\rho_1}q_{\rho_2}  \ldots q_{\rho_R} B^{\rho_1 \rho_2
			\ldots \rho_R}_{({\bf t})}  \right] 
		\, B^{\mu_1 \mu_2
			\ldots \mu_R}_{({\bf t})}\left( \delta^{\mu \nu} , \vec p_i \right), \quad
		i \not \in \{ r \}.
	\end{equation}
	Now, all tensor indices are carried by the metric tensor and external
	momenta  in $B^{\mu_1 \mu_2 \ldots \mu_R}_{({\bf t})}$, which are
	independent of the loop momentum. 
	The numerators of the integrals in the right-hand side are scalars, which we can
	now express as polynomials of the denominator factors,
	\begin{equation}
		q_{\rho_1}q_{\rho_2}  \ldots q_{\rho_R} B^{\rho_1 \rho_2
			\ldots \rho_R}_{({\bf t})}  =c_{(t)}^{(0)}  +
		\sum_{i \not \in \{ r \}}  c_{(t)}^{(i)} {\cal A}_i + \sum_{i,j \not \in
			\{ r \}}  c_{(t)}^{(i,j)} {\cal A}_i {\cal A}_j + \ldots 
	\end{equation}
	The above rewriting is possible because the number of independent
	scalar products with the loop-momentum $q$  which can appear in numerators equals the number of linearly independent denominator factors ${\cal A}_i,  \  i \not \in
	\{ r \}$. 
	After all, we have achieved to write a tensor integral in terms of scalar
	integrals,
	\begin{eqnarray}\label{eq:intermediatestep}
		&&
		\int d^Dq \frac{q^{\mu_1}q^{\mu_2}  \ldots  q^{\mu_R}}{\prod_{a \not \in \{ r\}}    {\cal A}_a^{d_a}} = 
		\sum_{{\bf t}} \, B^{\mu_1 \mu_2
			\ldots \mu_R}_{({\bf t})} \,
		\Bigg\{
		c_{(t)}^{(0)}
		\int d^Dq \frac{1}{\prod_{a \not \in \{ r\}}   {\cal
				A}_a^{d_a}}
		+\sum_{i \not \in \{ r\}}  c_{(t)}^{(i)}
		\int d^Dq \frac{ {\cal A}_i }{\prod_{a \not \in \{ r\}}   {\cal
				A}_a^{d_a}}
		\nonumber \\
		&&  \hspace{2cm}
		+\sum_{i,j \not \in \{ r\}}  c_{(t)}^{(i,j)}
		\int d^Dq \frac{ {\cal A}_i {\cal A}_j }{\prod_{a \not \in \{ r\}}   {\cal
				A}_a^{d_a}} + \ldots
		\Bigg\}
	\end{eqnarray} 
	In the first term of the right-hand side, we have an integral with the same number of propagators as in the tensor integral of the left-hand
	side and a constant ({\it i.e} loop-momentum independent) in the numerator. 
	In the remaining terms, we notice that numerator
	factors cancel against denominators.  
	After we carry out these
	cancellations of ${\cal A}_i$'s we obtain two classes of simpler integrals.
	\begin{enumerate}
		\item Integrals with constant numerators and fewer propagator
		denominators than the tensor integral in the left-hand side, 
		raised to positive or zero powers only. 
		\item Integrals with fewer denominator ${\cal A}_a$ factors than in
		the left-hand side but with some of them being still present in the numerator {(this happens for example when one of the ${\cal A}_i$ in the numerator of the right-hand side of \eqref{eq:intermediatestep} had $d_i=0$.)}.  These integral are of the form
		of Eq.~(\ref{eq:genNpoint}), with both positive and negative powers
		of propagators,  but with fewer denominators than what we started
		with.  Iterating the above steps as many times as necessary (the
		number of iterations is bounded by the rank of the original tensor
		integral) all integrals are eventually  made to have only
		propagator denominators and constant numerators.  
	\end{enumerate}
	
	After tensor reduction, all integrals have positive powers $\nu_i=d_i \geq 0$.  
	We can reduce all such powers to $\nu_i=1$ or $\nu_i=0$
	with the method of integration by parts, which we have described in
	detail for the cases of $N=2$ and $N=3$ earlier.  For a generic $N-$point one-loop integral, we can obtain $N$ integration by parts (IBP)  identities~\cite{Tkachov:1981wb,Chetyrkin:1981qh}, 
	{\begin{eqnarray}
			\label{eq:oneloopIBPstart}
			0&=&   \int d^Dq \; \vec{\nabla}_q  \cdot \frac{\vec q+\vec p_j}{\prod_{i=1}^N {\cal A}_i^{\nu_i}} \nonumber \\
			&=&  \int d^Dq \; \left( \prod_{i=1}^N \frac{1}{{\cal A}_i^{\nu_i}} \right)
			\left\{
			D  - \sum_{a=1}^{N} \nu_a \frac{ 2 \left( \vec q + \vec p_j \right) \cdot \left( \vec q + \vec p_a \right) }{ {\cal A}_a}
			\right\}
			\nonumber \\
			&=&  \int d^Dq \; \left( \prod_{i=1}^N \frac{1}{{\cal A}_i^{\nu_i}} \right)
			\left\{
			D  - \sum_{a=1}^{N} \nu_a \frac{ {\cal A}_a+{\cal A}_j - \left( p_a -p_j\right)^2 -M_a -M_j}{ {\cal A}_a}
			\right\}\ .
		\end{eqnarray}
	}
	For the vanishing of the total divergence integral, we need a
	convergent behaviour at infinity. 
	This is assumed due to dimensional
	regularization. 
	It is convenient to introduce the symbols, 
	\begin{equation}
		\label{eq:dmatrix}
		d_{ij} \equiv  \left( p_i -p_j\right)^2+M_i + M_j \, .  
	\end{equation}
	We can rewrite the system of IBP  equations~(\ref{eq:oneloopIBPstart}) in the form,  
	\begin{eqnarray}
		0=&& \int d^Dq \; \left( \prod_{i=1}^N \frac{1}{{\cal A}_i^{\nu_i}} \right)
		\left\{
		\sum_{a=1}^N d_{ja} \frac{\nu_a}{{\cal A}_a} +\left(D-\nu_j -\sum_{i=1}^N \nu_i \right) -\sum_{a \neq j} \, \nu_a \, 
		\frac{{\cal A}_j}{{\cal A}_a}
		\right\}. 
	\end{eqnarray}
	In a shorter notation, we write 
	\begin{equation}
		\label{eq:raise_a}
		\sum_{a=1}^N  d_{ja} \, \nu_a \, {\bf \hat a}^{+} =  \left(\nu_j
		+\sum_{i=1}^N \nu_i -D \right)  {\bf \hat 0}
		+ \sum_{a \neq j} \nu_a {\bf \hat a^+} \,
		{\bf \hat j^-} \, . 
	\end{equation}  
	where we have introduced raising (lowering) operators representing
	the original integral with a power of propagator increased (decreased) by one, 
	\begin{equation}
		{\bf \hat a}^{+}  \to 
		\int d^Dq \; \frac{1}{{\cal A}_a} \left( \prod_{i=1}^N
		\frac{1}{{\cal A}_i^{\nu_i}} \right)\ ,
		\quad
		{\bf \hat j}^{-}  \to 
		\int d^Dq \; {\cal A}_j \,  \left( \prod_{i=1}^N
		\frac{1}{{\cal A}_i^{\nu_i}} \right),  
	\end{equation}
	and a neutral operator which leaves the integral unchanged
	\begin{equation}
		{\bf \hat 0}  \to 
		\int d^Dq \;   \left( \prod_{i=1}^N
		\frac{1}{{\cal A}_i^{\nu_i}} \right)  \ .
	\end{equation}
	In the case of $\det(d_{ij}) \neq 0$, which is always the case for non-degenerate external momenta, we can use the inverse matrix $\tilde d_{ij}$, 
	\begin{equation}
		\sum_{j=1}^N {\tilde d}_{bj} \, d_{ja} = \delta_{ab},   
	\end{equation}
	to diagonalize the above system of difference equations,
	\begin{equation}
		\label{eq:IBPdiagonal}
		\nu_b \, {{\bf  \hat b}^{+}} = \sum_{j=1}^N {\tilde d}_{bj} \,
		\left[
		\left(\nu_j +\sum_{i=1}^N \nu_i -D \right) {\bf \hat 0}+ \sum_{a \neq j} \nu_a {\bf a^+} \, {\bf j^-}
		\right] \, . 
	\end{equation}
	Let us now consider positive non-zero integer
	powers of propagators $\nu_b = d_b > 0$. The integrals of the right-hand side are all simpler than the integral of the left-hand side. Specifically, the sum of powers of propagators  in the 
	${\bf \hat b}^{+}$ term of the left-hand side is $ \sum_{i \neq b} \nu_i  + (\nu_b +1) = \sum_i d_i+1$, while the {corresponding} sum for ${\bf \hat 0}$,  $ \sum_{i}\nu_i  = \sum_i d_i$ and   ${\bf a^+} \, {\bf j^-}$,  $ \sum_{i \neq a, j}\nu_i + (\nu_a+1) + (\nu_j-1)   = \sum_i d_i $ , in the right-hand side is lowered by a unit to $\sum_{i} d_i$. A sequential application of the recurrence identities of Eq.~(\ref{eq:IBPdiagonal}) will eventually reduce all the powers of propagators to $\nu_b \in \{ 0, 1\}$.
	In fact,  we have already seen this reduction mechanism at work for the triangle and bubble recursion relations (see the discussion around Eq.~\eqref{eq:rec_solved}) which are representative examples of the general one-loop case which we treat in this Section. 
	
	A further reduction can be now achieved by restricting ourselves to the physical limit of $D=3$ dimensions. 
	We find that only master integrals
	with at most three propagators, raised to unit powers, are truly independent. 
	This reduction has been shown in Ref.~\cite{vanNeerven:1983vr}, and we review it here, in a slightly modified derivation.
	
	Consider the case of the one-loop ``box'' integral with four propagators raised to a unit power,
	\begin{equation}
		I_4 \equiv  \int d^Dq \frac{1}{{\cal A}_1{\cal A}_2 {\cal A}_3 {\cal A}_4 }
	\end{equation}
	The three-dimensional part of the loop momentum $q$ can be decomposed as a superposition of the three independent linear combinations {$p_i, \; i=1\ldots 3$} of
	external momenta.
	Explicitly, this reads~\footnote{This equation can be derived by writing $\vec q=\sum_i c_i \, \vec p_i+\vec q_\perp$. Taking a dot product with $\vec p_j$, we get:
		\begin{equation}
			\vec p_j\cdot \vec q=\sum_i c_i \, \vec p_j\cdot \vec p_i\ .
		\end{equation}
		Multiplying both sides by $\Pi_{lj} $ and summing over $j$, we get
		\begin{equation}
			c_l=\sum_j\Pi_{lj}  \vec p_j\cdot \vec q\ ,
		\end{equation}
		reproducing our formula.}
	\begin{equation}
		\label{eq:normalbasis}
		q^\mu  = \sum_{l,j=1}^3  \left(   \vec q  \cdot \vec p_j \right)  \Pi_{jl}   \,  p_l^\mu + \vec q_\perp \, , 
	\end{equation}
	where $\Pi_{ik} $ is the inverse to the matrix $P_{ij} \equiv \vec p_i \cdot \vec p_j$ ($P_{ij}$ is non-singular given that the set of external momenta $\{\vec p_i\}$ are non-degenerate):
	\begin{equation}
		\label{eq:Pi}
		\sum_{k=1}^3 \Pi_{ik} \, \vec p_k \cdot \vec p_j  =\delta_{ij} \, ,  
	\end{equation} 
	and
	\begin{equation}
		\vec q_\perp \cdot \vec p_i = 0 \, . 
	\end{equation}
	Contracting both sides of Eq.~(\ref{eq:normalbasis}) with $q_\mu$,
	\begin{eqnarray}
		q^2 = \sum^3_{l,j}(  \vec q \cdot \vec p_j)\Pi_{jl}(  \vec p_l \cdot \vec q) + q^2_\perp
	\end{eqnarray}
	and recognizing that, since $p_4=0$, $q^2={\cal A}_4 - M_4$ and
	\begin{equation}
		\; 2 \vec q \cdot \vec p_j = {\cal A}_j -{\cal A}_4  -p_j^2 - M_j +M_4={\cal A}_j -{\cal A}_4  -\rho_j +\rho_4\ ,
	\end{equation}
	with
	\begin{equation}
		\rho_i  = p_i^2+M_i\ ,
	\end{equation}
	we can write
	\begin{eqnarray}
		{\cal A}_4 - \rho_4 - q^2_\perp &=& \frac{1}{2}\sum^3_{l,j}({\cal A}_j - {\cal A}_4 - \rho_j + \rho_4)\Pi_{jl}(\vec q  \cdot \vec p_l ) \nonumber \\
		&& \hspace{-2cm} =
		\frac{1}{2}\sum^3_{l,j}({\cal A}_j - {\cal A}_4 )\Pi_{jl}(\vec q  \cdot \vec p_l ) 
		-\frac{1}{2}\sum^3_{l,j}(  \rho_j - \rho_4)\Pi_{jl}(\vec q  \cdot \vec p_l ) 
		\nonumber \\
		&& \hspace{-2cm} =\frac{1}{2}\sum^3_{l,j}({\cal A}_j - {\cal A}_4 )\Pi_{jl}(\vec q  \cdot \vec p_l ) 
		-\frac{1}{4}\sum^3_{l,j}(  \rho_j - \rho_4)\Pi_{jl} ({\cal A}_l - {\cal A}_4 - \rho_l + \rho_4) \, . 
	\end{eqnarray}
	In the above manipulations, we aimed to express $q^2$ in terms of propagator denominators ${\cal A}_i$. However, we refrained from a complete substitution of all scalar products $\vec q \cdot \vec p_l$ in terms of ${\cal A}_l$ propagators, keeping the expression linear in ${\cal A}_i$. 
	This will soon prove be convenient for integration. Rearranging, we have
	\begin{eqnarray}
		\label{eq:neervenMiddle}
		&&{\cal A}_4 - \rho_4 + \frac{1}{4}\sum^3_{l,j}(\rho_j - \rho_4) \Pi_{ jl}({\cal A}_l - {\cal A}_4 - \rho_l + \rho_4)- q^2_\perp  \nonumber \\  && \hspace{1cm}
		=\frac{1}{2}\sum^3_{l,j}{\cal A}_{ l}\Pi_{lj}(\vec p_j \cdot \vec q) - \frac{{\cal A}_4}{2}\sum^3_{l,j}\Pi_{lj}(\vec p_j \cdot \vec q)     \nonumber \\
		&&
		\hspace{1cm}
		=\frac{1}{2}\sum^3_{l,j}{\cal A}_{ l}\Pi_{lj}(\vec p_j \cdot \vec q) - \frac{{\cal A}_4}{2}\sum^3_{l,j}\Pi_{lj}(\vec p_j \cdot \left( \vec q+ \vec p_1 \right) ) +  \frac{{\cal A}_4}{2}.
	\end{eqnarray} 
	In the last equality, we have used that $\sum^3_{l,j}\Pi_{lj}\vec p_1 \cdot \vec p_j = \sum^3_l \delta_{1l} = 1$. 
	We obtain the following relation of the propagators in the integral: 
	\begin{eqnarray}
		\label{eq:neervenA}
		&& \frac{ {\cal A}_4}{2} -\rho_4  + \frac 1 4 \sum_{i,j=1}^3 \left( \rho_i -\rho_4 \right) \Pi_{ij}  \left( {\cal A}_j - {\cal A}_4 -\rho_j+\rho_4 \right)  - \vec q_\perp^2
		= \frac{1}{2} \, \sum_{i,j=1}^3  {\cal A}_i   \, \Pi_{ij} \vec q \cdot \vec p_j
		\nonumber \\ && \hspace{2cm}
		- \frac{{\cal A}_4} {2} \sum_{i,j=1}^3 \Pi_{ij} \, \left( \vec q + \vec p_1\right) \cdot \vec p_j .
	\end{eqnarray}
	In the right-hand side, we have left some scalar products of the loop momentum without expressing them in terms of the propagator denominators ${\cal A}_i$.
	This is because, as we shall see next, they do not { contribute} to integration. 
	As a next step, we divide both sides of \eqref{eq:neervenA} with the
	product of ${\cal A}_1 {\cal A}_2 {\cal A}_3 {\cal A}_4 $ and integrate over the loop momentum $q$. 
	The right-hand side of the equation can be expressed as rank-one tensor integrals. 
	Performing a tensor reduction, as we described for
	tensor integrals at the start of this Section, we find that they vanish, as we now explain.
	Let us first consider explicitly the integral that originates from the last term of the right-hand side, which, after canceling the numerator factor ${\cal A}_4$ against a denominator,  reads
	\begin{eqnarray}
		&&\int d^Dq \,  \frac{
			\sum_{i,j=1}^3 \Pi_{ij} \, \left( \vec q + \vec p_1\right) \cdot \vec p_j
		}{ {\cal A}_1 {\cal A}_2 {\cal A}_3}
		\nonumber \\
		&&
		=  \int d^Dq^\prime \,  \frac{
			\sum_{i,j=1}^3 \Pi_{ij} \, \vec q^\prime \cdot \vec p_j
		}{ \left[ {\vec q^\prime}^2+M_1\right]  \left[ \left(\vec q^\prime+\vec p_2 -\vec p_1\right)^2+M_2\right]
			\left[ \left(\vec q^\prime+\vec p_3 -\vec p_1\right)^2+M_3\right]
		}
	\end{eqnarray}
	where, we have made a shift of integration variable $\vec q^\prime = \vec q +p_1$. 
	The external momenta entering the denominators are now the differences
	$\vec p_2 - \vec p_1$ and $\vec p_3 - \vec p_1$.
	Tensor reduction will project the numerator of the integral to
	\begin{eqnarray}
		\sum_{i,j=1}^3 \Pi_{ij} \, \vec q^\prime \cdot \vec p_j
		&\to& C_2  \sum_{i,j=1}^3 \Pi_{ij} \, \left( \vec p_2 -\vec p_1 \right) \cdot \vec p_j
		+C_3  \sum_{i,j=1}^3 \Pi_{ij} \, \left( \vec p_3 -\vec p_1 \right) \cdot \vec p_j
		\nonumber \\
		&=& C_2  \sum_{i=1}^3 \left( \delta_{i2} - \delta_{i1} \right) +C_3  \sum_{i=1}^3 \left( \delta_{i3} - \delta_{i1} \right) =0\ . 
	\end{eqnarray}
	Finally, the tensor reduction in the remaining integrals emerging from the right-hand side gives factors 
	proportional to  
	\begin{equation}
		\sum_{j \neq i} \Pi_{ij} \vec p_i \cdot  \vec  p_j  =\delta_{ij} =0\ ,
	\end{equation}
	as, for each $i$ the dependence on $p_i$ disappears from the integral. 
	Therefore, they vanish, too. 
	
	We are then left with the identity,
	\begin{eqnarray}
		\label{eq:neervenMain}
		&&  \int d^D q  \, \frac{
			-2 \rho_4 - \frac 1 2 \sum_{i,j=1}^3 \left(\rho_i - \rho_4 \right) \Pi_{ij} \left( \rho_j- \rho_4 \right)
		}
		{
			{\cal A}_1 {\cal A}_2 {\cal A}_3 {\cal A}_4
		} \,  =  2 \, 
		\int d^D q  \, \frac{
			\vec q_\perp^2
		}
		{
			{\cal A}_1 {\cal A}_2 {\cal A}_3 {\cal A}_4
		}
		\nonumber \\
		&& \hspace{2cm}
		- \frac 1 2 \int d^Dq \, \frac{  2{\cal A}_4 +
			\sum_{i,j=1}^3 \left( \rho_i -\rho_4 \right) \Pi_{ij} \left( {\cal A}_j -{\cal A}_4 \right)
		}{
			{\cal A}_1 {\cal A}_2 {\cal A}_3 {\cal A}_4
		} 
	\end{eqnarray}

	The integral with $\vec q_\perp^2$ is of order ${\cal O}(\epsilon)$ in $D=3-2\epsilon$ dimensions and drops out when we  take the exact limit of $D=3$. 
	Indeed, we can separate the integration measure into a three dimensional part and a $D-3=-2\epsilon$ part, 
	\begin{eqnarray}
		\int d^Dq \frac{q_\perp^2}{
			{\cal A}_1 {\cal A}_2 {\cal A}_3 {\cal A}_4 
		} &=&
		\int d^3 q  \, d^{-2\epsilon} q_\perp \frac{q_\perp^2}{
			{\cal A}_1 {\cal A}_2 {\cal A}_3 {\cal A}_4       }
	\end{eqnarray}
	The propagators ${\cal A}_i= q^2 + 2 \vec q \cdot p_i + \rho_i$ depend on the transverse momentum $\vec q_\perp$ only through its magnitude squared $\vec q_\perp^2$ in $q^2$, as the external momenta combinations $p_i$ are purely three dimensional and the scalar products $\vec q \cdot \vec p_l$ do not depend on $\vec q _\perp$. 
	Then, we are allowed to average over angles in the $-2\epsilon$-dimensional space, 
	\begin{eqnarray}
		\int d^{-2\epsilon} \vec q_\perp q_\perp^2  = \Omega_{-2\epsilon-1}
		\int d q_\perp q_\perp^{-2\epsilon+1} = \frac{\Omega_{-2 \epsilon-1}} {\Omega_{-2 \epsilon + 1}}  \int d^{-2 \left( \epsilon -1\right) } \vec q_\perp = \frac{-\epsilon}{\pi} \int d^{-2 \left( \epsilon -1\right) } \vec q_\perp\ ,  
	\end{eqnarray}
	using $\Omega_{n-1} = \frac{n}{2\pi}\Omega_{n + 1}$.

	The identity of Eq.~(\ref{eq:neervenMain}) is then a ``dimensional-shift'' of the box integral from $D=3-2 \epsilon$ dimensions to $D=5-2\epsilon$ dimensions~\cite{Bern:1993kr},
	\begin{eqnarray}
		\label{eq:dimshift}
		&&  \int d^D q  \, \frac{
			- 2 \rho_4 - \frac 1 2 \sum_{i,j=1}^3 \left(\rho_i - \rho_4 \right) \Pi_{ij} \left( \rho_j- \rho_4 \right)
		}
		{
			{\cal A}_1 {\cal A}_2 {\cal A}_3 {\cal A}_4
		} \,  =  -2 \epsilon \, 
		\int \frac{d^{D+2} q}{\pi}  \, \frac{
			1
		}
		{
			{\cal A}_1 {\cal A}_2 {\cal A}_3 {\cal A}_4
		}
		\nonumber \\
		&& \hspace{2cm}
		- \frac 1 2 \int d^Dq \, \frac{  2{\cal A}_4 +
			\sum_{i,j=1}^3 \left( \rho_i -\rho_4 \right) \Pi_{ij} \left( {\cal A}_j -{\cal A}_4 \right)
		}{
			{\cal A}_1 {\cal A}_2 {\cal A}_3 {\cal A}_4
		} 
	\end{eqnarray}
	The box integral is UV finite in $D<8$ dimensions, as can be seen with power counting for the ultraviolet degree of divergence. 
	Similarly,  IR power counting of the degree of divergence for potential soft or collinear singularities~\cite{Sterman:1978bi,Libby:1978qf,Collins:1989gx} shows that it is also IR  finite in $D>4$. 
	Therefore, the five dimensional box integral  in the right-hand side of Eq.~(\ref{eq:dimshift}) is finite. 
	As, it is multiplied with a factor of $\epsilon = \frac{3-D}{2}$, it drops out when integrals are computed through their finite parts in the expansion around $\epsilon=0$. 
	We then have the following identity,
	\begin{eqnarray}
		\label{eq:boxreduction}
		&&
		\left[
		-2 \rho_4 - \frac 1 2 \sum_{i,j=1}^3 \left(\rho_i - \rho_4 \right) \Pi_{ij} \left( \rho_j- \rho_4 \right)
		\right]
		\int d^D q  \, \frac{
			1
		}
		{
			{\cal A}_1 {\cal A}_2 {\cal A}_3 {\cal A}_4
		} \,  =  
		\nonumber \\
		&& \hspace{2cm}
		- \frac 1 2 \int d^Dq \, \frac{  2{\cal A}_4 +
			\sum_{i,j=1}^3 \left( \rho_i -\rho_4 \right) \Pi_{ij} \left( {\cal A}_j -{\cal A}_4 \right)
		}{
			{\cal A}_1 {\cal A}_2 {\cal A}_3 {\cal A}_4
		} +{\cal O}\left( \epsilon \right).
	\end{eqnarray}
	In the left-hand side we have isolated the three-dimensional one-loop
	box integral. 
	In the right-hand side, the numerator is a linear
	combination of the four propagators, yielding four {\it triangle}
	integrals. 
	Therefore, the one-loop box master integral is reduced to triangle master integrals in the physical case that we need to compute it through ${\cal O} (\epsilon^0)$.
	
	The above derivation is readily generalized~\cite{vanNeerven:1983vr}
	to all one-loop (pentagon, hexagon, $\ldots$)  $D$-dimensional master integrals.  
	Master integrals with $N > D=3$ external legs can always be expressed through their finite parts in terms of master integrals with at most three propagators.
	
	In this article, we have derived analytic expressions for the necessary
	tadpole, bubble and triangle master integrals. Together with the
	reduction identities of this Section, we have therefore presented all
	necessary ingredients for calculating generic $N-$point one-loop correlators in
	the EFTofLSS.

	\section*{Acknowledgments}
We wish to thank Ben Ruijl for help with recursion examples in Rust.

	\appendix	
	
	\section{Effect of changing the number of fitting functions}
	\label{sec:appfit}
	
	\begin{figure}[H]
		\centering
		\begin{subfigure}{0.5\textwidth}
			\centering
			\includegraphics[height=4.5cm]{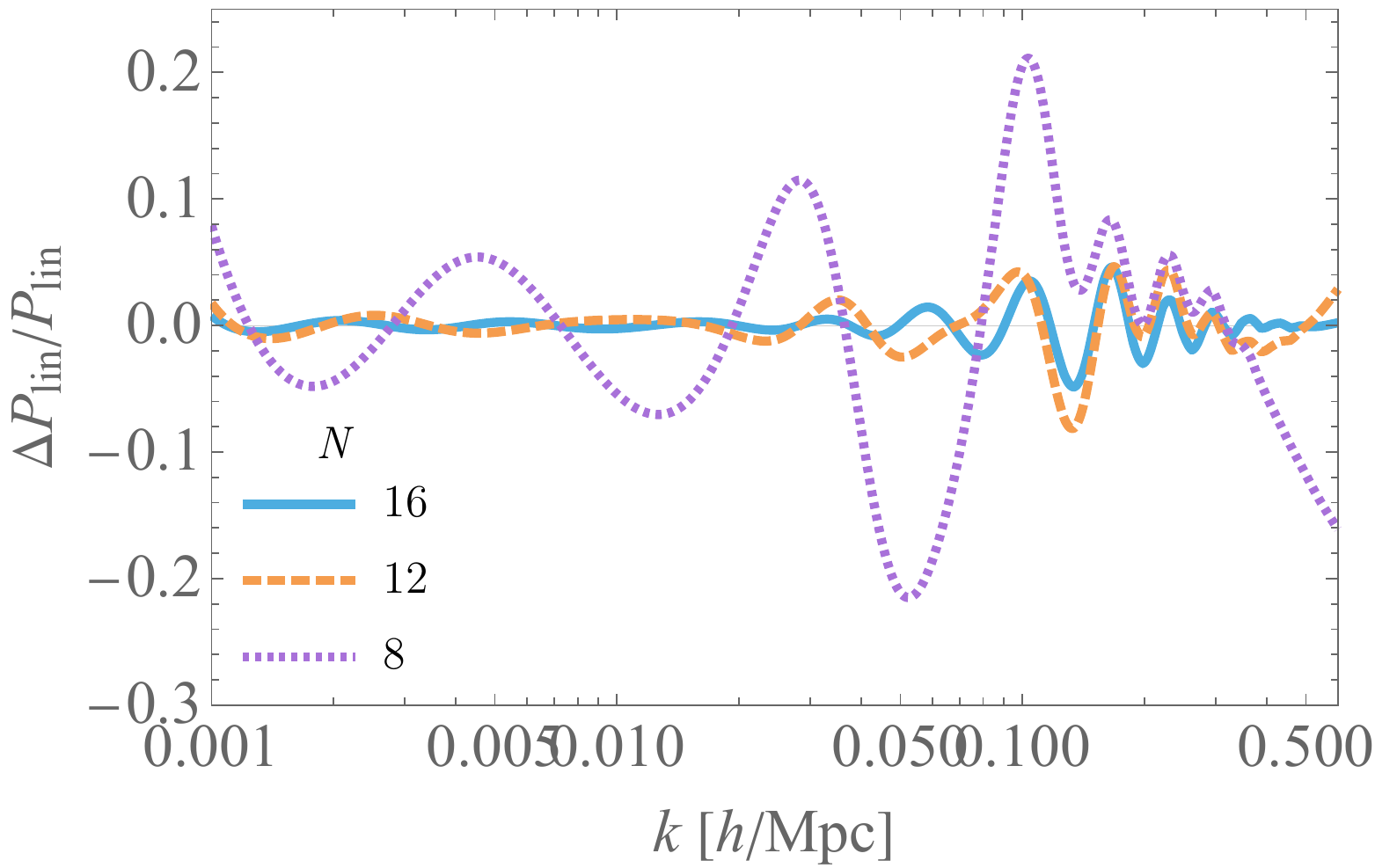}
		\end{subfigure}%
		\begin{subfigure}{0.5\textwidth}
			\centering
			\includegraphics[height=4.5cm]{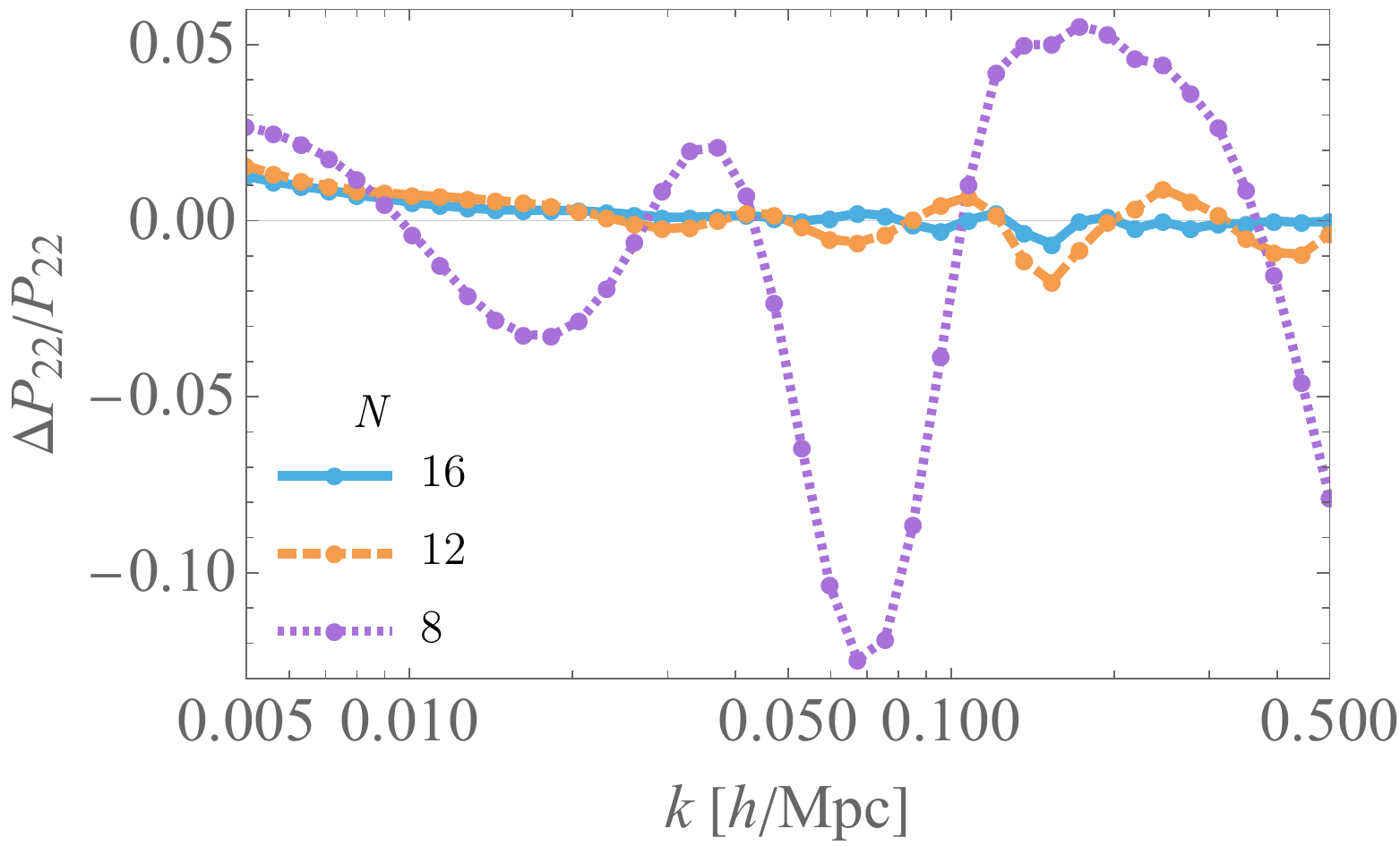}
		\end{subfigure}
		\caption{\label{fig:Pimax}Relative errors of $P_{\rm fit}$ to $P_{\rm lin}$ (left) and of analytically calculated $P_{22}$ to the numerical one (right). }
	\end{figure}
	
	\paragraph{Changing $N$, the number of fitting functions}
	In our fitting procedure, we can try to decrease even more the number of fitting functions $N$. In Fig.~\ref{fig:Pimax}, we show the effect of changing $N$ in $P_{\rm fit}$ and in the $P_{22}$ diagram.
	We can see that the accuracy becomes better as we increase the number of functions, as expected, {especially for the BAO wiggles. We also notice however} that for as low as 12 functions we can get a quite satisfactory fit. One can use this to make quick first estimates with less fitting functions that do not require so much precision. 
	
\begin{figure}[h]
		\centering
		\includegraphics[scale = 0.5]{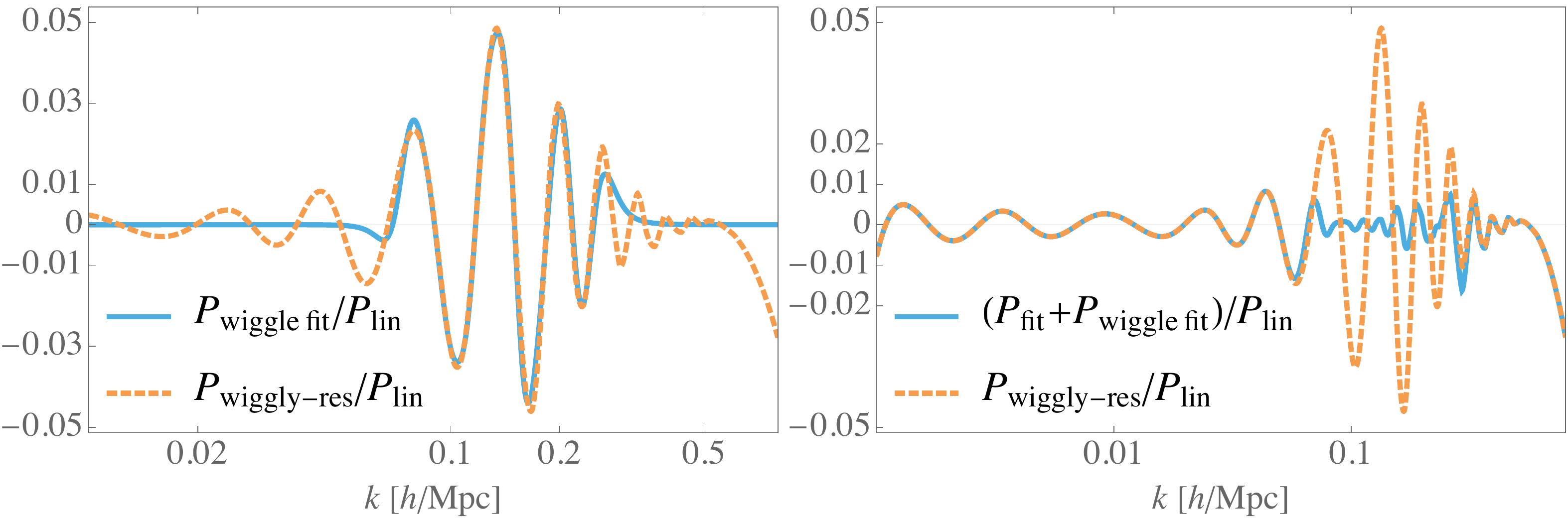}
		\caption{Left: Ratio plot of fitted $P_{\rm wiggle}/P_{\rm lin}$ using 12 BW functions and true wiggle $P_{\rm wiggle}/P_{\rm lin}$, Right: Ratio plot of fitted $P_{\rm fit}$ including wiggles and $P_{\rm fit}$ without wiggles. }
		\label{fig:wigglefitting}
	\end{figure}

	\paragraph{Adding Breit-Wigner functions to the fit} 
	{Our fitting procedure with $n=16$ does not capture very well the higher-$k$ BAO wiggles. This can be seen in Fig.~\ref{fig:wigglefitting}. As shown in Sec.~\ref{sec:P1loop}, this does not have significant effects on the accuracy of the procedure because the $P_{\rm fit}$ is only used inside the integrals, where the effect of the wiggles is smeared. Nevertheless, to improve the fitting residual in this regime, we can do the following. For clarity, we define the wiggly residual as 
		\begin{equation}
			P_{\rm wiggly-res.}(k) = P_{\rm lin}(k) - P_{\rm fit}(k).
		\end{equation}
		We then  can add Breit-Wigner (BW) functions with positive real masses {$m^2_{BW}$}. In practice, we add to our fitting basis a BW fitting basis as
		\begin{align}
			P_{\rm wiggle fit}(k) = \sum^n_{i = 1} \alpha_i \frac{k^2}{((k^2 - m^2_{BW, i})^2 + m^2_{BW, i}\delta BW^2_i)^2}\ ,
		\end{align}
		where $m_{BW, i}= k_{\rm max,BW} \exp (i/n)$, $\delta BW_i =\frac{0.02 k_{\rm max,BW}}{k_{\rm min,BW}}\exp (i/n)$ and $n = 12$. We fit the $P_{\rm wiggly-res.}(k) $ in the range of $k_{\rm min} = 0.02 h/{\rm Mpc}$ and $k_{\rm max} = 0.4 h/{\rm Mpc}$ using $k_{\rm min, BW} = 0.06 h/{\rm Mpc}$ and $k_{\rm max, BW} = 0.25 h/{\rm Mpc}$. The performance of our wiggle fits for $n = 12$ is shown in Fig.~\ref{fig:wigglefitting}.  The result seems satisfactory, and the procedure can be surely improved.
		
		In order to integrate our added BW fitting functions in the loop integrals, we notice that $m_{BW, i}$ in the BW functions are analogous to $k_{\rm peak}$ and $m_{BW, i}\delta BW_i$ to the {$k^2_{UV}$}} in our regular decomposition. So, we can define new $k^2_{\rm peak, BW}$ and  $M_{i, \rm BW}$ to put the new BW fitting functions in the standard form of our basis. The only  difference is that $k^2_{\rm peak} < 0$ and $m^2_{BW, i}> 0$, leading to $\Re M_{i, \rm BW} < 0$ in Eq.~\eqref{eq:Lfunc}. Thus when including BW functions to our fit, we need to consider triangle and bubble integrals with masses of negative real parts. We did not derive the resulting branch cut conditions due to negative real masses, but we expect it to be straightforward.
	
	\section{Check for systematic errors }
	\label{app:syst}
	{In this section, we provide several numerical checks of our code. All numerical integrations were performed with NIntegrate in Mathematica.}
	
	We show the comparison of the analytical evaluation of the master integrals to the numerical evaluation of the master integrals defined in Sec.~\ref{sec:l_func} {by integrating up to $q\to \infty$}, in Table~\ref{tab:compnummaster}. The evaluation of the master integrals given in the table below uses $M_1 = -k^2_{\rm peak, 1} + ik^2_{\rm UV, 1}, M_2 = -k^2_{\rm peak, 2} - ik^2_{\rm UV, 2}, M_3 = -k^2_{\rm peak, 3} + ik^2_{\rm UV, 3}$.
	
\begin{figure}[h]
		\centering
		\includegraphics[width=\textwidth]{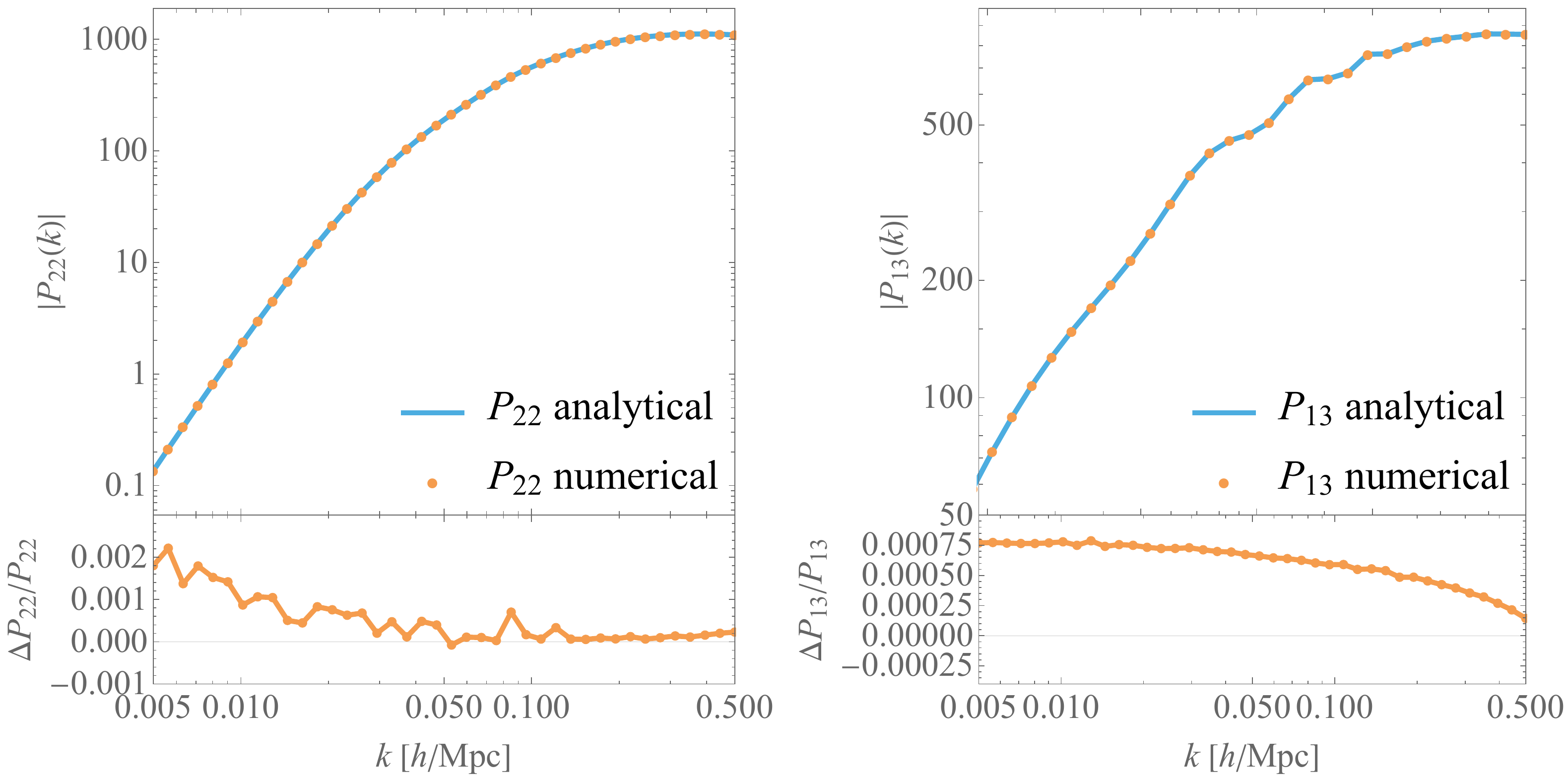}
		\caption{\label{fig:sysps}Two contributions for the one-loop power spectrum. Left: for $P_{22}$, comparison between analytical result, $\bar{P}_{22}$, and exact numerical result, using $P_{\rm fit}$ in both cases. Right: for  $P_{13}$, comparison between analytical result, $\bar{P}^{{\rm comp}}_{13}$, and exact numerical result, both using~$P_{\rm fit}$.}
	\end{figure}

\begin{figure}[h]
		\centering
		\includegraphics[width=\textwidth]{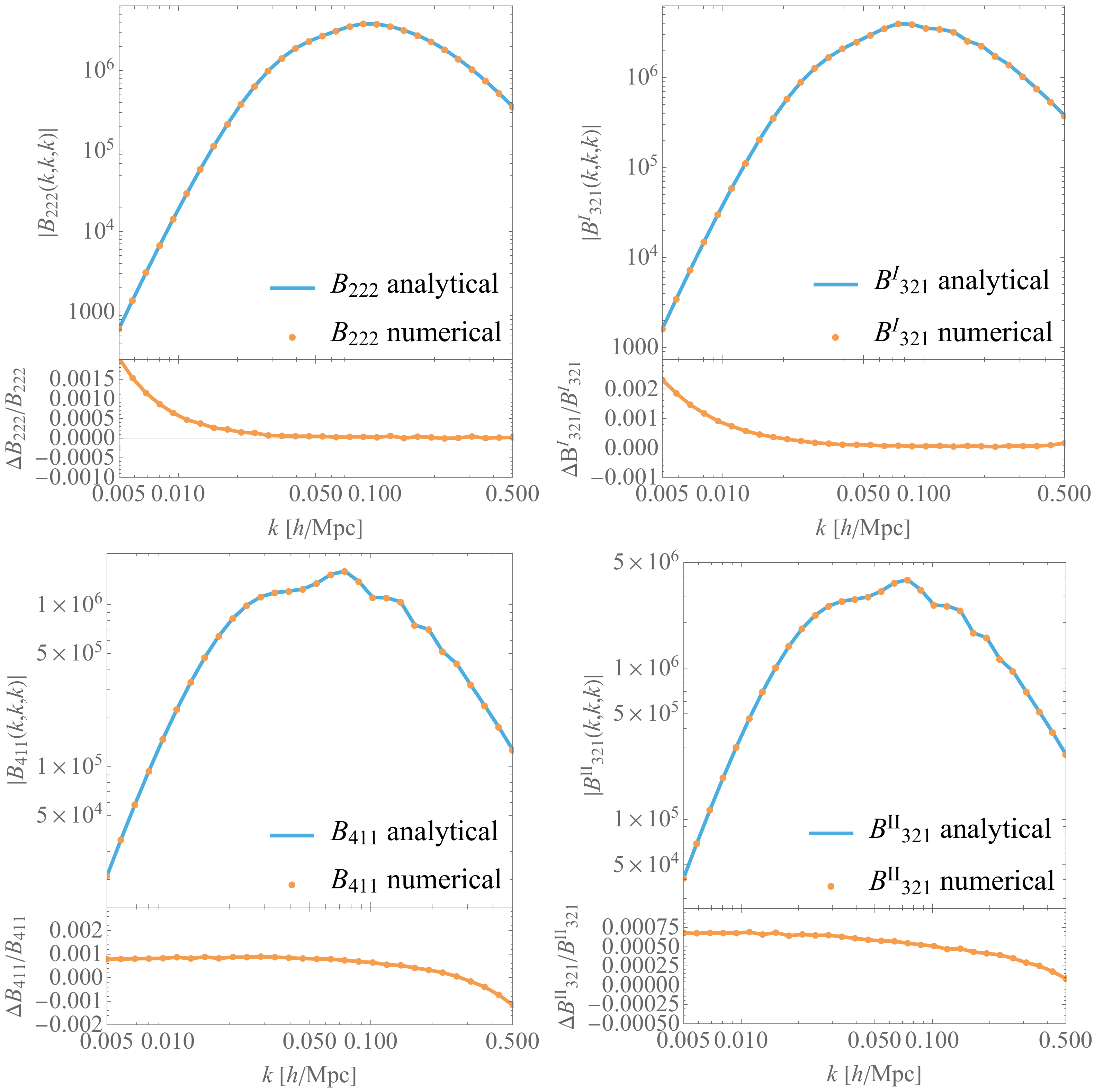}
		\caption{\label{fig:sysbs}Comparison of the analytical result with the numerical integration using $P_{\rm fit}$ for each one of the 4 diagrams contributing to the 1-loop bispectrum.}
	\end{figure}

	\begin{table}
		\centering
		\begin{tabular}{ccc}
			\hline
			$k=0.108 \hinvMpc$ & Analytic & Numerical \\
			\hline
			$\text{Tad}(M_1, n = 1, d = 3)$	& $4.06986 - 2.53169i$ & $4.06986 - 2.53169i$ \\
			$B_{\rm master}(k^2, M_1, M_2)$	& $4.44761 - 0.270344i$ & $4.44761 - 0.270343i$ \\
			$T_{\rm master}((0.8 k)^2, k^2, (1.2 k)^2, M_1, M_2, M_3)$	& $4.16315 - 0.364415 i$ & $4.16315 - 0.364415 i$ \\
			\hline
		\end{tabular}
		\caption{\label{tab:compnummaster} Comparison between analytical result and full numerical integration of momentum integrals of the master integrals defined in Sec.~\ref{sec:l_func} up to $q=\infty$ for $\text{Tad}, B_{\rm master}$, and $T_{\rm master}$ master integrals.}
	\end{table}
	To also ensure that our complete analytical integration procedure of the loop diagrams is correct, we now directly compare the result with numerical integration using the fitting power spectrum $P_{\rm fit}$.
	The numerical integration is performed, as in the case of the linear power spectrum, from $q_{\rm IR} = 1.0 \times 10^{-4} \hinvMpc$ to $q_{\rm UV} = 1.0 \hinvMpc$.
	We also perform a similar UV-correction to correctly account for the finite numerical integration region.
	The comparisons, in real space, are shown in Fig.~\ref{fig:sysps} for the 1-loop power spectrum diagrams and in Fig.~\ref{fig:sysbs} for the 1-loop bispectrum diagrams.
	We verify that our analytical procedure matches very well the numerical integration.
	The small discrepancies is probably mostly due to small residual IR and numerical inaccuracies.

	To further verify the correctness of our analytic integration, we show the results where the numerical integration up to $q\to\infty$ matches the analytical result for some specific cases in Table~\ref{tab:compnum}. 
		{
		\begin{table}
			\centering
			\begin{tabular}{ccc}
				\hline
				$\text{Diagram}(k_1,k_2,k_3)$ & Analytic & Numerical \\
				
				\hline
				$B_{222}(0.01829,0.024,0.02825)$ & $3.04388\times 10^5$ &  $3.04388\times 10^5$\\
				$B^I_{321}(0.01829,0.024,0.02825)$ & $8.04093\times 10^5$ &  $8.04092 \times 10^5$\\
				$B_{411}(0.01829,0.024,0.02825)$ & $-1.23417 \times 10^6$ &  $-1.23417 \times 10^6$\\
				\hline
			\end{tabular}
			\caption{\label{tab:compnum} Comparison between analytical result and full numerical integration of $P_{\rm fit}$ up to $q=\infty$ for some diagrams.{We demonstrate excellent agreement between numerical integration and our analytical procedure.}  }
		\end{table}
	}

	\section{Bubble master integral: mass conditions for branch cut}
	\label{sec:appmasscond}
	
	Let us check what are the conditions that $m_1$ and $m_2$ have to satisfy in order to have a branch cut crossing in Eq.~\eqref{eq:Bfinalsame}. 
	The conditions translate to $t_- > 0$ and $x_- \in ]0,1[$, where $t_-$ is given in Eq.~\eqref{eq:t} and $x_-$ in Eq.~\eqref{eq:xminus}. Solving $x_- > 0$ and $x_- < 1$ {and $t_- > 0$} yields, respectively:
	\begin{align}
		\label{eq:im2}
		& \Im(m_2) < -\frac{1}{2} (1 + \Re(m_1) - \Re(m_2)) \sqrt{\Delta(\Re(m_1), \Re(m_2))} \,, \\
		\label{eq:im1}
		& \Im(m_1) > \frac{1}{2} (1 + \Re(m_1) - \Re(m_2)) \sqrt{\Delta(\Re(m_1), \Re(m_2))}\,, \\
		\label{eq:appt}
		&{\Im(m_1)-\Im(m_2) - \sqrt{\Delta(\Re(m_1), \Re(m_2))} >0 \, } .
	\end{align}
	{As mentioned in the main text, in order for Eq.~\eqref{eq:appt} to be satisfied, one needs also $\Delta(\Re(m_1), \Re(m_2))>0$. Fig.~\ref{fig:contour} shows the region where this particular condition is satisfied.}
	We can now identify two cases:
	\begin{itemize}
		\item If $1 + \Re(m_1) - \Re(m_2)> 0$, then, defining $\kappa \equiv \frac{1}{2} (1 + \Re(m_1) - \Re(m_2)) \sqrt{\Delta(\Re(m_1), \Re(m_2))} >0$, we have $\Im(m_1) > \kappa > 0$ and $\Im(m_2) < -\kappa < 0$.
		Since we are now considering the case where the two masses have the same imaginary part sign, the inequalities cannot be simultaneously satisfied. 
		
		\item If $1 + \Re(m_1) - \Re(m_2) < 0$, then, defining $\kappa \equiv -\frac{1}{2} (1 + \Re(m_1) - \Re(m_2)) \sqrt{\Delta(\Re(m_1), \Re(m_2))} >0$, we have $\Im(m_1) > -\kappa$ and $\Im(m_2) < \kappa$. It is therefore possible to satisfy the above inequalities and thus to get branch cut crossings. 
		
	\end{itemize}

	\section{Relating the integrals using contour integration}
	\label{sec:appcontour}
	We noticed that at the end of the day, we obtained the same expression as Eq.~\eqref{eq:Bfinalsamecut} for the case of equal sign imaginary parts of the mass and opposite sign given by Eq.~\eqref{eq:Bfinalcut}.  
	Something is hinting that there may be a simpler way to relate the case where the masses have the same sign of the imaginary part and where they have opposite signs. 
	Indeed, we can prove using contour integration that the two integrals are closely related.
	Recall that in the case of opposite imaginary parts, we take $\Im m_1 > 0$ and $\Im m_2 < 0$.
	
	First, let us call $z_+$ and $z_-$ the roots of the second order polynomial $x (1-x) + m_1 x + m_2 (1-x)$.
	Then, we can write Eq.~\eqref{eq:Bdiffsign2} as 
	\begin{equation}
		\label{eq:Bmasterepsilon}
		B_{\rm master}(k^2,M_1,M_2) = {\frac{\sqrt{\pi}}{k}}\left(\int_0^{\frac12 - \frac{1}{\epsilon}}  \frac{dx}{\sqrt{(z_+ - x)(x - z_-)}} + \int_{\frac12 + \frac{1}{\epsilon}}^{1}  \frac{d\hat{x}}{\sqrt{(z_+ - x)(x - z_-)}} + {\pi} \right)\,.
	\end{equation}
	
	We shall now prove that, if $\Im(m_1)>0$, $\Im(m_2)<0$, and $\Re(m_i)>0$, then $\Im(z_+)>0$ and $\Im(z_-)>0$. 
	First, observe that  
	\begin{equation}
		\label{eq:zpm1}
		\begin{split}
			& z_+ + z_- = 1 + m_1 - m_2
			\Rightarrow 
			\begin{cases}
				\Re (z_+) + \Re (z_-) = 1 + \Re (m_1) - \Re (m_2)\,,\\
				\Im (z_+) + \Im (z_-) = \Im (m_1) - \Im (m_2)\,,
			\end{cases}
		\end{split}
	\end{equation}
	and
	\begin{equation}
		\label{eq:zpm2}
		\begin{split}
			& z_+ z_- = -m_2
			\Rightarrow 
			\begin{cases}
				\Re (z_+ z_-) = -\Re (m_2)\,, \\
				\Im (z_+ z_-) = -\Im(m_2)\,.
			\end{cases}
		\end{split}
	\end{equation}
	Now, we have  
	\begin{align}
		\Im (z_+ z_-) = -\Im(m_2) &\Leftrightarrow  \Im(z_+) \Re(z_-) + \Im(z_-) \Re(z_+) = -\Im(m_2)\\
		\label{eq:imzpzm}
		& \Leftrightarrow  \Re(z_-) = \frac{-\Im(m_2) - \Im(z_-) \Re(z_+)}{\Im(z_+)}\,.
	\end{align}
	So, using Eqs.~\eqref{eq:zpm1} and \eqref{eq:imzpzm}, we get 
	\begin{align}
		&\Re(z_+) = \frac{\Im(z_+) (1 + \Re(m_1-m_2)) + \Im(m_2)}{\Im(z_+-z_-)} \,,\\
		&\Re(z_-) = \frac{-\Im(z_-) (1 + \Re(m_1-m_2)) - \Im(m_2)}{\Im(z_+-z_-)} \,.
	\end{align}
	Now, using Eq.~\eqref{eq:zpm2}, we have 
	\begin{equation}
		\begin{split}
			&\Re(z_+ z_-) = - \Re(m_2) \Leftrightarrow  \Re(z_+) \Re(z_-) - \Im(z_+) \Im(z_-) = -\Re(m_2)\\
			\Leftrightarrow &(\Im(z_+) (1 + \Re(m_1-m_2)) + \Im(m_2)) (-\Im(z_-) (1 + \Re(m_1-m_2)) - \Im(m_2))\\
			&- \Im(z_+) \Im(z_-)\Im(z_+-z_-)^2  = -\Re(m_2) \Im(z_+-z_-)^2\,.
		\end{split}
	\end{equation}
	This equation can be simplified using $\Im(z_+-z_-)^2 = \Im(z_+ + z_-)^2 - 4 \Im(z_+) \Im(z_-)$ together with Eq.~\eqref{eq:zpm1}, yielding 
	\begin{align}
		& - \Im(z_+) \Im(z_-)((1 + \Re(m_1-m_2))^2 + \Im(z_+-z_-)^2) - \nonumber \\
		&\Im(m_2) (1 + \Re(m_1-m_2)) (\Im(z_+)+\Im(z_-)) -\Im(m_2)^2 \nonumber \\
		&= -\Re(m_2) (\Im(z_+ + z_-)^2- 4 \Im(z_+) \Im(z_-))  \\
		\Leftrightarrow & - \Im(z_+) \Im(z_-)((1 + \Re(m_1-m_2))^2 + \Im(m_1 - m_2)^2- 4 \Im(z_+)\Im(z_-) + 4 \Re(m_2)) - \nonumber \\
		&\Im(m_2) (1 + \Re(m_1-m_2)) (\Im(m_1 - m_2)) -\Im(m_2)^2  \nonumber \\
		&= -\Re(m_2) \Im(m_1 - m_2)^2\\
		\Leftrightarrow &  \Im(z_+) \Im(z_-)((1 + \Re(m_1-m_2))^2 + \Im(m_1 - m_2)^2- 4 \Im(z_+)\Im(z_-) + 4 \Re(m_2))  \nonumber \\
		& = \Re(m_2) \Im(m_1 - m_2)^2 \nonumber \\
		&- \Im(m_2) (1 + \Re(m_1-m_2)) \Im(m_1 - m_2) -\Im(m_2)^2\\
		\Leftrightarrow &  \Im(z_+) \Im(z_-)((1 + \Re(m_1-m_2))^2 + \Im(m_1 - m_2)^2 + 4 \Re(m_2)) \nonumber \\
		& = \Re(m_2) \Im(m_1 - m_2)^2 \nonumber  \\
		&- \Im(m_2) (1 + \Re(m_1-m_2)) \Im(m_1 - m_2) -\Im(m_2)^2 + 4 (\Im(z_+)\Im(z_-))^2 \\
		\Leftrightarrow &  \Im(z_+) \Im(z_-)((1 + \Re(m_1-m_2))^2 + \Im(m_1 - m_2)^2 + 4 \Re(m_2)) \nonumber \\
		& = -\Im(m_1) \Im(m_2) + \Im(m_1 - m_2) (\Re(m_2) \Im(m_1) - \Re(m_1) \Im(m_2)) \nonumber\\
		&+ 4 (\Im(z_+)\Im(z_-))^2\,.
	\end{align} 
	Therefore, since each term on the right hand side is positive since $\Im m_1 >0$, $\Im m_2 < 0$, and we use masses that have $\Re m_i >0$, we have 
	\begin{equation}
		\label{eq:x1x2same}
		\begin{split}
			\Im(z_+)\Im(z_-) &\left[\Im(m_1 - m_2)^2 + (1 + \Re(m_1-m_2))^2 + 4 \Re(m_2) \right] > 0 \,,
		\end{split}
	\end{equation}
	so if $\Re(m_i)>0$, then $\Im(z_+)\Im(z_-) > 0$. 
	Combined with $\Im(z_+ + z_-)=\Im[m_1-m_2]>0$, it implies that $\Im(z_+)>0$ and $\Im(z_-)>0$. 
	
	Thus, we can separate the square roots in the two terms (using Eq.~\eqref{eq:ssigncondition}, which gives $s(z_+ - x, x - z_-) = 1$ as $\textrm{sign}\Im (z_+ - x) = -\textrm{sign}\Im (x - z_-)$):
	\begin{equation}
		\sqrt{(z_+ - x)(x - z_-)} = \sqrt{z_+ - x}\sqrt{x - z_-}\,.
	\end{equation}
	Let us now define $f(z) = \sqrt{z_+ - z}\sqrt{z - z_-}$ for complex $z$. 
	According to our definition of the square root, the function $f(z)$ has two horizontal branch cuts in the upper imaginary plane, but in the lower imaginary plane $f$ is analytic. 
	\begin{figure}[h]
		\centering
		\includegraphics[width = 0.9 \textwidth]{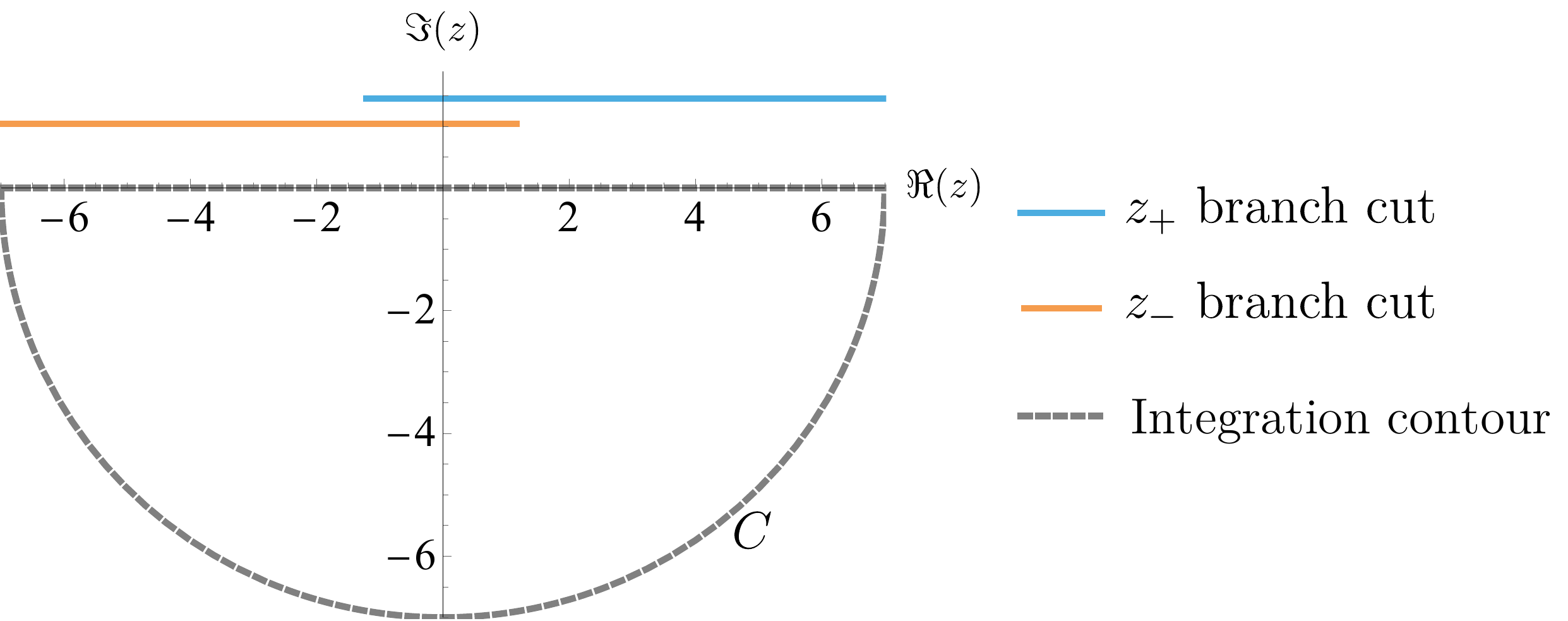}
		\caption{Branch cuts of $f(z)$ and integration contour used in Eq.~\eqref{eq:contourint}. In this plot we used $m_1 = 2+2\,i$ and $m_2 = 3-0.5 \,i$.}
		\label{fig:intcontourB}
	\end{figure}
	Thus, by using a contour constituted of the real line and a lower semi-circle $C$, as shown in Fig.~\ref{fig:intcontourB}, we find that
	\begin{align}
		\label{eq:contourint}
		&	{\int_{-\infty}^{\infty}  \frac{dx}{\sqrt{z_+ - x}\sqrt{x - z_-}} +\lim_{\rho\to \infty} \int_0^\pi d\theta\frac{(-i)\rho e^{-i \theta }}{\sqrt{z_+-\rho e^{-i \theta }} \sqrt{\rho e^{-i \theta }-z_-}}=0}\ , \\
		\Rightarrow &\int_{-\infty}^{\infty}  \frac{dx}{\sqrt{z_+ - x}\sqrt{x - z_-}} + \int_0^\pi d\theta\frac{(-i) e^{-i \theta }}{\sqrt{-e^{-i \theta }} \sqrt{e^{-i \theta }}} = 0\ , \\
		\Rightarrow & \int_{-\infty}^{\infty}  \frac{dx}{\sqrt{z_+ - x}\sqrt{x - z_-}} = \pi \ , \\
		\Rightarrow & \int_{0}^{1}  \frac{dx}{\sqrt{z_+ - x}\sqrt{x - z_-}} = \pi + \int_{0}^{-\infty}  \frac{dx}{\sqrt{z_+ - x}\sqrt{x - z_-}} + \int_{\infty}^{1}  \frac{dx}{\sqrt{z_+ - x}\sqrt{x - z_-}}\,,
	\end{align}
	and therefore, when $\Im(m_1)>0$, $\Im(m_2)<0$, and $\Re(m_i)>0$, we obtain for Eq.~\eqref{eq:Bmasterepsilon} and taking the limit $\epsilon \to 0$: 
	\begin{equation}
		B_{\rm master}(k^2,M_1,M_2) ={\frac{\sqrt{\pi}}{k}} \,\int_0^{1}  \frac{dx}{\sqrt{z_+ - x}\sqrt{x - z_-}} = {\frac{\sqrt{\pi}}{k}} \,\int_0^{1}  \frac{dx}{\sqrt{(z_+ - x)(x - z_-)}}\,,
	\end{equation}
	which is exactly Eq.~\eqref{eq:Bsame}.
	Note that the requirement $\Re(m_i) > 0$ is important in this case. In general, there could be a crossing of a branch cut between $x=0$ and $x=1$ if $\Im(m_1)>0$ and $\Im(m_2)<0$. 
	So we conclude that the two cases of equal sign and opposite sign for the imaginary parts of the masses are indeed identical.

	\section{Triangle master integral for masses with a general real part}
	\label{sec:alternativeT}
	With the inclusion of extra Breit-Wigner functions to our fit to better capture the BAO wiggles in $P_{\rm fit}$, we will encounter negative real masses in our integrals since $\Re M_{\rm BW} < 0$.
	We now provide an alternative derivation of the triangle master integral without assuming positive real part of the masses.
	
	\subsection{Triangle master integral for masses with the same imaginary part sign and general real part.}
	\label{sec:trianglesame}
	First let us consider the case of all three masses $M_1, M_2$ and $M_3$ having positive imaginary parts. Using Schwinger parametrization, the momentum integral becomes, 
	\begin{align}
		\begin{split}
			T_{\rm master}(\dots) =& \frac{i}{\pi^{3/2}}\int_0^{+\infty}  ds_1 ds_2 ds_3{\int \frac{d^3\vec q}{\pi^{3/2}}} (1+i\epsilon_1)(1 + i\epsilon_2)(1 + i\epsilon_3)\\
			& \times \exp\left[ i\left( (\vec{k}_1 - \vec{q})^2 + M_1
			\right)(1 + i\epsilon_1)s_1 + i(q^2 + M_2)(1 + i\epsilon_2)s_2 \right.\\
			& \hspace{5cm} \left. + i\left( (\vec{k}_2 + \vec{q})^2 + M_3 \right)(1 + i\epsilon_3)s_3
			\right]\,.
		\end{split}
	\end{align}
	Expanding the exponent yields 
	\begin{align}
		\begin{split}
			&  (1 + i\epsilon_1)s_1(k^2_1 + q^2 - 2\vec{k}_1\cdot \vec{q} + M_1) + (1 + i\epsilon_2)s_2(q^2 + M_2) \\
			& \qquad \qquad \qquad \qquad \qquad \qquad \qquad + (1 + i\epsilon_3)s_3(k^2_2 + q^2 + 2\vec k_2\cdot \vec q +
			M_3) \\
			&  = S\left( \vec q - \frac{\tilde{s}_1\vec k_1 - \tilde{s}_3\vec k_2}{S} \right)^2  + \frac{\tilde{s}_1\tilde{s}_3 k^2_3 + \tilde{s}_1\tilde{s}_2k^2_1 + \tilde{s}_2\tilde{s}_3k^2_2}{S} + \tilde{s}_1 M_1 + \tilde{s}_2 M_2 + \tilde{s}_3 M_3\,,
		\end{split}
	\end{align}
	where $\tilde{s}_i = s_i(1 + i\epsilon_i)$, $S = \tilde{s}_1 + \tilde{s}_2 + \tilde{s}_3$, and we used $\vec{k}_1 +
	\vec{k}_3 + \vec{k}_3 = 0$.
	Since $\Im(S) > 0$, we can do the Gaussian integration in $\vec{q}$, obtaining 
	\begin{align}
		T_{\rm master} = -\frac{1}{(-i)^{1/2}}\int_0^{+\infty} ds_1 ds_2 ds_3\frac{e^{i I}}{S^{3/2}}\,,
	\end{align}
	where $I = (\tilde{s}_1\tilde{s}_2k^2_1 + \tilde{s}_3\tilde{s}_1k^2_3 + \tilde{s}_2\tilde{s}_3k^2_2)/{S} + \tilde{s}_1M_1 + \tilde{s}_2M_2 + \tilde{s}_3M_3$.
	Changing integration variables $s_1 = \tau x_1, s_2 =
	\tau x_2, s_3 = \tau(1 - x_1 - x_2) = \tau x_3$, where $\tau \in [0,+\infty[$, $x_1 \in [0,1]$, $x_2 \in [0,1]$, and $x_1 + x_2 < 1$ {and $x_3=1-x_2-x_1$}, and observing that the Jacobian is $\tau^2$, we obtain
	\begin{align}
		\label{eq:tau}
		T_{\rm master} = -\frac{1}{(-i)^{1/2}}\int_0^1 dx_1 dx_2 \int_0^{+\infty}d\tau \tau^{1/2}e^{i\tau \tilde{I}}\,,
	\end{align}
	where we defined 
	\begin{equation}
		\tilde{I} = \frac{\tilde{x}_1 \tilde{x}_2 k^2_1 + \tilde{x}_2 \tilde{x}_3 k^2_2 + \tilde{x}_1 \tilde{x}_3 k^2_3 }{1 + i (x_1 \epsilon_1 + x_2 \epsilon_2 + x_3 \epsilon_3)} + \tilde{x}_1 M_1 + \tilde{x}_2 M_2 + \tilde{x}_3 M_3\,,
	\end{equation}
	and used $\tilde{x}_i \equiv x_i (1 + i \epsilon_i)$. 		
	
	To perform the $\tau$ integration, we must have $\Im(\tilde{I})>0$.
	This condition is satisfied if any $\Im(M_i)>0$, and we can safely send $\epsilon_i \to 0$. 
	However, we still have to check that $\Im(\tilde{I})>0$ even when the masses are all real and positive. 
	Note that this case is equivalently discussed also in Sec.~\ref{sec:triangle_master}.
	We can check that it is indeed the case. 
	Setting $\epsilon_i \to \epsilon$ and using real masses, we obtain (using $x_1+x_2+x_3=1$) 
	\begin{equation}
		\Im(\tilde{I}) = \epsilon \left(M_1 x_1 + M_2 x_2 + M_3 x_3 + k_1^2 x_1 x_2 +  k_2^2 x_2 x_3 + k_3^2 x_1 x_3    \right)\,,
	\end{equation}
	which is always positive if all masses are real and positive (which is the case in our parametrization).  	
	Performing the $\tau$ integration in Eq.~\eqref{eq:tau}, we then obtain
	\begin{align}
		\label{eq:Tbefore}
		T_{\rm master} = \frac{\sqrt{\pi}}{2}\int_0^1 dx_1 \int_0^{1-x_1} dx_2\tilde{I}^{-3/2}\,.
	\end{align}
	At this point, we can safely set all $\epsilon_i \to 0$, and $\tilde{I}$ becomes simply $\tilde{I} = x_1 x_2 k^2_1 + x_2 x_3 k^2_2 + x_1 x_3 k^2_3 + x_1 M_1 + x_2 M_2 + x_3 M_3$.	
	This expression is the same as the one in Eq.~\eqref{eq:Tbeforesame}, so this case gives the same result as the case discussed in the main text where all masses have a positive real part. 
	Note that this is true even including branch cut crossings. 
	In fact, the only other branch cut crossings one could have would be in the integrand of Eq.~\eqref{eq:Tbefore}, but since the imaginary part of $\tilde{I}$ is always positive, as explained above, there are no branch cut crossings in that integrand.
	The branch cuts of the resulting $\arctan$ are treated exactly the same way as described in Sec.~\ref{sec:Fintcut}.

	\subsection{{Triangle master integral for masses of different imaginary part sign and general real part.}}
	\label{sec:trianglediff}
	
	Let us consider the case of two masses $M_1$ and $M_2$ having positive imaginary parts and $M_3$ having a negative imaginary part:
	\begin{align}
		\begin{split}
			T_{\rm master}(\dots) &= -i\int ds_1 ds_2 ds_3 \int \frac{d^3\vec{q}}{\pi^{3/2}} (1 + i\epsilon_1)(1 + i\epsilon_2)(1 - i\epsilon_3)  \\
			& \times \exp\left[ i\left( (\vec{k}_1 - \vec{q})^2 + M_1\right)(1 + i\epsilon_1)s_1 + i\left( q^2 + M_2 \right)(1 + i\epsilon_2)s_2 \right.\\
			&\hspace{5cm} \left. - i\left( (\vec{k}_2 + \vec{q})^2 +
			M_3 \right)(1 - i\epsilon_3)s_3 \right]\,.
		\end{split}
	\end{align}
	Expanding the exponent (factoring out an $i$), we get 
	\begin{align}
		\begin{split}
			&  (1 + i\epsilon_1)s_1(k^2_1 + q^2 - 2\vec{k}_1\cdot \vec{q} + M_1) + (1 + i\epsilon_2)s_2(q^2 + M_2) \\
			& \qquad \qquad \qquad \qquad \qquad \qquad \qquad - (1 - i\epsilon_3)s_3(k^2_2 + q^2 + 2\vec k_2\cdot \vec q +
			M_3) \\
			&  = S_-\left( \vec{q} - \frac{\tilde{s}_1\vec k_1 + \tilde{s}_3\vec k_2}{S_-} \right)^2  + \frac{\tilde{s}_1\tilde{s}_2k^2_1 - \tilde{s}_2\tilde{s}_3k^2_2 - \tilde{s}_1\tilde{s}_3 k^2_3}{S_-} + \tilde{s}_1 M_1 + \tilde{s}_2 M_2 - \tilde{s}_3 M_3\,,
		\end{split}
	\end{align}
	where $\tilde{s}_i = s_i(1 + i\epsilon_i)$ for $i=\{1,2\}$ and $\tilde{s}_3 = s_3(1 - i\epsilon_3)$, $S_- = \tilde{s}_1 + \tilde{s}_2 - \tilde{s}_3$, and we used $\vec{k}_1 +
	\vec{k}_3 + \vec{k}_3 = 0$.
	Since $\Im(S_-) > 0$, we can do the Gaussian integration in $\vec{q}$, obtaining 
	\begin{align}
		T_{\rm master} = \frac{1}{(-i)^{1/2}}\int_0^{+\infty} ds_1 ds_2 ds_3\frac{e^{i I}}{S_-^{3/2}}\,,
	\end{align}
	where $I = (\tilde{s}_1\tilde{s}_2k^2_1 - \tilde{s}_3\tilde{s}_1k^2_3 - \tilde{s}_2\tilde{s}_3k^2_2)/{S_-} + \tilde{s}_1M_1 + \tilde{s}_2M_2 + \tilde{s}_3M_3$.
	As before, changing integration variables $s_1 = \tau x_1, s_2 =
	\tau x_2, s_3 = \tau(1 - x_1 - x_2) = \tau x_3$, where $\tau \in [0,+\infty[$, $x_1 \in [0,1]$, $x_2 \in [0,1]$, $x_1 + x_2 < 1$, and $x_3=1-x_2-x_1$, and observing that the Jacobian is $\tau^2$, we obtain
	\begin{align}
		\label{eq:taudiff}
		T_{\rm master} = \frac{1}{(-i)^{1/2}}\int_0^1 dx_1 \int_0^{1-x_1} dx_2 \int_0^{+\infty}d\tau
		\frac{\tau^{1/2}}{\left(\tilde{x}_1+ \tilde{x}_2 - \tilde{x}_3\right)^{3/2}}
		e^{i\tau \tilde{I}}\,,
	\end{align}
	where we defined 
	\begin{equation}
		\tilde{I} = \frac{\tilde{x}_1 \tilde{x}_2 k^2_1 - \tilde{x}_2 \tilde{x}_3 k^2_2 - \tilde{x}_1 \tilde{x}_3 k^2_3}{\tilde{x}_1 + \tilde{x}_2 - \tilde{x}_3} + \tilde{x}_1 M_1 + \tilde{x}_2 M_2 - \tilde{x}_3 M_3 \,,
	\end{equation}
	and used $\tilde{x}_i \equiv x_i (1 + i \epsilon_i)$ for $i=\{1,2\}$ and $\tilde{x}_3 \equiv x_3 (1 - i \epsilon_3)$.
	To perform the $\tau$ integration, we must have $\Im(\tilde{I})>0$. 
	There are two interesting limits to check for the sign of $\Im(\tilde{I})$. 
	The first is when $\epsilon_i\to 0$, with $x_1+x_2\neq1/2$.
	This gives
	\begin{equation}\label{eq:limitim0}
		\left.\Im(\tilde{I})\right|_{x_1+x_2\neq1/2} = x_1 \Im(M_1) + x_2 \Im(M_2) - x_3 \Im(M_3) + \mathcal{O}(\epsilon)\,,
	\end{equation}
	where we set $\epsilon_i \to \epsilon$, as we do for the rest of this section. 
	If $\Im(M_1)>0$ or $\Im(M_2)>0$ or $\Im(M_3)<0$, then, assuming the terms in $\epsilon$ are not enhanced anywhere in the integration region, we can  choose infinitesimal $\epsilon > 0$ while keeping $\Im(\tilde{I})>0$.
	If all the masses are real, then we use the approach for the same sign of the imaginary part outlined in Sec.~\ref{sec:trianglesame}, which is valid when the masses are all real (and, in that case, they must be positive, which is always the case in our parametrization).
	
	As mentioned, the limit in (\ref{eq:limitim0}) does not explore the case in which there might exist some terms in  $\Im(\tilde{I})$ that, though suppress by $\epsilon$, are enhanced by a divergent contribution. 
	Indeed, inspection of $\tilde{I}$ shows that the limit $x_2=\frac{1}{2}-x_1$ is singular for $\epsilon\to 0$. In this case, by setting  $x_2=\frac{1}{2}-x_1$, one can easily see that the imaginary of $\tilde I$ part is proportional to $1/\epsilon$, for $\epsilon\to0$, but it is indeed positive:
	\begin{equation} 
		\label{eq:limitim2}
		\left.\Im(\tilde{I})\right|_{x_2+x_1=1/2} =\frac{k^2_2(1-2x_1) + 2x_1(k^2_3 + k^2_1(2x_1 -1))}{4\epsilon} + C\,,
	\end{equation}
	where $C$ is bounded by a constant for any value of $x_1$, and where positivity of the first term can be proven in the following way
	\begin{align}
		\begin{split}
			k^2_2(1-2x_1) + 2x_1(k^2_3 + &k^2_1(2x_1 -1)) = (1-2x_1)(k^2_2 - 2x_1 k^2_1 - k^2_3) + k^2_3\\
			&= (1-2x_1)(k^2_1 - 2x_1 k^2_1 + 2\vec{k}_1 \cdot \vec{k}_3) + k^2_3\\
			&= (1-2x_1)^2k^2_1 + 2\vec{k}_1 \cdot \vec{k}_3(1-2x_1) + k^2_3\\
			&= (1-2x_1)^2k^2_1 + (\vec{k}_1(1-2x_1) + \vec{k}_3)^2 - k^2_1(1-2x_1)^2\\
			&=(\vec{k}_1(1-2x_1) + \vec{k}_3)^2 \geq 0\, ,
		\end{split}
	\end{align}
	where in the second line we use $k^2_2 = k^2_1 + k^2_3 + 2\vec{k}_1\cdot\vec{k}_3$. 
	
\begin{figure}[h]
		\centering
		\includegraphics[width = 0.5 \textwidth]{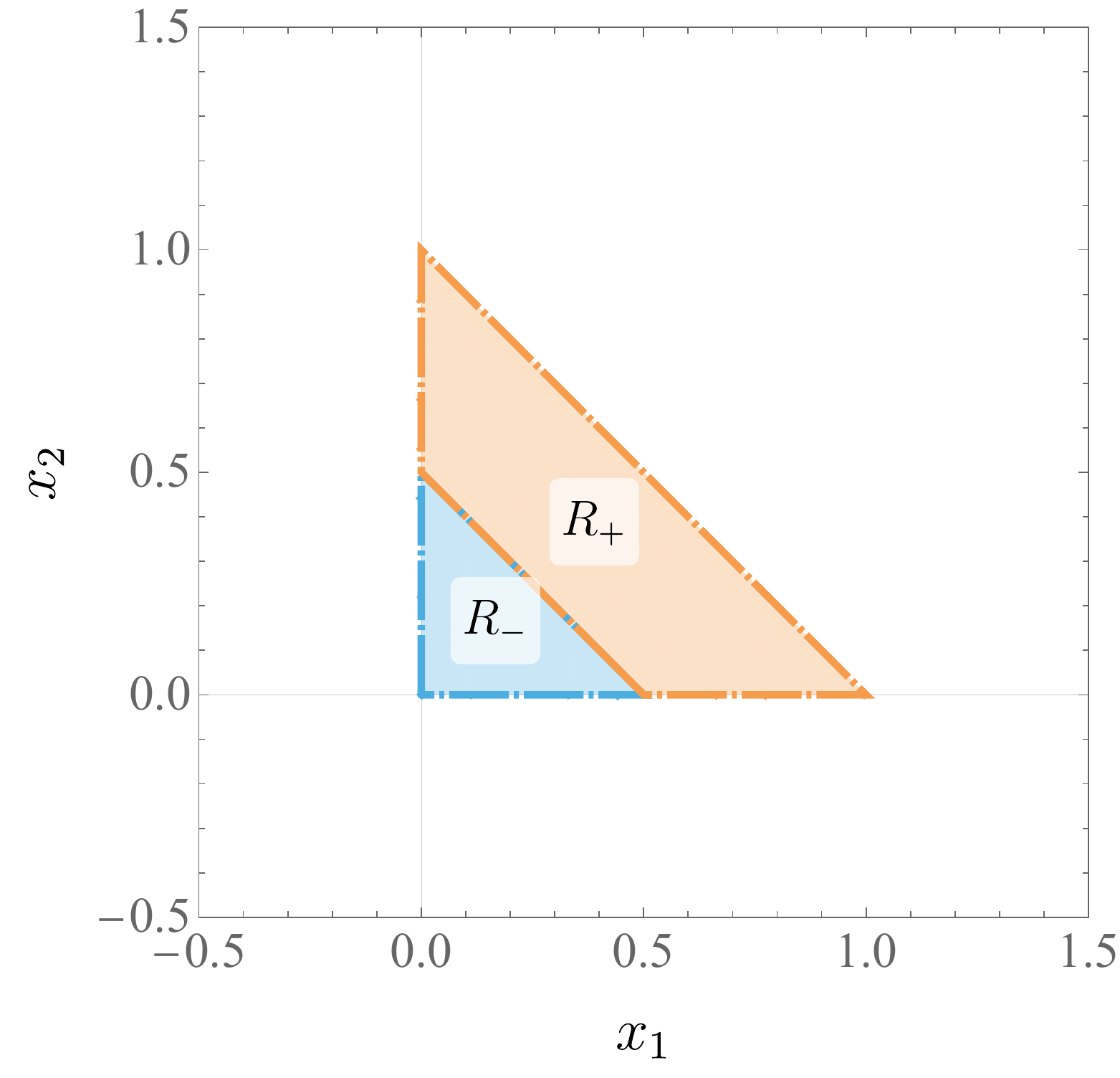}
		\caption{Integration region of $x_1$ and $x_2$ split into the regions $R_-$ and $R_+$.}
		\label{fig:boundpm}
	\end{figure}

	We can now do the $\tau$ integration in Eq.~\eqref{eq:taudiff}, obtaining
	\begin{align}
		\label{eq:Tbeforediff}\nonumber
		T_{\rm master} &= \frac{1}{(-i)^{1/2}}\frac{\sqrt{\pi}}{2}\int dx_1 dx_2 \frac{1}{\left(2(x_1 + x_2) - 1 + i\epsilon\right)^{3/2} (-i \tilde{I})^{3/2}} \\
		&= -\frac{\sqrt{\pi}}{2}\int dx_1 dx_2 \frac{1}{\left(2(x_1 + x_2) - 1 + i\epsilon\right)^{3/2}\tilde{I}^{3/2}}\,,
	\end{align}
	where we recall that we use $\epsilon_i \to \epsilon$, and $\tilde{I}$ simplifies to
	\begin{equation}
		\tilde{I} = \frac{\tilde{x}_1 \tilde{x}_2 k^2_1 - \tilde{x}_2 \tilde{x}_3 k^2_2 - \tilde{x}_1 \tilde{x}_3 k^2_3}{2(x_1 + x_2) - 1 + i\epsilon} + x_1(1+ i\epsilon) M_1 + x_2(1+ i\epsilon) M_2 - x_3(1 - i\epsilon) M_3\,.
	\end{equation}		
	The integration region in the $x_1,x_2$ plane is just the triangle defined by the three points $(0,0)$, $(1,0)$, and $(0,1)$. 
	The integrand diverges for the segment ($0<x_1<1/2,x_2=1/2-x_1$), which divides the triangle in two parts.
	Let us call them $R_-$ and $R_+$, corresponding to $x_1 + x_2 <1/2$ and $x_1 + x_2 > 1/2$, respectively. 
	These two regions are shown in Fig.~\ref{fig:boundpm}.
	
	We can simplify the integrand in Eq.~\eqref{eq:Tbeforediff} for each region $R_-$ and $R_+$ separately.
	First, note that for $a>0$ and $\Im(z)>0$, $s(a \pm i \epsilon, z) = 1$  and $s(-a \pm i \epsilon, z) = \mp 1$, where $\epsilon$ is arbitrarily small and positive.  
	We can also separate the integrand as 
	\begin{equation}
		\begin{split}
			\frac{1}{\left(2(x_1 + x_2) - 1 + i\epsilon\right)^{3/2}\tilde{I}^{3/2}}& = \\
			&\frac{1}{\left(2(x_1 + x_2) - 1 + i\epsilon\right)\tilde{I}} \times \frac{1}{\left(2(x_1 + x_2) - 1 + i\epsilon\right)^{1/2}\tilde{I}^{1/2}}\,,
		\end{split}
	\end{equation}
	where we remind the reader that $a^{3/2}=(a^{1/2})^3$.
	On the right hand side, the denominator of first term is given by
	\begin{equation}
		\begin{split}
			& \tilde{x}_1 \tilde{x}_2 k^2_1 - \tilde{x}_2 \tilde{x}_3 k^2_2 - \tilde{x}_1 \tilde{x}_3 k^2_3 +\\
			&+(2(x_1 + x_2) - 1 + i\epsilon)\left(x_1(1+ i\epsilon) M_1 + x_2(1+ i\epsilon) M_2 - x_3(1 - i\epsilon) M_3\right)\,,
		\end{split}
	\end{equation}
	in both regions.
	Concerning the second term on the right hand side, we wish to combine the two square roots into one, but the result depends on the region according to the sign of the function $s(,)$. 
	In $R_+$ we have $2(x_1 + x_2) - 1>0$ and $\Im(\tilde{I})>0$, so we get $s(2(x_1 + x_2) - 1 + i\epsilon,\tilde{I}) =1$, so the denominator is given by
	\begin{equation}
		\begin{split}
			& [\tilde{x}_1 \tilde{x}_2 k^2_1 - \tilde{x}_2 \tilde{x}_3 k^2_2 - \tilde{x}_1 \tilde{x}_3 k^2_3 +\\
			&+(2(x_1 + x_2) - 1 + i\epsilon)\left(x_1(1+ i\epsilon) M_1 + x_2(1+ i\epsilon) M_2 - x_3(1 - i\epsilon) M_3\right)]^{1/2}\,.
		\end{split}
	\end{equation}
	In $R_-$ we have $2(x_1 + x_2) - 1<0$ and $\Im(\tilde{I})>0$, so we get $s(2(x_1 + x_2) - 1 + i\epsilon,\tilde{I}) = -1$, so the denominator is given by (note the minus sign)
	\begin{equation}
		\begin{split}
			& -[\tilde{x}_1 \tilde{x}_2 k^2_1 - \tilde{x}_2 \tilde{x}_3 k^2_2 - \tilde{x}_1 \tilde{x}_3 k^2_3 +\\
			&+(2(x_1 + x_2) - 1 + i\epsilon)\left(x_1(1+ i\epsilon) M_1 + x_2(1+ i\epsilon) M_2 - x_3(1 - i\epsilon) M_3\right)]^{1/2}\,.
		\end{split}
	\end{equation}
	We can then safely send $\epsilon \to 0$ in each region, obtaining
	\begin{equation}
		\begin{split}
			\label{eq:Tregulardiff}
			T_{\rm master} &= -\frac{\sqrt{\pi}}{2}\left(\int_{R_+} - \int_{R_-}\right) \\
			&\frac{1}{[x_1 x_2 k^2_1 - x_2 x_3 k^2_2 - x_1 x_3 k^2_3 +(2(x_1 + x_2) - 1)\left(x_1 M_1 + x_2 M_2 - x_3 M_3\right)]^{3/2}} \,,
		\end{split}
	\end{equation}
	where the integrand is regular everywhere in the region of integration.	
	
\begin{figure}[h]
		\centering
		\includegraphics[width = 0.5 \textwidth]{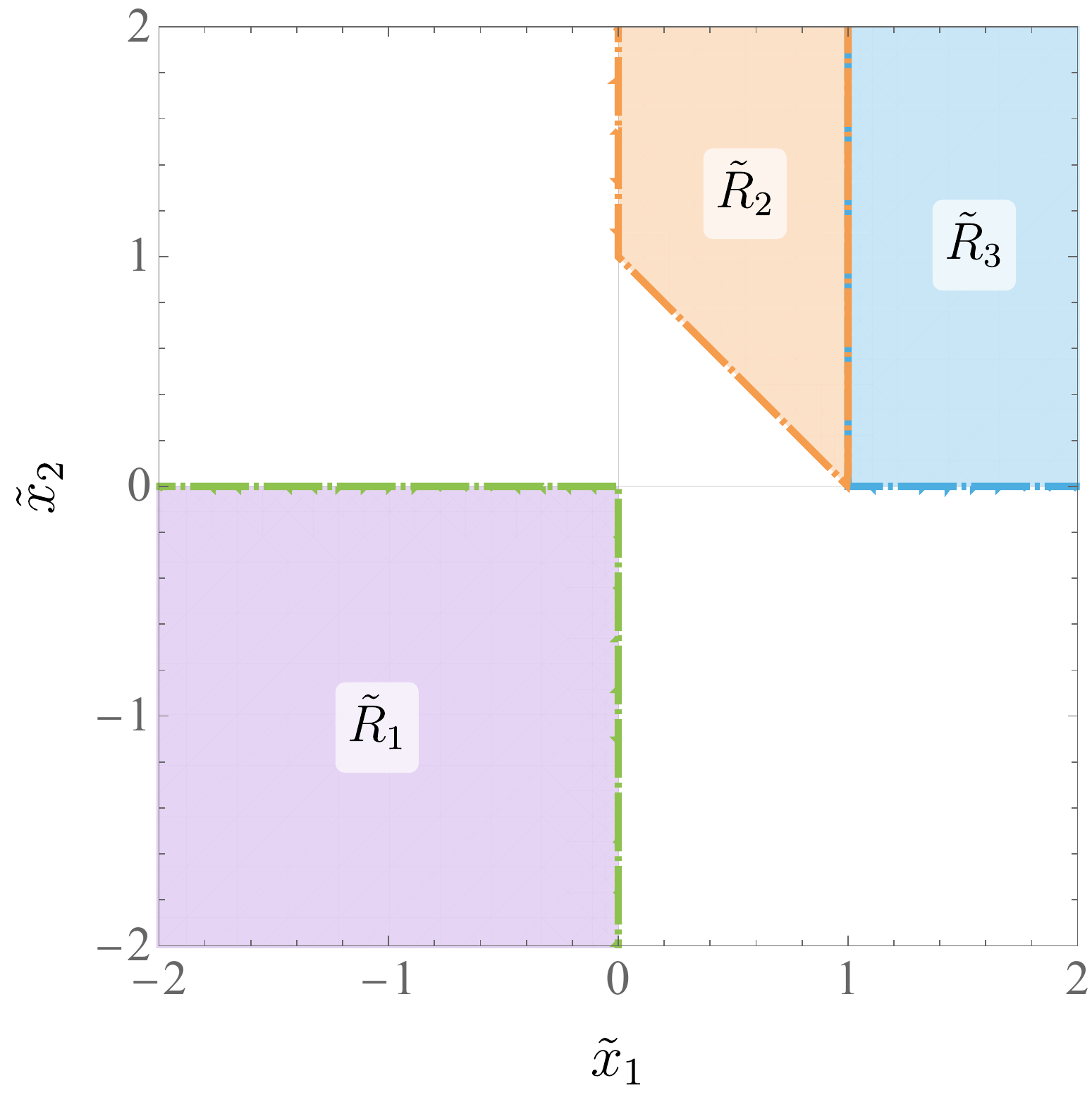}
		\caption{Integration region of $\tilde{x}_1$ and $\tilde{x}_2$ split into the regions $\tilde{R}_1$, $\tilde{R}_2$ and $\tilde{R}_3$.  
		}
		\label{fig:bound}
	\end{figure}

	We can then make the following substitutions, $\tilde{x}_1 = \frac{x_1}{2(x_1 + x_2) - 1}, \tilde{x}_2 = \frac{x_2}{2(x_1 + x_2) - 1}$ (note that the Jacobian is $|\frac{1}{(1-2(\tilde{x}_1 + \tilde{x}_2))^3}|$).
	Note that these variables are different from those defined in Eq.~\eqref{eq:taudiff}, and in particular they are real.
	The new region of integration is $\{\tilde{x}_1 \tilde{x}_2 > 0\} \setminus \{(\tilde{x}_1>0) \wedge (\tilde{x}_2>0) \wedge (\tilde{x}_1 + \tilde{x}_2 < 1)\}$, that is the first and third quadrants except the triangle whose vertices are (0,0), (0,1) and (1,0)~(\footnote{
		One can see this by analyzing how the boundaries of the original triangle transform under the change of variables. 
		The segment ($x_1 = 0$, $0<x_2<1/2$) maps to ($\tilde{x}_1 = 0$, $-\infty<\tilde{x}_2<0$).
		The segment ($x_1 = 0$, $1/2<x_2<1$) maps to ($\tilde{x}_1 = 0$, $1<\tilde{x}_2<\infty$). Similarly for $x_2=0$ (as the region of integration and the change of variables are symmetric in $x_1\leftrightarrow x_2,\ \tilde x_1\leftrightarrow \tilde x_2 $). 
		The hypotenuse ($0<x_1<1$, $x_2=1-x_1$) maps to itself. 
		Finally, the hypotenuse that splits $R_-$ and $R_+$ ($0<x_1<1/2$, $x_2=1/2-x_1$)  is mapped to $\tilde x_1=-\infty,\ \tilde x_2=-\infty$ if approaching the hypotenuse from $R_-$, and to $\tilde x_1=+\infty,\ \tilde x_2=+\infty$  if approaching it from $R_+$.
	}
	).  
	The integration region is shown in Fig.~\ref{fig:bound}.  		
	
	First, let us see how {the region of integration} $R\equiv R_+\cup R_-$ in Eq.~\eqref{eq:Tregulardiff} is transformed:
	\begin{equation}
		\begin{split}
			R \equiv& \left(\int_{R_+} - \int_{R_-}\right)\\
			= & \left( \int_0^1  \int_{1-\tilde{x}_1}^{+\infty}  + \int_{1}^{+\infty}  \int_{0}^{+\infty} - \int_{-\infty}^0  \int_{-\infty}^0 \right) d\tilde{x}_1 d\tilde{x}_2\left|\frac{1}{(2(\tilde{x}_1 + \tilde{x}_2)-1)^3}\right| \\
			= & \left(\int_{\tilde{R}_2}  + \int_{\tilde{R}_3}  - \int_{\tilde{R}_1}\right) d\tilde{x}_1 d\tilde{x}_2\left|\frac{1}{(2(\tilde{x}_1 + \tilde{x}_2)-1)^3}\right| \,,
		\end{split}
	\end{equation}
	where $\int_{\tilde{R}_1} = \int^0_{-\infty} \int^0_{-\infty}d\tilde{x}_1d\tilde{x}_2$, $\int_{\tilde{R}_2} = \int^1_{0} \int^{+\infty}_{1 - \tilde{x}_1 }d\tilde{x}_1d\tilde{x}_2$, and $\int_{\tilde{R}_3} = \int^{+\infty}_{1} \int^{+\infty}_{0}d\tilde{x}_1d\tilde{x}_2$.
	The integration regions of $\tilde{R}_1$, $\tilde{R}_2$ and $\tilde{R}_3$ are shown in Fig.~\ref{fig:bound}. 
	In  particular, $R_-$ is mapped to $\tilde{R}_1$ and $R_+$ is mapped to $\tilde{R}_2 \cup \tilde{R}_3$ {(we have split it into two regions $\tilde{R}_2$ and $\tilde{R}_3$ so that the boundary of integration can be easily expressed)}.

	Having seen how the integration region transforms under the change of variables, let us see how the integrand in Eq.~\eqref{eq:Tregulardiff} is modified.
	Applying the change of variables, we obtain for the integrand denominator
	\begin{equation}
		\begin{split}
			&x_1 x_2 k^2_1 - x_2 x_3 k^2_2 - x_1 x_3 k^2_3 +(2(x_1 + x_2) - 1)\left(x_1 M_1 + x_2 M_2 - x_3 M_3\right)\\
			& =\frac{\tilde{x}_1 \tilde{x}_2 k^2_1 + \tilde{x}_2 \tilde{x}_3 k^2_2 + \tilde{x}_1 \tilde{x}_3 k^2_3 +\tilde{x}_1 M_1 + \tilde{x}_2 M_2 + \tilde{x}_3 M_3}{(2(\tilde{x}_1 + \tilde{x}_2) - 1)^2}\\
			&=\frac{\hat{I}}{(2(\tilde{x}_1 + \tilde{x}_2) - 1)^2}\,,
		\end{split}
	\end{equation}
	where $\hat{I}$ is given by
	\begin{equation}
		\hat{I} \equiv k_1^2 \tilde{x}_1 \tilde{x}_2 + k_2^2 \tilde{x}_2 \tilde{x}_3 +k_3^2 \tilde{x}_1 \tilde{x}_3 + \tilde{x}_1 M_1 + \tilde{x}_2  M_2 + M_3 \tilde{x}_3\,,
	\end{equation}
	and $\tilde{x}_3 \equiv 1 - \tilde{x}_1 -\tilde{x}_2 $. 
	The new expression for $T_{\rm master}$ becomes then,
	\begin{align}
		\begin{split}
			T_{\rm master} 
			&=  -\frac{\sqrt{\pi}}{2}\left(\int_{\tilde{R}_2}+\int_{\tilde{R}_3} - \int_{\tilde{R}_1}\right) d\tilde{x}_1 d\tilde{x}_2 \left|\frac{1}{(1-2(\tilde{x}_1 + \tilde{x}_2))^3}\right|
			\left(\frac{(2(\tilde{x}_1 + \tilde{x}_2) - 1)^2}{\hat{I}}\right)^{3/2} \\
			&=  -\frac{\sqrt{\pi}}{2}\left(\int_{\tilde{R}_2}+\int_{\tilde{R}_3} - \int_{\tilde{R}_1}\right) d\tilde{x}_1 d\tilde{x}_2 
			\left(\frac{1}{\hat{I}}\right)^{3/2} 
			=  \frac{\sqrt{\pi}}{2}\left(\int_{\tilde{R}_1} - \int_{\tilde{R}_2} - \int_{\tilde{R}_3}\right) \hat{I}^{-3/2} \,,
		\end{split}
	\end{align}	
	where in the first step we used Eq.~\eqref{eq:srealposcondition} as $(2(\tilde{x}_1 + \tilde{x}_2) - 1)^2$ is real and positive, so $\sqrt{(2(\tilde{x}_1 + \tilde{x}_2) - 1)^2}=|2(\tilde{x}_1 + \tilde{x}_2) - 1|$.

\begin{figure}[h]
		\centering
		\includegraphics[scale = 0.5]{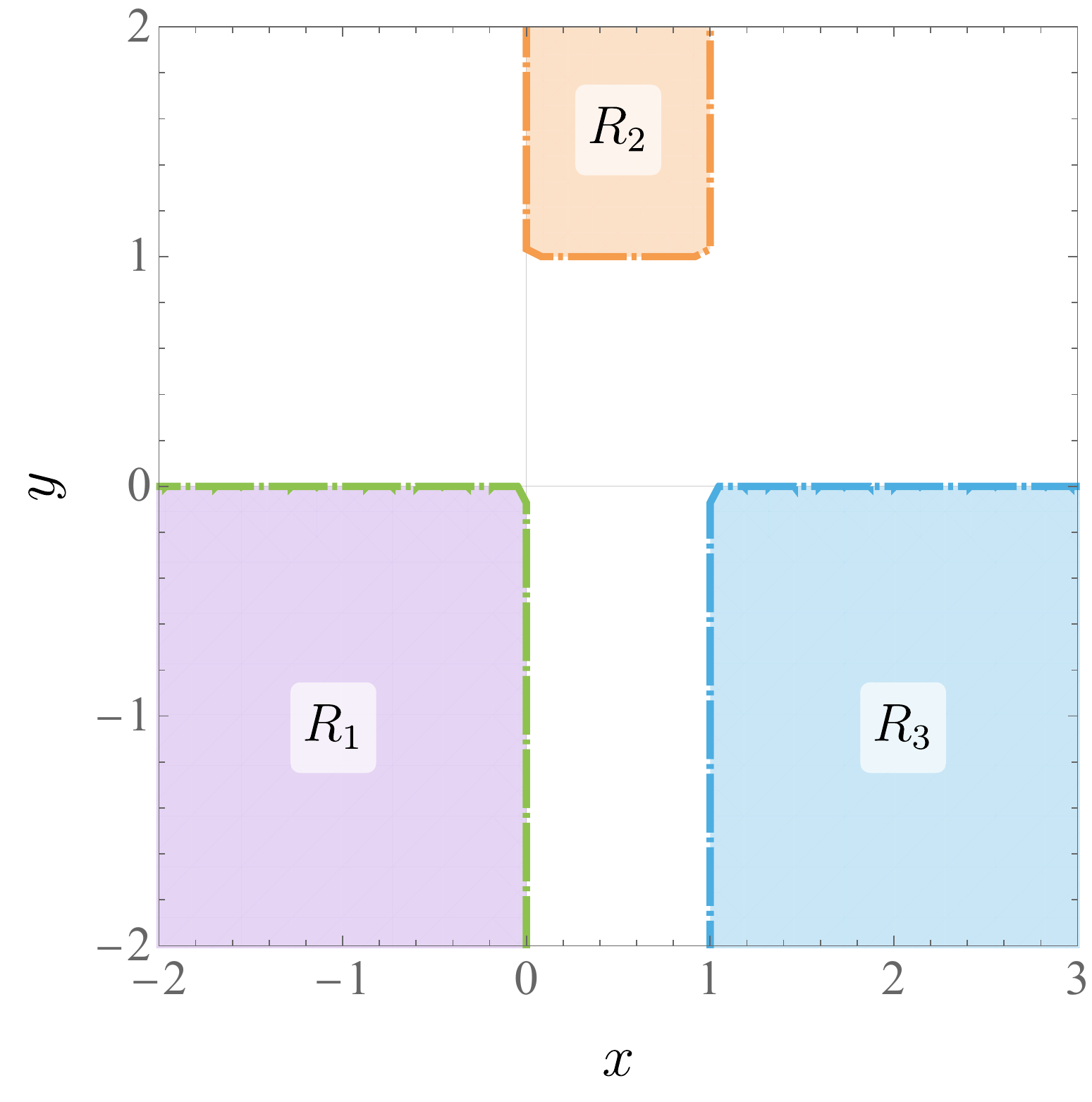}
		\caption{Integration region of $x$ and $y$ split into the regions $R_1$, $R_2$ and $R_3$. }
		\label{fig:boundxy}
	\end{figure}

\begin{figure}[h]
		\centering
		\includegraphics[width = 0.5 \textwidth]{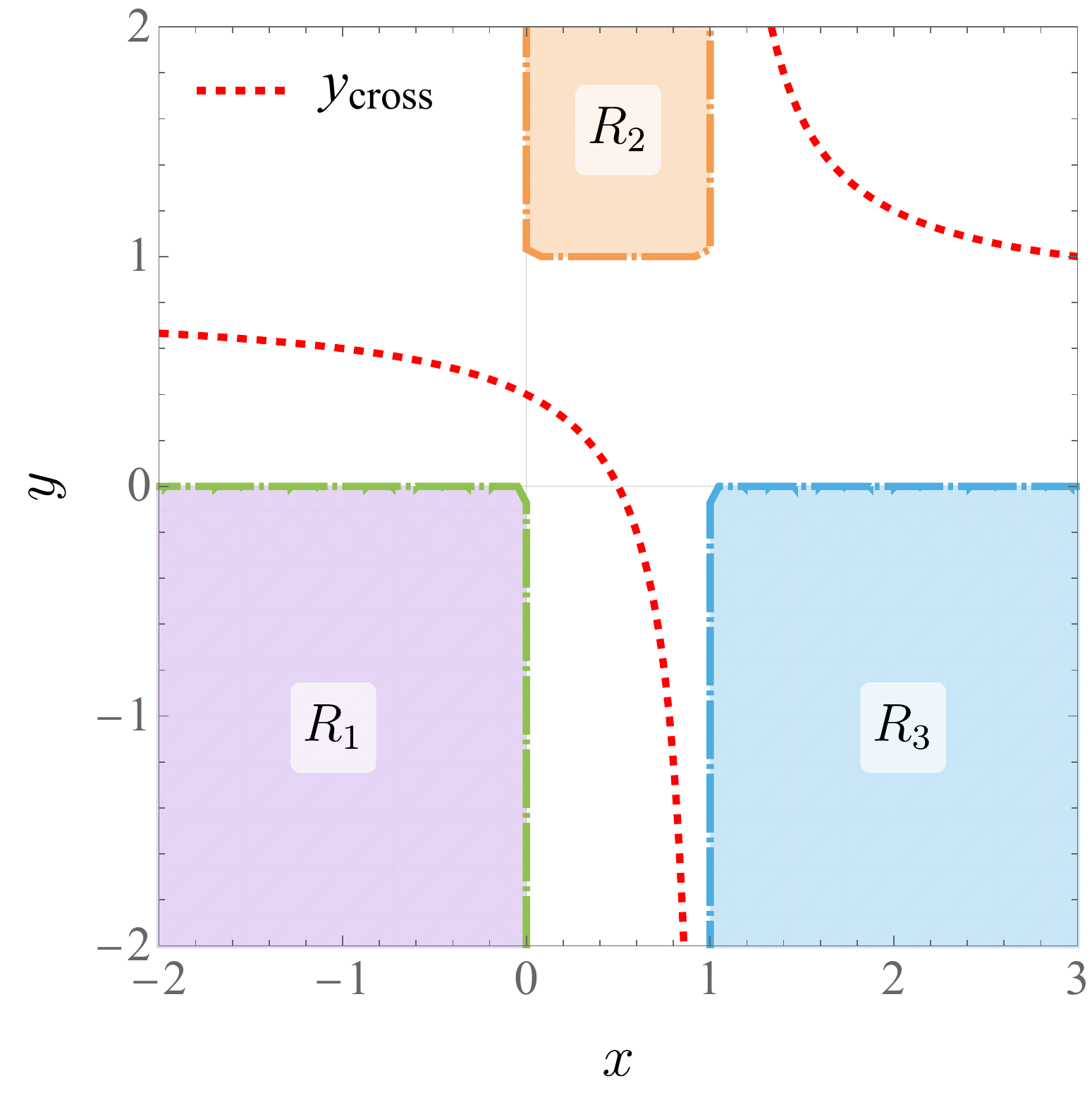}
		\caption{
			Integration regions $R_1$, $R_2$, and $R_3$, along with the line $y_{\rm cross}$ for masses $M_1 = 1.1+0.2\,i$, $M_2 = 0.3+0.3\,i$, and $M_3 = 0.1-0.2\,i$. We can see that $R_2$ and $R_3$ are always in the same branch. $R_1$, on the other side, can be in a different branch.}
		\label{fig:R123cross}
	\end{figure}

	We can go further and do another change of variables: $\tilde{x}_1 = x$, $\tilde{x}_2 = (1-x)y$, whose Jacobian is $1-x$.
	Let us separate the integration region in three parts, one in the third quadrant, and another two in the first quadrant.
	In the third quadrant, we have  
	\begin{equation}
		\int_{\tilde{R}_1} = \int_{-\infty}^0 d\tilde{x}_1 \int_{-\infty}^0 d\tilde{x}_2 = \int_{-\infty}^0 dx (1-x)\int_{-\infty}^0 dy  \,,
	\end{equation}
	and in the first quadrant we have 
	\begin{equation}
		\int_{\tilde{R}_2} = \int_0^1 d\tilde{x}_1 \int_{1-\tilde{x}_1}^{+\infty} d\tilde{x}_2 = \int_0^1 dx \int_1^{+\infty} dy (1-x) \,,
	\end{equation}
	and 
	\begin{equation}
		\int_{\tilde{R}_3} = \int_1^{+\infty} d\tilde{x}_1 \int_{0}^{+\infty} d\tilde{x}_2 = \int_1^{+\infty} dx \int_{-\infty}^0 dy |1-x| = -\int_1^{+\infty} dx \int_{-\infty}^0 dy (1-x) \,.
	\end{equation} 
	The integration regions in terms of $x$ and $y$ are shown in Fig.~\ref{fig:boundxy}. 
	So, we get for $T_{\rm master}$: 
	\begin{equation}
		\begin{split}
			\label{eq:Tdiff}
			T_{\rm master} &=  \frac{\sqrt{\pi}}{2}\left(\int_{-\infty}^0 dx \int_{-\infty}^0 dy 
			-\int_0^1 dx \int_{1}^{+\infty}dy
			+ \int_{1}^{+\infty} dx \int_{-\infty}^{0}dy\right) (1-x)\hat{I}^{-3/2} \\
			&=\frac{\sqrt{\pi}}{2} \left(\int_{R_1} - \int_{R_2} + \int_{R_3} \right) (1-x)\hat{I}^{-3/2}\, ,
		\end{split}
	\end{equation}
	where we redefined $\int_{R_1} = \int^0_{-\infty} \int^0_{-\infty}dx dy$, $\int_{R_2} = \int^1_{0} \int^{+\infty}_{1} dx dy$, and $\int_{R_3} = \int^{+\infty}_{1} \int^{0}_{-\infty}dx dy$ and $\hat{I}$ is given in terms of $x$ and $y$ by
	\begin{equation}
		\begin{split}
			\hat{I} =& M_1 x +  M_2 y (1-x) + M_3 (1-y) (1-x) + k^2_1 x y (1-x) + k^2_2 (1-x)^2 y (1-y) \\
			&+ k^2_3 x (1-x)(1-y) \,.
		\end{split}
	\end{equation}		
	Let us now briefly check that the sign of $\Im(\hat{I})$ is constant in each integration region separately (so that the integrand does not have any branch cut crossings).
	For that, we analyze when $\Im(\hat{I})=0$, which happens when:
	\begin{equation}
		\label{eq:ycond}
		\begin{split}
			y_{\rm cross} &= \frac{x \Im(M_1) + (1-x) \Im(M_3)}{(-1+x)\Im(M_2 - M_3)} \\
			& = \frac{\Im(M_1-M_3)}{\Im(M_2 - M_3)} + \frac{1}{x-1} \frac{\Im(M_1)}{\Im(M_2 - M_3)}\, .
		\end{split}
	\end{equation}
	So since $\Im M_1>0$, $\Im M_2>0$ and $\Im M_3<0$, {we find that, by looking at the first line in (\ref{eq:ycond})}, if $x<0$ the crossing happens at $y_{\rm cross}>0$, which means there is no crossing in $R_1$. 
	Likewise, {we find that, by looking at the second line in (\ref{eq:ycond})}, if $x>1$, the crossing also happens at $y_{\rm cross}>0$, which means there is no crossing in $R_3$. 
	Let us now look at $R_2$.
	Notice that $y_{\rm cross}$ in Eq.~\eqref{eq:ycond} is a monotonically decreasing function of $x$ for $x < 1$ and for $x>1$. 
	By looking at the first line of Eq.~\eqref{eq:ycond}, we notice that at $x=0$, $y_{\rm cross} = \frac{-\Im(M_3)}{\Im(M_2 - M_3)} < 1$ given that $\Im M_2 > 0$ and $\Im M_3 <0$. 
	Hence, in the region $R_2$ where $y > 1$ there is no crossing as $y_{\rm cross} < 1$ for $0<x<1$. 
	This is expected since we have specifically chosen regions of integration where no crossing can occur.
	In summary, the sign of $\Im(\hat{I})$ is constant in each integration region $R_1$, $R_2$, and $R_3$, so the integrand does not cross any branch cut.
	Furthermore, by looking at the relative positions of $R_2$, $R_3$, and the hyperbole $y_{\rm cross}$ as shown in Fig.~\ref{fig:R123cross}, we can verify that we can always go from $R_2$ to $R_3$ without crossing $y_{\rm cross}$. Therefore, the sign of $\Im(\hat{I})$ in $R_2$ is equal to its value in $R_3$.
	Furthermore, the sign of $\Im(\hat{I})$ in the $R_1$ region will always be the opposite than in the region $R_2$ and $R_3$.
	
	Now that we are confident that no branch cut crossings occur in $R_1$, $R_2$ and $R_3$, we can proceed to evaluate each of these regions separately.
	We can define the integrals in each integration region as
	\begin{align}
		I_1 &=  \frac{\sqrt{\pi}}{2}\int_{R_1}(1-x)\tilde{I}^{-3/2}\, ,\\
		I_2 &= \frac{\sqrt{\pi}}{2}\int_{R_2}(1-x)\tilde{I}^{-3/2}\, ,\\
		I_3 &=   
		\frac{\sqrt{\pi}}{2}\int_{R_3}(1-x)\tilde{I}^{-3/2}\, ,\\
		\label{eq:TmasterI123}
		I_1 +I_3 - I_2 &= T_{\rm master}\,.
	\end{align}
	
	Examining the integrands in each of the regions, and performing the indefinite $y$ integration, we obtain
	\begin{align}
		\label{eq:gen_integrand}\nonumber
		&\frac{\sqrt{\pi}}{2}\int dx \,dy \frac{1-x}{(-a y^2 + by + c)\sqrt{-a y^2 + by + c}}\, \\
		&={\frac{\sqrt{\pi}}{2}} \int dx\frac{N_1 x + N_0}{\sqrt{R_2 x^2 + R_1 x + R_0}(S_2 x^2 + S_1 x + S_0)}\, ,
	\end{align}
	where the terms $a,b,c$ and $R_0,R_1,R_2,S_0,S_1,S_2, N_0,N_1$ are defined respective in Eqs.~\eqref{eq:Tsame} and \eqref{eq:trimaster_params},
	and will be evaluated at the $y-$integration limits, as we will more explicitly indicate below.
	Note that the integrand in Eq.~\eqref{eq:gen_integrand} contains $\sqrt{\tilde{I}}$, and so does not cross any branch cut in each integration region as $\textrm{sign}\Im(\tilde{I})$ is constant as we have shown.
	So, to calculate the $y$ integral for a general $x$ in each region, it suffices to take the difference in the $y$ antiderivative.
	We are clearly obtaining the same indefinite integrals as in the main section, and so we then define the indefinite integral of $F_{\rm int}$ (see \eqref{eq:Fintexpr}), $W_{\rm int}$:
	\begin{align}
		\label{eq:Wint}\nonumber
		W_{\rm int}(x, z_+, z_-, x_0) &= \frac{\sqrt{\pi}}{2}\int dx \frac{1}{\sqrt{-R_2(z_+ - x)(x - z_-)}(x - x_0)}\, \\
		& = s(z_+ -x, x - z_-)\frac{\sqrt{\pi}}{\sqrt{|R_2|} }\frac{\arctan\left(\frac{\sqrt{z_+-x}\sqrt{x_0 - z_-}}{\sqrt{x_0-z_+}\sqrt{x-z_-}}\right)}{\sqrt{x_0 - z_+}\sqrt{x_0 - z_-}}\, .
	\end{align}
	Thus the indefinite integral of Eq.~\eqref{eq:gen_integrand} is given by,  
	\begin{align}
		\begin{split}\label{eq:Wint2}
			{\frac{\sqrt{\pi}}{2}}\int dx\frac{N_1 x + N_0}{\sqrt{R_2 x^2 + R_1 x + R_0}(S_2 x^2 + S_1 x + S_0)} &= c_1 W_{\rm int}(x, z_+, z_-, x_+)+ c_2 W_{\rm int}(x, z_+, z_-, x_-) \\
			&{\equiv}\tilde{W}_{\rm int}(x,z_+(y), z_-(y))\, ,
		\end{split}
	\end{align}
	where we remind the definitions of $c_1,c_2$ and $x_\pm$: $c_1 = \frac{N_0+N_1 x_+}{S_2(x_+ - x_-)}$, and $c_2 = - \frac{N_0+N_1 x_-}{S_2(x_+ - x_-)}$ and $x_{\pm} = -\frac{S_1}{2S_2} \pm \frac{\sqrt{S^2_1 - 4 S_0S_2}}{2S_2}$. 
	All the parameters, except for $x_\pm$, in general depend on $y$.  
	To simplify the notation, we will write $\tilde{W}_{\rm int}(x,y) \equiv \tilde{W}_{\rm int}(x,z_+(y), z_-(y))$.
	$\tilde{W}_{\rm int}(x,y)$ is then the full expression of the indefinite integral.
	More explicitly, to calculate the integral in a generic rectangular region $R$ delimited by $x_a<x<x_b$ and $y_a<y<y_b$, we do the following:  
	\begin{equation}
		\begin{split}
			{\frac{\sqrt{\pi}}{2}}&\int_{x_a}^{x_b} dx \int_{y_a}^{y_b} dy (1-x)\tilde{I}^{-3/2} = {\frac{\sqrt{\pi}}{2}}\int_{x_a}^{x_b} dx \left.\frac{N_1 x + N_0}{\sqrt{R_2 x^2 + R_1 x + R_0}(S_2 x^2 + S_1 x + S_0)}\right|_{y_a}^{y_b} \\
			&= \left. \left(\left.\tilde{W}_{\rm int}(x,y) \right|_{y_a}^{y_b}\right)\right|_{x_a}^{x_b}{ - \,  {\rm discontinuities}} \\
			&= \left. \left(\tilde{W}_{\rm int}(x,y_b)-\tilde{W}_{\rm int}(x,y_a) \right)\right|_{x_a}^{x_b}{ - \,  {\rm discontinuities}}\\
			&= \tilde{W}_{\rm int}(x_b,y_b)-\tilde{W}_{\rm int}(x_b,y_a) -\tilde{W}_{\rm int}(x_a,y_b) + \tilde{W}_{\rm int}(x_a,y_a) { - \,  {\rm discontinuities}}\,,
		\end{split}
	\end{equation}
	where we used the notation $f(x)|_{x=a}^{x=b} = f(b) - f(a)$.
	The evaluation of $I_{1,2,3}$ in terms of $W_{\rm int}$ can then be decomposed as, 
	\begin{align}
		I_1 &= \left.\tilde{W}_{\rm int}(x,y=0)\right|^{x=0}_{x=-\infty} -\left.\tilde{W}_{\rm int}(x,y=-\infty)\right|^{x=0}_{x=-\infty}{ - \,  {\rm discontinuities}}   \, , \\
		I_2 &= \left.\tilde{W}_{\rm int}(x,y=+\infty)\right|^{x=1}_{x=0} -\left.\tilde{W}_{\rm int}(x,y=1)\right|^{x=1}_{x=0} { - \,  {\rm discontinuities}}\, ,\\
		I_3 &= \left.\tilde{W}_{\rm int}(x,y=0)\right|^{x=+\infty}_{x=1} -\left.\tilde{W}_{\rm int}(x,y=-\infty)\right|^{x=+\infty}_{x=1}{ - \,  {\rm discontinuities}}\, .
	\end{align}
	Next, we define the limits of the function $W_{\rm int}$ which we will use to evaluate the integration regions $I_1$, $I_2$, and $I_3$.
	
	\paragraph{Limits of $W_{\rm int}$:}
	
	The regions $I_1$ and $I_3$ require, in particular, the evaluation of $W_{\rm int}(x = \pm \infty, z_+, z_-, x_0)|^{y = 0}$. 
	To simplify our result, notice that the $W_{\rm int}$ function can be rewritten as, 
	\begin{align}
		W_{\rm int}(x, z_+,z_-, x_0) = \frac{\sqrt{\pi}}{\sqrt{|R_2|}}\frac{\sqrt{z_+ - x}\sqrt{x - z_-}}{\sqrt{(z_+ - x)(x-z_-)}}\frac{\arctan\left(\frac{\sqrt{z_+-x}\sqrt{x_0 - z_-}}{\sqrt{x_0-z_+}\sqrt{x-z_-}}\right)}{\sqrt{x_0 - z_+}\sqrt{x_0 - z_-}}\, ,
	\end{align}
	where we have used $\sqrt{(z_+ - x)(x-z_-)} = s(z_+ - x, x - z_-)\sqrt{z_+ - x}\sqrt{x - z_-}$. 
	Taking the limit of the redefined $W_{\rm int}$ function, we find 		
	\begin{align}
		W_{\rm int}(x = \pm \infty, z_+,z_-, x_0)\left.\right|^{y = 0} &= -{\left. \frac{i\sqrt{\pi}\left(\log(1-\frac{\sqrt{x_0 - z_-}}{\sqrt{x_0 - z_+}})-\log(1+\frac{\sqrt{x_0 - z_-}}{\sqrt{x_0 - z_+}})\right)}{2\sqrt{|R_2|}\sqrt{x_0 - z_-}\sqrt{x_0 - z_+}}\right|^{y = 0} }\, ,
	\end{align}
	such that $\tilde{W}_{\rm int}(x=\pm\infty,y = 0) = c_1 W_{\rm int}(x=\pm\infty,z_+,z_-, x_+)\left.\right|^{y = 0} + c_2 W_{\rm int}(x=\pm\infty,z_+,z_-, x_-)\left.\right|^{y = 0}$. 
	
	In the following limits of $\tilde{W}_{\rm int}$, which involve $y\to\pm \infty$, we derive the expressions by the following procedure.
	First take the $y$-limit of the {$x$-integrand},  
	\begin{align}
		U &= \lim_{y \to \pm \infty}{\frac{\sqrt{\pi}}{2}}\frac{N_1 x + N_0}{\sqrt{R_2 (x - z_+)(x - z_-)}(S_2(x - x_+)(x - x_-))}\, ,\\
		&= 
		\begin{cases}
			\mp \frac{2 i k_2\sqrt{\pi}}{S_2(x-x_+)(x-x_-)} & \text{if } x < 1\\
			\pm \frac{2 i k_2\sqrt{\pi}}{S_2(x-x_+)(x-x_-)} & \text{if } x \geq 1
		\end{cases}\, ,
	\end{align}
	then we integrate in $x$~(\footnote{We adopt the notation where $f(x,y)\left.\right|^{x = x_a}_{x = x_b}|^{y = y_a} = f(x_a,y_a) - f(x_b, y_a)$.}),
	\begin{align}
		\tilde{W}_{\rm int}\left.\right|^{x = x_a}_{x = x_b}\left.\right|^{y = \pm \infty} = \int^{x_a}_{x_b}dx \, U,
	\end{align}
	where we have omitted the arguments of $\tilde{W}_{\rm int}$ for simplicity.
	The indefinite integral of $U$ can be calculated using 
	\begin{equation}
		\label{eq:antiU}
		\int \frac{dx}{(x-x_+)(x-x_-)} = \frac{\log(x-x_+) - \log(x-x_-)}{x_+ - x_-}\,.
	\end{equation}
	In the $I_1$ region we evaluate $\tilde{W}_{\rm int}\left.\right|^{x = 0}_{x = -\infty}\left.\right|^{y = -\infty}$, which is given by, paying attention to branch cut crossings,  
	\begin{align}
		\tilde{W}_{\rm int}(x, y =-\infty)\left.\right|^{x = 0}_{x = -\infty} = -\frac{2i\sqrt{\pi}k_2(\log(x_-) - \log(x_+))}{S_2(x_+ - x_-)}\, .
	\end{align}
	In the $I_2$ region we evaluate $\tilde{W}_{\rm int}(x, y = \infty)\left.\right|^{x = 1}_{x = 0}$, which is given by,  
	\begin{align}
		\tilde{W}_{\rm int}(x, y = +\infty)\left.\right|^{x = 1}_{x = 0} &= {-}
		\frac{2 i k_2\sqrt{\pi}(\log(-x_-) - \log(-x_+)+\log(1-x_+)-\log(1-x_-))}{S_2(x_+ - x_-)}\, .
	\end{align}
	In the $I_3$ region we also evaluate $W_{\rm int}(x, y = -\infty)\left.\right|^{x = \infty}_{x = 1}$, given by, 
	\begin{align}
		\tilde{W}_{\rm int}&(x, y = -\infty)\left.\right|^{x = \infty}_{x = 1} = {-}\frac{2 i \sqrt{\pi}k_2(\log(1-x_-)-\log(1-x_+))}{S_2(x_+ - x_-)}\, .
	\end{align}
	The other values of $\tilde W$ at finite values of $y$ and $x$ that are needed to evaluate $I_1$, $I_2$, and $I_3$ can be directly evaluated, and thus, we can calculate the value of $T_{\rm master}$ using Eq.~\eqref{eq:TmasterI123}.		
	
	However, before we proceed to evaluate each region, one must also take into account the branch cut crossing in the $\arctan$ of the $\tilde{W}_{\rm int}$ function similar to what is {described} in Sec.~\ref{sec:triangle_master}. 
	The procedure will be exactly the same except instead taking the integration regions of $I_1$, $I_2$ and $I_3$ and analyzing the branch cut crossings in each region. 
	
	We do not present a complete study, but include an example to highlight the branch cut crossing features, which are similar to Sec.~\ref{sec:Fint}. We leave a detailed implementation to future work.
	For definiteness, consider the masses $M_1 = -5 + 0.5 i, M_2 = -1.5 + 2.3 i$ and $M_3 = -0.7 -0.6i$ and the momenta $k^2_1 =  k^2_2 = k^2_3 = 0.1$. 
	Let us show how to take the branch cuts into account in the following terms~\footnote{
		These are the only terms that can have a branch cut crossing, since the other terms, that have infinite $y$, have already been considered and treated above. 
		In fact, the approach for finite and infinite $y$ is different.
		For finite $y$ (and infinite $x$ for example), we calculate the limit of the $\arctan$ in $\tilde{W}_{\rm int}$, so we need to consider the branch cut crossing of the $\arctan$.
		For infinite $y$ we are not taking the limit of the $\arctan$. Instead, we first take the limit of the integrand itself in $y$, followed by integration in $x$ (so it is no longer related to the $\arctan$).
		In fact, for those terms, we provided directly the definite integrals.		
	}: 
	\begin{itemize}
		\item $\left.\tilde{W}_{\rm int}(x,y=0)\right|^{x=0}_{x=-\infty}$ that appears in $I_1$, 
		\item $\left.\tilde{W}_{\rm int}(x,y=1)\right|^{x=1}_{x=0}$ that appears in $I_2$,
		\item $\left.\tilde{W}_{\rm int}(x,y=0)\right|^{x=+\infty}_{x=1}$ that appears in $I_3$.
	\end{itemize}
	Note, as from Eq.~\eqref{eq:Wint2}, that each $\tilde{W}_{\rm int}$ above contains two $\arctan$ terms that need to be checked for branch cut crossings. 
	These terms correspond to $x_0 = x_+$ and $x_0=x_-$ in Eq.~\eqref{eq:Wint}.
	
\begin{figure}
		\centering
		\begin{subfigure}{.5\textwidth}
			\centering
			\includegraphics[width = 0.9\textwidth]{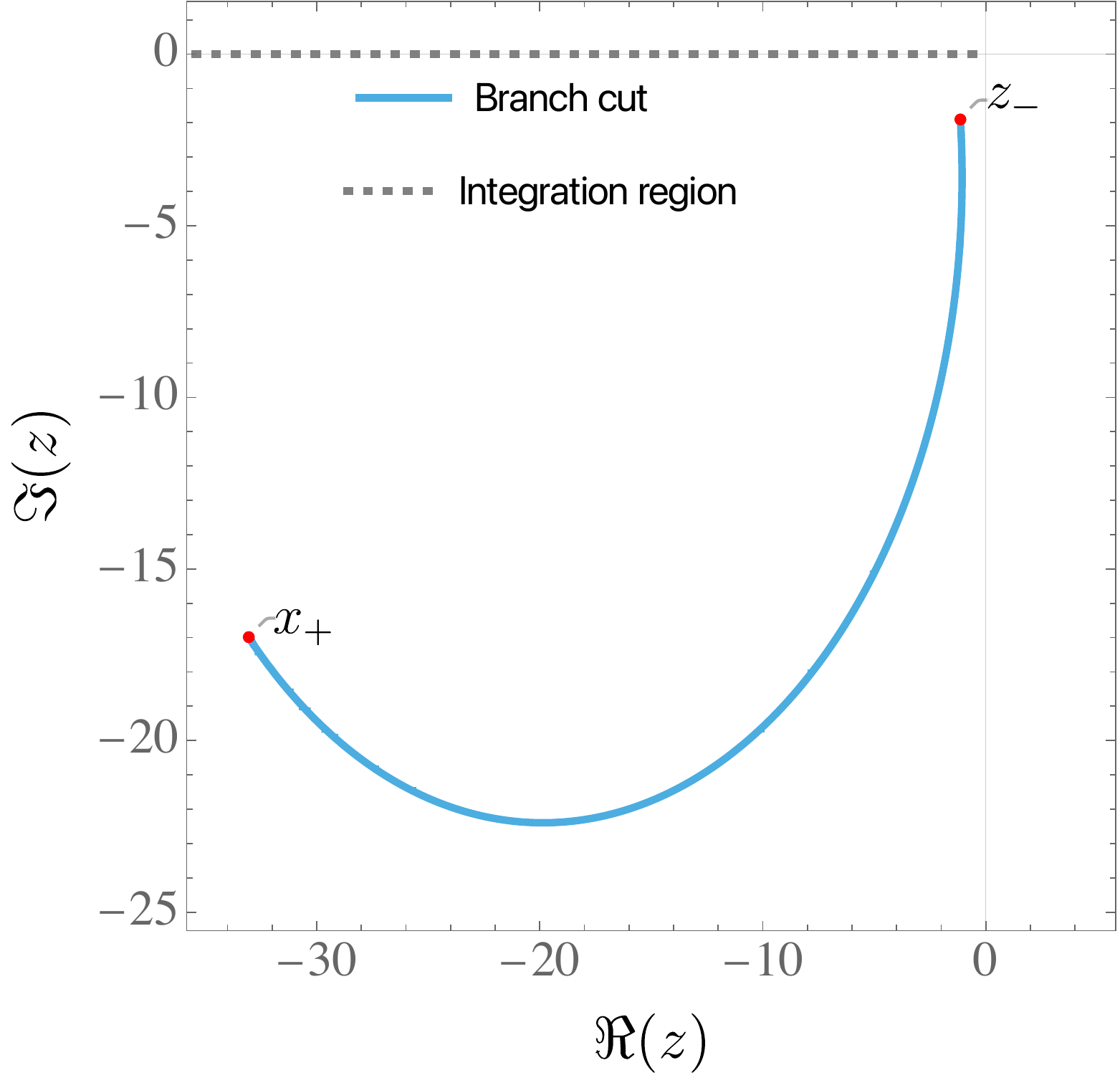}
			\label{fig:R1_xp}
		\end{subfigure}%
		\begin{subfigure}{.5\textwidth}
			\centering
			\includegraphics[width = 0.9\textwidth]{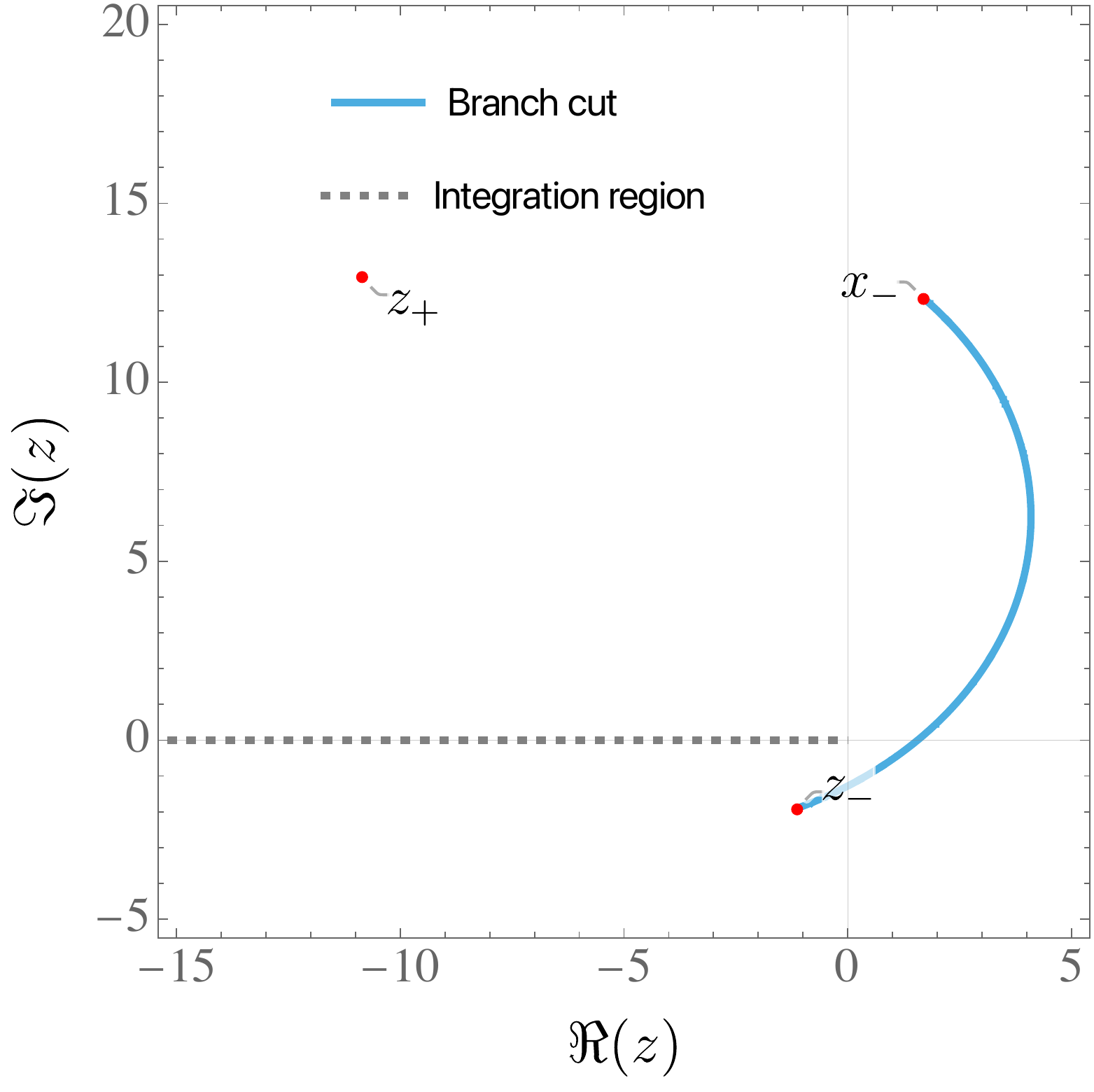}
			\label{fig:R1_xm}
		\end{subfigure}
		\caption{Arc described by $B \leq 0$ defined in Eq.~\eqref{eq:B} for $R_1$ region with left: $x_0 = x_+$ and right: $x_0 = x_-$. The expressions for $x_+$ and $x_-$ are given in Eq.~\eqref{eq:Tsamefinal}.}
		\label{fig:R1_branch}
	\end{figure}

\begin{figure}
		\centering
		\begin{subfigure}{.5\textwidth}
			\centering
			\includegraphics[scale = 0.45]{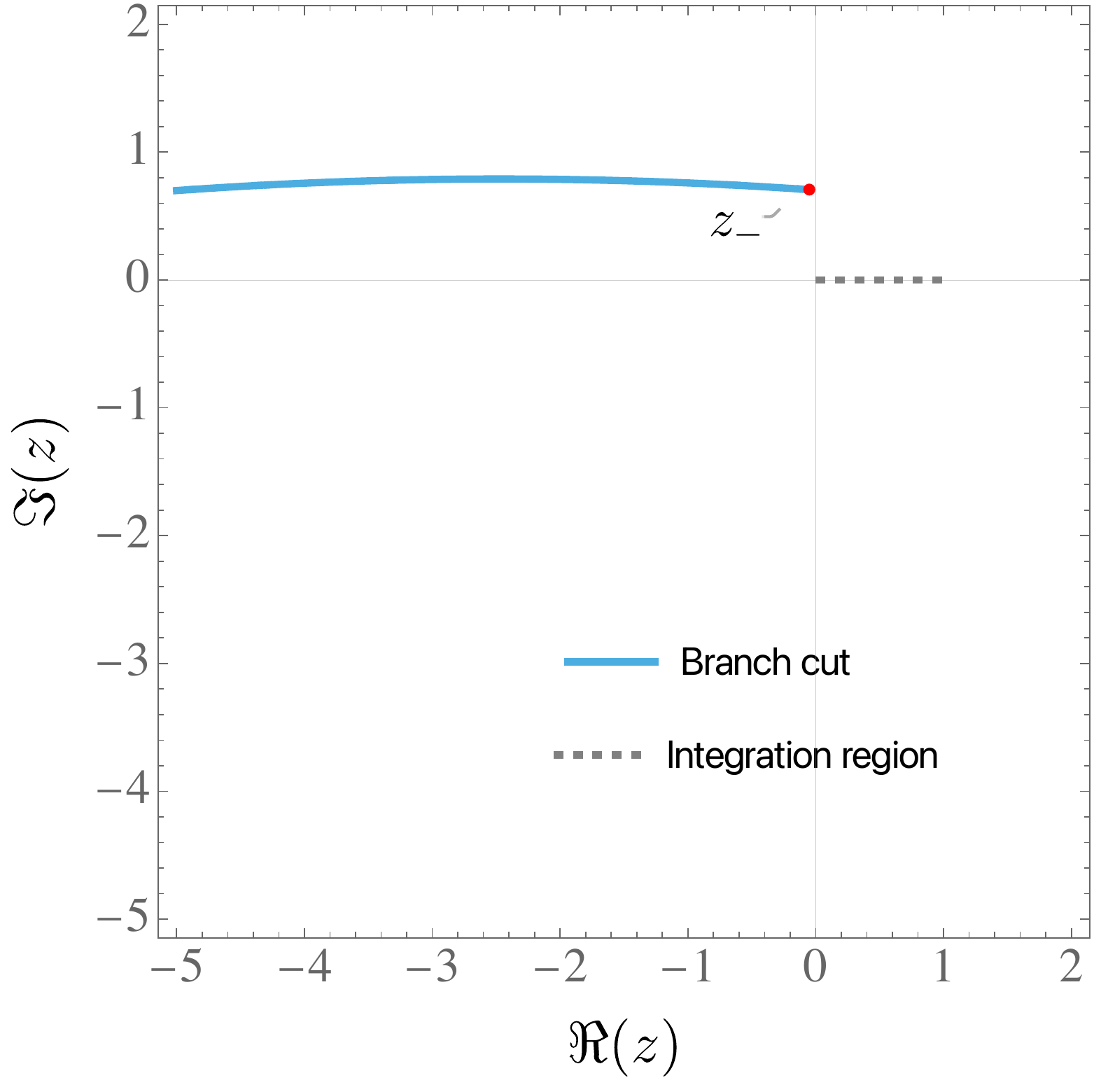}
			\label{fig:R2_xp}
		\end{subfigure}%
		\begin{subfigure}{.5\textwidth}
			\centering
			\includegraphics[scale = 0.45]{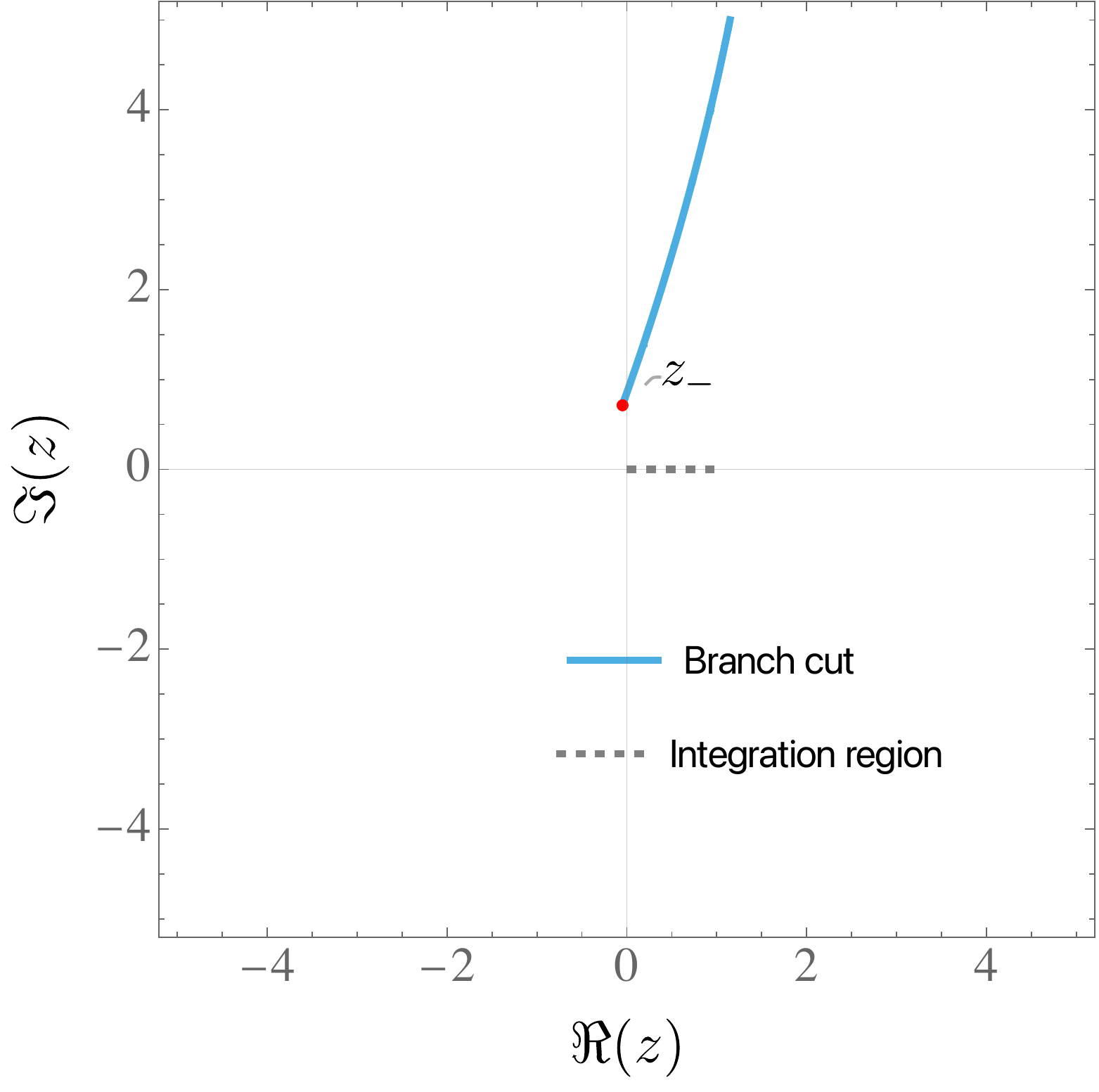}
			\label{fig:R2_xm}
		\end{subfigure}
		\caption{Arc described by $B \leq 0$ defined in Eq.~\eqref{eq:B} for $R_2$ region with left: $x_0 = x_+$ and right: $x_0 = x_-$. The expressions for $x_+$ and $x_-$ are given in Eq.~\eqref{eq:Tsamefinal}.}
		\label{fig:R2_branch}
	\end{figure}

\begin{figure}
		\centering
		\begin{subfigure}{.5\textwidth}
			\centering
			\includegraphics[scale = 0.45]{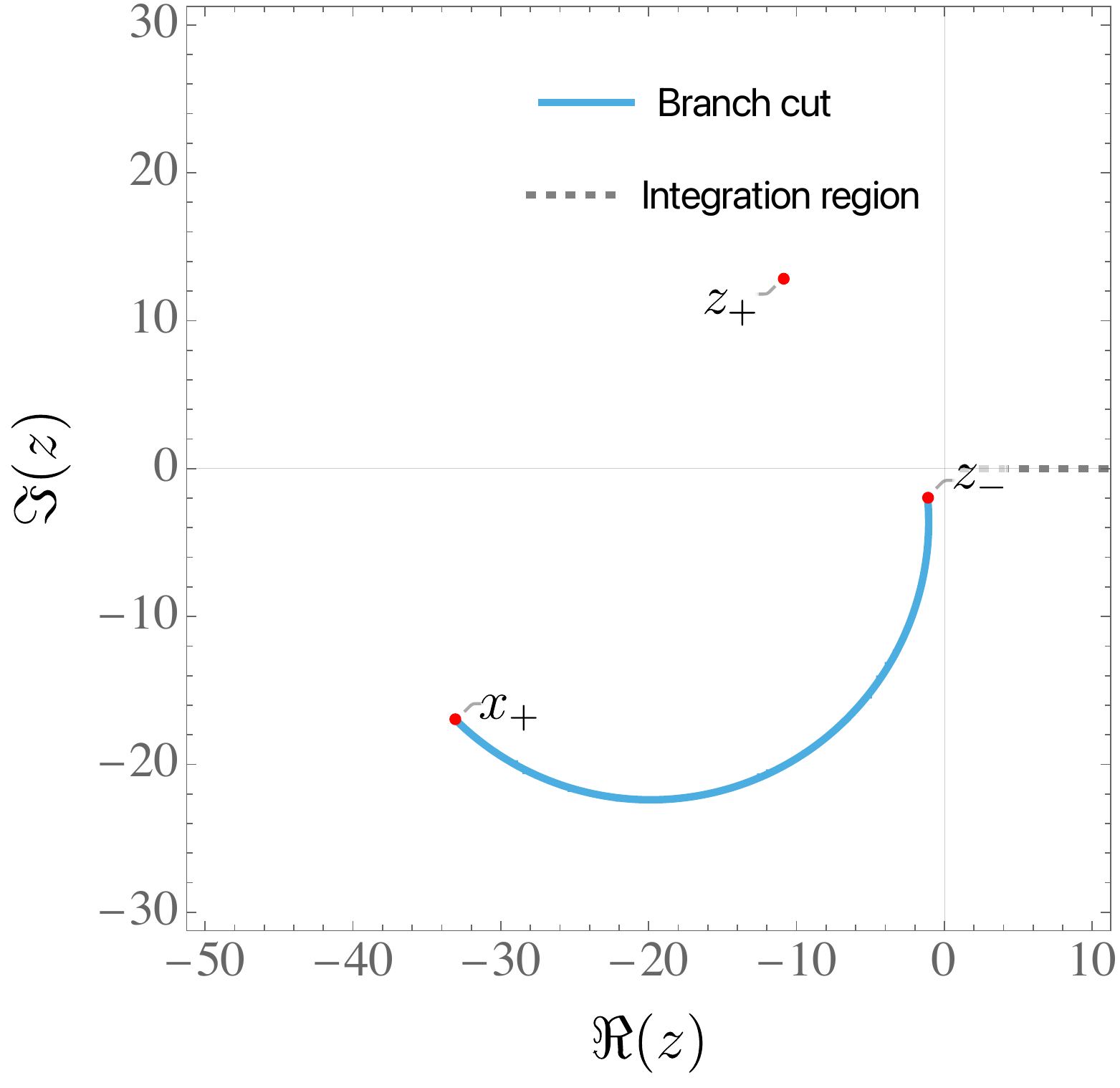}
			\label{fig:R3_xp}
		\end{subfigure}%
		\begin{subfigure}{.5\textwidth}
			\centering
			\includegraphics[scale = 0.45]{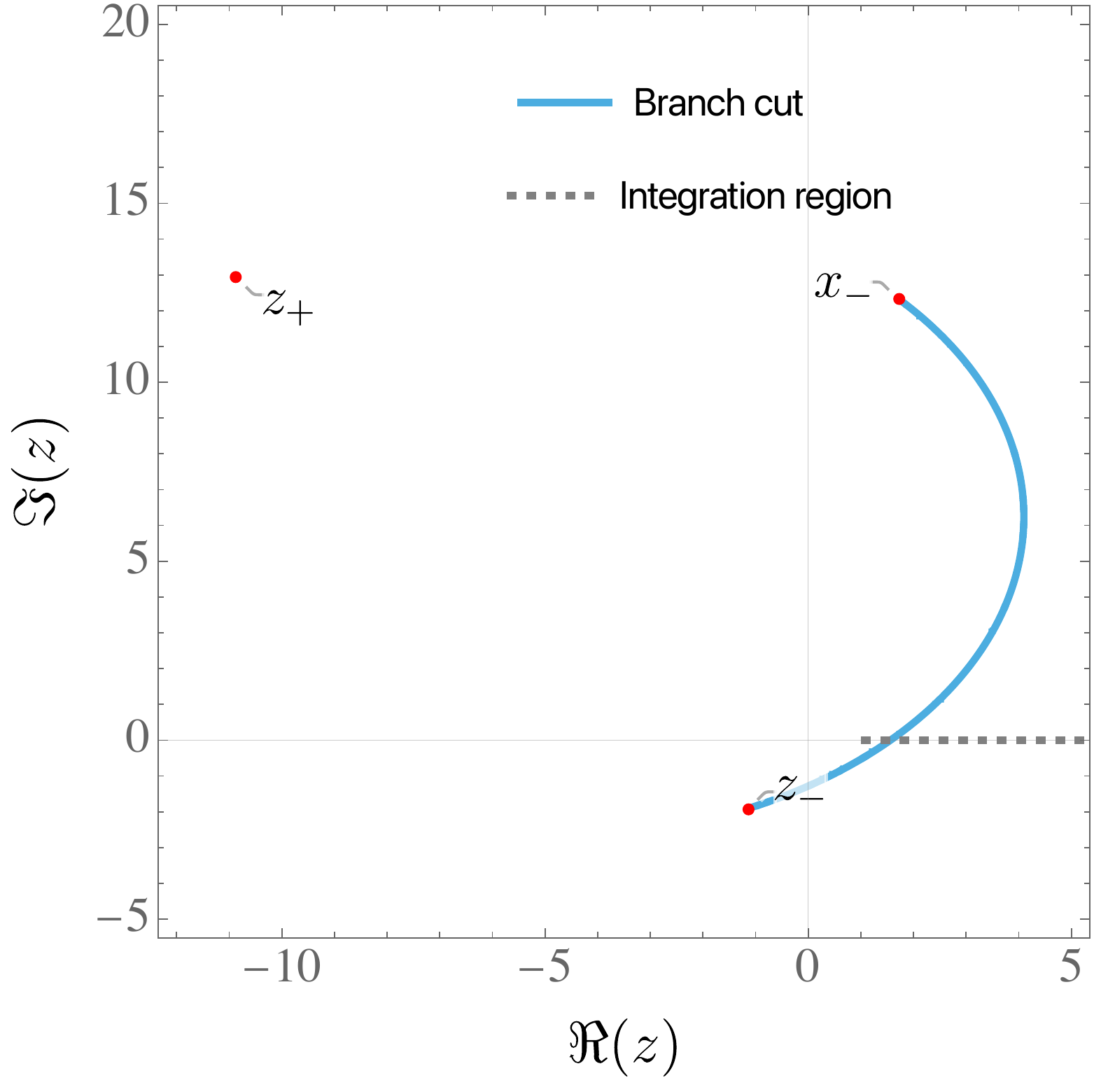}
			\label{fig:R3_xm}
		\end{subfigure}
		\caption{Arc described by $B \leq 0$ defined in Eq.~\eqref{eq:B} for $R_3$ region with left: $x_0 = x_+$ and right: $x_0 = x_-$. The expressions for $x_+$ and $x_-$ are given in Eq.~\eqref{eq:Tsamefinal}.}
		\label{fig:R3_branch}
	\end{figure}

	The arc defined by Eq.~\eqref{eq:B} describing the $\arctan$ branch cut for each term and the corresponding $x$ integration region are shown in Figs.~\ref{fig:R1_branch}, \ref{fig:R2_branch}, and \ref{fig:R3_branch}.
	We evaluate $z_+, z_-, x_+, x_-$ at $y = 0$ in Figs.~\ref{fig:R1_branch} and \ref{fig:R3_branch}, and at $y=1$ in Fig.~\ref{fig:R2_branch}. 	
	Notice that only in the right figure of Fig.~\ref{fig:R3_branch}, corresponding to $x_0 = x_-$, we have an intersection between the integration region and the arc. 
	Plugging in the values for the masses, we find $\Re dA/dx=-0.0069$, where $A$ is the argument of the $\arctan$ defined in Eq.~\eqref{eq:Adef}.
	Since in this case $\Re dA/dx < 0$, as explained in Sec.~\ref{sec:Fintcut}, we must add $\pi$ to the $\arctan$ of $c_2 W_{\rm int}(x, z_+, z_-, x_-)$ in the calculation of $\left.\tilde{W}_{\rm int}(x,y=0)\right|^{x=+\infty}_{x=1}$. 
	A similar study can be done for checking when $B=0$, where $B=A^2 + 1$. 
	We leave a complete study and an automation of the procedure to future work.
	
	\paragraph{Relating to case of two negative imaginary masses and one positive.} 
	
	Without loss of generality, we can consider the case of two negative imaginary masses and one positive to be the case of $\tilde{M}_1 = M_1^*$, $\tilde{M}_2 = M_2^*$ and $\tilde{M}_3 = M_3^*$ where $\Im(M_1) > 0$, $\Im(M_2) > 0$ and $\Im(M_3) < 0$. Taking $\tilde{M}_1$, $\tilde{M}_2$ and $\tilde{M}_3$ to be the new masses, we then have $\Im(\tilde{M}_1) < 0$, $\Im(\tilde{M}_2) < 0$, and $\Im(\tilde{M}_3) > 0$. It then follows that
	\begin{align}
		\begin{split}
			&\int_q\frac{1}{\left((\vec{k}_1 - \vec{q})^2 + \tilde{M}_1 \right)\left(\vec{q}^2 + \tilde{M}_2 \right)\left((\vec{k}_2 + \vec{q})^2 + \tilde{M}_3 \right)} \\
			&\qquad \qquad = \left(\int_q\frac{1}{\left((\vec{k}_1 - \vec{q})^2 + M_1 \right)\left(\vec{q}^2 + M_2 \right)\left((\vec{k}_2 + \vec{q})^2 + M_3 \right)} \right)^*\, .
		\end{split}
	\end{align}
	We have derived above,
	\begin{align}
		T_{\rm master}{(M_1,M_2,M_3)} = \int_q\frac{1}{\left((\vec{k}_1 - \vec{q})^2 + M_1 \right)\left(\vec{q}^2 + M_2 \right)\left((\vec{k}_2 + \vec{q})^2 + M_3 \right)}\, ,
	\end{align}
	for $\Im(M_1) > 0$, $\Im(M_2) > 0$ and $\Im(M_3) < 0$. Hence, it follows that
	\begin{align}
		\int_q\frac{1}{\left((\vec{k}_1 - \vec{q})^2 + \tilde{M}_1 \right)\left(\vec{q}^2 + \tilde{M}_2 \right)\left((\vec{k}_2 + \vec{q})^2 + \tilde{M}_3 \right)} = (T_{\rm master}{(\tilde{M}_1^*,\tilde{M}_2^*,\tilde{M}_3^*)})^*.
	\end{align}

	\bibliographystyle{JHEP} 
	\bibliography{references}	
	
\end{document}